\documentclass[11pt]{cernrep}
\usepackage{epsf}
\usepackage{psfig}
\usepackage{graphicx}
\usepackage{here}
\usepackage{rotating}
\usepackage{amssymb}
\usepackage{times}

\begin{document}

\def\la{\mathrel{\mathpalette\fun <}}
\def\ga{\mathrel{\mathpalette\fun >}}
\def\fun#1#2{\lower3.6pt\vbox{\baselineskip0pt\lineskip.9pt
\ialign{$\mathsurround=0pt#1\hfil##\hfil$\crcr#2\crcr\sim\crcr}}}  
\def\lrang#1{\left\langle#1\right\rangle}

\title{HARD PROBES IN HEAVY ION COLLISIONS AT THE LHC:\\
JET PHYSICS}
\author{
{\bf Convenors}: R.~Baier$^{6}$, A.~Morsch$^{9}$, I.P.~Lokhtin$^{8}$, 
X.N.~Wang$^{15}$, U.A.~Wiedemann$^{4}$\\
{\bf Editor}: U.A.~Wiedemann$^{4}$\\
{\bf Contributors}: A.~Accardi$^{1,2}$, F.~Arleo$^{3}$, N.~Armesto$^{4,5}$, 
R.~Baier$^{6}$, D.~d'Enterria$^{1}$, 
R.J.~Fries$^{7}$, O.~Kodololva$^{8}$,
I.P.~Lokhtin$^{8}$, A.~Morsch$^{9}$, A.~Nikitenko$^{10}$,
S.~Petrushanko$^{8}$, J.W.~Qiu$^{11}$, C.~Roland$^{12}$, C.A.~Salgado$^{4}$,
L.I.~Sarycheva$^{8}$, S.V.~Shmatov$^{13}$,
A.M.~Snigirev$^{8}$, S.~Tapprogge$^{14}$, I.~Vardanian$^8$, I.~Vitev$^{11}$, 
R.~Vogt$^{15,16}$, E.~Wang$^{17}$, X.N.~Wang$^{15}$, U.A.~Wiedemann$^{4}$, 
P.I.~Zarubin$^{13}$, B.~Zhang$^{17}$}

\institute{$^1$ Department of Physics, Columbia University, New York, USA \\
           $^2$ Institut f\"ur Theoretische Physik der Universit\"at
                Heidelberg, Germany\\
           $^3$ ECT$^*$ and INFN, Trento, Italy\\
           $^4$ Theory Division, CERN, Gen\`eve, Switzerland\\
           $^5$ Departamento de F\'{\i}sica, Universidad de C\'ordoba, Spain\\
           $^{6}$ Fakult\"at f\"ur Physik, Universit\"at Bielefeld, 
                   Bielefeld, Germany\\
           $^7$ Physics Department, Duke University, Durham, NC, USA \\
           $^{8}$ Institute of Nuclear Physics, Moscow State University,
                  Moscow, Russia\\
           $^{9}$ Experimental Physics Division, CERN, Gen\`eve, Switzerland\\
           $^{10}$ Imperial College, London, UK\\ 
           $^{11}$ Department of Physics and Astronomy, Iowa State University, 
                   Ames, Iowa, USA \\
           $^{12}$ Massachusetts Institute of Technology, Cambridge,
                   MA 02139-4307 USA\\
           $^{13}$ Joint Institute of Nuclear Research, Dubna, Russia\\
           $^{14}$ HIP, Helsinki, Finland   \\ 
           $^{15}$ Nuclear Science Division, Lawrence Berkeley National 
                   Laboratory, Berkeley, USA\\
           $^{16}$ Physics Department, University of California, Davis, 
                   California, USA\\
           $^{17}$ Institute of Particle Physics, Huazhong Normal 
                   University, Wuhan 430079, China
}
\maketitle
\begin{abstract}
We discuss the importance of high-$p_T$ hadron and jet
measurements in nucleus-nucleus collisions at the LHC.
\end{abstract}

\newpage
\tableofcontents
\newpage
\section{INTRODUCTION}
{\em U.A. Wiedemann}

This report summarizes the current understanding of how 
the production of high-$p_T$ partons in
nucleus-nucleus collisions at the LHC can be used as a ``hard probe'',
i.e. as a diagnostic tool 
for the produced QCD matter either in thermalized 
(quark gluon plasma) or in other non-equilibrated but dense
forms.

The production of high-$p_T$ partons (observed as high-$p_T$
hadrons or jets) involves a ``hard'' perturbative scale 
$Q \gg \Lambda_{\rm QCD}$. This report mainly considers the case when
this scale is harder than any momentum scale characterizing 
the medium produced in the nucleus-nucleus collision. Momentum scales 
proposed to characterize the medium  (such as the initial temperature 
$T$ or the saturation momentum $Q_s$) may be perturbatively large themselves, 
in which case $Q \gg T, Q_s$ and $ T, Q_s \gg \Lambda_{\rm QCD}$. 
This restriction
aims at insuring that {\it hard} partonic production processes are not 
part of the ``bulk matter'': they occur in the primary partonic collisions 
on temporal and spatial scales $\Delta\tau \sim 1/Q$, $\Delta r \sim 1/Q$
which are sufficiently small to be unaffected by the properties of 
the produced matter. This makes them promising candidates of 
processes whose primary partonic production process is unaffected by the
presence of a medium, while the development of the final
(and possibly initial) state parton shower leaving (entering)
the hard partonic subprocess is sensitive to the medium. 
If collinear factorization is applicable in nucleus-nucleus
collisions, then inclusive cross sections of high-$p_T$ partons 
measured in proton-proton collisions or calculated in perturbative
QCD can be used as benchmark against which one can search for the 
actual signals and properties of the hot and dense matter created 
in nucleus-nucleus collisions at the LHC.

{\bf Section~\ref{sec2}} discusses benchmark calculations for
jet spectra and identified high-$p_T$ hadronic spectra in nucleus-nucleus 
collisions at LHC, calculated in the framework of collinear factorized 
QCD. The question to what extent collinear factorization can be expected
to hold in nucleus-nucleus collisions, and how its validity can
be tested experimentally, is discussed in chapter \cite{pAwriteup}
of this workshop report.

{\bf Section~\ref{sec3}} addresses the main theoretical arguments
for strong final state medium-modifications in
jet production at the LHC. A jet is the hadronized remnant of 
a final state parton shower related to a produced highly 
virtual parton. Sections~\ref{sec31} and ~\ref{sec32} focus 
mainly on the modification of this final state parton shower 
due to multiple parton scattering in a spatially extended 
dense medium. The current understanding of the additional 
medium-induced radiative contributions and the transverse 
momentum broadening of the parton shower is discussed. 
Section~\ref{sec33} compares radiative and collisional
modifications of the parton shower. Section~\ref{sec34}
discusses calculations of the main observables in which
these medium-modifications are expected to show up. 
Related results obtained in the formalism of medium-enhanced 
higher twist expansion are summarized in section~\ref{sec35}.
Finally, the section~\ref{sec36} on other potentially 
large medium-modifications discusses medium effects which may
become important at moderate scales, $Q$ up to $\sim$ 10 GeV. 

Once medium-modifications of hard probes are determined, the 
question arises to what extent the properties of the hot and dense
matter produced in nucleus-nucleus collisions differ from those of 
normal cold nuclear matter. To this end, Section~\ref{sec314} 
summarizes what is known about the medium-modifications
of hard probes in cold nuclear matter. 
 
The remainder of this report summarizes the experimental
situation. {\bf Section~\ref{sec4}} gives a short overview
of the data available for Au+Au collisions at $\sqrt{s} = 200$
GeV from the Relativistic Heavy Ion Collider RHIC at Brookhaven.
Several measurements at RHIC indicate that strong 
final state medium-modifications of the hadron production
in central Au-Au collisions persist at RHIC up to the
highest transverse momenta explored ($p_T < 15$ GeV). 
This further supports the theoretical expectations of strong 
final state medium-modifications in nucleus-nucleus collisions
at LHC where a much wider range in transverse energy is
experimentally accessible. Finally, {\bf Section~\ref{sec5}}
discusses the current status of how the LHC experiments 
ALICE, ATLAS and CMS will measure jets and their medium-modifications
in the high-multiplicity background of a heavy ion collision.

In summary:
\begin{itemize}
 \item
Jets and high-$p_T$ hadrons are the most abundant
hard probes produced in nucleus-nucleus collisions at LHC. 
Within one month of running at design luminosity and in the
absence of strong medium-modifications, jet spectra
up to at least $E_T = 200$ GeV and leading hadron
spectra up to $p_T = 100$ GeV are accessible.
$\longrightarrow$ Section~\ref{sec2}.
\item
Both theoretical arguments ($\longrightarrow$ Section~\ref{sec3})
and data at lower center of mass energy 
($\longrightarrow$ Section~\ref{sec4})
suggest large medium-modifications of hadronic high-$p_T$
spectra in nucleus-nucleus collisions at LHC.
\item
A variety of theoretical approaches (comparison with benchmark 
calculations $\longrightarrow$ Section~\ref{sec2} ) and 
experimental techniques (comparison
to benchmark measurements, dependence of medium-modifications
on nuclear geometry $\longrightarrow$ Section~\ref{sec346}
, etc.) are available to quantify the
medium-dependence of jet production at LHC. The mutual
consistency of these different approaches is a prerequisite
for any characterization of the produced hot and dense 
medium from the medium-dependence of hard probes 
(see also Ref.~\cite{HQwriteup,Photonwriteup}). We
emphasize that data from p-A collisions at the LHC are an important part
of this program~\cite{pAwriteup}. 
\end{itemize}

\section{BENCHMARK CROSS SECTIONS}
\label{sec2}

\subsection{Jet and Dijet Rates in Nucleus-Nucleus Collisions}
{\em A. Accardi, N. Armesto, I. P. Lokhtin}
\label{sec21}

Jet studies will play a central role as a proposed signature of the
formation of QGP in AB collisions. Energy loss of energetic partons 
inside a medium where color charges are present, the so-called jet 
quenching \cite{Baier:2000mf}, has been suggested to behave very
differently in cold nuclear matter and in QGP. It has been postulated 
as a tool to
probe the properties of this new state of matter
\cite{Gyulassy:2001nm,Wang:2002ri,Salgado:2002cd,Vitev:2002pf}.

On the other hand, jet calculations at NLO have been successfully
confronted with experimental data in hadron-hadron
collisions \cite{Affolder:2001fa}. Monte Carlo codes have become available:
among them, we will use that of
\cite{Frixione:1995ms,Frixione:1997np,Frixione:1997ks} adapted to
include isospin effects and modifications of nucleon pdf inside nuclei,
see the section on jet and dijet rates in pA collisions~\cite{pAwriteup} 
for more information.
Here we will present the results of 'initial' state effects, i.e. no
energy loss of any kind will be taken into account. These results can be
considered as the reference, hopefully to be tested in pA, whose failure
should indicate the presence of new physics.
As in pA collisions,
we will work in the LHC lab frame, which for symmetric AB collisions
coincides with the center-of-mass one, and the accuracy of our
computations, limited by CPU time, is the same as in the pA case:
\begin{itemize}
\item For the transverse energy distributions, 2~\% for the lowest
and 15~\% for the highest $E_T$-bins.
\item For the pseudorapidity distributions, 3~\%.
\item For the dijet distributions of the angle between the two hardest
jets, 20~\% for the least populated and 3~\% for the most populated
bins.
\end{itemize}
All the energies will be given per
nucleon and, in order to compare with the pp case, all cross sections
will be presented per nucleon-nucleon pair, i.e. divided by AB.
 
Unless explicitly stated and as in the pA case,
we will use as nucleon pdf MRST98 central
gluon \cite{Martin:1998sq} modified inside nuclei using the EKS98
parameterizations \cite{Eskola:1998iy,Eskola:1998df},
a factorization scale equal to the
renormalization scale $\mu=\mu_F=\mu_R=E_T/2$, with $E_T$ the total
transverse energy of all the jets in the generated event,
and for jet reconstruction we will employ
the $k_T$-clustering algorithm \cite{Catani:1993hr,Ellis:tq} with $D=1$.
The kinematic regions we
are going to consider are the same as in the pA case:
\begin{itemize}
\item $|\eta_i|<2.5$, with $\eta_i$ the pseudorapidity of the jet; this
corresponds to
the acceptance of the central part of the CMS detector.
\item $E_{Ti}>20$ GeV in the pseudorapidity distributions, with
$E_{Ti}$ the transverse energy of the jet; this will ensure the validity
of perturbative QCD.
\item $E_{T1}>20$ GeV and $E_{T2}>15$ GeV for the $\phi$-distributions,
with $E_{T1}$ ($E_{T2}$) the transverse energy of the hardest
(next-to-hardest) jet entering the CMS acceptance, and $\phi$ the angle
between these two jets.
\end{itemize}
For more information, we refer to the chapter~\cite{pAwriteup} of 
this report. The centrality dependence is not studied in
either contribution.

\begin{figure}
\begin{center}
\includegraphics[width=12.5cm,clip=,bbllx=0pt,bblly=30pt,bburx=560,bbury=545pt]{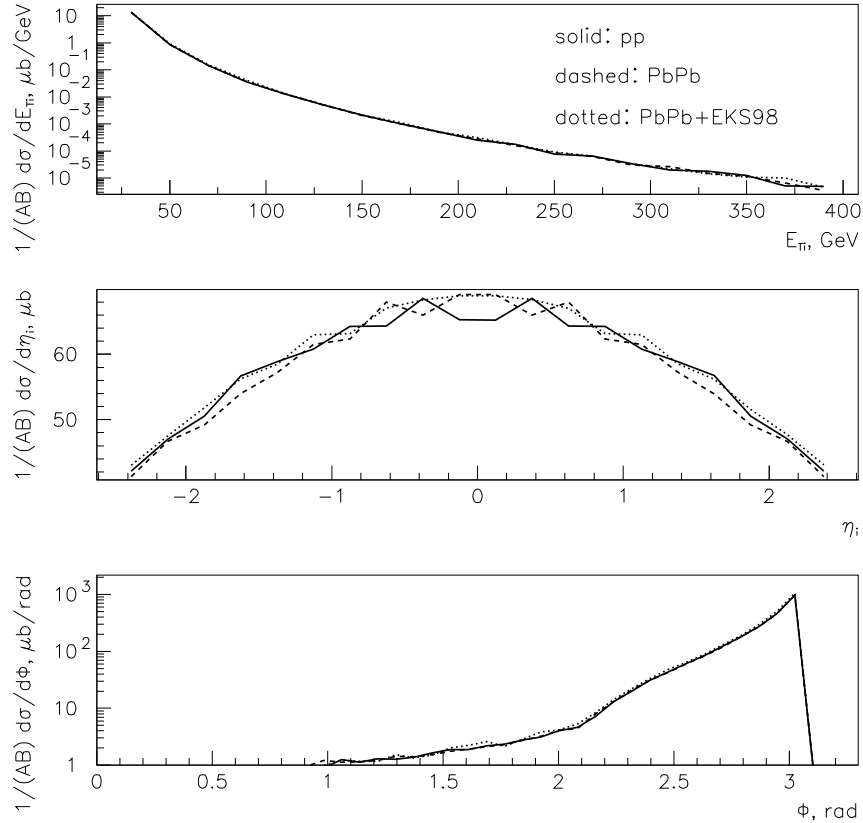}
\caption{Isospin and nuclear pdf dependence of jet cross sections (pp results:
solid lines; PbPb results without modification of nucleon pdf inside
nuclei: dashed lines; PbPb results with EKS98 modification of nucleon
pdf inside nuclei: dotted lines) versus
transverse energy of the jet (for $|\eta_i|<2.5$, upper plot) and
pseudorapidity of the jet (for $E_{Ti}> 20$ GeV, middle plot), and
dijet cross sections (lower plot)
versus angle between the two hardest jets for $E_{T1}>20$ GeV,
$E_{T2}>15$ GeV and $|\eta_1|,|\eta_2|<2.5$, for collisions at 5.5 TeV.
Unless otherwise stated default options are
used, see text.}
\label{abfig1}
\end{center}
\end{figure}

The words of caution about our results which were given in the pA
Section are even more relevant in AB collisions, as our ignorance on soft
multiparticle production in this case is even larger than in pA
collisions. For
example, the number of particles produced at midrapidity in a central
PbPb collision at the LHC may vary as much as a factor 3
\cite{Armesto:2000xh,aliceppr}
among different
models which, in principle, are able to reproduce the available
experimental data on multiplicities at SPS, RHIC and TeVatron.
Therefore, these
issues of the underlying event \cite{Huston:zr,Field:2000dy}
and multiple hard parton scattering
\cite{Accardi:ur,Accardi:2001ih,Ametller:1985tp,Abe:1997xk,Abe:1997bp}
demand extensive Monte Carlo studies including full detector simulation.
Preliminary analysis, based on the developed sliding window-type jet
finding algorithm (which subtracts the large background from the
underlying event) and full GEANT-based simulation of the CMS
calorimetry,
shows that even in the worst case of central PbPb collisions with
maximal
estimated charged particle density at mid-rapidity $dN^{\pm}/dy|_{y=0}=8000$,
jets can be reconstructed with almost 100~\% efficiency, low noise and
satisfactory energy and spatial resolution starting from $E_{Ti} \sim 100$
GeV (see the Section on Jet Detection at CMS).
In the case of more realistic, lower multiplicities,
the minimal threshold for adequate jet reconstruction could even decrease.

As in the pA case, see the previously mentioned section on pA
collisions,
the influence of disconnected collisions on jet
production in AB collisions may be studied using
simple estimates on the number $\langle n \rangle$ of
nucleon-nucleon collisions involved in the production of jets with
$E_{Ti}$ greater than a given $E_{T0}$, which can be obtained in the Glauber
model \cite{Capella:1981ju,Braun:me} in the optical approximation: 
$\langle n \rangle
(b,E_{T0})=ABT_{AB}(b)\sigma(E_{T0})/\sigma_{AB}(b,E_{T0})$, with $b$ the
impact parameter, $T_{AB}(b)=\int d^2s T_{A}(s)T_{B}(b-s)$
the convolution of the nuclear profile functions of projectile and
target normalized to unity, $\sigma(E_{T0})$ the
cross section for production of jets with
$E_{Ti}$ greater than $E_{T0}$ in pp collisions, and
$\sigma_{AB}(b,E_{T0})=1-[1-T_{AB}(b)\sigma(E_{T0})]^{AB}$. Taking
$\sigma(E_{T0})=294$, 0.463
and 0.0012 $\mu$b as representative values in PbPb collisions at 5.5 TeV
for $E_{T0}=20$, 100 and 200 GeV respectively (see results
in Fig.
\ref{abfig2-4} below), the number
of nucleon-nucleon collisions involved turns out to be respectively
3.5, 1.0 and 1.0 
for minimum bias collisions (i.e. integrating numerator and
denominator in $\sigma_{pA}(b,E_{T0})$ between $b=0$ and $\infty$),
while for central collisions (integrating between $b=0$ and 1 fm) the
numbers are 8.9, 1.0 and 1.0 respectively.
So, in AB collisions at LHC
energies the contribution of multiple hard scattering coming from
different nucleon-nucleon collisions seems to be negligible for
transverse energies of the jets greater than
$\sim 100$ GeV, while for $E_{Ti}$ smaller
than $\sim 50$ GeV this effect
might need to be taken into account more
carefully in our computations.
\vspace*{-1cm}
\begin{figure}
\begin{center}
\includegraphics[width=12.5cm,bbllx=0pt,bblly=30pt,bburx=555,bbury=535pt]{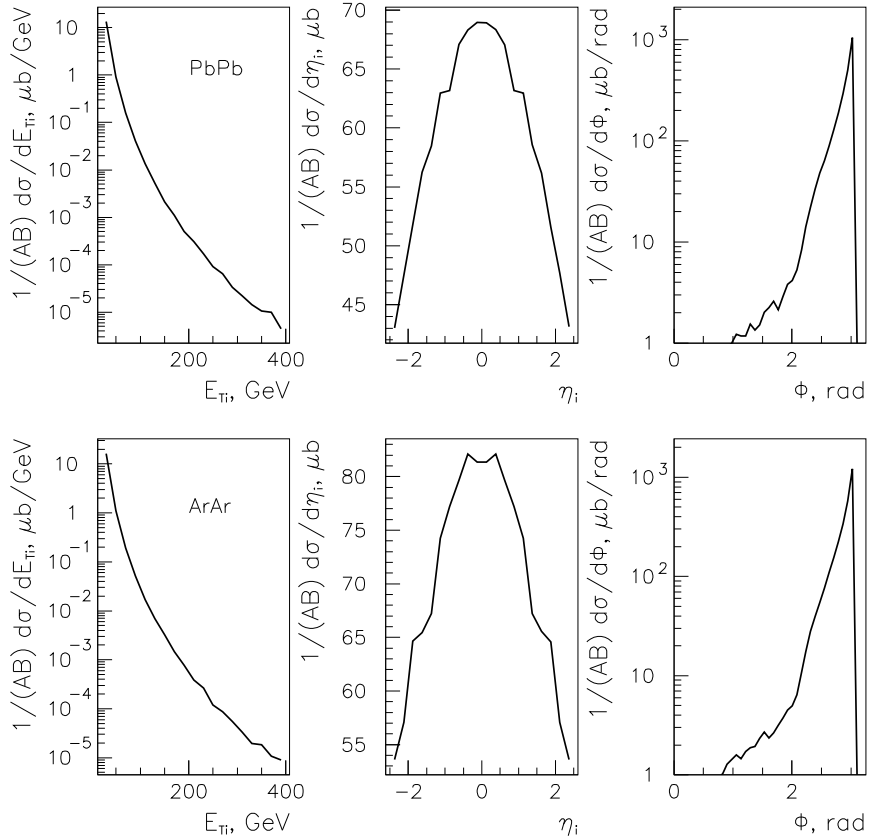}
\caption{Jet cross sections
versus
transverse energy of the jet (for $|\eta_i|<2.5$, plots on the left) and
pseudorapidity of the jet (for $E_{Ti}> 20$ GeV, plots in the middle),
and dijet cross sections
versus angle between the two hardest jets for $E_{T1}>20$ GeV,
$E_{T2}>15$ GeV and $|\eta_1|,|\eta_2|<2.5$ (plots on the right), for PbPb
collisions at 5.5 TeV (upper plots) and ArAr collisions at 6.3 TeV
(lower plots). Default options are
used, see text.}
\label{abfig2-4}
\end{center}
\end{figure}

\subsubsection{Uncertainties}

Uncertainties on the renormalization/factorization scale, on the
jet reconstruction algorithm and on nucleon pdf, have been discussed
in the mentioned Section on pA collisions and show very similar features
in the AB case, so we will discuss them no longer. Here we will focus,
see Fig.~\ref{abfig1},
on isospin effects (obtained from the
comparison of pp and PbPb without any modifications of nucleon pdf
inside nuclei at the same energy per nucleon, 5.5 TeV) and on
the effect of modifications of nucleon pdf inside nuclei,
estimated by using EKS98 \cite{Eskola:1998iy,Eskola:1998df} nuclear 
corrections.

On the transverse momentum distributions isospin effects are
negligible, while effects of EKS98 result in a $\sim 3$~\% increase. On
the pseudorapidity distributions, isospin effects apparently tend to
fill a small dip at $\eta\simeq 0$ present in the pp distribution,
while EKS98 results in some increase, but nevertheless
effects never go beyond 5~\% and are not very significant when statistical
errors are considered. On the dijet angular distributions, isospin
effects are negligible while EKS98 produces an increase of order 10~\%
at maximum.

\subsubsection{Results}

The expected LHC luminosities in different collisions
are collected into Table 1, 
and also shown are
$\cal{L}\times{\rm (1 \ month)}$ in units of $\mu{\rm b}/(AB)$.  Using this
Table and the cross sections for inclusive one-jet production and
dijet production in the Figures, it is possible to extract the number of
expected jets (Figs.~\ref{abfig1} and~\ref{abfig2-4})
or dijets (Fig.~\ref{abfig2-4}) in different ranges
of the kinematic variables.
For example, examining the
solid line in Fig.
\ref{abfig2-4} (upper-left) one can expect, within the
pseudorapidity region we have considered, the following number of
jets per month in PbPb collisions at 5.5 TeV with a luminosity of $5
\cdot 10^{26}$ cm$^{-2}$s$^{-1}$: $2.2\cdot
10^{7}$ jets  with $E_{Ti}\sim 50$ GeV (corresponding to a cross section
of 1 $\mu$b/(AB)), and $2.2\cdot 10^3$ jets with $E_{Ti}\sim 250$ GeV
(corresponding to a cross section of $10^{-4}$ $\mu$b/(AB)).

A detailed study of jet quenching
\cite{Baier:2000mf,Gyulassy:2001nm,Wang:2002ri,Salgado:2002cd,Vitev:2002pf}
and of associated characteristics as jet
profiles, which should be sensitive to radiation from the jet
\cite{Baier:1999ds}, should be feasible with
samples of $\sim 10^3$ jets. 
Looking at the results given in Fig.~\ref{abfig2-4}, it
becomes evident that, from a theoretical point of view, 
the study of such samples should be possible up to a transverse energy
$E_{Ti}\sim 275$ GeV
with a run of 1 month at the considered luminosity: indeed, from Table 1,
$10^3$ jets for PbPb would correspond to a cross section of $4.5\cdot
10^{-5}$ $\mu$b/(AB), which in Fig.~\ref{abfig2-4} (upper-left) cuts the
curve at $E_{Ti}\sim 275$ GeV.

The centrality
dependence of the observables has not been examined
due to our poor knowledge of the
centrality behavior of the modification of nucleon pdf inside nuclei;
if this behavior becomes clear in future experiments at eA colliders
\cite{Arneodo:1996qa,eacoll,Abramowicz:2001qt}, such study would become 
very useful \cite{Lokhtin:2000wm}.
In any case, a variation of
nuclear sizes should allow a systematic study of the
dependence of jet spectra on the size and energy density of the produced
plasma.

\begin{table}[t]
\label{armestoab:table1}
\begin{center}
\begin{tabular}{|c|c|c|c|}
\hline
Collision & $E_{cm}$ per nucleon (TeV) & $\cal{L}$
(cm$^{-2}$s$^{-1}$) & Number of jets/events per month per $\mu$b/(AB) \\
\hline
ArAr & 6.3 & $10^{29}$ & $1.6\cdot 10^8$ \\
\hline
ArAr & 6.3 & $3\cdot 10^{27}$ & $4.8\cdot 10^6$ \\
\hline
PbPb & 5.5 & $5\cdot 10^{26}$ & $2.2\cdot 10^7$ \\
\hline
\end{tabular}
\end{center}
{\small Table 1:
Luminosities $\cal{L}$ in units of cm$^{-2}$s$^{-1}$ and
${\cal L}\times 10^6$~s in units of $\mu{\rm b}/AB$ for different
collisions at the LHC. The numbers of expected jets and dijets
in a certain kinematic range are obtained by multiplying the latter
column by the cross sections given in Figs.~\ref{abfig1},~\ref{abfig2-4}
(jets) and~\ref{abfig2-4} (dijets).}
\end{table}

\subsection{Benchmark Particle Cross Sections}
\label{sec22}
{\em Ivan Vitev} 

Hadron production in leading order pQCD is reviewed. 
The shape of the single inclusive particle spectra is well 
described for $p_T \geq 2-3$~GeV at  center of mass energies 
from $20$~GeV to  $2$~TeV.  The phenomenological  
K-factor is found to decrease systematically with $\sqrt{s}$. 
For ultra-relativistic heavy ion reactions
the calculation is augmented with the effects of initial multiple
parton scattering and final state radiative energy loss.    
Baseline CERN-LHC predictions for hadron production in $p+p$ 
and suppression in central $Pb+Pb$ reactions at 
$\sqrt{s} = 5.5$~TeV are given in comparison to the  
corresponding results at BNL-RHIC  and CERN-SPS energies. 
  
The purpose of this section is to present a {\em lowest order} (LO) 
analysis of inclusive hadron production up to the Tevatron 
energies and discuss hadron differential cross sections
and composition  at the LHC. 

\subsubsection{Hadroproduction in Factorized pQCD}
\label{sec221}

The standard  factorized pQCD hadron production formalism 
expresses the differential 
hadron cross  section in $N+N \rightarrow h+X$  as a convolution of 
the measured parton distribution functions (PDFs)  
$f_{\alpha/N}(x_\alpha,Q_\alpha^2)$  for the interacting 
partons ($\alpha = a,b$), 
with the fragmentation function  (FFs) $D_{h/c}(z,Q^2_c)$ for
the leading scattered parton  $c$  into a hadron of flavor $h$ and the
parton-parton differential cross sections for the elementary sub-process 
$d\sigma^{(ab \rightarrow cd)}/d\hat{t}$: 
\begin{eqnarray} 
E_{h}\frac{d\sigma^{NN}}{d^3p} &=&
K_{NLO}   \sum_{abcd}\,  \int\limits_0^1  dz_c  
\int\limits_{x_{a  \min}}^1 \int\limits_{x_{b  \min}}^1dx_a  dx_b \;
f_{a/p}(x_a,Q^2_a) f_{b/p}(x_b,Q^2_b) \nonumber \\
&& \times \; D_{h/c}(z_c,{Q}_c^2) 
 \frac{\hat{s}}{\pi z^2_c} \frac{d\sigma^{(ab\rightarrow cd)}}
{d{\hat t}} \delta(\hat{s}+\hat{u}+\hat{t}) \; .
\label{hcrossec}
\end{eqnarray}
A list of the lowest order partonic cross sections can be found 
in~\cite{Owens:1986mp}. In Eq.~(\ref{hcrossec}) $x_a, x_b$ are the
initial  momentum  fractions  carried  by the interacting partons 
and  $z_c=p_h/p_c$  is  the momentum fraction of the observed hadron. 
$K_{NLO}$ is a phenomenological factor that is meant to account 
for next-to-leading order (NLO) corrections. It is $\sqrt{s}$ 
and scale dependent  and takes typical values $ \simeq 1-4$.
One usually finds that Eq.~(\ref{hcrossec}) over-predicts 
the  curvature  of the inclusive hadron spectra 
$ | \partial_{p_T} d\sigma^h |$  at transverse momenta $p_T \leq 4$~GeV. 
This can be partly corrected by the introduction of a small 
intrinsic (or primordial) $k_T$-smearing of partons, transversely 
to  the collision axis, and generalized parton distributions
$\tilde{f}_\alpha(x,k_T,Q^2)$  motivated by the pQCD initial state 
radiation. For the corresponding modification of the kinematics
in (\ref{hcrossec}) in addition to the $\int d^2 k_T^a 
\int  d^2 k_T^b \,(\cdots) $ integrations  see~\cite{Owens:1986mp}.
The generalized parton distributions are often approximated as
\begin{equation}
\tilde{f}_\alpha(x,k_T,Q^2) \approx f_\alpha(x,Q^2) g(k_T), \quad 
g(k_T) = \frac{e^{-k_T^2/\langle {k}_T^2 \rangle}}
{\pi\langle {k}_T^2 \rangle } \; ,
\label{gassm}
\end{equation}
where the width  $\langle {k}_T^2 \rangle$ of the Gaussian enters as a 
phenomenological parameter.

\vskip 0.5cm
\begin{figure}[htb!]
\begin{center} 
\hspace*{-0.2in} 
\includegraphics[height=3.2in,width=2.6in,bbllx=90,bblly=10,bburx=600,bbury=700,clip=,angle=-90]{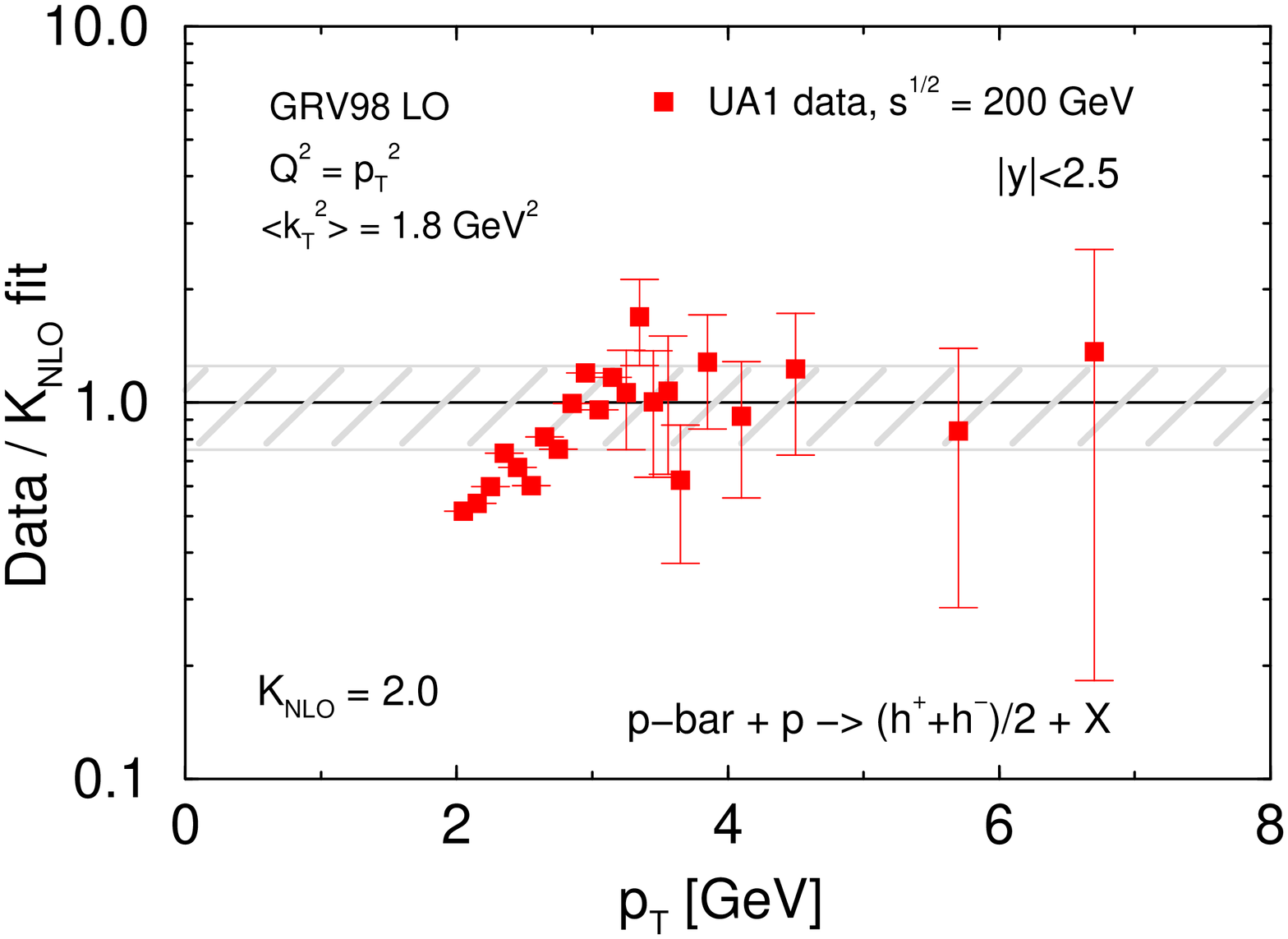}
\includegraphics[height=3.2in,width=2.6in,bbllx=90,bblly=10,bburx=600,bbury=700,clip=,angle=-90]{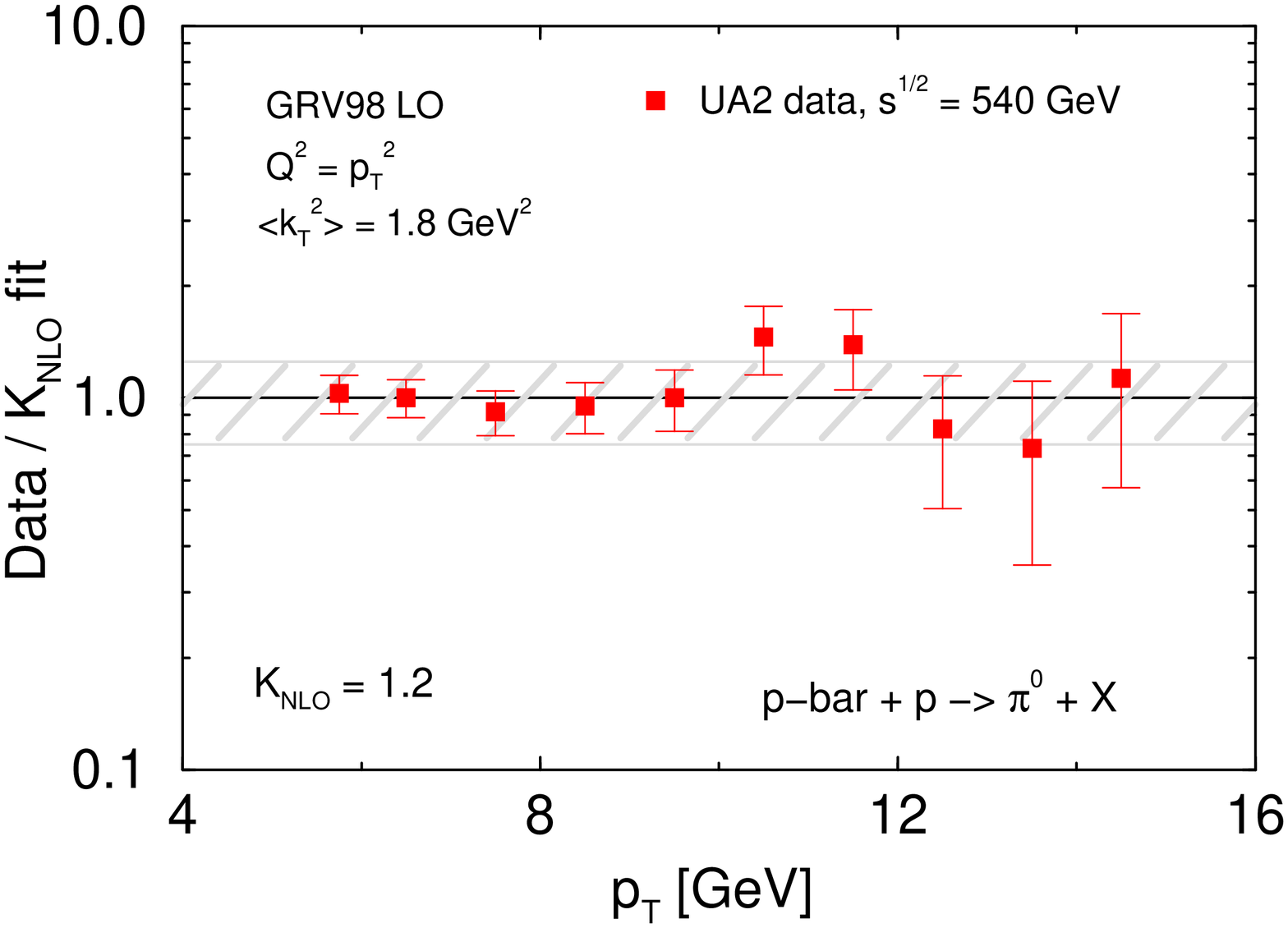}
\hspace*{-0.2in} 
\includegraphics[height=3.2in,width=2.6in,bbllx=90,bblly=10,bburx=600,bbury=700,clip=,angle=-90]{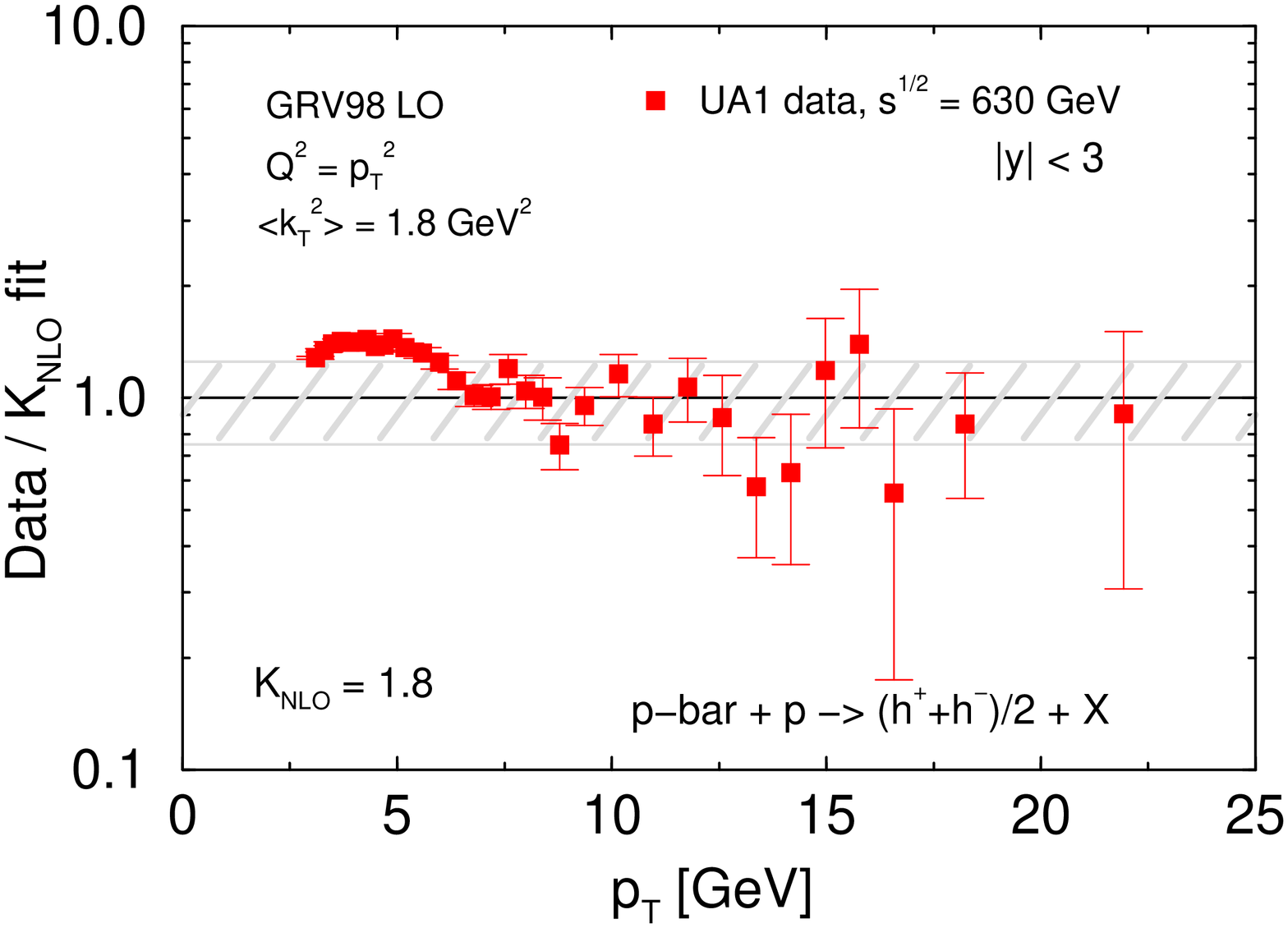}
\includegraphics[height=3.2in,width=2.6in,bbllx=90,bblly=10,bburx=600,bbury=700,clip=,angle=-90]{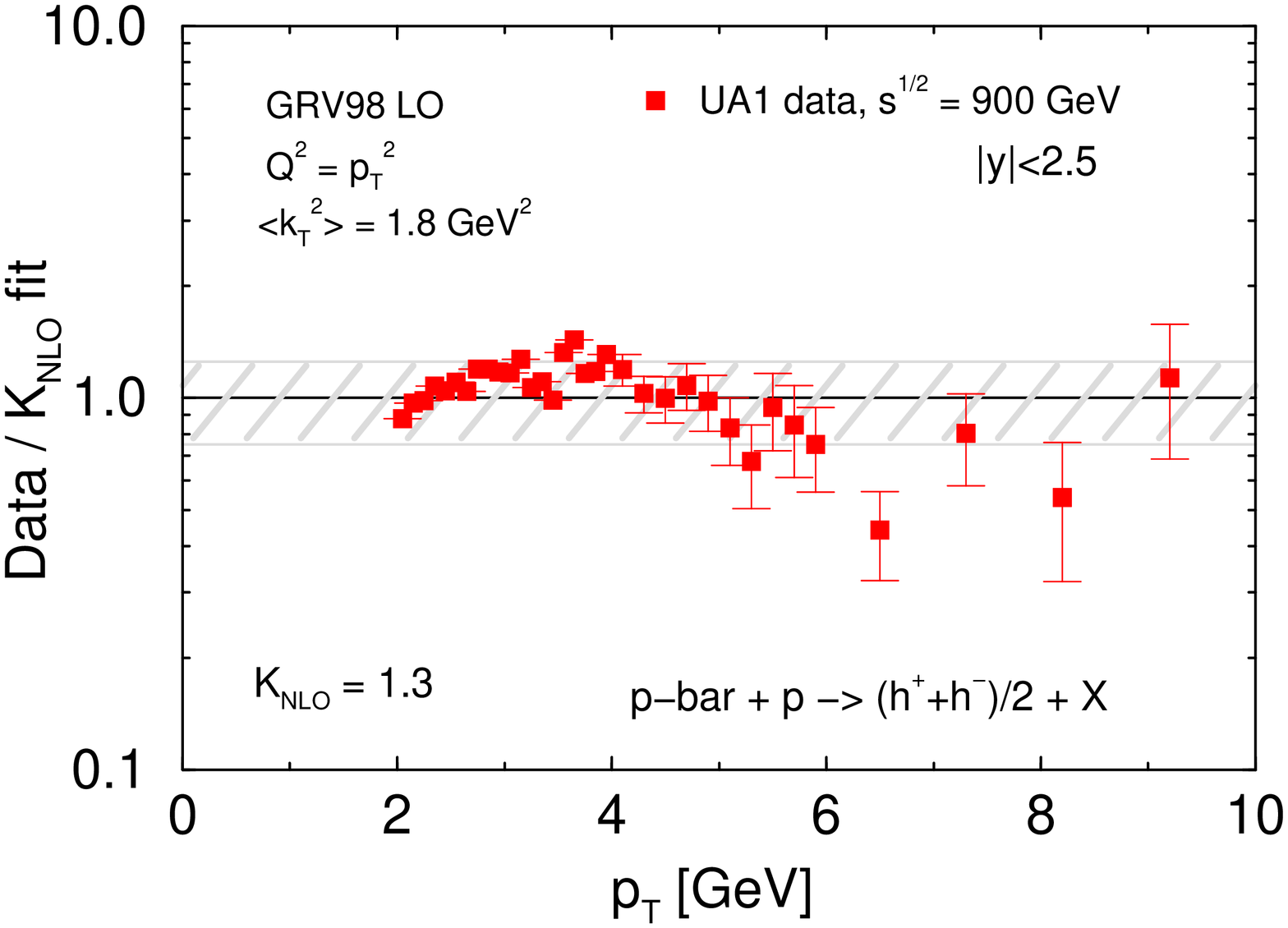}
\hspace*{-0.2in} 
\includegraphics[height=3.2in,width=2.6in,bbllx=90,bblly=10,bburx=600,bbury=700,clip=,angle=-90]{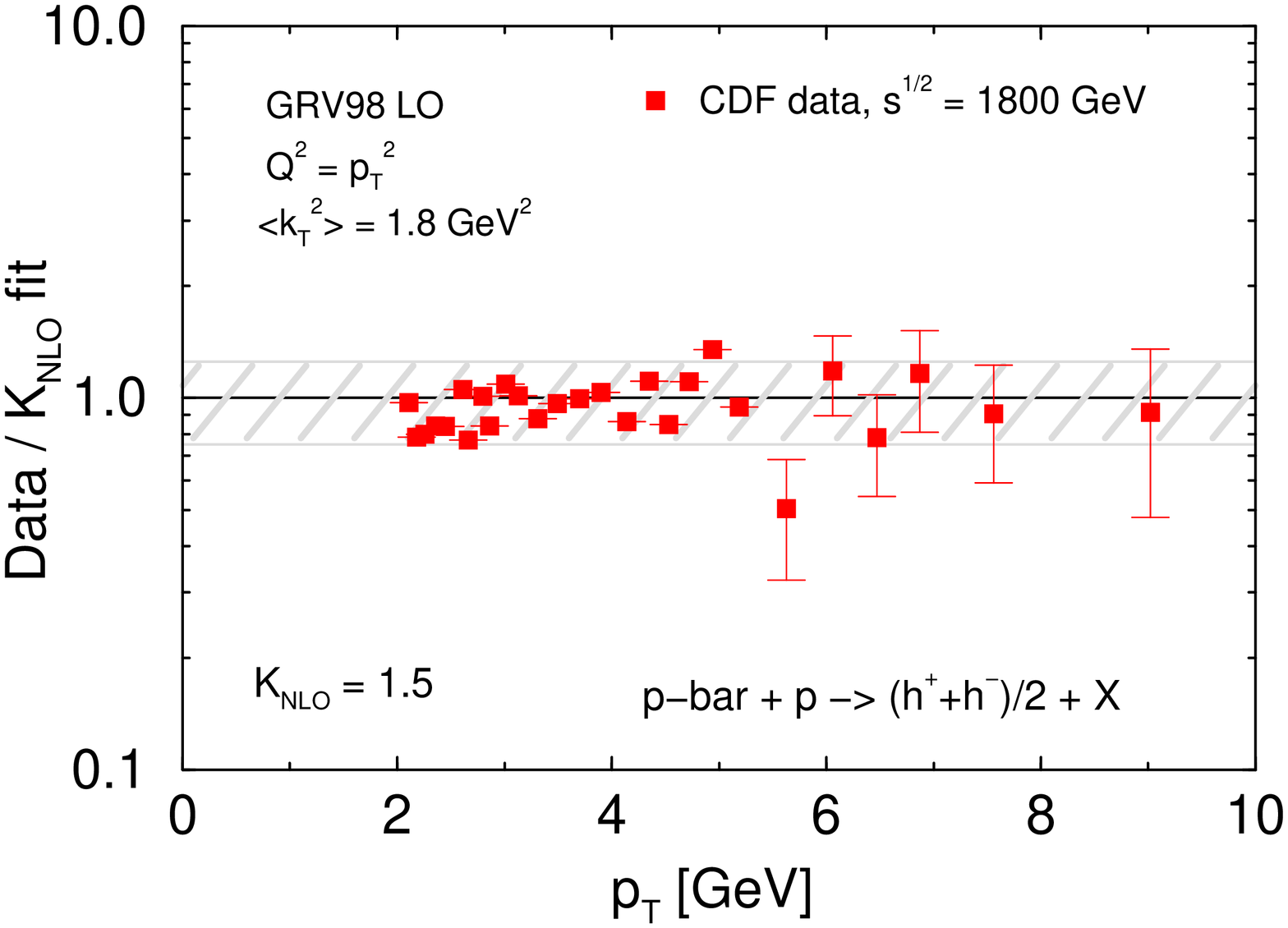}
\includegraphics[height=3.2in,width=2.6in,bbllx=90,bblly=10,bburx=600,bbury=700,clip=,angle=-90]{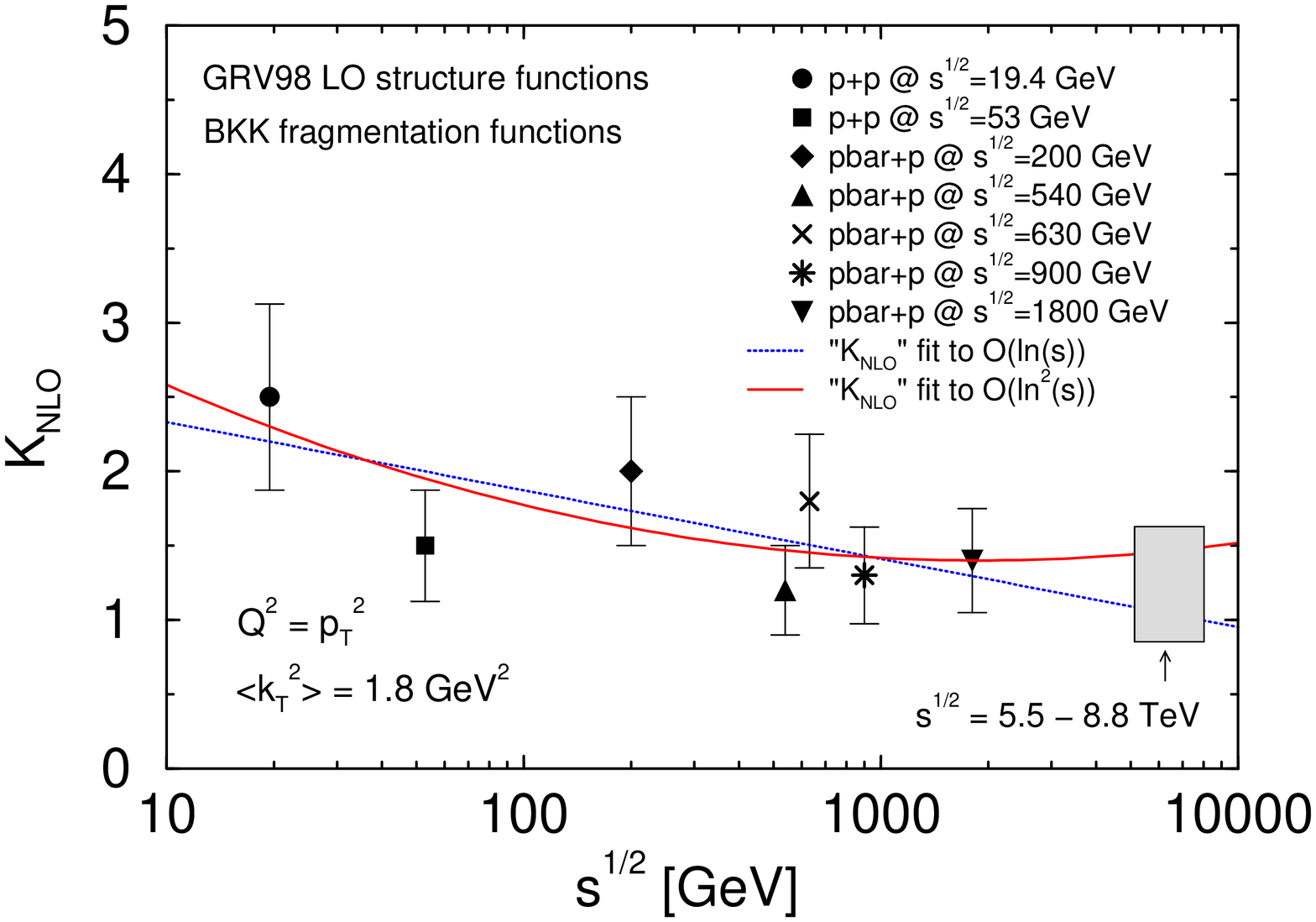}
\vspace*{-0.1in}
\caption{\small  Extracted  $K_{NLO}$  from comparison of LO 
pQCD calculation to data~\protect{\cite{Antreasyan:cw,Alper:nv,Albajar:1989an,Banner:1984wh,Bocquet:1995jr,Abe:yu}  } at and about mid-rapidity  in the 
range $2\leq p_T \leq 25$~GeV. A systematic decrease of $K_{NLO}$ 
with $\sqrt{s}$ is observed and illustrated in the bottom right panel. The 
projected 50\% uncertainty at $\sqrt{s}= 5.5 -8.8$~TeV is also shown. }     
\label{lhc-h:fig1}
\end{center} 
\end{figure}
Perturbative QCD fits to data~\cite{Cronin:zm,Straub:xd,Antreasyan:cw,Alper:nv,Albajar:1989an,Banner:1984wh,Bocquet:1995jr,Abe:yu}  
use  different coupled choices for  $K_{NLO}$ and 
$\langle {k}_T^2 \rangle$ and  the extracted values are thus not 
directly comparable. However, similar agreement 
between data and theory at the level of spectral shapes and  the $\sqrt{s}$ 
dependence of the corrective factors discussed above is found.
In~\cite{Zhang:2001ce} the factorization and fragmentation scales 
were set to $Q_{PDF}=p_T/2$  and  $Q_{FF}=p_T/2z_c$ and no $K_{NLO}$ 
factors were employed.  The extracted  $\langle {k}_T^2 \rangle$  
decreases from $2.7$~GeV$^2$  at $\sqrt{s}\simeq 50$~GeV to  
$0.75$~GeV$^2$ at  $\sqrt{s}\simeq 2$~TeV. Alternatively, 
in~\cite{Eskola:2002kv} no primordial $k_T$-smearing was used and the 
scales in the calculation were fixed to be  $Q_{PDF}=Q_{FF}=p_T$. The 
deduced  $K_{NLO}$ decreases from  $\sim 6$ at $\sqrt{s}\simeq 50$~GeV 
to $\sim 1.5$ at   $\sqrt{s}\simeq 2$~TeV.

In the fits shown in Fig.~\ref{lhc-h:fig1} we have used  the GRV98 LO 
PDFs~\cite{Gluck:1998xa} and the BKK LO FFs~\cite{Binnewies:1994ju}. 
Proton+antiproton fragmentation has been  parameterized  as 
in~\cite{Wang:1998bh}, inspired from PYTHIA~\cite{Sjostrand:1993yb}   
results.  A fixed $\langle {k}_T^2 \rangle_{pp}=1.8$~GeV$^2$ has been 
employed, leading to  a $K_{NLO}$ parameter  that  naturally exhibits a 
smaller variation with $\sqrt{s}$. A $\pm 25 \%$ error band 
about the $K_{NLO}$ value, fixed by the requirement to match the moderate- 
and high-$p_T$  behavior of the data, is also shown. The  
fragmentation and factorization scales were fixed as 
in~\cite{Eskola:2002kv}. In the lower right  panel the systematic 
decrease of the next-to-leading order K-factor is
presented. Two fits to $K_{NLO}$ have been used: linear  
$K_{NLO} = 2.7924-0.0999 \ln s $ and quadratic   
$K_{NLO} = 3.8444-0.3234 \ln s + 0.0107 \ln^2 s$ in  $\ln s$.  
For center of mass energies up to 1~TeV the two parameterization  differ
by less than 15\% but this difference is seen to grow to 30\%-50\% at 
$\sqrt{s}=5-10$~TeV.

\vskip 0.5cm
\begin{figure}[htb!]
\begin{center} 
\hspace*{-0.2in}
\includegraphics[height=6.2in,width=3.6in,bbllx=90,bblly=0,bburx=600,bbury=800,clip=,angle=-90]{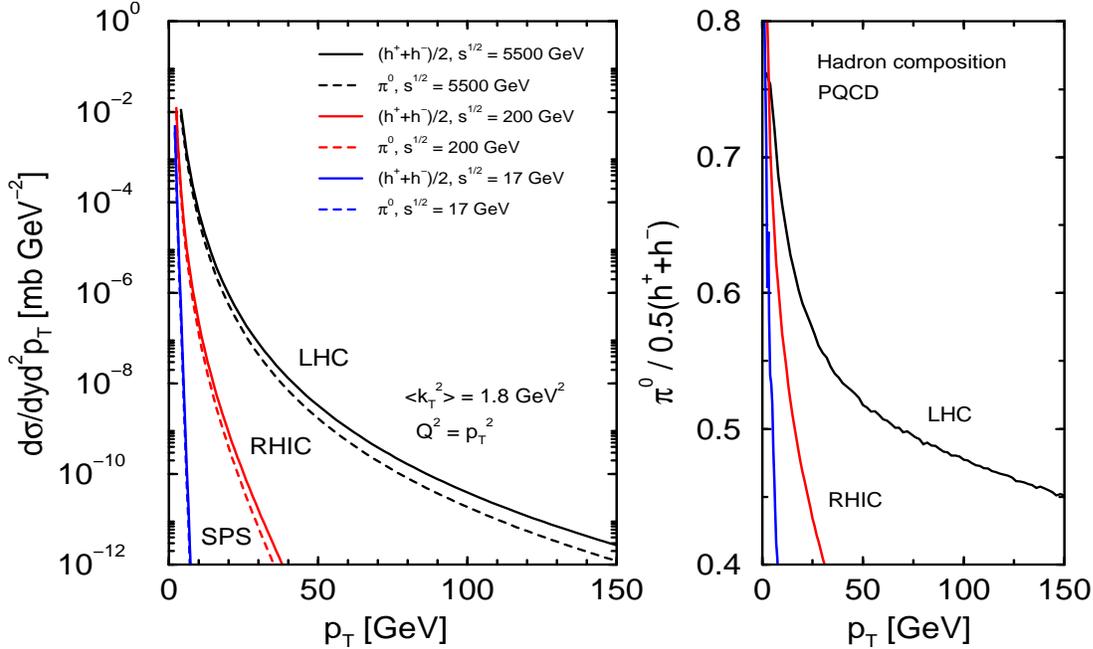}
\vspace*{-0.1in}
\caption{\small The predicted  LO  differential  cross  section 
$d \sigma^{pp} / dy d^2 p_T$ for inclusive neutral pion 
and charged hadron  production at midrapidity $y=0$ in 
$p+p$ ($\bar{p}+p$) reactions  is shown for $\sqrt{s}=17, 200,$ 
and  $5500$~GeV. The ratio of neutral pions to inclusive charged 
hadrons versus  $p_T$  is given in the right panel.} 
\label{lhc-h:fig2}
\end{center} 
\end{figure}

In Fig.~\ref{lhc-h:fig2} the predicted transverse momentum 
distribution of neutral pions and inclusive charged hadrons is shown,  
corresponding to the quadratic in $\ln s$ fit to $K_{NLO}$ for 
energies typical of SPS, RHIC, and the LHC. The 
{\em significant} hardening of the spectra with $\sqrt{s}$ has 
two important consequences for $p+A$ and $A+A$ collisions: a notably 
reduced sensitivity to initial state kinematic effects 
(smaller Cronin) and larger variation of the manifested final-state 
multi-parton scattering (energy loss)  with $p_T$~\cite{Vitev:2002pf}. 
We have also investigated the effect of isospin asymmetry between 
$p+p$ and $p+\bar{p}$ reactions in $\pi^0$ and $h^+ + h^-$  
production and found it to be small. More quantitatively,  at 
$\sqrt{s}=5.5$~TeV  the fractional difference 
$|d\sigma^{\bar{p}p}-d\sigma^{pp}|/d\sigma^{pp}$ 
varies from 2.5\% at $p_T = 5$~GeV to 4.8\% at $p_T = 150$~GeV. 
This is insignificant as compared to the projected 50\% 
uncertainty that comes from the extrapolation of 
$K_{NLO}$ in LO calculations (see Fig.~\ref{lhc-h:fig1}) or the 
choice of scale in NLO calculations. A recent study  
showed {\em no} deviation from DGLAP evolution
at $Q^2=10$~GeV$^2$  down to $x=10^{-5}$ in $N+N$ 
reactions~\cite{Eskola:2002yc}. The nuclear 
amplification effect $\propto A^{1/3} \simeq 10$ for a large nucleus 
is still insufficient  to enable measurements of high initial 
gluon density QCD at RHIC,  but will play an important role 
at the LHC.

\subsubsection{Perturbative QCD Hadron Composition}
\label{sec222}

The predicted hadron composition in $p+p$ ($\bar{p}+p$) reactions is plotted
in the right panel of  Fig.~\ref{lhc-h:fig2}. The proton+kaon
fraction is seen to increase systematically with $p_T$
($x_T = 2p_T/\sqrt{s}$) and is reflected in the decreasing 
$\pi^0/0.5(h^+ + h^-)$. At RHIC and LHC energies this ratio becomes 
$\sim 0.5$ at  $p_T \simeq 15$~GeV and  $p_T \simeq 75$~GeV, respectively.     
At transverse momenta $p_T \simeq 2-4$~GeV the contribution of baryons and
kaons to $h^+ + h^-$ is  $\leq 20\%$. This is significantly 
smaller compared to  data on $N+N$ reactions,
 with the discrepancy being amplified  in central $A+A$. 
Possible explanations include: enhanced  baryon production  
via topological gluon configurations (junctions)  and its interplay 
with jet quenching~\cite{Wang:xy,Gyulassy:1993hr} in 
$A+A$~\cite{Vitev:2001zn,Vitev:2002wh}, hydrodynamic 
transverse flow~\cite{Teaney:2001av},  uncertainty of the 
fragmentation functions $D_{p/c}(z_c,Q^2)$ into protons and 
antiprotons~\cite{Zhang:2002py}, and quark recombination driven 
by unorthodox (extracted) parton distributions inside 
nuclei~\cite{Hwa:2002tu}. The approaches in 
Refs.~\cite{Vitev:2001zn,Vitev:2002wh,Teaney:2001av} also address the 
centrality dependence of the baryon/meson ratios in heavy ion collisions 
at RHIC.  In~\cite{Vitev:2002wh} in has been shown 
that similar nuclear enhancement is
expected in $\Lambda, \bar{\Lambda}$ production (as compared to kaons). 
The combined low-$p_T$ baryon  enhancement and the growth of the 
non-pionic hadron fraction in the 
perturbative regime may lead to an approximately constant  pion to 
charged hadron ratio in the full measured $p_T$ region  at RHIC at 
$\sqrt{s}_{NN}=200$~GeV.  We propose that the LHC may play 
a critical role in resolving the mystery of enhanced baryon production 
in $A+A$ through the significantly larger experimentally 
accessible $p_T$ range. Effects associated  with baryon transport  and 
transverse flow are not expected to extend  beyond $p_T=10-15$~GeV and may 
result in a detectable minimum of the baryon/meson ratio versus
$p_T$  before a secondary subsequent rise. On the other hand, 
fragmentation functions (possibly enhanced at 
large $z_c$ relative to current  parameterizations) are expected 
to exhibit a much more monotonic behavior.

\section{FINAL STATE EFFECTS IN DENSE AND HOT MATTER}
\label{sec3}

{\bf Bjorken argument}
In August 1982 J.~D.~Bjorken published a preprint {\cite{Bjorken:1982tu}}
on "Energy Loss of Energetic Partons in Quark-Gluon Plasma:
Possible Extinction of High $p_{\perp}$ Jets in Hadron-Hadron Collisions",
in which he discussed that high energy quarks  and gluons propagating through
quark-gluon plasma (QGP) suffer differential energy loss, and where he
further  pointed
out that as an interesting signature events may be observed in which the
hard collisions may occur such that one jet is escaping without
 absorption and the other is fully absorbed.

The arguments in this work have been based on elastic scattering of high 
momentum partons from quanta in the QGP, with a resulting 
("ionization") loss $-dE/dz \simeq \alpha_s^2 \sqrt{\epsilon}$,
with $\epsilon$ the energy density of the QGP. The loss turns out to be less 
than the string tension of $O(1~\rm{GeV/fm})$ \cite{Thoma:1995ju} .

However, as in QED, bremsstrahlung is another important source of
energy loss \cite{Gyulassy:1993hr}. Due to multiple (inelastic) scatterings 
and induced gluon radiation high momentum jets and leading large 
$p_{\perp}$ hadrons become depleted, quenched \cite{Gyulassy:1990ye} 
or may even become extinct. In \cite{Baier:1994bd} 
it has been shown that a genuine pQCD process (Fig.~\ref{fig:rad})
is responsible for the dominant loss: after the gluon is radiated off
the energetic parton it suffers multiple scatterings in the medium.
Indeed, further studies by 
\cite{Baier:1996kr,Baier:1996sk,Zakharov:1996fv,Zakharov:1997uu,Baier:1998kq,Zakharov:1998sv,Zakharov:1999zk,Zakharov:2000iz,Wiedemann:1999fq,Wiedemann:2000ez,Wiedemann:2000za,Wiedemann:2000tf,Gyulassy:2000fs,Gyulassy:2000er}
support this observation. For reviews, 
see~\cite{Baier:2000mf,Gyulassy:2003mc,Kovner:2003zj}.

\subsection{Radiative Energy Loss and Medium-Induced Gluon Radiation}
\label{sec31}
\subsubsection{Qualitative Arguments}
\label{sec311}
{\em R. Baier, U.A. Wiedemann}

\begin{figure}[htb]
\hspace*{3cm}
\psfig{figure=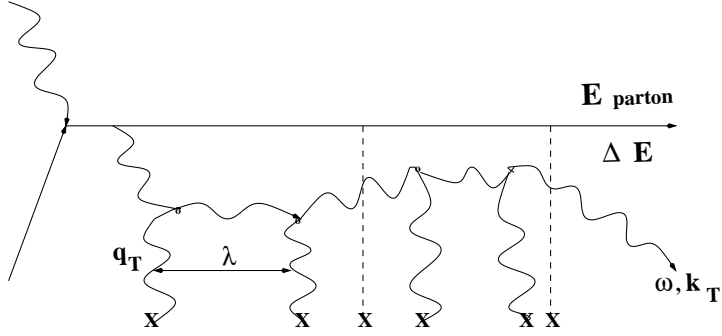,angle=-90,width=9.5cm}
\caption{Typical gluon radiation diagram }
\label{fig:rad}
\end{figure}

After its production in a hard collision, the energetic parton
radiates a gluon which both traverse a finite size $L$ medium.  
Due to its non-abelian nature and its interaction with the medium, 
this gluon follows a zig-zag path (Fig.~\ref{fig:rad}),
with a  mean free path $\, \lambda~ > ~1/\mu~$, which is  the
range of screened multiple gluon interactions. 
%
%
We estimate the medium-induced gluon radiation in two different limits,
requiring that the gluon is emitted from the hard parton if it 
picks up sufficient transverse momentum to decohere from the partonic
projectile.

\underline{In the multiple soft scattering limit}, the average phase 
$\varphi$ accumulated by the gluon due to multiple scattering is
\begin{equation}
  \varphi = \Bigg\langle \frac{k_T^2}{2\omega}\, \Delta z \Bigg\rangle
  \sim \frac{\hat{q}\, L}{2\omega} L = \frac{\omega_c}{\omega}\, .
  \label{2.5}
\end{equation}
Here, the medium dependence is controlled by the transport coefficient 
\begin{equation}
 \hat{q} \simeq \mu^2/\lambda \simeq \rho 
        \int~d^2 q_{\perp}~q^2_{\perp}~ d\sigma/d^2 q_{\perp} \, ,
\label{qhat}
\end{equation}
where $\rho $ is the density of the medium (a nucleus, or  partons)
and $\, \sigma $ the cross section of the gluon-medium interaction.
According to (\ref{2.5}), gluons are emitted from a hard parton 
traversing a finite path length $L$ in the medium, if the phase
$\varphi > 1$. Thus, their energy has to be smaller than
the ``characteristic gluon frequency''
\begin{equation}
  \omega_c = \frac{1}{2}\, \hat{q}\, L^2\, .
  \label{2.6}
\end{equation}
For an estimate of the shape of the energy distribution,
consider the number $N_{\rm coh}$ of scattering centers which 
add coherently in the gluon phase (\ref{2.5}), 
$k_T^2 \simeq N_{\rm coh}\, \mu^2$. Based on expressions
for the coherence time of the emitted gluon, 
$t_{\rm coh} \simeq \frac{\omega}{k_T^2} \simeq 
\sqrt{\frac{\omega}{\hat{q}}}$
and $N_{\rm coh} = \frac{t_{\rm coh}}{\lambda} = 
\sqrt{\frac{\omega}{\mu^2\, \lambda}}$, one estimates for the
gluon energy spectrum per unit path length
\begin{equation}
  \omega \frac{dI}{d\omega\, dz} \simeq 
  \frac{1}{N_{\rm coh}}\, 
  \omega \frac{dI^{\rm 1\, scatt}}{d\omega\, dz} \simeq
  \frac{\alpha_s}{t_{\rm coh}}
  \simeq \alpha_s\, \sqrt{\frac{\hat{q}}{\omega}}\, .
  \label{2.7}
\end{equation}
Again, this $1/\sqrt{\omega}$-energy dependence of the
medium-induced non-abelian gluon energy spectrum is
expected for sufficiently small $\omega < \omega_c$.
The average energy loss of the parton 
(in the limit $E_{parton} \rightarrow \infty$)
due to gluon radiation with a spectrum $\frac{\omega dI}{d \omega}$
is then determined by the characteristic gluon energy
$~\omega_c$ as follows,  
\begin{equation}
\Delta E = \int^{\omega_c}
\frac{\omega dI}{d \omega}~{d \omega} \simeq \alpha_s ~\omega_c \, \, , 
\label{loss}
\end{equation}
It shows a characteristic quadratic dependence on the in-medium
pathlength. The medium-induced BDMPS gluon spectrum
(valid for finite size $\, L >> \lambda \, $ and for soft gluon energies 
$\,\, \, \omega_{GB}=
\hat{q} \lambda^2 < \omega << E_{parton} \, \rightarrow \infty $)
with the characteristic behavior:
$\, \,  \frac{\omega dI}{d\omega} \simeq
 \alpha_s ~\sqrt{\frac{\omega_c}{\omega}} \, $ , 
 $\, \omega < \omega_c~$ is suppressed by $1/N_{coh}$ for
$\omega > \omega_{GB}$ with
respect to the incoherent Gunion-Bertsch spectrum. For comparison with 
QED the Landau-Pomeranchuk-Migdal~\cite{Landau:gr,Landau:um,Migdal:1956tc}
(LPM) suppressed photon spectrum behaves as 
$ \frac{\omega dI}{d\omega} \simeq \sqrt{\omega}$.

\underline{Opacity expansion:} we turn now to the limiting
case in which the radiation pattern results from an incoherent
superposition of very few $n_0L$ single hard scattering processes
positioned within path length $L$. Consider a hard partonic projectile
which picks up a single transverse momentum $\mu$ by interacting with a 
single hard scatterer. An additional gluon of energy $\omega$ decoheres
from the projectile wave function if its typical formation
time $\bar{t}_{\rm coh} = \frac{2\omega}{\mu^2}$ is smaller than the
typical distance $L$ between the production point of the parton
and the position of the scatterer. The relevant phase is 
\begin{equation}
  \gamma = \frac{L}{\bar{t}_{\rm coh}} \equiv \frac{\bar{\omega}_c}{\omega}
  \, ,
  \label{2.16}
\end{equation}
which indicates a suppression of gluons with energy $\omega$
larger than the characteristic gluon energy 
\begin{equation}
  \bar\omega_c = \frac{1}{2} \mu^2\, L\, .
  \label{2.17}
\end{equation}
The gluon energy spectrum per unit path length can be estimated
in terms of the coherence time $\bar{t}_{\rm coh}$ and of the
average number $n_0\, L$ of scattering centers contributing
incoherently
\begin{equation}
  \omega \frac{dI^{N=1}}{d\omega\, dz} \simeq 
  (n_0\, L)\, \frac{\alpha_s}{\bar{t}_{\rm coh}}
  \simeq (n_0\, L)\, \alpha_s\, \frac{\mu^2}{\omega}\, .
  \label{2.18}
\end{equation}
This is the typical $1/\omega$-dependence of the non-abelian
gluon radiation spectrum in the absence of LPM-type
destructive interference effects. It will result again in
a quadratic $L$-dependence of the average energy loss~\cite{Gyulassy:2000fs}.

\subsubsection{Quantitative Results}
\label{sec312}

{\em C.A. Salgado, U.A. Wiedemann}

There are several calculations of the inclusive energy distribution of
medium-induced gluon radiation from Feynman multiple scattering diagrams \cite{Baier:1996kr,Baier:1996sk,Zakharov:1996fv,Zakharov:1997uu,Baier:1998kq,Zakharov:1998sv,Zakharov:1999zk,Zakharov:2000iz,Wiedemann:1999fq,Wiedemann:2000ez,Wiedemann:2000za,Wiedemann:2000tf,Gyulassy:2000fs,Gyulassy:2000er}. They can
be obtained as particular limiting cases of the following compact
expression~\cite{Wiedemann:2000za,Wiedemann:2000tf}
\begin{eqnarray}
  \omega\frac{dI}{d\omega}
  &=& {\alpha_s\,  C_R\over (2\pi)^2\, \omega^2}\,
    2{\rm Re} \int_{\xi_0}^{\infty}\hspace{-0.3cm} dy_l
  \int_{y_l}^{\infty} \hspace{-0.3cm} d\bar{y}_l\,
   \int d{\bf u}\,  \int_0^{\chi \omega}\, d{\bf k}_T\, 
  e^{-i{\bf k}_T\cdot{\bf u}}   \,
  e^{ -\frac{1}{2} \int_{\bar{y}_l}^{\infty} d\xi\, n(\xi)\,
    \sigma({\bf u}) }\,
  \nonumber \\
  && \times {\partial \over \partial {\bf y}}\cdot
  {\partial \over \partial {\bf u}}\,
  \int_{{\bf y}=0}^{{\bf u}={\bf r}(\bar{y}_l)}
  \hspace{-0.5cm} {\cal D}{\bf r}
   \exp\left[ i \int_{y_l}^{\bar{y}_l} \hspace{-0.2cm} d\xi
        \frac{\omega}{2} \left(\dot{\bf r}^2
          - \frac{n(\xi) \sigma\left({\bf r}\right)}{i\, \omega} \right)
                      \right]\, .
    \label{2.1}
\end{eqnarray}
Here, ${\bf k}_T$ denotes the transverse momentum of the emitted gluon.
The limit $k_T = \vert{\bf k}_T\vert < \chi\, \omega$
on the transverse phase space allows to
discuss gluon emission into a finite opening angle $\Theta$,
$\chi = \sin\Theta$. For the
full angular integrated quantity, $\chi = 1$. 
The radiation of hard quarks or gluons differs by 
the Casimir factor $C_R = C_F$ or $C_A$, respectively.

The two-dimensional transverse coordinates ${\bf u}$, ${\bf y}$
and ${\bf r}$ emerge in the derivation of (\ref{2.1}) as distances
between the positions of projectile components in the amplitude
and complex conjugate amplitude. The longitudinal coordinates
$y_l$, $\bar{y}_l$ integrate over the ordered longitudinal
gluon emission points in amplitude and complex conjugate amplitude,
which emerge in time-ordered perturbation theory. These
internal integration variables play no role in the following 
discussion. They are
explained in more detail in Ref. \cite{Wiedemann:2000za}.

\begin{figure}[htb]
\begin{minipage}[t]{78mm}
\includegraphics[width=8cm]{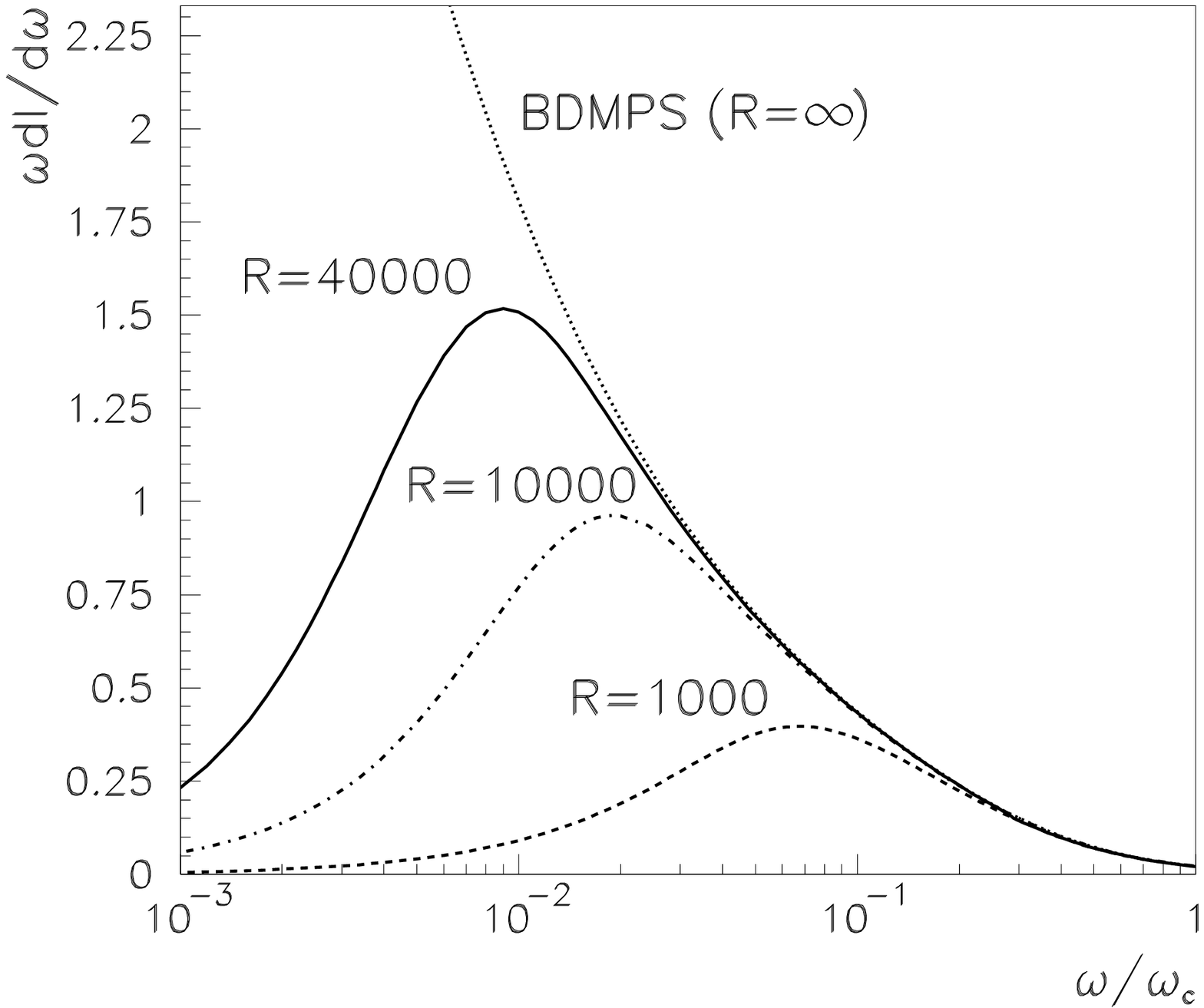}
\caption{The medium-induced gluon energy distribution 
$\omega \frac{dI}{d\omega}$ in the multiple soft scattering
approximation for different values of the 
kinematic constraint $R = \omega_c\, L$.}
\label{bdmpscomp}
\end{minipage}
\hspace{\fill}
\begin{minipage}[t]{78mm}
\includegraphics[width=7.5cm]{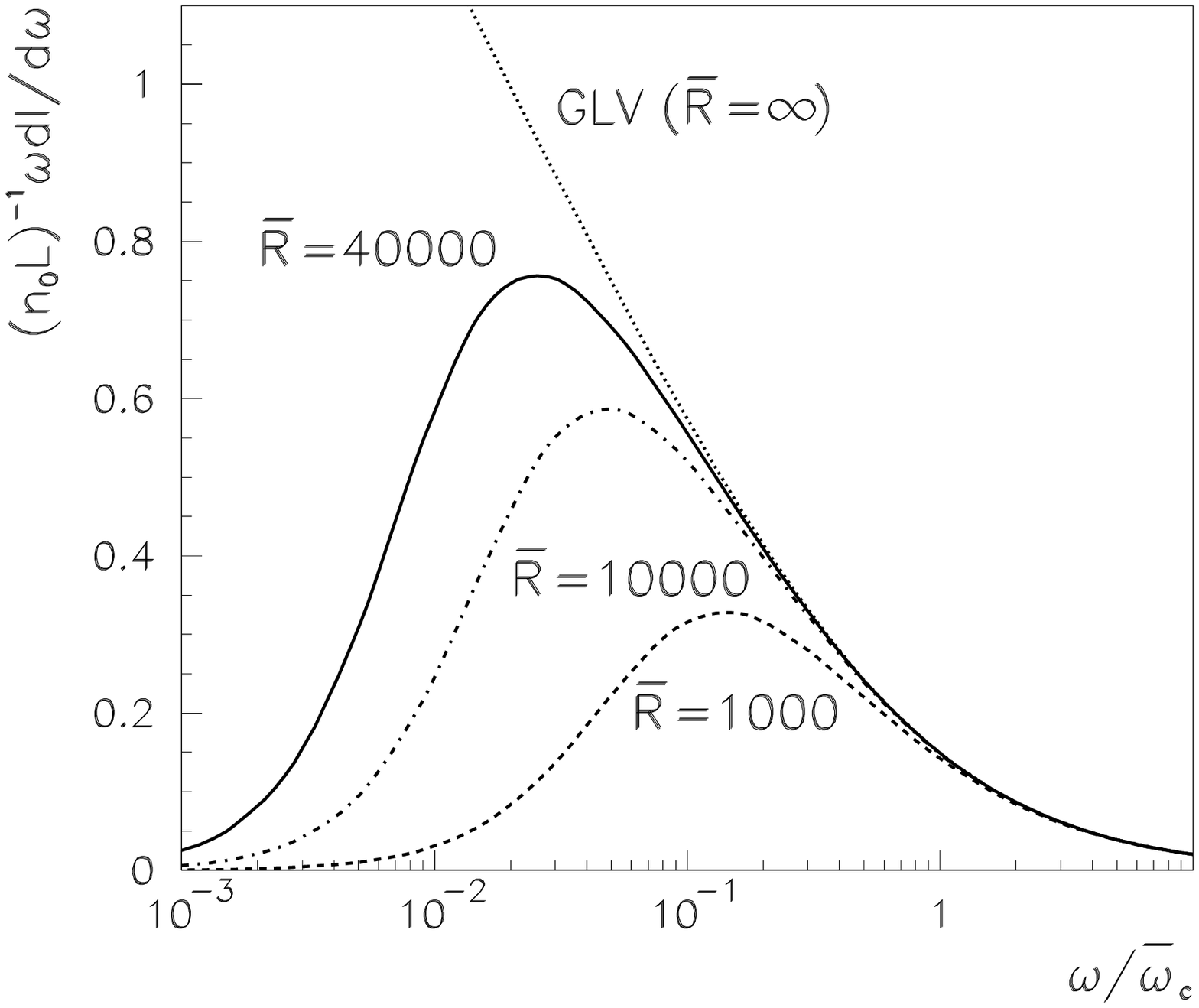}
\caption{The medium-induced gluon energy distribution 
$\omega \frac{dI}{d\omega}$ for a hard quark in the 
$N=1$ opacity expansion, calculated for different
values of the kinematic constraint $\bar{R}$.}
\label{neq1}
\end{minipage}
\end{figure}

The properties of the medium enter Eq.~(\ref{2.1}) in terms of the product 
of the time-dependent density $n(\xi)$ of scattering centers times 
the strength of a single elastic scattering $\sigma({\bf r})$. 
This dipole cross section $\sigma({\bf r})$ is given 
in terms of the elastic high-energy 
cross section $\vert a({\bf q})\vert^2$ of a single scatterer,
\begin{eqnarray}
 \sigma({\bf r}) = 2 \int \frac{d{\bf q}}{(2\pi)^2}\,
                    \vert a({\bf q})\vert^2\, 
                    \left(1 - e^{i{\bf q}\cdot {\bf r}}\right)\, .
 \label{2.2}
\end{eqnarray}
The full expression (\ref{2.1}) has been studied in two limiting
cases:
\begin{enumerate}
  \item \underline{Multiple soft scattering limit}\\
  For arbitrary many soft scattering centers, the projectile performs a 
Brownian motion in transverse momentum. This dynamical limiting
case can be studied in the saddle point approximation of the
path-integral (\ref{2.1}), using\cite{Zakharov:1996fv,Zakharov:1998sv}
\begin{eqnarray}
  n(\xi)\, \sigma({\bf r}) \simeq \frac{1}{2}\, \hat{q}(\xi)\, {\bf r}^2\, .
  \label{2.3}
\end{eqnarray}
Here, $\hat{q}(\xi)$ is the transport coefficient\cite{Baier:1996sk} 
which characterizes the medium-induced transverse momentum squared 
$\langle q_T^2\rangle_{\rm med}$ transferred to the projectile 
per unit path length $\lambda$ (for details and numerical estimates,
see section~\ref{sec314}). For a static medium, the transport
coefficient is time-independent,
\begin{equation}
  \hat{q} = \frac{\langle q_T^2\rangle_{\rm med}}{\lambda}\, .
  \label{2.4}
\end{equation}
In the approximation (\ref{2.3}),
the path integral in (\ref{2.1}) is equivalent 
to that of a harmonic oscillator. Technical details
of how to evaluate (\ref{2.1}) are given in Ref.~\cite{Kovner:2003zj}.
In the multiple soft scattering approximation, the gluon 
energy distribution (\ref{2.1}) depends on the characteristic
gluon energy $\omega_c$ and a dimensionless parameter $R$,
\begin{equation}
  R = \omega_c\, L\, .
\end{equation}
The spectrum is shown in Fig.~\ref{bdmpscomp}. 
For the case of medium showing one-dimensional Bjorken
expansion, $R$ can be related to the initially produced
gluon density~\cite{Salgado:2002cd}. In the limit $R\to \infty$,
the full spectrum (\ref{2.1}) reduces to a compact 
analytical expression first derived in  \cite{Baier:1996sk},
\begin{equation}
  \lim_{R\to \infty}\, 
   \omega \frac{dI}{d\omega} =
   \frac{\alpha_s C_R}{\pi}\, 
   \ln \Bigg [
   {\cosh^2\,\sqrt{\frac{\omega_c}{2\omega}}\,
    - \sin^2\,\sqrt{\frac{\omega_c}{2\omega}} }
   \Bigg ] \, .
   \label{2.9}
\end{equation}
In the limit of large and small gluon energies, this 
expression coincides with the qualitative expectations:
it shows a characteristic $1/\sqrt{\omega}$-energy 
dependence for small $\omega$ which is suppressed above the 
characteristic gluon frequency $\omega_c$:
\begin{eqnarray}
   \lim_{R\to \infty}\, 
   \omega \frac{dI}{d\omega} \simeq 
           \frac{2\alpha_s C_R}{\pi} 
          \left\{ \begin{array} 
                  {r@{\qquad  \hbox{for}\quad}l}                 
                  \sqrt{\frac{\omega_c}{2\, \omega}}
                  & \omega < \omega_c\, , \\ 
                  \frac{1}{12} 
                  \left(\frac{\omega_c}{\omega}\right)^2
                  & \omega > \omega_c \, .
                  \end{array} \right.
  \label{2.10}
\end{eqnarray}
  \item \underline{Opacity Expansion}\\
In the opacity 
expansion\cite{Wiedemann:2000za,Gyulassy:2000fs,Gyulassy:2000er}, 
the path integral
\begin{eqnarray}
 &&{\cal K}({\bf r}(y_l),y_l;{\bf r}(\bar{y}_l),\bar{y}_l|\omega)
  = \int {\cal D}{\bf r}
   \exp\left[ \int_{y_l}^{\bar{y}_l} d\xi
        \left(i\frac{\omega}{2} \dot{\bf r}^2
          - \frac{1}{2}  n(\xi) \sigma\left({\bf r}\right) \right)
                      \right]
  \label{eq55}
\end{eqnarray}
in (\ref{2.1}) is expanded in powers of the dipole cross 
section.
%
%
%
%
To first order, the entire medium-dependence comes from the interaction 
of the hard parton with a single static scattering center, multiplied
by the number $n_0L = L/\lambda$ of scattering centers along the
path. Modeling the single scatterer by a Yukawa potential with 
Debye screening mass $\mu$, one finds~\cite{Salgado:2003gb}
\begin{eqnarray}
  \omega \frac{dI^{N=1}}{d\omega} &=& 2 \frac{\alpha_s\, C_R}{\pi}\,
   (n_0L)\, \gamma\,   
  \int_0^\infty  dr\,  \frac{r - sin(r)}{r^2}
                 \nonumber \\   
  && \times
  \left( \frac{1}{r + \gamma} - 
         \frac{1}{\sqrt{( (\bar{R}/2\gamma) + r + \gamma)^2
                       - 4 r\bar{R}/2\gamma}}\right)\, .
  \label{2.19}
\end{eqnarray}
This result is also obtained to leading order in opacity from
the reaction operator approach (for details, see section~\ref{sec313}).
The energy distribution (\ref{2.19}) depends via the phase factor 
$\gamma = \frac{\bar{\omega}_c}{\omega}$ on the characteristic
gluon energy $\bar{\omega}_c$ in (\ref{2.17}), and on the 
dimensionless quantity 
$\bar{R} = \bar\omega_c\, L$. The energy distribution (\ref{2.19})
is plotted in Fig.\ref{neq1}. 
In the limit $\bar{R} \to \infty$, the 
characteristic $1/\omega$-energy dependence of the estimate (\ref{2.18})
is recovered for sufficiently large gluon energies 
$\omega > \bar\omega_c$,
\begin{eqnarray}
   \lim_{\bar{R}\to \infty}\, 
   \omega \frac{dI^{N=1}}{d\omega} &=& 
   2\, \frac{\alpha_s\, C_R}{\pi}\, \left( n_0\, L\right)\,
   \gamma\,   
  \int_0^\infty  dr\, \frac{1}{r + \gamma}\,  
                  \frac{r - sin(r)}{r^2}\, 
   \nonumber \\
   &\simeq & 
   2\, \frac{\alpha_s\, C_R}{\pi}\, \left( n_0\, L\right)\,
          \left\{ \begin{array} 
                  {r@{\qquad  \hbox{for}\quad}l}
                  \log \left[ \frac{\bar\omega_c}{\omega}\right]
                  & \bar\omega_c > \omega\, ,\\ 
                  \frac{\pi}{4}\, \frac{\bar\omega_c}{\omega}
                  & \bar\omega_c < \omega\, .  
                  \end{array} \right.
  \label{2.21}
\end{eqnarray}
\end{enumerate}
%
%
In both limiting cases, the multiple soft and the single hard
scattering limit, the gluon energy distributions show similar
dependencies on the gluon energy and the dimensionless ``kinematic
constraint'' $R = \omega_c\, L$ ($\bar{R} = \bar{\omega}_c\, L$).
In the opacity expansion, one additional model parameter enters
since one specifies both the average momentum transfer $\mu$ per
scattering as well as the average number $n_0\, L$ of scattering
centers involved. One can establish, however, a numerical relation
between  transport 
coefficient, Debye screening mass and opacity, for which both 
approximations lead to comparable results~\cite{Salgado:2003gb}.

\subsubsection{Energy Loss in the Reaction Operator Approach}
\label{sec313}
{\em I.~Vitev}

In this section we review the finite opacity  
GLV approach~\cite{Gyulassy:2000er,Gyulassy:2000fs} 
to the computation of the induced gluon radiative energy loss  
in dense nuclear matter. This calculational framework is well 
equipped for practical 
applications~\cite{Vitev:2002pf,Vitev:2001zn,Gyulassy:2000gk}
and underlies the numerical results presented in sections~\ref{sec343} 
and~\ref{sec346}. This section also complements the discussion
and numerical results on the opacity expansion in section~\ref{sec312}.

We first discuss some of the important physical constraints in the case 
of nucleus-nucleus collisions that lead to the development of the Reaction 
Operator approach:

\begin{itemize} 

\item A prerequisite for the consistency of all current theoretical 
approaches to non-abelian jet energy loss is the path (or time) ordering 
of the exchanged gluons between the propagating jet+induced gluon  system 
the dense nuclear matter. This approximation holds as long as the range of 
the typical scattering in the medium, $R= \mu^{-1}$, is much smaller 
than the mean free path $\lambda_g$, in which case diagrams with crossed 
gluon exchanges are suppressed by 
$\sim e^{-\mu \lambda_g}$~\cite{Baier:1996kr,Gyulassy:1999zd}
(see also section~\ref{sec311}). This 
condition by itself puts a theoretical constraint on the applicability of 
the large number of scattering centers limit for the case of heavy ion 
reactions where the typical size of the medium $\langle L \rangle 
\sim 5$~fm. The conclusion is reinforced by the fact that the hot and
dense quark-gluon plasma, that is expected to be created in energetic
$A+A$ reactions, expands in the final state, which leads to additional 
growth of the mean free path.      

 \item  The inherent dynamical nature of heavy ion reactions and 
the final state expansion of the system require  careful treatment of 
the interference phases along the eikonal line of jet+gluon propagation 
that are the basis of the LPM  destructive interference 
effect~\cite{Landau:gr,Landau:um,Migdal:1956tc}  in QCD. 
Symmetry arguments for the exchange gluons {\em do not} apply  since
their collective properties, e.g. the Debye screening mass, and  
correspondingly the transport coefficient $\hat{q}=\mu^2/\lambda_g$ are
strongly position dependent. Most of the contribution to the 
interference phases is accumulated in the early stages of jet 
propagation. The explicit solution in the GLV reaction operator approach 
for the  medium induced bremsstrahlung spectrum  keeps the exact position 
and  momentum  information in the jet and gluon phases and 
propagators, see Eq.~(\ref{difdistro}). 
This particular formulation of the problem of multiple 
radiative parton scattering can  therefore answer the important 
question about  the {\em rate} at which the transition to the asymptotic 
large number  of interactions limit might takes 
place~\cite{Gyulassy:2000er,Gyulassy:2000fs}.
        
\item  For phenomenological 
applications~\cite{Vitev:2002pf,Vitev:2001zn,Gyulassy:2000gk} 
the Reaction Operator solution 
for the radiative spectrum $\omega dN^g / d \omega$  and the mean 
energy loss $\langle \Delta E  \rangle $  does implement finite
kinematic bounds (e.g.  $ \mu < k_T < \omega$, $ \omega_{pl} 
 \sim \mu  < \omega < E$ and 
$q_T^2 < s/4$ )~\cite{Gyulassy:2000er,Gyulassy:2000fs}, where
the effective parton mass is determined by the 
medium properties~\cite{Braaten:1989kk,Blaizot:2000fc}. In the infrared 
and collinear regions $\mu$ plays the a role similar to the mass 
of a heavy quark~\cite{Dokshitzer:2001zm,Djordjevic:2003qk,Djordjevic:2003be}.
The finiteness of the available phase space is particularly important at 
RHIC energies, but this also holds true at LHC energies for 
$p_T < 20 - 50$~GeV. The analytic solutions discussed in the previous 
sections and in the sections that follow relax the kinematic constraints. 
It is difficult to {\em a posteriori} adequately account for  
the overestimate of the available phase space even on average. Instead,    
the integration limits have to be imposed directly in the full
solution Eq.~(\ref{difdistro}).

\item   The dominance of the lowest order terms in the opacity expansion 
series, which has been demonstrated in~\cite{Gyulassy:2000er,Gyulassy:2000fs},
is not unique to the gluon bremsstrahlung problem. For example, a  perturbative
expansion of nuclear enhanced  power  corrections in  the  number  of 
quark-nucleon scatterings in DIS on nuclei was recently computed and 
resummed~\cite{Qiu:2003vd}. The full solution is well approximated by the first
few terms of the series for a wide parameter range.   
\end{itemize}

\begin{figure}
\hspace*{0.2cm}
  \includegraphics[width=2.9in,height=2.2in]{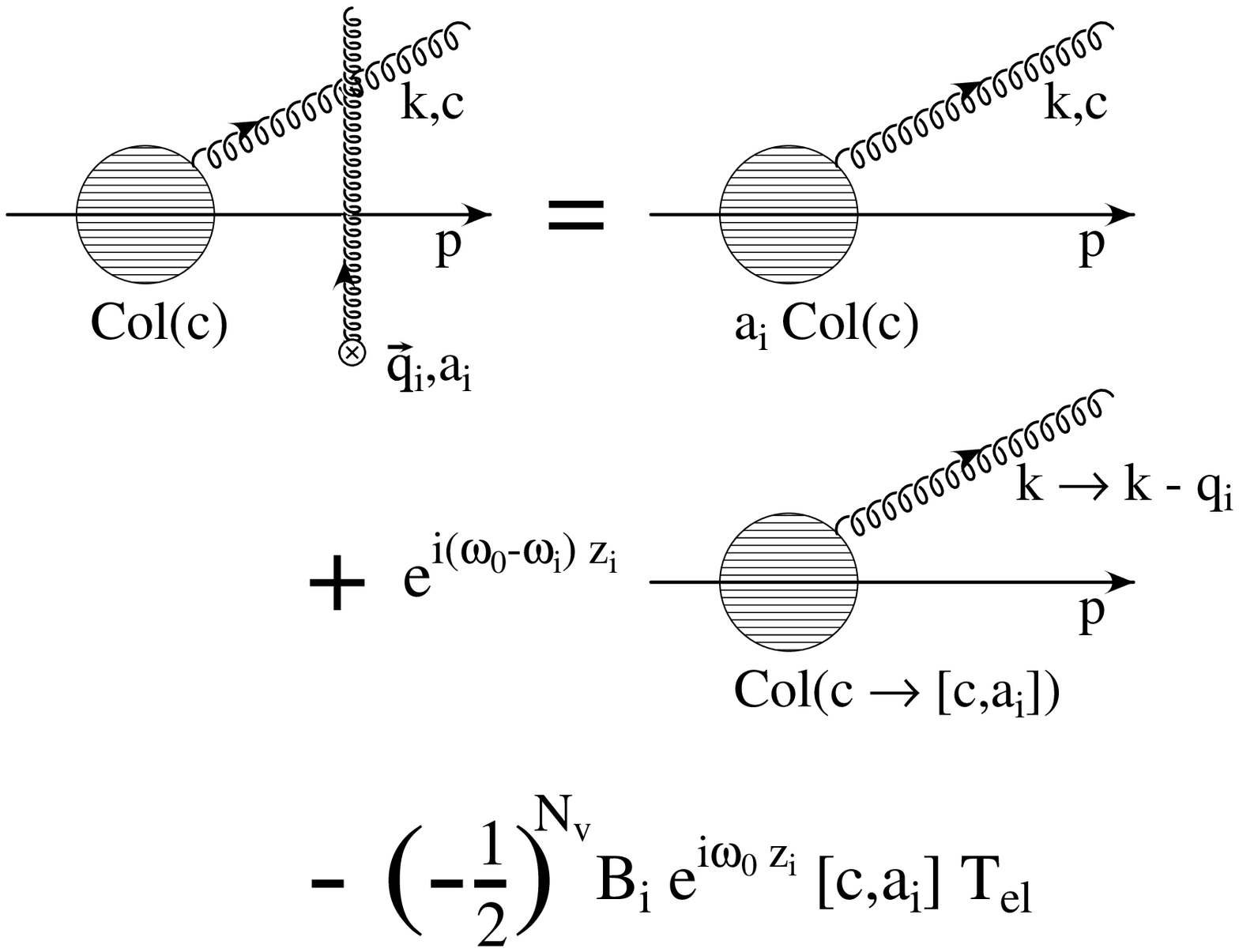}
\hspace*{0.3cm}
  \includegraphics[width=2.9in,height=2.2in]{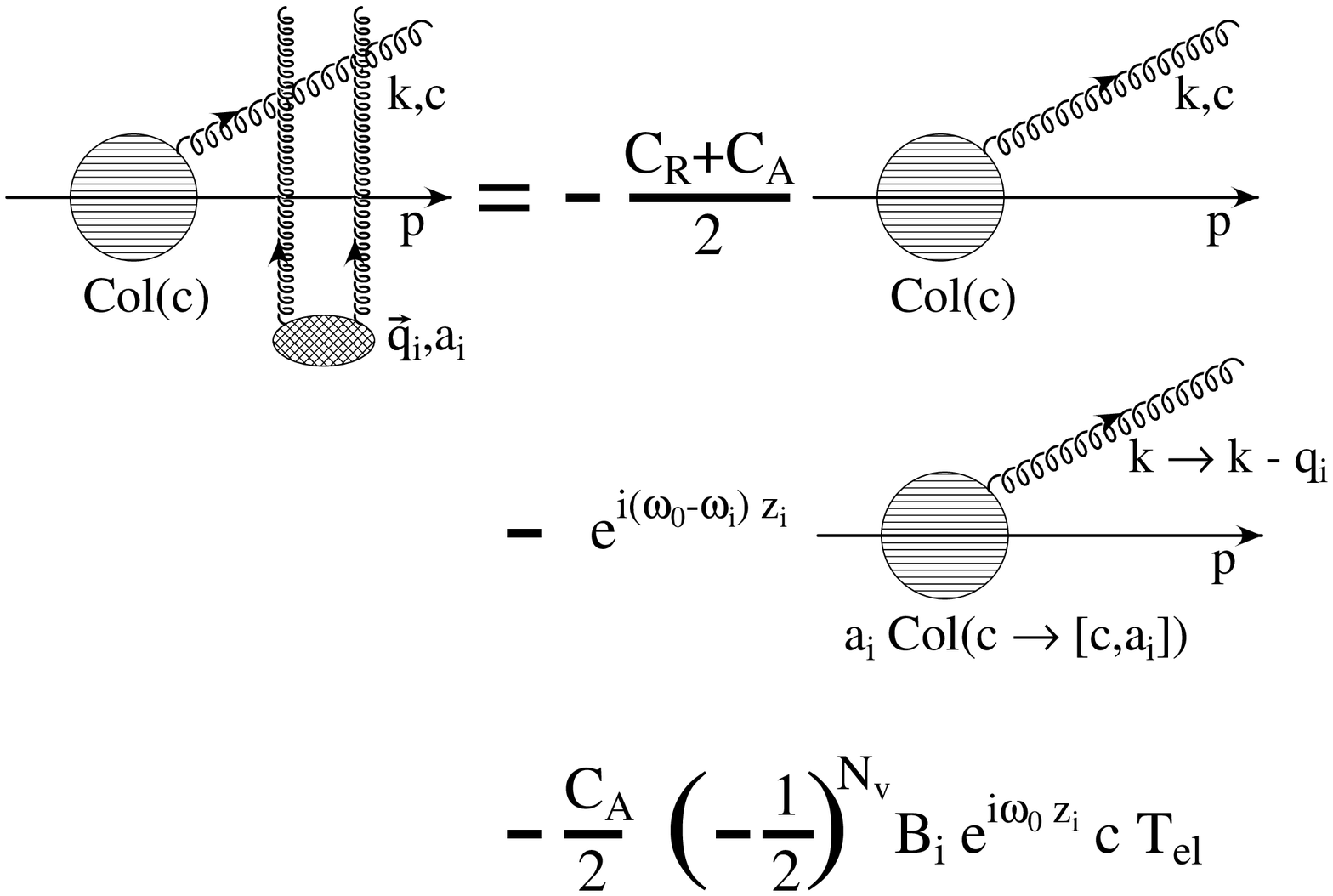}
\vspace*{0.3cm}
  \caption{Left panel: diagrammatic representation of the action of the 
direct  insertion  operator  $\hat{D}_i$ (single hit) at position $z_i$  
on a jet+gluon state described by an amplitude  ${\cal A}$. The 
generated  kinematic modification, color factors or color rotation, 
symmetry factors and phases from the energy difference before and after
the momentum exchange are explicitly shown. 
Right panel: diagrammatic representation of the action of the 
direct  insertion  operator  $\hat{V}_i$ (double hit) at position $z_i$.
Figure is adapted from Ref.~\cite{Gyulassy:2000er}.}
\label{fig3:dv}
\end{figure}

A powerful way to address multiple interactions of systems traversing 
abelian and non-abelian media is to decompose the complex 
multi-parton dynamics in a product of basic operator insertions that 
represent the  interaction  with a single scattering 
center~\cite{Gyulassy:2000er,Gyulassy:2000fs,Qiu:2003vd,Qiu:2003pm,Gyulassy:2002yv},
 subject  to the assumption that the mean free path of the propagating system 
significantly exceeds the  range of the scattering potential as discussed 
above. Let ${\cal A}_{i_1\cdots i_{n-1}}(x,{\bf k},c)$ be the amplitude of the propagating 
jet+gluon system that has already undergone $n-1$ scatterings where 
$x=k^+/p^+ \approx \omega/ E$ is the gluon momentum fraction, ${\bf k}$ 
is its transverse momentum and $c$ -- its color matrix. When the composite
system passes by a scattering center  it can miss, which is 
which leaves its amplitude unchanged ($\hat{\bf 1}$). If the system exchanges
a single momentum with the scattering, left panel of Fig.~\ref{fig3:dv}, 
there corresponding modification of the color and kinematics of its
amplitude  is given by
the direct insertion operator $\hat{D}$ and reads:
\begin{eqnarray} 
& &  \hat{D}_n {\cal A}_{i_1\cdots i_{n-1}}(x,{\bf k},c) 
 = {a_n} {\cal A}_{i_1\cdots i_{n-1}}(x,{\bf k},c) +  
e^{i(\omega_0-\omega_n)z_n }  
{\cal A}_{i_1\cdots i_{n-1}}(x,{\bf k}- {{\bf q}_{n}},{[c,a_n]}) 
\nonumber \\[1.ex] 
&&
-{\left(-{\textstyle\frac{1}{2}} \,\right )^{N_v({\cal A}_{i_1\cdots i_{n-1}})}} {\bf B}_n \,  
e^{i \omega_0 z_n} {[c,a_n]} \,  T_{el}({\cal A}_{i_1\cdots i_{n-1}}) \;\; ,  
\label{dop}
\end{eqnarray} 
where ${\bf B}_n \, = 
{\bf H}-{\bf C}_n={\bf k}/{\bf k}^2-({\bf k}-{\bf q}_n)/({\bf k}-{\bf q}_n)^2)$ 
is the so-called Bertsch-Gunion amplitude 
for producing a gluon with transverse momentum ${\bf k}$ 
in an isolated single collision 
with scattering center $n$. The momentum transfer to the jet is 
${\bf q}_n$. 
The notation $\omega_n=({\bf k}-{\bf q}_n)^2/2\omega$ is 
for a gluon with energy $\omega$  ($\omega_0={\bf k}^2/2\omega$)
and $a_n$ is the color matrix in the $d_R$ dimensional 
representation of the jet with color Casimir $C_R$. 
$N_v=\sum^{n-1}_{m=1}\delta_{i_m,2}$  
counts the number of virtual interactions 
in ${\cal A}_{i_1\cdots i_{n-1}}$.   $T_{el}({\cal A}_{i_1\cdots i_{n-1}})$ is the elastic color factor associated with  
all $n-1$ momentum transfers from the medium to the jet line.
Similarly, for the case of two momentum transfers given by the 
virtual insertion operator $\hat{V}$, right panel of Fig.~\ref{fig3:dv},  
\begin{eqnarray}
&&\hat{V}_n {\cal A}_{i_1\cdots i_{n-1}}(x,{\bf k},c)
= - \frac{C_R+C_A}{2} {\cal A}_{i_1\cdots i_{n-1}}(x,{\bf k},c)
 -  e^{i(\omega_0-\omega_n)z_n } a_n
{\cal A}_{i_1\cdots i_{n-1}}(x,{\bf k}- {\bf q}_{n}, [c,a_n])
\nonumber \\[.5ex]
&& - \left(-{\textstyle\frac{1}{2}}\,\right)^{N_v}
\frac{C_A}{2} \, {\bf B}_n \,
e^{i \omega_0 z_n} c \, a_{n-1}^{i_{n-1}}\cdots a_1^{i_1} \;\;.
\label{vidamit}
\end{eqnarray}   

To build one power of the elastic scattering cross section, or equivalently
one power of opacity $\chi = \bar{n} = L / \lambda_g$, two gluon exchanges 
at a fixed position $z_n$ are needed. Therefore, it is easy to 
see that the basic operator unit  that represents  one additional scattering
with the medium -- the GLV Reaction Operator -- has the form:  
\begin{equation} 
\hat{R}_n  = \hat{D}_n^\dagger 
\hat{D}_n+\hat{V}_n+\hat{V}_n^\dagger \;\;, 
\label{reacop} 
\end{equation} 
where $\hat{D}_n$, $\hat{V}_n$ are defined in Eqs.(\ref{dop},\ref{vidamit}).
The full solution~\cite{Gyulassy:2000er} for the medium induced 
gluon radiation  off jets produced in a hard collisions inside  the  
nuclear  medium of length $L$ and {\em to all orders} 
in the correlations  between the  multiple  scattering  centers  
is computed  via the iterative action of the Reaction Operator
on an initial condition given by the vacuum bremsstrahlung and 
averaging over the momentum transfers and the positions of the 
scattering center, respectively,
\begin{eqnarray}
 \sum\limits_{n=1}^\infty  x\frac{dN^{(n)}}{dx\, d^2 {\bf k}}   
& = & \frac{C_R \alpha_s}{\pi^2} \sum\limits_{n=1}^{\infty} 
 \prod_{i=1}^n \int\limits_0^{L-\sum_{a=1}^{i-1} \Delta z_a } 
 \frac{d \Delta z_i }{\lambda_g(i)} 
 \int \prod_{i=1}^n         \left(d^2{\bf q}_{i} \, 
\left[ |\bar{v}_i({\bf q}_{i})|^2 - \delta^2({\bf q}_{i}) \right]\right) 
\,  \nonumber \\[1.ex] 
&\;& \times 
\left( -2\,{\bf C}_{(1, \cdots ,n)} \cdot 
\sum_{m=1}^n {\bf B}_{(m+1, \cdots ,n)(m, \cdots, n)} 
\right . \nonumber \\[1.ex] &\;&  \left. \times \;
\left[ \cos \left (
\, \sum_{k=2}^m \omega_{(k,\cdots,n)} \Delta z_k \right)
-   \cos \left (\, \sum_{k=1}^m \omega_{(k,\cdots,n)} \Delta z_k \right)
\right]\; \right) \;, \quad \qquad  
\label{difdistro} 
\end{eqnarray}
where $\sum_2^1 \equiv 0$ is understood. In (\ref{difdistro})  
${\bf C}_{(m, \cdots ,n)} =  \frac{1}{2} \nabla_{{\bf k}} 
\ln ({\bf k} - {\bf q}_m - \cdots  - {\bf q}_n )^2$,   
${\bf B}_{(m+1, \cdots ,n)(m, \cdots, n)} = 
{\bf C}_{(m+1, \cdots ,n)} - {\bf C}_{(m, \cdots ,n)}$ are the 
color current propagators, $\omega_{(m,\cdots,n)}^{-1}  = 
2 x E / |C_{(m,\cdots,n)}^2 | $ are formation times, and 
$ \Delta z_k = z_k - z_{k-1} $ are the separations of subsequent 
scattering centers. The momentum transfers ${\bf q}_i$ are distributed 
according to a normalized  elastic scattering cross section  
$|\bar{v}_i({\bf q}_{i})|^2 = 
\sigma_{el}^{-1}d \sigma_{el} / d^2 {\bf q}_i  = 
\mu^2/\pi ({\bf q}_i^2+\mu^2)^2 $ for the color-screened Yukawa type 
and  the  radiative spectrum can   
be evaluated from  (\ref{difdistro}) for any initial nuclear geometry 
with an arbitrary  subsequent  dynamical evolution 
of the matter density. It is this stage of the calculation, 
Eq.~(\ref{difdistro}), at which the
finite kinematic constraints have to be imposed for the remaining 
${\bf q},{\bf k}$ and $x$ integrals.

\begin{figure}
\begin{center}
  \includegraphics[width=3.2in,height=4.2in,angle=-90]{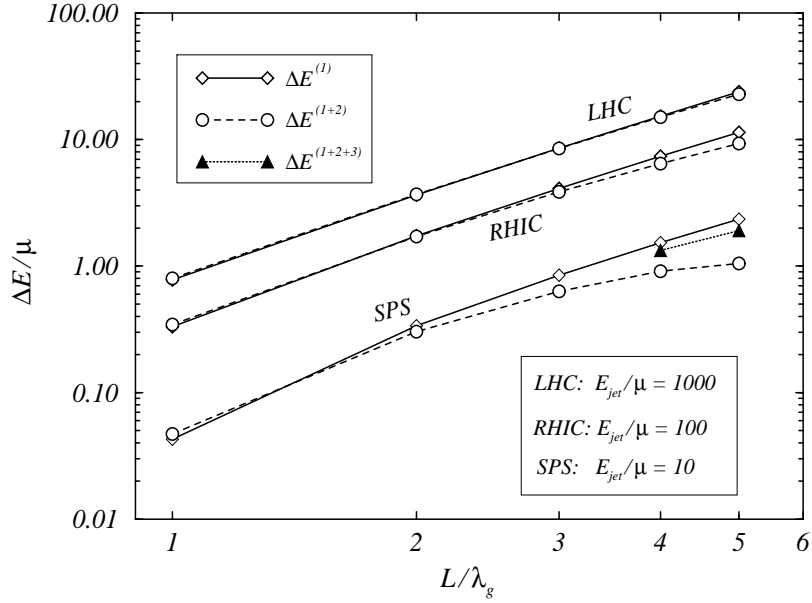}
\end{center}
  \caption{The radiative energy loss of a quark jet
        with energy $E_{jet}=5,50,500$~GeV  (at SPS, RHIC, LHC) 
        is plotted as a function of 
        the opacity $L/\lambda_g$. for a static medium  
($\lambda_g=1$~fm, $\mu=0.5$~GeV). Solid curves show the first order 
in opacity results. The  dashed curves show
results up to second order in opacity, and two third order
results are shown by solid triangles for SPS energies. 
Figure is adapted from Ref.~\cite{Gyulassy:2000er}.}
\label{fig3:conv}
\end{figure}

As argued above, the explicit all order solution for the double
differential radiative spectrum, Eq.~(\ref{difdistro}), provides 
an unambiguous way to study  the convergence of the opacity 
series and the relative importance of its terms. It can be seen from
Fig.~\ref{fig3:conv} that at large jet energies  the lowest order 
correlation between the jet production point one of the 
scatterings that follow gives the dominant  contribution to the 
non-abelian energy loss. It also gives the  quadratic  
dependence of $\Delta E$  on the size of the plasma, $\Delta E \propto L^2$  
for {\em static} media~\cite{Gyulassy:2000fs}. For {\em realistic} 
plasmas higher order opacity  corrections may become important only 
for large number of scatterings $n \geq 5$ and small jet energies 
$E \sim 5-10$~GeV.

\begin{figure}
  \includegraphics[width=2.5in,height=3.1in,angle=-90]{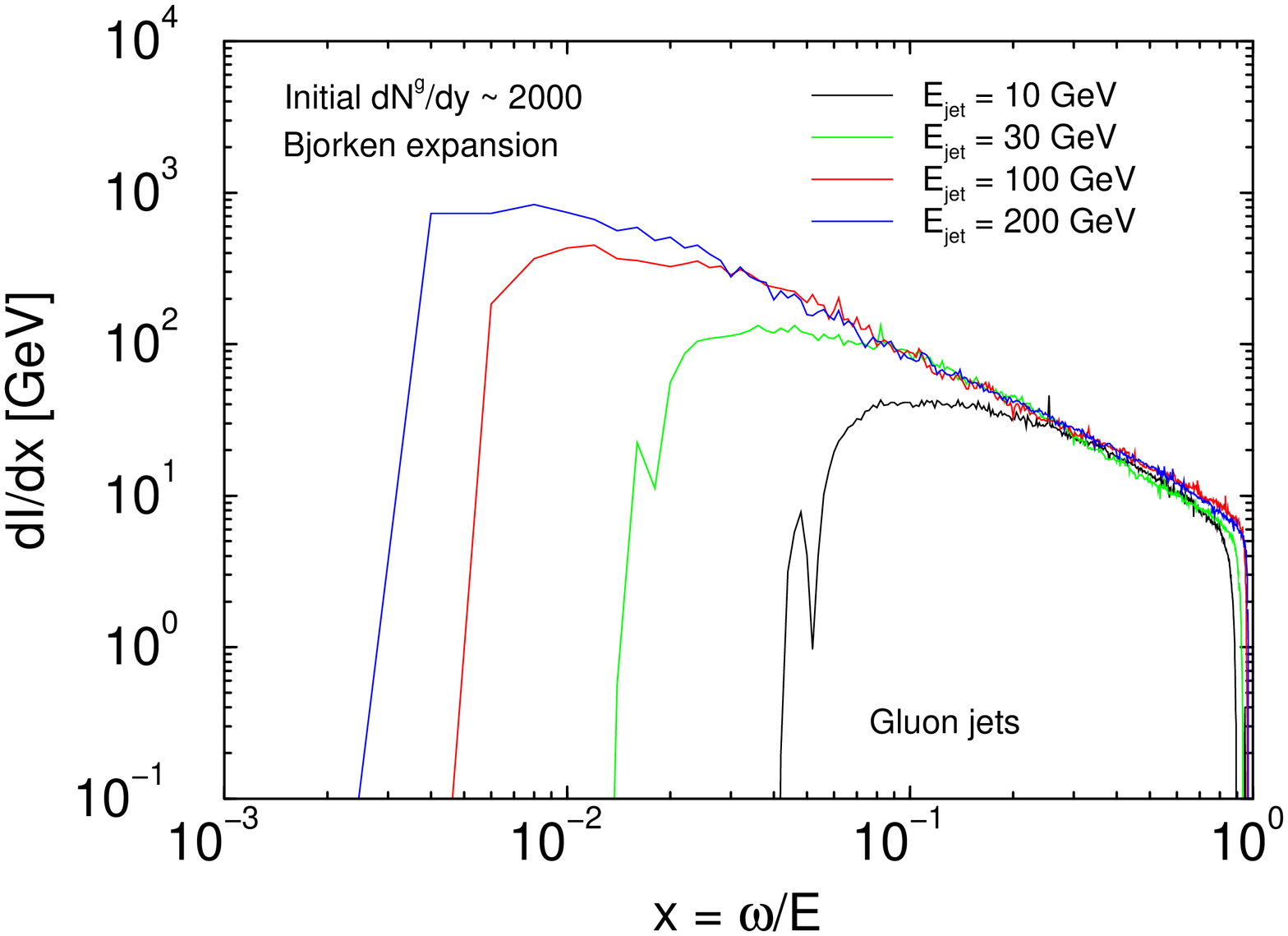}
\hspace*{0.2cm}
  \includegraphics[width=2.5in,height=3.1in,angle=-90]{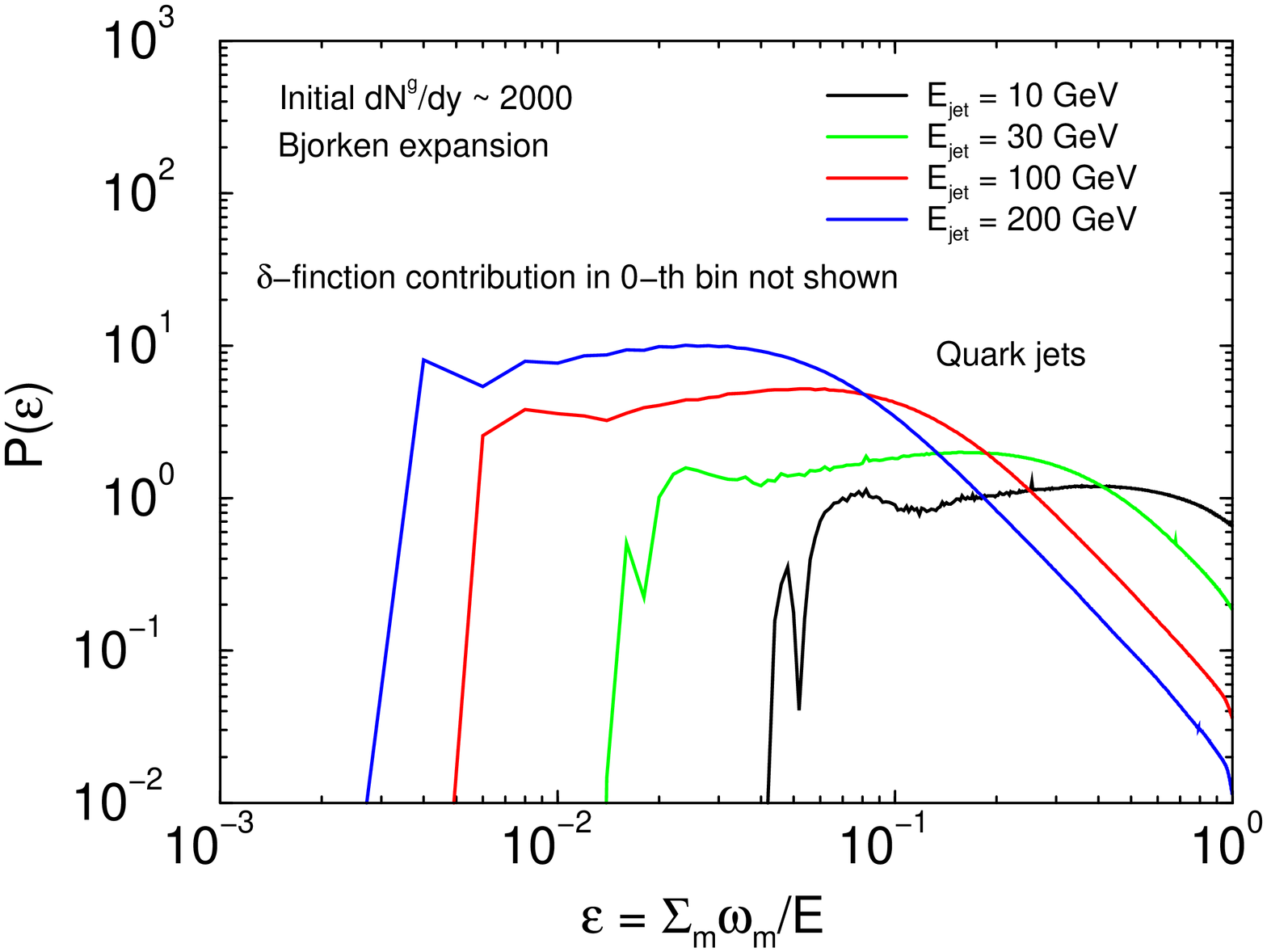}
  \caption{Left panel: the radiative spectrum $dI/dx$ computed  numerically
up to third order in opacity from Eq.~(\ref{difdistro}) in a dynamical
Bjorken expansion scenario with a gluon plasma rapidity density 
$dN^g/dy = 2000$.  Right panel: the fractional energy loss 
probability distribution $P(\epsilon = \sum_n \omega_n/E)$  
of fractional  radiative energy loss computed as 
in~\cite{Gyulassy:2001nm}. The no-radiation $\delta$-function contribution
is not shown.}  
\label{fig4:fs}
\end{figure}

Despite the dominance of the first order in the opacity 
expansion~\cite{Gyulassy:2000fs}, to improve the numerical accuracy 
for small parton energies we include corrections up to third order 
in $\chi$. The left panel of Fig.~\ref{fig4:fs} shows the radiation
intensity $dI/dx$ with an infrared cut-off  at small  $x = \omega / E$ 
given by the plasmon mass $\mu$.     
The dynamical expansion of the bulk soft matter is assumed to be 
of Bjorken type. The medium induced radiative energy loss is 
proportional to the density of the scattering centers in the medium
and for the cases of 1+1D and 1+3D expansions it has been shown 
that a useful to drive the calculation by
$dN^g/dy$~\cite{Gyulassy:2000gk,Gyulassy:2001kr} since the gluon
rapidity density can be related to the hadron multiplicities and the 
number of participants in $A+A$ collisions. 
The right panel of Fig.~\ref{fig4:fs} shows the probability distribution 
of fractional energy loss $\epsilon = \sum_n \omega_n/E$,   
numerically computed as in~\cite{Gyulassy:2001nm} from the gluon number 
distribution $dN(x,E)/dx$, in the  Poisson 
approximation~\cite{Gyulassy:2001nm,Baier:2001yt,Salgado:2002cd,Arleo:2002kh}
  of  independent gluon emission
\begin{equation}
P(\epsilon,E)=\sum_{n=0}^\infty P_n(\epsilon,E)\;, \quad
P_{n+1}(\epsilon,E) =  \frac{1}{n+1} \int_{x_0}^{1-x_0} dx_n \; 
\frac{dN(x_n,E)}{dx}  P_n(\epsilon-x_n,E) \;. 
\label{poisiter}
\end{equation}
The number of radiated gluons $\langle N^g(E) \rangle$  is finite and 
small which leads to an explicit finite  no-radiation contribution
 $P_0(\epsilon,E)=
e^{-\left\langle N^g(E) \right\rangle }\delta(\epsilon)$.

{\bf Analytic examples in the one scattering center limit}

In the case of 1+1D Bjorken longitudinal expansion with initial plasma 
density $\rho_0$ and formation time  $\tau_0$, i.e.   
 $\rho(\tau)=\rho_0\left({\tau_0}/{\tau}\right)$, 
it is possible to obtain  a closed form analytic 
formula~\cite{Gyulassy:2000gk}
under the strong asymptotic no kinematic bounds assumption. 
For a hard jet penetrating the quark-gluon plasma 
the LPM effect originates from the formation physics 
function  defined in~\cite{Gyulassy:2000gk} as 
$f(x,\tau) = \int_0^\infty {du} \,  
\left[ 1 -   \cos \left (\,u Z(x,\tau) \right) \right]/[{u(1+u)} ]$. 
With $Z(x,\tau)=(\tau-\tau_0)\mu^2(\tau)/2 x E$ being the local formation 
physics  parameter,  two simple analytic limits  apply: 
for $x \gg x_c = L\mu^2(L)/2 E$, in which case  
the formation length is large compared to the size of the medium,  
the small $Z(x,\tau)$ limit applies, leading to  $f(Z) \approx \pi Z / 2 $. 
The interference pattern along the  gluon path becomes important and accounts 
for the the non-trivial dependence of the energy loss on $L$.   
When $x \ll x_c$, i.e. the formation length  is small compared to the plasma 
thickness, one gets $f(Z) \approx \log Z$. The bremsstrahlung intensity
distribution reads:  
\begin{eqnarray}
&&{  \frac{dI^{(1)}}{dx} }
=  \frac{9 C_R \pi \alpha_s^3}{4} \frac{1}{A_T} 
\frac{dN^{g}}{dy}   
\times \left\{  \begin{array}{ll} 
\displaystyle    \frac{ L  }{x}  + \cdots \,, \quad  &
 \displaystyle x \gg \frac{\mu^2(L) L}{2 E}    \\[2ex]
\displaystyle \frac{ 6 E  }{\pi \mu^2(L)} 
\, \ln \frac {\mu^2(L) L}{2 x E} 
 + \cdots \,, \quad  &
 \displaystyle  x \ll \frac{\mu^2(L) L}{2 E}   
\end{array} \right. \; . 
\label{boostsp}
\end{eqnarray}
where $C_R=4/3 \; (3)$ for quark (gluon) jets. In Eq.~(\ref{boostsp}) 
$A_T$ is the transverse size of  the medium, e.g. 
$A_T = \pi R^2$ for central nucleus-nucleus collisions.  
The mean  energy loss (to first order in $\chi$) integrates to 
\begin{eqnarray}
&&  \Delta E^{(1)} =  
 \frac{ 9 C_R \pi \alpha_s^3  }{4} 
 \frac{1}{A_T} \frac{dN^{g}}{dy} \, L 
 \; \left( \ln   \frac{2 E}{\mu^2 L}  + 
\frac{3}{\pi} + \cdots \right) \;.
\label{rap-de}
\end{eqnarray}
We emphasize the linear rather than quadratic dependence 
of the energy loss on the size of the medium~\cite{Gyulassy:2000gk} 
in the Bjorken expansion case. 
The logarithmic enhancement with energy comes from the  
$ x_c < x < 1$ region~\cite{Gyulassy:2000fs}. 
In the case of sufficiently large jet energies  
($E\rightarrow \infty$) this term dominates.  

In the reaction operator approach,
medium induced radiative energy loss in transversely expanding plasma is
discussed in~\cite{Gyulassy:2001nm}.
To derive an analytic expression taking transverse flow into account, 
we consider an asymmetric expanding sharp {\em elliptic}  
density profile the surface of which is defined by  
$x^2(R_x+v_x \tau^{-2} + y^2 (R_y+v_y \tau)^{-2} = 1$.
The area of this elliptic transverse profile increases with time, $\tau$,  
as  $A_T(\tau)=\pi(R_x+v_x \tau) (R_y+v_y \tau)$. 
A short calculation for the $\propto \ln E$  term in the opacity series 
leads to 
\begin{eqnarray}  
\Delta E^{(1)}(\phi_0)   
  &\approx&  \frac{9}{4} \, \frac{ C_R\alpha_s^3}{R_x R_y}\frac{dN^g}{dy}  
\; \frac{\log   \frac{1 + a_x \tau(\phi_0)}{1+ a_y \tau(\phi_0)}}{a_x-a_y}  
\; \log \frac{2E}{\mu^2 L}  \;\; , 
\label{deaz}   
\end{eqnarray}
where $a_x=v_x/R_x, a_y=v_y/R_y$. 
This expression is a central result for transversely expanding media
and provides a simple analytic  generalization 
that interpolates between pure Bjorken 1+1D expansion for small 
$a_{x,y} \tau$, and 3+1D expansion at large $a_{x,y} \tau$. 
In the special  case  of  pure  Bjorken (longitudinal) expansion 
with $v_x=v_y=0$  Eq.~(\ref{deaz}) reduces to  Eq.~(\ref{rap-de})  with
$A_T = \pi R_x R_y$.
We also note that for a jet originating near the center of the medium  and  
{\em fully penetrating} the plasma the enhanced escape time  
due to expansion $\tau=R/(1-v_T)$ compensates for the 
$1/(1+v_T\tau/R)$ dilution factor. Therefore, in this isotropic case, 
the extra dilution due to transverse expansion 
 has in fact little or no effect of the total energy loss 
$\Delta E^{(1)}_{1D}(b=0\; {\rm fm})  
\approx \Delta E^{(1)}_{3D}( b=0\; {\rm fm})$,
modulo logarithmic factors which become sizable  only for large $v_T$.
An important consequence of our finding is that the  inclusive 
azimuthally averaged jet quenching pattern in central  collisions is 
approximately independent of the transverse flow.

\subsubsection{Estimates for Cold Nuclear Matter Transport Coefficients}
\label{sec314}
{\em F. Arleo}

The modification of high-$p_T$ hadro- and jet production due to multiple
medium-induced interactions depends on the spatial extension of the medium, 
and on the probability and strength of the multiple scatterings which the 
hard partons suffer. To characterize medium-modifications of high-$p_T$ 
jets produced in nucleus-nucleus collisions, and to relate them to
the properties of hot and dense QCD matter produced in the collision
region, knowledge about the multiple scattering strength of cold nuclear 
matter is a baseline of obvious importance. 
Several parameterizations, suited for different processes,
have been proposed to characterize this strength of multiple scattering 
which a hard parton undergoes while propagating through cold nuclear 
matter. Here we summarize the information currently available
from theoretical predictions, as well as from data analysis of
processes in which incoming or outgoing quarks propagate
through nuclear matter, and we comment on the relation between 
different parameterizations.

There are two parameters often used for the characterization of the
strength of multiple scattering. 
\begin{itemize}
\item BDMPS transport coefficient $\hat{q}$\\
In the approach of Baier, Dokshitzer, Mueller, Peign\'e, and Schiff 
(BDMPS)~\cite{Baier:1996sk,Baier:1996kr,Baier:1998kq}, this 
transport coefficient of the medium is given by 
$\hat{q} = \mu^2/\lambda_{\rm mfp}$. Here $\mu$ is the typical 
transverse momentum exchanged with the medium and $\lambda_{\rm mfp}$ 
denotes the parton mean free path in the medium. In the following, 
$\hat{q}$ will refer to the transport coefficient of a propagating gluon. 

The transport coefficient $\hat{q}$ is a measure of the scattering
strength of the QCD medium. It is related to the local density of 
color charges. In the BDMPS framework, $\hat{q}$ is related to the 
elastic scattering cross section $\sigma$ of a parton on a scattering 
center in the medium, see Eq.~(\ref{qhat}). For
cold nuclear matter, it is given by the gluon density of the 
nucleon (see Appendix~B of~\cite{Baier:1996sk}). This allows for a 
simple expression of the gluon transport coefficient in terms of the 
gluon distribution $x G(x)$ and the nuclear density $\rho$,
\begin{equation}
\hat{q} = \frac{4\,\pi^2\alpha_s\,N_c}{N_c^2-1}\,\rho\,x\,G(x,Q^2)\, .
\end{equation}
Taking $\alpha_s \simeq 1/2$, $\rho \simeq 0.16$~fm$^{-3}$ and $x G(x) 
\simeq 1$, Baier et al. have estimated the value for cold nuclear
matter, $\hat{q} \,=\, 0.045\,\,\,\mathrm{GeV^2/fm}$. Since also
the saturation momentum $Q_s$ of gluons for central gluon-nucleus 
(radius $R_A$)  collisions at small $x$ is given in terms
of the gluon distribution function, there is a 
linear relation between $Q_s$ and $\hat{q}$,~\cite{Baier:2002tc}
\begin{equation}
 Q_s^2  = 2 R_A~ \hat{q} \, .
\end{equation}
\item LQS scale parameter $\lambda$\\
In the perturbative QCD approach developed by Luo, Qiu, and Sterman 
(LQS)~\cite{Luo:ui}, $\lambda$ denotes the strength of twist-4 
matrix elements. These determine the strength of double parton
scattering, see section~\ref{sec351}. 

To determine a numerical estimate, LQS calculated the momentum imbalance 
of di-jets in photoproduction on nuclei. This is defined as
\begin{equation}
  \Delta \langle k_T^2 \rangle \,=\, \left(p_{\perp_1}^2 + p_{\perp_2}^2 
  \right) \,\times\, \sin \Delta \Phi\, ,
\end{equation}
where $p_{\perp_i}$ denote the transverse momenta of the partons and 
$\Delta \Phi$ the angle between the two corresponding jets. Their 
analysis assumes that the nuclear enhancement seen in the data is 
due to the rescattering of one of the produced partons (either quark 
or gluon). Assuming only the rescattering of the quark when $x$ is not 
too small, they can rewrite the momentum imbalance as
\begin{equation}
  \Delta \langle k_T^2 \rangle 
  = C_R\,\pi^2\,\alpha_s\,\lambda^2\,A^{1/3}\, .
  \label{lqskt}
\end{equation}
Comparing with the measurements (see Table 2) 
reported at Fermilab by the E683 
collaboration,
\begin{equation}
    \Delta \langle k_T^2 \rangle \,
    =\, 2\times 0.216 \ln A \sim 2\times 0.216 A^{1/3}\, ,
\end{equation}
they extract $\lambda^2  \simeq 0.1$~GeV$^2$. Assuming moreover the 
non-perturbative scale $\lambda$ to be greater than $\Lambda_{QCD}$, they 
conclude $\lambda^2 \,=\, 0.05 - 0.1 \,\,\, \mathrm{GeV^2}$.
The original LQS estimate is based on rescattering of partons in 
the final state. X.F. Guo~\cite{Guo:1998rd} gave a different estimate
based on the $\langle k_T^2 \rangle$ of Drell-Yan pairs produced 
in $h$-A collisions. This observable shows a nuclear enhancement due
to the multiple scatterings of the incoming quark entering the nucleus,
\begin{equation}\label{eq:ktbroadening1}
\Delta \langle k_T^2 \rangle =  \langle k_T^2 \rangle_{h\,A}  - 
\langle k_T^2 \rangle_{h\,N} \, .
\end{equation}
In a calculation~\cite{Guo:1998rd} to leading order in $\alpha_s$ but
taking into account nuclear enhanced power corrections, this quantity
was shown to be proportional to the four-parton 
correlation function in nuclei, $T_q(x, A)$, given in the Luo, Qiu, and 
Sterman model by
\begin{equation}
  T_q(x, A) \,=\, \lambda^2 \,A^{1/3}\,f_q^A(x, A) \, ,
\end{equation}
which depends linearly on the length $\sim A^{1/3}$ covered by the hard 
incoming quark, and where $\lambda$ is the unknown non-perturbative scale. 
Assuming only one quark flavor to contribute to the DY process, the LO 
$\langle k_T^2 \rangle$ broadening~(\ref{eq:ktbroadening1}) reads
\begin{equation}\label{eq:ktbroadening2}
  \Delta \langle k_T^2 \rangle \,=\, \left(\frac{4\,\pi^2\,
  \alpha_s}{3}\right)\,\lambda^2\,A^{1/3}\, .
\end{equation}
Experimentally, the $\langle k_T^2 \rangle$ broadening has been 
measured by the NA10 and E772 collaborations in pion and proton induced 
reactions on nuclei respectively, see Table 2. The value 
\begin{equation}
 \lambda^2 \,=\, 0.01 \,\,\, \mathrm{GeV^2}
\end{equation}
has been extracted from a comparison of Eq.~(\ref{eq:ktbroadening2}) with 
these data. This is a factor at least 5 smaller than the original LQS 
estimate. 
It may come from the strong interference beyond the leading order for the
initial-state interactions~\cite{Qiu:2001hj}. 
\end{itemize}

The following discussion focuses on these two parameters. Also, we 
shall express numerical estimates in terms of the mean energy loss 
per unit length, $-dE/dz$. Other 
parameterizations in the literature can be related to them. For example,
the product $n_0\, C$ is used in the works of 
Zakharov~\cite{Zakharov:1998sv} and 
Wiedemann~\cite{Wiedemann:2000za,Wiedemann:2000tf} on
parton energy loss. It denotes the product of the density $n_0$ of 
charges in the medium times their scattering strength $C$, and can
be expressed in terms of the
BDMPS transport coefficient,  $\hat{q} = 2\,n_0 C$.  

\def\ce#1{\centerline{#1}}
\begin{center}
\begin{tabular}{|p{0.4cm}|p{3.3cm}||p{3.6cm}|p{1.6cm}|p{5.0cm}|}
\hline
\multicolumn{2}{|c||} ~ & \ce{Observable} & \ce{Data} & \ce{Reaction} \\
\hline
\hline
&Vasiliev {\it et al.} (E866) \cite{Vasilev:1999fa} & \ce{Drell-Yan} 
\ce{$x_1$ dependence} & \ce{E866} & \ce{$p$(800 GeV)- Be, Fe, W}  \\
\cline{2-5}& Johnson {\it et al.} (E772) \cite{Johnson:2000ph} & 
\ce{Drell-Yan} \ce{$x_1$ and $M$ dependence} & \ce{E772}\ce{E866} & 
\ce{$p$(800 GeV)- D, C, Ca, Fe, W}  \ce{$p$(800 GeV)- Be, Fe, W}\\
\cline{2-5}&Arleo \cite{Arleo:2002ph} & \ce{Drell-Yan} \ce{$x_1$ dependence} 
& \ce{NA3} & \ce{$\pi^-$(150 and 280 GeV)- $p$, Pt}  \\
\cline{2-5}
\rule{1mm}{0pt}
\begin{rotate}{90}%
\hspace*{5mm} {\Large incoming $q$}
\end{rotate}\rule{1mm}{0pt}
&Guo \cite{Guo:1998rd}& \ce{Drell-Yan} \ce{$k_T$ broadening} & 
\ce{NA10} \ce{E772} & \ce{$\pi^-$ (140 and 280 GeV) - D, W} 
\ce{$p$ (800 GeV) - C, Ca, Fe, W} \\
\cline{2-5}
&Gyulassy, Vitev \cite{Vitev:2002pf}& \ce{Hadroproduction} 
\ce{Cronin enhancement}  
& \ce{E300} \ce{E605} & \ce{$p$ (400 and 800 GeV) - Be, W} \\
\hline
\hline
&Luo, Qiu, Sterman \cite{Luo:ui} & \ce{Photoproduction} \ce{dijet momentum} 
\ce{imbalance} & \ce{E683} & \ce{$\gamma$ - $p$, D, Be, C, Al, Cu, Sn, Pb} 
\ce{${\sqrt s} =  21$~GeV} \\
\cline{2-5}
\rule{1mm}{0pt}
\begin{rotate}{90}%
\hspace*{-3.5mm} {\large outgoing $q$}
\end{rotate}\rule{1mm}{0pt}
&Wang, Wang \cite{Wang:2002ri} & \ce{DIS fragmentation} \ce{functions} & 
\ce{HERMES} & \ce{$e$- D, N, Kr $\sqrt{s}= 7.2$~GeV}  \\
\hline
\end{tabular}
\end{center}
\vskip 0.5cm
{\small Table 2: Summary of various data analysis to extract the amount 
of quark energy 
loss in nuclear matter.}

%
\noindent
\paragraph{Relation between BDMPS transport coefficient, LQS scale
parameter and mean energy loss}
The BDMPS transport coefficient $\hat{q}$ and the LQS scale parameter 
$\lambda$ can be connected assuming the $k_T$ broadening and the 
dijet momentum imbalance to be directly comparable~\cite{Baier:1996sk}. 
In the BDMPS framework, the broadening of an incoming parton (with 
color $C_R$) is given by the transport coefficient and the 
length it has traveled though the medium,
\begin{equation}\label{eq:ktbdmps}
\langle k_T^2 \rangle_{\mathrm{BDMPS}} = \frac{C_R}{C_A}\,\hat{q}\,L.
\end{equation}
In the LQS approach to jet broadening~\cite{Luo:ui}, it is given by
Eq.~(\ref{lqskt}).
Comparing (\ref{eq:ktbdmps}) and (\ref{lqskt}), a simple expression 
between $\hat{q}$ and $\lambda$ is found~\cite{Baier:1996sk}
\begin{eqnarray}
\hat{q} &=& C_A\,\pi^2\,\alpha_s\,\frac{A^{1/3}}{L}\,\lambda^2\\
&=& \frac{4}{3}\,C_A\,\pi^2\,\alpha_s\,\times\,\left(\frac{1}{r_0}\right)
\times\,\lambda^2
\end{eqnarray}
where $L= 3/4 \, r_0\,A^{1/3}$ with $r_0 \simeq 1.2$~fm. The transport 
coefficient can also
be related to the mean energy loss per unit length $-dE/dz$.
This parameter depends linearly on the length $L$ of the medium
and is therefore proportional to the atomic mass number $A^{1/3}$
~\cite{Baier:1998kq}
\begin{eqnarray}\label{eq:BDMPSwlout}
  \left(-\frac{dE}{dz}\right)_{\mathrm{out}} 
  &=& \frac{1}{4}\,\alpha_s\,C_R\,\hat{q}\,L 
  \\
  &=& \frac{1}{4}\,\alpha_s^2 \,C_A\,C_R\,\pi^2\,A^{1/3}\,\lambda^2
\end{eqnarray}
for an outgoing parton, while
\begin{eqnarray}\label{eq:BDMPSwlin}
  \left(-\frac{dE}{dz}\right)_{\mathrm{in}} &=& \frac{1}{12}\,
  \alpha_s\,C_R\,\hat{q}\,L 
  \\
  &=& \frac{1}{12}\,\alpha_s^2 \,C_A\,C_R\,\pi^2\,A^{1/3}\,\lambda^2
\end{eqnarray}
in the case of partons approaching the medium.

For a numerical comparison of estimates for $-dE/dz$ with $\hat{q}$ 
and $\lambda$, we use the mean energy loss $-dE/dz$ in a $L= 5$~fm 
nucleus, and obtain the following relations
\begin{equation}
  \frac{1}{3} \left(-\frac{dE}{dz}\right)_{\mathrm{out}} = 
  \left(-\frac{dE}{dz}\right)_{\mathrm{in}} \,\left[\mathrm{GeV/fm}\right] 
  = 1.39 \,\hat{q} \,
  \left[\mathrm{GeV^2/fm}\right] = 22.8 \,\lambda^2  \,
  \left[\mathrm{GeV^2}\right]\, ,
\end{equation} 
where $\alpha_s \simeq 0.5$ and $C_R = 4/3$. 

\label{tab:results}
\begin{center}
\begin{tabular}{|p{0.4cm}|p{3.2cm}||p{2.7cm}|p{2.7cm}|p{2.7cm}|p{2.7cm}|}
\hline
\multicolumn{2}{|c||} ~ & \ce{$\hat{q}$} & \ce{$\lambda^2$} & 
\ce{$\left(-dE/dz\right)_{\mathrm{in}}$} & 
\ce{$\left(-dE/dz\right)_{\mathrm{out}}$}\\
\hline
&Brodsky, Hoyer \cite{Brodsky:1992nq} & \ce{$\le 0.72$} & \ce{$\le 0.022$} & 
\ce{$\le {\bf 0.5}$}& \ce{$\le {\bf 0.5}$} \\
\cline{2-6}
\rule{1mm}{0pt}
\begin{rotate}{90}%
\hspace*{-0.4cm} {\Large theory}
\end{rotate}\rule{1mm}{0pt}
&Baier {\it et al.} \cite{Baier:1996sk} & {\bf \ce{0.045}} & \ce{0.0029} & 
\ce{0.063}& \ce{0.19} \\
\hline
\hline
\cline{2-6}&Vasiliev {\it et al.} (E866) \cite{Vasilev:1999fa} &
\ce{${\bf \le 0.24}$}&\ce{$\le 0.014$}&\ce{$\le 0.33$}&\ce{$\le 0.99$} \\
\cline{2-6}&Johnson {\it et al.} (E772) \cite{Johnson:2000ph} & \ce{$2.0\pm 
0.3 \pm 0.4$} & \ce{$0.12\pm 0.02 \pm 0.02$} & \ce{${\bf 2.7 \pm 0.4 \pm 0.5}$} 
& \ce{$8.2 \pm 1.1 \pm 1.5$} \\
\cline{2-6}&Arleo \cite{Arleo:2002ph} & \ce{${\bf 0.14 \pm 0.11}$} & 
\ce{$0.009 \pm 0.007$} & \ce{$0.20 \pm 0.15$}  & \ce{$0.60 \pm 0.45$}\\
\cline{2-6}&Guo \cite{Guo:1998rd} & \ce{$0.16$} & \ce{${\bf 0.01}$} & 
\ce{$0.23$} & \ce{$0.69$} \\
\cline{2-6}&Gyulassy, Vitev \cite{Vitev:2002pf} & \ce{{\bf  0.1}} & \ce{$0.0061$} & 
\ce{$0.14$} & \ce{$0.42$} \\
\cline{2-6}&Luo, Qiu, Sterman \cite{Luo:ui} & \ce{$0.82$ -- $1.6$} & 
\ce{${\bf 0.05}$ -- ${\bf 0.1}$} & \ce{$1.14$ -- $2.28$} & \ce{$3.4$ -- $6.8$} \\
\cline{2-6}
\rule{1mm}{0pt}
\begin{rotate}{90}%
\hspace*{1.cm} {\Large data analysis}
\end{rotate}\rule{1mm}{0pt}
&Wang, Wang \cite{Wang:2002ri} & \ce{$0.12$} & \ce{$0.0073$} & \ce{$0.17$} & 
\ce{${\bf 0.5}$}\\
\hline
\end{tabular}
\end{center}
\vskip 0.5cm
{\small Table 3: Compilation of the different estimates for the magnitude 
of quark energy loss, given either in terms of $\hat{q}$, $\lambda$, or 
the mean energy loss per unit length for an incoming 
$\left(-dE/dz\right)_{\mathrm{in}}$ and outgoing quark 
$\left(-dE/dz\right)_{\mathrm{out}}$. The correspondence between the 
variables has been tempted and is explained in the text. In bold are the 
original estimates given by the various groups.}


{\bf Comparison of estimates for multiple
scattering in cold nuclear matter}

\begin{center}
\begin{figure}\label{resarleo}
\includegraphics[width=14.9cm]{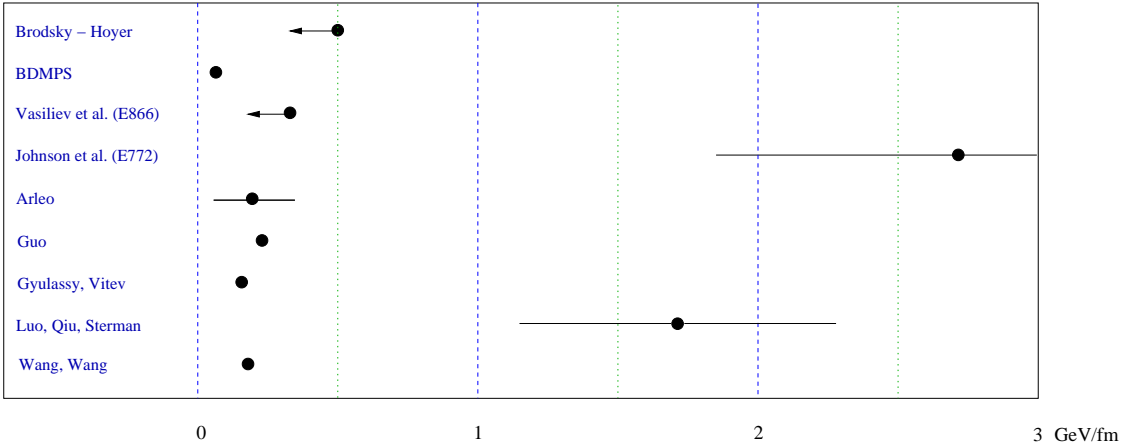}
\caption{Compilation of the different estimates for the magnitude of an 
incoming quark mean energy loss per unit length, 
$\left(-dE/dz\right)_{\mathrm{in}}$, in a $L = 5$~fm 
nucleus (see text and Table 3).}
\end{figure}
\end{center}

The results of different numerical estimates for the scattering
properties of cold nuclear matter are tabulated in 
Table 3 and summarized in Fig.~\ref{resarleo}. 
In the last subsection, the origin of the 
estimates of Baier {\it et al.}, Luo, Qiu and Sterman, as well as 
X.F. Guo have been reviewed already. Here, we explain how the other
estimates were obtained and we give arguments for the discrepancies
between different  estimates.

\noindent
{\bf Theoretical arguments} ~\cite{Brodsky:1992nq,Baier:1996sk}
The arguments leading Baier {\it et al.} to relate the energy
loss of cold nuclear matter to the gluon distribution in the
nucleon are discussed above. Without comparison to experimental
data, S.~Brodsky and P.~Hoyer had suggested an upper bound for
parton energy loss~\cite{Brodsky:1992nq}. It is based on the
argument that the minimum longitudinal momentum transfer to 
the hard parton due to gluon radiation is set by the uncertainty 
principle, $\Delta p_z \,\times\,L > 1$, 
where $L$ is the distance between two scattering centers. Let $\Delta E$ 
be the energy carried away by the emitted gluon and $k_T$ its 
transverse momentum. The longitudinal momentum transfer then reads 
$\Delta p_z \simeq k_T^2 /\,2\Delta E$, leading to
\begin{equation}
\frac{k_T^2}{2\,\Delta E}\,\times\,L \,>\, 1.
\end{equation}
While this expression comes directly from the Heisenberg principle, a 
similar expression has been explicitly derived by Brodsky and Hoyer 
in a simpler QED model. The maximal radiative energy loss for partons 
(note that it should apply to both quarks and gluons) now amounts to
\begin{eqnarray}\label{eq:BH_bound}
- \frac{dE}{dz}\,&<&\,\frac{k_T^2}{2}
\end{eqnarray}
where $k_T^2$ can be related to the typical transverse momentum
which partons acquire in the medium. Brodsky and Hoyer mainly emphasize
that this radiative energy loss is not proportional to the energy
of the scattering particle. They point out that a previous analysis
of Drell-Yan data \cite{Gavin:1991qk} based on 
$- \frac{dE}{dz} \propto E$ violates the uncertainty principle at 
large $E$. With the estimate $k_T^2 = 0.1$~GeV$^2$ for partons 
traversing cold nuclear matter, Brodsky and Hoyer arrive at
$-\frac{dE}{dz} \,\le\, 0.5 \,\,\,\mathrm{GeV/fm}$, taking into account 
a similar energy loss as~(\ref{eq:BH_bound}) due to elastic scattering. 
Clearly, this upper bound depends on the choice of $k_T^2$ and is violated
if  $k_T^2$ turns out to be larger. Thus, the 
analysis of Brodsky and Hoyer does not constrain the 
scattering properties of cold nuclear matter, but it constrains the
energy dependence of $- \frac{dE}{dz} $ in the ultra-relativistic
limiting case.

{\bf Estimates for outgoing quarks} ~\cite{Luo:ui,Wang:2002ri}
The original estimate for the LQS scale parameter $\lambda$ in 
cold nuclear matter is based on the transverse momentum broadening
of di-jets measured in photoproduction on nuclei.
Another analysis of the energy loss of outgoing quarks in cold 
nuclear matter was given by E.~Wang and X.-N.~Wang~\cite{Wang:2002ri}. 
In their approach, the 
multiple scattering of the produced quarks escaping the nucleus 
modifies the fragmentation functions in nuclei, $D(z, Q^2, A)$. 
The strength of this modification (here denoted 
$\tilde{C}$) is again related to nuclear enhanced twist-4 
parton correlation functions.

The HERMES collaboration measured hadron production in $e$-A collisions 
on D, N, and Kr targets (${\sqrt s = 7.2}$~GeV) as a function of the 
virtual photon energy $\nu$ and the momentum fraction carried by the 
produced hadron, $z$. These measurements give a direct access to the 
nuclear dependence of the fragmentation functions. Comparing with these 
preliminary data, E.~Wang and X.-N.~Wang found a good agreement provided 
that
\begin{equation}
  \tilde{C} \alpha_s^2 \simeq 0.00065 \,\,\, \mathrm{GeV^2}.
\end{equation}
They translate this quantity into a mean energy loss of
\begin{equation}
 \left(-\frac{dE}{dz}\right)_{\mathrm{out}} \,=\, 0.5 \,\,\, \mathrm{GeV/fm}
\end{equation}
for a $L = 5$~fm nucleus.

{\bf Estimates for incoming quarks} 
~\cite{Guo:1998rd,Vasilev:1999fa,Johnson:2000ph,Johnson:2001xf,Arleo:2002ph}
Several works have attempted to parametrize multiple scattering 
effects of cold nuclear matter from Drell-Yan measurements in 
hadron-nucleus collisions. The estimate of X.F. Guo based on the
nuclear enhanced transverse momentum broadening of Drell-Yan
pairs was reviewed. Three other groups estimated
the parton energy loss from the $x_1$-dependence of
Drell-Yan data. 

The data analysis of  
M.A. Vasiliev {\it et al.} (E866)~\cite{Vasilev:1999fa}
is based on Drell-Yan measurements in 800 GeV proton induced reactions on 
Beryllium, Iron, and Tungsten targets. The 
data cover a wide range in the momentum fraction of the projectile 
parton, $x_1$, integrated over the invariant mass interval 
$4 < M < 8.4$~GeV. The data analysis assumes~\cite{Vasilev:1999fa}
that the multiple scatterings of the incoming (anti)quark 
in the nucleus shift the momentum fraction of the hard parton on
average by a 
\begin{equation}\label{eq:shift}
  \Delta x_1 = \frac{\kappa}{s}\,A^{2/3}\, ,
\end{equation}
where $\kappa$ parametrizes the strength of the energy loss. The 
leading-order Drell-Yan cross section 
\begin{equation}\label{eq:dyxs}
\sigma^{p\,A}(x_1) \sim f^p(x_1 + \Delta x_1) \times f^A(x_2) \times 
\hat{\sigma}
\end{equation}
is then fitted to the data with $\kappa$ considered as a free parameter. 
The effects of nuclear shadowing $f^A(x_2) \ne f^p(x_2)$ in~(\ref{eq:dyxs}) 
have been taking into account using the EKS98 
parameterization~\cite{Eskola:1998df}.

The amount of quark energy loss was found to be negligible in these data 
sets, with a one-standard deviation upper limit $\kappa < 0.10$~GeV$^2$. 
Assuming the length covered by the incoming parton to be given by the 
nuclear radius, $L = 3/4\,r_0\,A^{1/3}$, one can relate easily the 
$\kappa$ parameter to the BDMPS transport coefficient. Using 
(\ref{eq:shift}) in (\ref{eq:BDMPSwlin}), it is given by
\begin{equation}\label{eq:qkappa}
\hat{q}=\frac{16\,\kappa}{m_p\,r_0^2}
\end{equation}
with $r_0 \simeq 1.2$~fm. The upper limit extracted by this group amounts to 
\begin{equation}
 \hat{q} \,\le\, 0.237 \,\,\,\mathrm{GeV^2/fm}.
 \label{smallupper}
\end{equation}

M. Johnson {\it et al.} (E772)~\cite{Johnson:2000ph,Johnson:2001xf}
performed a different analysis of the E772 and E866 Drell-Yan
measurements. They raised the point that E772 Drell-Yan data have 
also been used in the EKS98 analysis. In principle, the small upper
bound Eq.~(\ref{smallupper}) could originate from an erroneous attribution
of the sizable nuclear dependence to shadowing effects.
The authors ~\cite{Johnson:2000ph,Johnson:2001xf} then attempt to
disentangle the effects of quark energy loss and nuclear shadowing
on the basis of a light-cone formulation of the Drell-Yan process 
which allows to calculate shadowing corrections. For the E772 and
E866 data sets, they extract
\begin{equation}
 \left(-\frac{dE}{dz}\right)_{\mathrm{in}} \,=\, 2.73 \pm 0.37 
 \pm 0.5 \,\,\,\mathrm{GeV/fm}
\end{equation}
where the errors are statistics and systematics respectively. This
value is an order of magnitude larger than any other estimate
for cold nuclear matter and depends entirely on the
validity of a theoretical light-cone calculation of nuclear 
shadowing, whose uncertainties are difficult to evaluate. 

As illustrated by the two analysis above, the poorly known 
shadowing corrections in the Drell-Yan process render the 
extraction of quark energy loss from Drell-Yan data difficult.
F. Arleo~\cite{Arleo:2002ph} emphasized that Drell-Yan 
production in pion induced reactions at lower beam energy is much 
less sensitive to shadowing effects~\cite{Arleo:2002ph} mainly for
two reasons. First, the pion beam favors the fusion of valence quarks 
for which shadowing corrections are 
well fixed from DIS measurements only. Moreover, the low beam energy 
probes a target momentum fraction range $x_2 \sim 0.1$ where shadowing 
is known to be small.

The Drell-Yan cross section has been computed in the QCD-improved parton 
model to leading-order. In these calculations, the energy loss $\epsilon$
is modeled by the BDMPS probability distribution~\cite{Baier:2001yt} 
$P(\epsilon)$ which was determined from the medium-induced BDMPS
gluon energy distribution. The cross section reads
\begin{equation}\label{eq:dyxs2}
  \sigma^{\pi^-\,A}(x_1) \sim f^{\pi^-}(x_1 + \epsilon / E_\pi) \times 
  f^A(x_2) \times \hat{\sigma} \times P(\epsilon)\, ,
\end{equation}
In the fit of  Eq.~(\ref{eq:dyxs2}) to NA3 data, 
the transport coefficient was considered a free parameter. It has 
been found that
\begin{equation}
 \hat{q} \,=\, 0.144 \pm 0.108 \,\,\,\mathrm{GeV^2/fm}
\end{equation}
which corresponds to a mean energy loss per unit length
\begin{equation}
   \left(-\frac{dE}{dz}\right)_{\mathrm{in}} \,=\, 0.20 \pm 0.15 
   \,\,\,\mathrm{GeV/fm}
\end{equation}
in a large ($L = 5$~fm) nucleus.

An alternative way to determine the nuclear matter transport coefficient is 
to analyze the multiple scattering and associated transverse momentum
broadening of incoming partons. The Cronin effect observed in $p+A$  
reactions relative to the Glauber-scaled $p+p$ 
result~\cite{Cronin:zm,Straub:xd,Antreasyan:cw} has been analyzed 
by Gyulassy and Vitev~\cite{Vitev:2002pf} in the framework of 
multiple initial state scatterings of partons in cold nuclei.
Parton broadening due to random elastic scatterings is computable 
from~\cite{Gyulassy:2002yv}. The possibility of hard 
fluctuations along the projectile path leads to a power law tail of 
the $k_T$ distribution that enhances $ \langle \Delta  k^2_T \rangle$ 
beyond the naive Gaussian random walk result $ \mu^2 L/\lambda$.  
For a high energy parton with transverse  momentum $p_T$ produced
in a $p+A$ reaction $Q_{max}^2 \sim p_T^2$.  
The Cronin effect is modeled  by using 
\begin{equation} 
\langle k_T^2 \rangle_{pA} -  \langle k_T^2 \rangle_{pp}  \approx  
\frac{\mu^2}{\lambda} L_A \,  \ln(1 +  c \, p_T^2/\mu^2 )  \, .  
\label{cron} 
\end{equation}  
Calculations are consistent with the energy,
$\sqrt{s} = 27.4, 38.8$~GeV, and $p_T$ dependence observed in 
$p+W/p+Be$  reactions with parameters of the nuclear medium set as follows: 
$c/\mu^2=0.18 \, /{\rm GeV}^{2}$ and   
$\mu^2 / \lambda=0.05$~GeV$^2$/fm. 
The corresponding transport coefficient and initial state  
mean energy loss per unit length using Eq.~(10) are:
\begin{equation} 
\hat{q} \simeq 0.1 \; {\rm GeV}^2/{\rm fm} 
\label{qgv} 
\end{equation} 
and    
\begin{equation} 
   \left(-\frac{dE}{dz}\right)_{\mathrm{in}} \simeq 
   0.14  \; {\rm GeV}/{\rm fm} 
   \label{dedzgv} 
\end{equation} 
for a quark jet approaching the nucleus. The uncertainties in the above 
estimates are correlated with the uncertainties in the fragmentation 
functions.

\subsubsection{The Transport Coefficient $\hat{q}$ for a Hot and
Expanding Medium}
\label{sec315}
{\em R. Baier, U.A. Wiedemann}

\begin{figure}[h]\epsfxsize=8.7cm
\centerline{\epsfbox{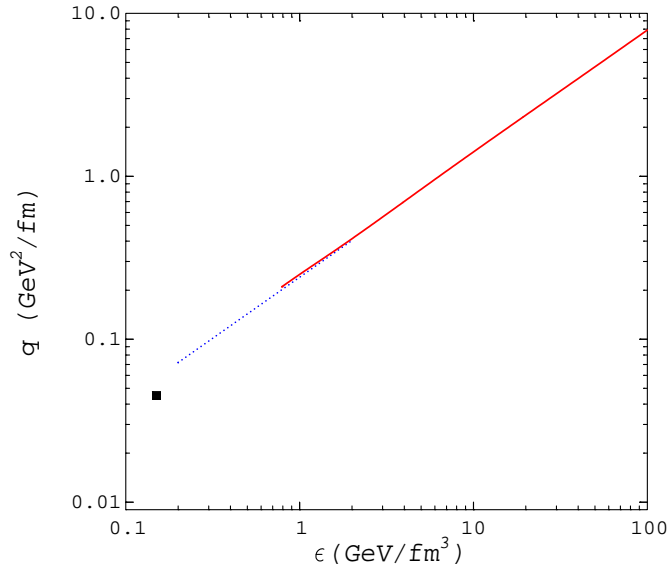}}
\caption{Transport coefficient as a function of energy density
for different media: cold, massless hot pion gas (dotted) 
and (ideal) QGP (solid curve). Figure taken from \cite{Baier:2002tc} 
}\label{fig:transportcoeff}
\end{figure}

Various data on DIS electron-nucleus and hadron-nucleus collisions
indicate that the multiple scattering properties of cold nuclear 
matter can be described by a cold nuclear matter transport coefficient
$\hat{q} \vert_{\rm{nuclear~ matter}} < 0.5-1 $ GeV$^2$/fm. This
information is compiled in the previous section~\ref{sec314}. For hot
equilibrated matter, the estimated dependence of $\hat{q}$  on the 
energy density $\epsilon$ is shown in Fig.~\ref{fig:transportcoeff}. For
example, for a quark gluon plasma, the number density is translated 
into $\epsilon$ as $\rho(T) \sim T^3 \sim \epsilon^{3/4}$. A "smooth" 
increase of $\hat{q}$ with increasing $\epsilon$ is observed, such
that 
\begin{equation}
\hat{q} \vert_{\rm{QGP}} \gg \hat{q} \vert_{\rm{nuclear~ matter}} \, .
\end{equation}
%
%
The QCD phase transition near $\epsilon \simeq 1~ {\rm {GeV/fm^3}}$ 
\cite{Karsch:2001vs} leaves no structure in the $\epsilon$-dependence of
$\hat{q}$. In contrast, plotting $\hat{q}$ versus temperature, one
would find a sharp increase of the transport coefficient at the
critical temperature. 

What matters in practice for jet quenching in a heavy ion collision 
is {\it for how long} the transport coefficient takes values which 
are significantly above the cold nuclear matter reference value.

In nucleus-nucleus collisions at collider energies, the produced
hard partons propagate through a rapidly expanding medium. The
density of scattering centers and thus the transport coefficient
$\hat{q}(\tau)$ is expected to reach a 
maximal value $\hat{q}_d$ around the plasma formation time $\tau_0$,
and then decreases with time $\tau$
rapidly due to the strong longitudinal and 
- to a lesser extent - transverse expansion, 
\begin{equation}
  \hat{q}(\tau) = \hat{q}_d \left( \frac{\tau_0}{\tau} \right)^\alpha\, .
  \label{4.27}
\end{equation}
Here, $\alpha = 0$ characterizes the static medium discussed
above. The value $\alpha =1$ corresponds to a one-dimensional, 
boost-invariant longitudinal expansion and approximates the 
findings of hydrodynamical simulations. The formation time $\tau_0$ 
of the medium may be set by
the inverse of the saturation scale $p_{\rm sat}$~\cite{Eskola:1999fc} and 
is $\approx$ 0.2 fm/c at RHIC and $\approx$ 0.1 fm/c at LHC. Since the
time difference between the formation of the hard parton and the
formation of the medium bulk is irrelevant for the evaluation of
the radiation spectrum (\ref{2.1}), one can replace in (\ref{2.1}) 
the production time $\tau_0$ of the parton by $0$.
%
\begin{figure}[t]\epsfxsize=14.7cm
\centerline{\epsfbox{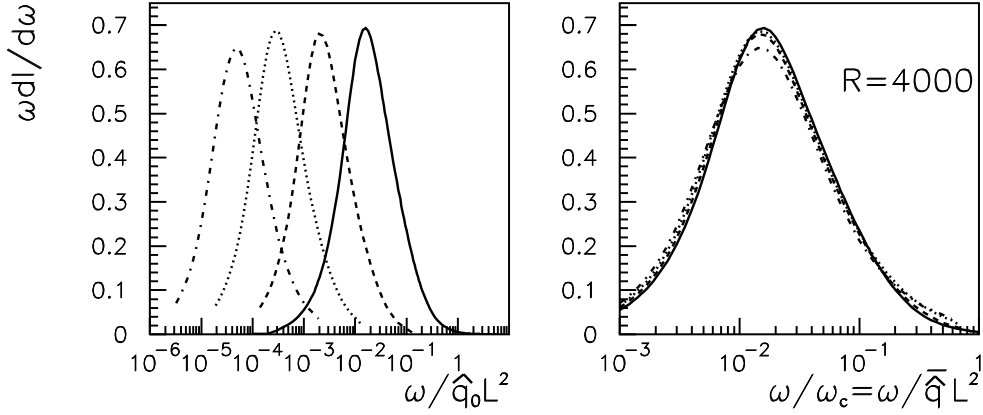}}
\caption{LHS: The medium-induced gluon energy distribution 
for expanding collision regions 
(\protect\ref{4.27}) with expansion parameter $\alpha = $
0, 0.5, 1.0 and 1.5. The value of the transport coefficient
$\hat{q}_0$ is taken at initial time $\tau_0$.
RHS: The same gluon radiation spectrum with parameters
rescaled according to (\ref{4.32}). Figure taken from
\protect\cite{Salgado:2003gb}.
}\label{fig4}
\end{figure}
 
For a dynamically evolving medium of the type (\ref{4.27}),
the path-integral (\ref{eq55}) in (\ref{2.1}) is the path integral of 
a 2-dimensional harmonic oscillator with time dependent (imaginary) 
frequency $\Omega^2 (\tau)\equiv {\hat{q}(\tau)\over i2\omega}$
and {\it mass} $\omega$~\cite{Baier:1998yf}. In this way, one
can calculate the medium-induced gluon energy distribution (\ref{2.1})
for a dynamically expanding medium~\cite{Baier:1998yf}. The result is 
shown in Fig.\ref{fig4}. The radiation spectrum 
$\omega \frac{dI}{d\omega}$ satisfies a simple scaling law which 
relates the radiation spectrum of a dynamically expanding
collision region to an equivalent static scenario. The 
linearly weighed line integral \cite{Salgado:2002cd}
\begin{eqnarray}
  \overline{\hat{q}} & =  & \frac{2}{L^2}\int_{\tau_0}^{\tau_0+L} d\tau\, 
  \left( \tau - \tau_0\right)\, \hat{q}(\tau) 
  \label{4.32}\\
& \simeq & \frac{2}{2 - \alpha} ~\hat{q}(L) \, \, 
\, {\rm for} \, \, \,  \tau_0 \rightarrow 0  ,
\end{eqnarray}
defines the transport coefficient of the equivalent static
scenario. 
The linear weight in (\ref{4.32}) implies that scattering centers
which are further separated from the production point of the
hard parton are more effective in leading to partonic energy
loss. In contrast to earlier believe that parton energy loss
is most sensitive to the hottest and densest initial stage of
the collision, this implies for a dynamical expansion following
Bjorken scaling [$\alpha = 1$ in Eq.~(\ref{4.27})] that all 
time scales contribute equally to the average transport coefficient. 
This makes partonic energy loss a valuable tool for the measurement
of the quark-gluon plasma lifetime.

\subsubsection{Angular Dependence of Radiative Energy Loss}
\label{sec316}
{\em R. Baier, U.A. Wiedemann}

Hard jets, when produced in a heavy ion collision, will be measured in a 
typical calorimeter experiment within an angular cone of opening angle
$\theta_{{\rm cone}}$. Here we summarize what is known about the 
angular dependence of medium-induced radiative energy loss.

\paragraph{Average energy loss}

\begin{figure}[h]\epsfxsize=8.7cm 
\centerline{\epsfbox{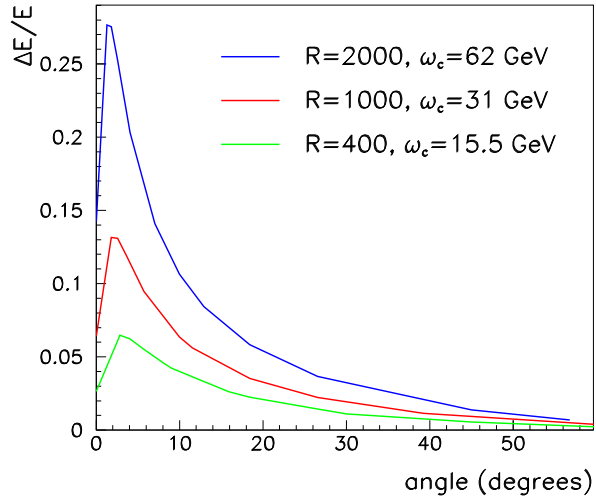}}
\caption{The fraction of the total radiative 
energy loss $\Delta E/E$ emitted outside a jet cone of fixed angle 
$\Theta$. This calculation is for a jet of total energy $E=100$ GeV
and $R = \frac{1}{2} \hat{q}\, L^3$, $\omega_c = \hat{q}\, L^2$.
}\label{angularcarlos}
\end{figure}

The mean energy loss due to gluons, induced by the medium and radiated 
outside the cone, has been investigated in \cite{Baier:1999ds} 
and more recently in \cite{Wiedemann:2000tf,Baier:2001qw,Salgado:2003gb}.
The calculations are based on the integrated energy loss 
{\it outside} an angular cone of opening angle $\theta_{{\rm cone}}$,
\begin{equation}\label{eq:3.1}
\Delta E (\theta_{{\rm cone}}) = \int^\infty_0 \, d\omega\,
\int^\pi_{\theta_{{\rm cone}}} \, \frac{\omega dI}{d\omega d\theta} 
d\theta\, .
\end{equation}
Fig.~\ref{angularcarlos} shows numerical results for this 
angular dependence obtained from evaluating Eq.~(\ref{2.1}) in 
the multiple soft scattering approximation with different values
for the BDMPS transport coefficient $\hat{q}$ and the in-medium
pathlength $L$.
Here, $\frac{\Delta E}{E}(\Theta)$ does not decrease monotonously with 
increasing $\Theta$ but has a maximum at finite jet opening angle. The 
reason is that the
radiative energy loss outside a cone angle $\Theta$ receives
additional contributions from the Brownian ${\bf k}_T$-broadening
of the standard DGLAP vacuum radiation,
\begin{equation}
        \frac{1}{{\bf k}^2_T} \longrightarrow
         \frac{1}{({\bf k}_T+ {\bf q}_T)^2}
        \label{4.18}
\end{equation}
 This redistribution in
transverse momentum space does not affect the total 
${\bf k}_T$-integrated yield $ \frac{\Delta E}{E}(\Theta=0)$, 
but shows up as soon as a finite cone size is chosen. Thus, 
strictly speaking, the total ${\bf k}_T$-integrated radiative 
energy loss $\frac{\Delta E}{E}(\Theta=0)$
is not the upper bound for the radiative energy loss outside a
finite jet cone angle $ \frac{\Delta E}{E}(\Theta)$. 
%
%

To sum up, there is a simple physical reason for the non-monotonic
behavior of $\Delta E (\theta_{{\rm cone}})$ as a function of the
jet cone, namely the redistribution of the vacuum component of
gluon radiation in transverse phase space. However, the size of the
effect remains unclear. First, one may expect that the region of 
small $\theta_{{\rm cone}}$ is dominated by higher order QCD 
contributions which are not yet taken into account. Second, the
effect shown in Fig.~\ref{angularcarlos} is seen in the multiple
soft scattering approximation of Eq.~(\ref{2.1}). However, in 
the single hard scattering approximation, $\Delta E (\theta_{{\rm cone}})$
decreases monotonously~\cite{Salgado:2003gb}. Irrespective of these
differences for small opening angle, it is worth emphasizing that
both approximations agree quantitatively for 
$\theta_{{\rm cone}} > 10^\circ$~\cite{Salgado:2003gb}.

\paragraph{Universality of angular dependence}

\begin{figure}[htb!]
\begin{center} 
\includegraphics[height=2.6in,width=2.9in,angle=0]{newinsert.epsi}
\hspace*{0.1in}
\includegraphics[height=3.0in,width=3.1in,angle=0]{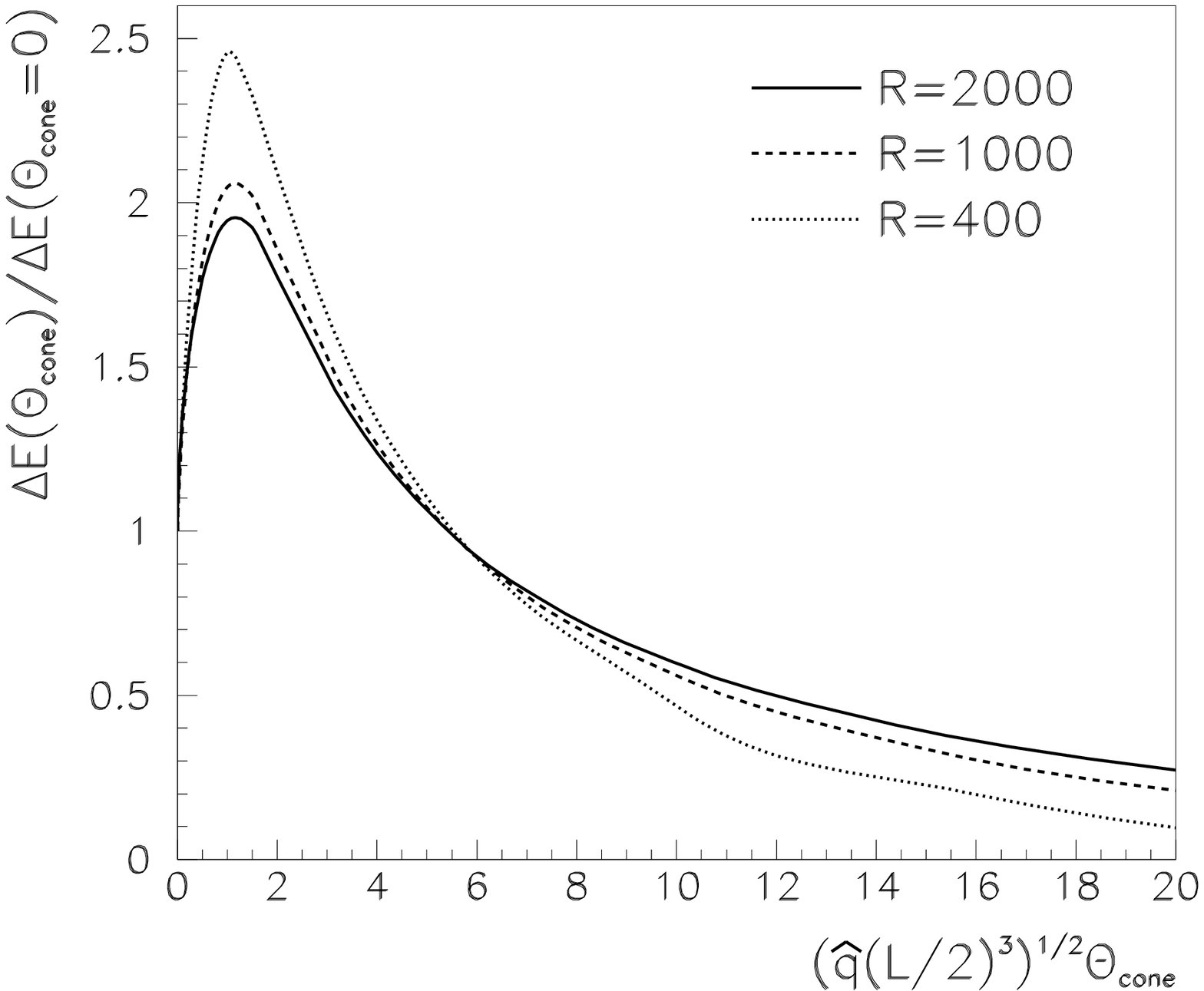}
\caption{The mean energy loss radiated outside an opening cone
$\theta_{{\rm cone}}$ normalized to the total average energy loss.
LHS: in the BDMS calculation~\cite{Baier:2001qw}, the ratio is universal,
depending on a single dimensionless variable only. RHS: if the
kinematic constraints in transverse phase space are taken into 
account, deviations from the universal behavior remain 
small~\cite{Salgado:2003gb}.}
\label{angledep}
\end{center} 
\end{figure}

The integrated mean loss $\Delta E (\theta_{{\rm cone}})$, normalized to 
$\Delta E \equiv \Delta E (\theta_{{\rm cone}}=0)$, is defined by 
\begin{equation}  \label{eq:15b}
  R (\theta_{{\rm cone}} ) = 
  \frac{\Delta E (\theta_{{\rm cone}})}{\Delta E}
\end{equation}
with $R (\theta_{{\rm cone}}=0 ) = 1.$.
In the BDMPS limiting case, the
total energy loss $\Delta E$ depends only on the characteristic 
gluon energy, see Eq.~(\ref{loss}).
In this limit, it was observed that the ratio $R (\theta_{{\rm cone}} )$
depends on a single dimensionless variable, 
which includes all the medium dependent parameters, 
namely~\cite{Baier:1999ds}
\begin{equation}  \label{eq:15}
  R (\theta_{{\rm cone}} ) 
  = R ( c (L) \theta_{{\rm cone}})\, , 
\end{equation}
where
\begin{equation}   \label{eq:16}
c^2 (L) = \frac{N_c}{2C_F} \hat q \,\,(L/2)^3\, . 
\end{equation}
The typical dependence on $\theta_{{\rm cone}}$ is shown in 
Fig.~\ref{angledep}. 
When comparing hot and cold QCD matter we recall that
for fixed in-medium pathlength $L$,
\begin{equation}\label{eq:3.13}
c (L) \Big|_{{\rm HOT}} \gg c(L) \big|_{{\rm COLD}}. 
\end{equation}
This is a consequence of the temperature dependence of the
BDMPS transport coefficient $\hat{q}$.
$R (\theta_{{\rm cone}})$ is also universal in the sense that it is 
the same for an energetic quark and for an energetic gluon jet. 
In a recent study\cite{Salgado:2003gb}, the ratio $R (\theta_{{\rm cone}} )$
was calculated in the presence of kinematic constraints on the
transverse phase space. Small deviations from the universal 
behavior (\ref{eq:15}) were observed, but for all practical
purposes these are negligible.
%

\subsection{Multiple Gluon Emission and Quenching Weights}
\label{sec32}
{\em R. Baier, U.A. Wiedemann}

Irrespective of the number of additionally radiated gluons, 
what matters for the medium modification of hadronic observables
is how much {\it additional} energy $\Delta E$ is radiated off a 
hard parton. In this section, we first discuss the so called
quenching weight which is the probability distribution
$P(\Delta E)$ of the additional medium-induced
energy loss. For independent gluon emission, this probability
is the normalized sum of the emission probabilities
for an arbitrary number of $n$ gluons which carry away a total
energy $\Delta E$:\cite{Baier:2001yt}
\begin{eqnarray}
  P(\Delta E) = \sum_{n=0}^\infty \frac{1}{n!}
  \left[ \prod_{i=1}^n \int d\omega_i \frac{dI(\omega_i)}{d\omega}
    \right]
    \delta\left(\Delta E - \sum_{i=1}^n \omega_i\right)
    e^{- \int d\omega \frac{dI}{d\omega}}\, .
   \label{5.1}
\end{eqnarray}
In general, the quenching weight (\ref{5.1}) has a discrete and a 
continuous part,\cite{Salgado:2002cd}
\begin{equation}
  P(\Delta E) = p_0\, \delta(\Delta E) + p(\Delta E)\, .
   \label{5.2}
\end{equation}
The discrete weight $p_0$ emerges as a consequence of a finite
mean free path. It determines the probability that 
no additional gluon is emitted due to in-medium scattering 
and hence no medium-induced energy loss occurs. 

In order to determine the discrete and continuous part of
(\ref{5.2}), it is convenient to rewrite Eq.~(\ref{5.1}) 
as a Laplace transformation~\cite{Baier:2001yt}
\begin{eqnarray}
  P(\Delta E) &=& \int_C \frac{d\nu}{2\pi i}\, {\cal P}(\nu)\,
  e^{\nu\Delta E}\, ,
  \label{5.3}\\
  {\cal P}(\nu) &=& \exp\left[ -\int_0^\infty
    d\omega\, \frac{dI(\omega)}{d\omega}\,
    \left(1- e^{-\nu\, \omega}\right)\right]\, .
  \label{5.4}
\end{eqnarray}
Here, the contour $C$ runs along the imaginary axis with
${\rm Re}\nu = 0$. 

For the further discussion, it is useful to treat the
medium-induced gluon energy distribution $\omega \frac{dI}{d\omega}$ 
in Eq.~(\ref{2.1}) explicitly as the medium modification of a
``vacuum'' distribution \cite{Salgado:2003gb}
\begin{equation}
 \omega \frac{dI^{(tot)}}{d\omega} = 
 \omega \frac{dI^{(vac)}}{d\omega} +
 \omega \frac{dI}{d\omega}
 \label{5.5}
\end{equation}
From the Laplace transform (\ref{5.3}), one finds the 
total probability
\begin{equation}
  P^{(tot)}(\Delta E) = \int_0^\infty  d\bar{E} \, 
  P(\Delta E - \bar{E}) \, 
   P^{(vac)}(\bar{E})\, . 
   \label{5.6}
\end{equation} 
This probability $P^{(tot)}(\Delta E)$ is normalized to unity and it is 
positive definite. In contrast, the medium-induced modification of this 
probability, $P(\Delta E)$, is a generalized probability. It can take
negative values for some range in $\Delta E$, as long as its normalization
is unity,
\begin{equation}
  \int_0^\infty  d\bar{E} \,  P(\bar{E})
  = p_0 + \int_0^\infty  d\bar{E} \,  p(\bar{E}) = 1\, .
  \label{5.7}
\end{equation}
We now discuss separately the properties of the discrete contribution
$p_0$ and the continuous one $p(\bar{E})$.

\begin{figure}[h]\epsfxsize=8.7cm
\centerline{\epsfbox{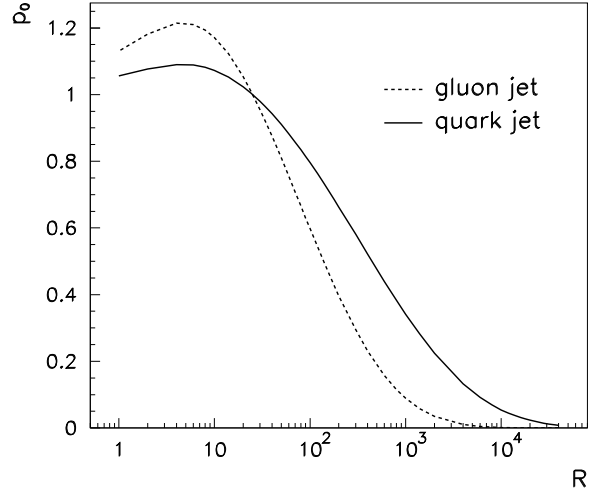}}
\caption{The discrete part $p_0$ of the quenching weight calculated
in the multiple soft scattering limit as a function of $R$.
Figure taken from \protect\cite{Salgado:2003gb}. 
}\label{fig7}
\end{figure}
\paragraph{Discrete part of the quenching weight}
The discrete part of the quenching weight
can be expressed in terms of the total gluon multiplicity,
\begin{equation}
  p_0 = \lim_{\nu \to \infty}\, {\cal P}(\nu) = 
  \exp \left[ -N(\omega = 0)\right]\, ,
  \label{5.8}
\end{equation}
where the multiplicity $N(\omega)$ of gluons with energy larger 
than $\omega$ emerges by partially integrating the exponent of
(\ref{5.4}),
\begin{equation}
  N(\omega) \equiv \int_\omega^\infty d\omega'\,
                    \frac{dI(\omega')}{d\omega'}\, .
  \label{5.9}
\end{equation}
For the limiting case of infinite in-medium pathlength, the
total multiplicity $N(\omega)$ diverges and the discrete 
part vanishes. In general, however, $p_0$ is finite. 
A typical dependence of $p_0$ on model parameters is
shown in Fig.~\ref{fig7} for the radiation spectrum
calculated in the multiple soft scattering limit.
A qualitatively similar behavior is found in the 
opacity expansion. 
Remarkably, $p_0$ can exceed unity for some parameter
range, since the medium modification $\omega \frac{dI}{d\omega}$
to the radiation spectrum (\ref{5.5}) can be negative.  
The value $p_0> 1$ then compensates a predominantly 
negative continuous part $p(\Delta E)$ and satisfies 
the normalization (\ref{5.7}). It indicates
a phase space region at very small transverse momentum, 
into which {\it less} gluons 
are emitted in the medium than in the vacuum. 
This effect is more pronounced for gluons than for quarks. 

\begin{figure}[t]\epsfxsize=12.7cm
\centerline{\epsfbox{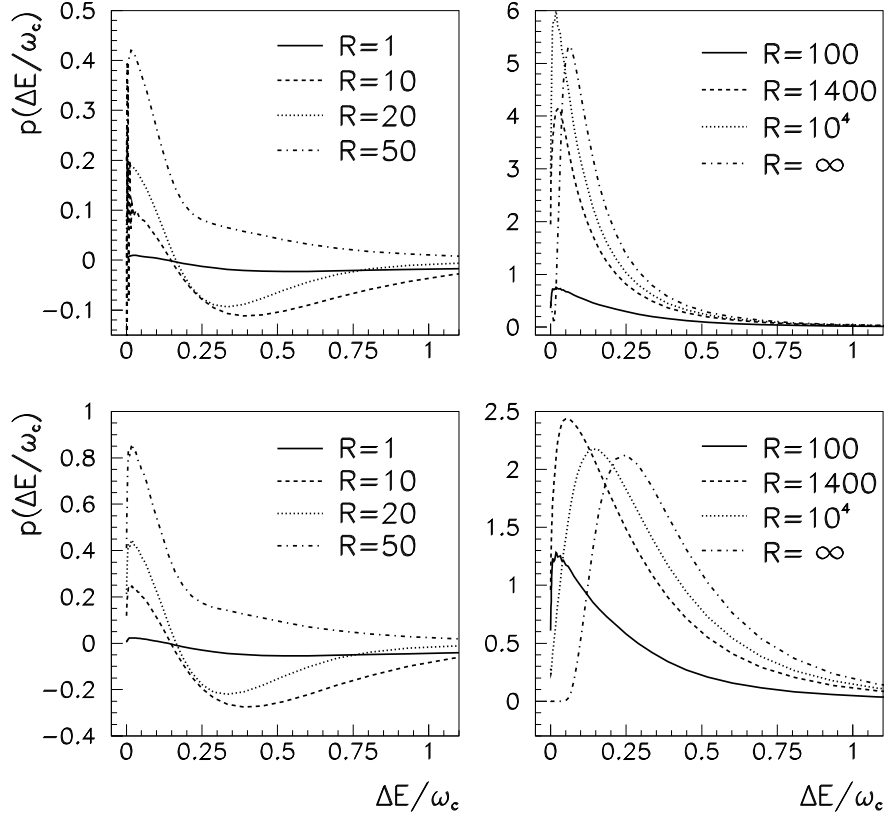}}
\caption{The continuous part of the quenching weight 
(\protect\ref{5.2}), calculated in the multiple soft scattering limit
for a hard quark (upper row) or hard gluon (lower row).
Figure taken from \protect\cite{Salgado:2003gb}.
}\label{fig8}
\end{figure}

\paragraph{Continuous part of the quenching weight}
The continuous part $p(\Delta E)$ of the probability distribution 
(\ref{5.2}) is shown in  Fig.~\ref{fig8} calculated in the
multiple soft scattering limit. In the opacity expansion,
it looks qualitatively similar. With increasing density of
scattering centers (i.e. increasing $R = \frac{1}{2}\hat{q}L^3$)
the probability of loosing a significant energy fraction
$\Delta E$ increases. Also, since the interaction between
partonic projectile and medium are larger for a hard gluon 
than a hard quark, the energy loss is larger for gluons.
This can be seen in Fig~\ref{fig8} from the larger width 
of $p(\Delta E)$ for the gluonic case. 
Finally, as expected from the normalization
condition (\ref{5.7}), the continuous part $p(\Delta E)$ shows
predominantly negative contributions for the parameter range for
which the discrete weight $p_0$ exceeds unity. 
%

In the multiple soft scattering limit and for infinite pathlength,
the quenching weight was found to be fit very well by a two-parameter 
log-normal distribution\cite{Arleo:2002kh}. 
Analytically, an estimate of the quenching 
weight can be obtained~\cite{Baier:2001yt} in the limit $R\to \infty$ 
from the small-$\omega$ approximation 
$\omega \frac{dI}{d\omega} \propto \frac{1}{\sqrt{\omega}}$ in 
the multiple scattering limit
\begin{equation}
  P^{\rm approx}_{\rm BDMS}(\epsilon)
  = \sqrt{\frac{a}{\epsilon^3}} 
     \exp\left[-\frac{\pi\, a}{\epsilon} \right]\, ,
     \qquad \hbox{\rm where}\, \, 
     a=\frac{2\, \alpha_s^2\, C_R^2}{\pi^2}\omega_c\, .
  \label{5.10}
\end{equation}
This approximation is known to capture \cite{Salgado:2002cd} the 
rough shape of the probability distribution for large system size, but 
it has an unphysical large $\epsilon$-tail with infinite first moment 
$\int d\epsilon\, \epsilon\, P^{\rm approx}_{\rm BDMS}(\epsilon)$.
Remarkably, Equation (\ref{5.10}) provides a semi-quantitative 
understanding of the degree of partonic energy loss shown in
Fig.~\ref{fig8}. In particular, comparing for $R\to\infty$ the 
maxima of the curves in Fig.\ref{fig8}
for quarks and gluons, one finds a displacement by a factor 
$\approx$ 5. This agrees well with the square of the
relative Casimir factors $(C_A/C_F)^2$
by which the maximum of $\epsilon  P^{\rm approx}_{\rm BDMS}(\epsilon)$ 
changes.

\subsection{Collisional versus Radiative Energy Loss}
\label{sec33}
{\em I. Lokhtin, A.M. Snigirev}

The collisional energy loss due to elastic scattering with high-momentum 
transfer have been originally estimated by Bjorken in~\cite{Bjorken:1982tu}, 
and recalculated later in~\cite{Mrowczynski:da,Thoma:1991em} taking also 
into account the loss with low-momentum transfer dominated by the 
interactions with plasma collective modes. 
The method for quantum field-theoretic calculating energy loss in the low  
exchange momentum region of the collisions (screening effect in the plasma) 
have been developed by Braaten and 
Pisarski~\cite{Braaten:1989kk,Braaten:1989mz,Frenkel:br}. 
It allows one to calculate the hard thermal loop (HTL) corrections to the 
propagator of the exchanged gluon in the $Qq \rightarrow Qq$ and the 
$Qg \rightarrow Qg$ processes. 

The average collisional energy loss per mean free path $\lambda$ 
can be written as~\cite{Bjorken:1982tu}
\begin{equation}
  \frac{dE_{\rm coll}}{dz} = 
  \frac{\Delta E_{\rm coll}}{\lambda} = 
  \int d^3k\, \rho(k) \int dt\, 
   \frac{d\sigma}{dt} \, \frac{t}{2k}
  \simeq \frac{1}{4T \sigma\, \lambda} 
  \int\limits_{\displaystyle\mu^2_D}^
  {\displaystyle 3T E / 2}dt\frac{d\sigma}{dt}t
  \, .
\end{equation}
Here, $\lambda = 1/\int d^3k \rho(k) \sigma$, and $\rho(k)$
defines the thermal density of partonic scatterers (the sum over
which is implicit). The factor
$\frac{t}{2k}$ denotes the energy transfer per scattering 
times the flux factor of the incident participating partons
\cite{Baier:2000mf}.
The dominant contribution to the differential cross section 
$d\sigma / dt$ for scattering of a parton with energy $E$ off the 
"thermal" partons with energy $3T \ll E$ 
at temperature $T$ is taken to be the LO perturbative scattering
cross section~\cite{Bjorken:1982tu}
\begin{equation} 
\frac{d\sigma}{dt} \cong \frac{2\pi\alpha_s^2(t)}{t^2}\, . 
\end{equation} 
The integrated 
parton scattering cross section $\sigma$ is regularized by the 
Debye screening mass squared 
$\mu_D^2 \cong 4 \pi \alpha_s T^2 (1 + N_f / 6)$.

There are marked differences between collisional and radiative
energy loss. For collisional energy loss, the scattering centers
act incoherently. The value $\Delta E_{coll}$ is independent 
of in-medium path length, and it depends only logarithmically on the
initial parton energy. It is determined mainly by the medium 
temperature~\cite{Bjorken:1982tu}
\begin{equation} 
 \frac{dE_{\rm coll}}{dz} \propto \alpha_s^2\, T^2\, \ln E/T\, . 
\end{equation} 
The dependence of the total collisional energy loss on in-medium
pathlength can be weaker than linear for an expanding medium and it
is linear for a static one.

The angular-integrated radiative energy loss of a high energy projectile 
parton is known to dominate over the collisional energy loss by up to an 
order of magnitude~\cite{Gyulassy:1993hr,Wang:1994fx,Mustafa:1997pm}. However, the 
angular dependence of the lost (i.e. redistributed) energy is very
different for both mechanisms. With increasing parton energy, the maximum 
of the angular distribution of 
bremsstrahlung gluons shifts towards the direction of the parent parton. 
This means that measuring the jet energy as a sum of the energies of 
final hadrons moving inside an angular cone with a given finite size 
$\theta_0$ will allow the bulk of the gluon radiation to belong to the 
jet and thus the major fraction of the initial parton energy to be 
reconstructed. Therefore, the medium-induced radiation mainly 
softens the particle energy distributions inside the jet, and 
increases the multiplicity of secondary particles. Only to a lesser degree 
does it affect the total jet energy. It is important to notice that the 
coherent 
Landau-Pomeranchuk-Migdal radiation induces a strong dependence of the 
radiative energy loss of a jet (but not a leading parton) on the jet 
angular cone size~\cite{Lokhtin:1998ya,Baier:1998yf,Wiedemann:2000za,Wiedemann:2000tf,Gyulassy:2000fs}. 

On the other hand, collisional energy loss turns out to be practically 
independent of jet cone size, because the bulk of "thermal" particles 
knocked out of the dense matter by elastic scatterings fly away in 
almost transverse direction relative to the jet axis~\cite{Lokhtin:1998ya}. 
In fact, in relativistic kinematics, $E \gg m_0 = 3\, T$, in the rest 
system of 
the target with effective mass $m_0$ we get the following estimate for 
the transverse $p_T^{t,i}$ and longitudinal $p_L^{t,i}$ momenta of the 
incident and ``thermal'' particles: $p_T^t \simeq \sqrt{t}$, 
$p_L^t \simeq t / 2m_0$; 
$p_T^i \simeq -\sqrt{t}$, $p_L^i \simeq E - t / 2m_0$. 
Scattering angle $\theta_i$ of the incident parton vanishes in the 
relativistic limit,  $\tan{\theta_i} =  p_T^i / p_L^i \simeq \sqrt{t} / E
\rightarrow 0$. The scattering angle  $\theta_t$ of a struck ``thermal'' 
particle with respect to the initial direction of the fast parton can be 
estimated as  $\tan{\theta_t} = p_T^t /p_L^t \simeq 2 m_0 / \sqrt{t}$.  
The minimal and maximal values of $\tan{\theta_t}$ are 
$\tan{\theta^{max}_t} \simeq 2 m_0 / \mu_D$ and $\tan{\theta^{min}_t} 
\simeq 2 m_0 / \sqrt{0.5 m_0 E}$ respectively. 
It is straightforward to evaluate the average 
$\left \langle \tan{\theta_t}\right\rangle$ as~\cite{Lokhtin:1998ya}     
\begin{equation}  
  \left \langle \tan{\theta_t} \right\rangle  
  = \left \langle \frac{2m_0}{\sqrt{t}} \right\rangle 
  \simeq \frac{6T}{\sigma} 
  \int\limits_{\displaystyle \mu^2_D}^
  {\displaystyle 3T E / 2} dt \frac{d\sigma}{dt} \frac{1}{\sqrt{t}}.  
\end{equation} 
Neglecting a weak $\alpha_s (t)$ dependence we obtain 
$\left \langle \tan{\theta_t}\right\rangle \simeq \frac{2}{3} 
\tan{\theta^{max}_t} \simeq 4m_0 / 3\mu_D$. 
Substituting  $\mu_D$,
we arrive at  $<\theta_t> \sim 60^0$ for $T > 200$~MeV.  
This value exceeds typical cone sizes $\theta_0\sim 10^0 - 30^0$ 
used to experimentally define hadronic jets.  
This means that the major part of ``thermal'' particles will fly outside the 
cone of the jet and thus cause the ``jet energy loss''. 
%
\begin{figure}[htb]\epsfxsize=7.7cm
\centerline{\epsffile{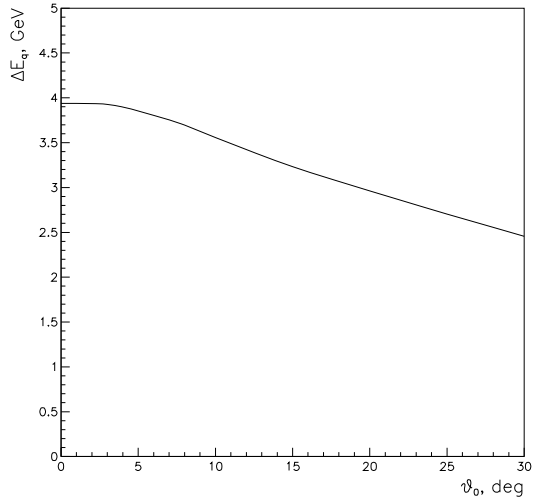}}
\vskip -1cm
\caption{Angular dependence of the collisional energy loss for a 
100 GeV quark-initiated jet, according to Ref.\cite{Lokhtin:1998ya}. 
In comparison to the
radiative energy loss shown in Fig.\ref{angularcarlos}, the contribution 
is relatively 
small at small angles, but cannot be neglected at large angles.}  
\label{lokhtinangular}
\end{figure}
%
This study indicates that radiative energy loss indeed dominates the
medium-dependence of jets for small cone opening angles $\theta_0$. 
However, collisional energy loss may have a significant contribution 
to jet quenching for larger cone opening angles $\theta_0$,
see Fig.\ref{lokhtinangular}. 

Here, we have considered only 
massless partons propagating through a dense QCD-matter. Although a 
full description of the coherent gluon radiation from a massive color 
charge is still lacking, finite quark mass effects are expected to lead 
to a relative suppression of medium-induced radiation of heavy 
(especially $b$) quarks~\cite{Dokshitzer:2001zm}. 
In this case the influence of collisional energy loss on experimental 
observables of "heavy quark quenching" (such as high-mass dileptons 
and secondary charmonium production) can be comparable with the effect 
of medium-induced radiation~\cite{Lokhtin:1998ya}, see 
section~\ref{sec344}. 


\subsection{Observable Consequences of Radiative Energy Loss}
\label{sec34}
%
\subsubsection{$E_T$-Distributions and $p_T$-Spectra: 
Quenching Weights}
\label{sec341}
{\em R. Baier, U.A. Wiedemann}

Assume that a hard parton looses an additional energy fraction
$\Delta E$ while escaping the collision region. 
The medium-dependence of the corresponding inclusive transverse momentum
spectra can be characterized in terms of the quenching 
factor $Q$ \cite{Baier:2001yt}
\begin{eqnarray}
 Q(p_T)&=&
{{d\sigma^{\rm med}(p_T)/ dp^2_T}\over
{d\sigma^{\rm vac}(p_T)/ dp^2_T}}=
\int d{\Delta E}\, P(\Delta E)\left(
{d\sigma^{\rm vac}(p_T+\Delta E)/ dp^2_T}\over
{d\sigma^{\rm vac}(p_T)/ dp^2_T}\right)
\nonumber\\[2mm]
&\simeq&  \int d{\Delta E}\, P(\Delta E)\,  
    \left({p_T\over p_T+\Delta E}\right)^n\, ,
 \label{5.11}
\end{eqnarray}
where $P(\Delta E)$ is the quenching weight given in Eq.~(\ref{5.1}).
%
\begin{figure}[h]\epsfxsize=12.7cm
\centerline{\epsfbox{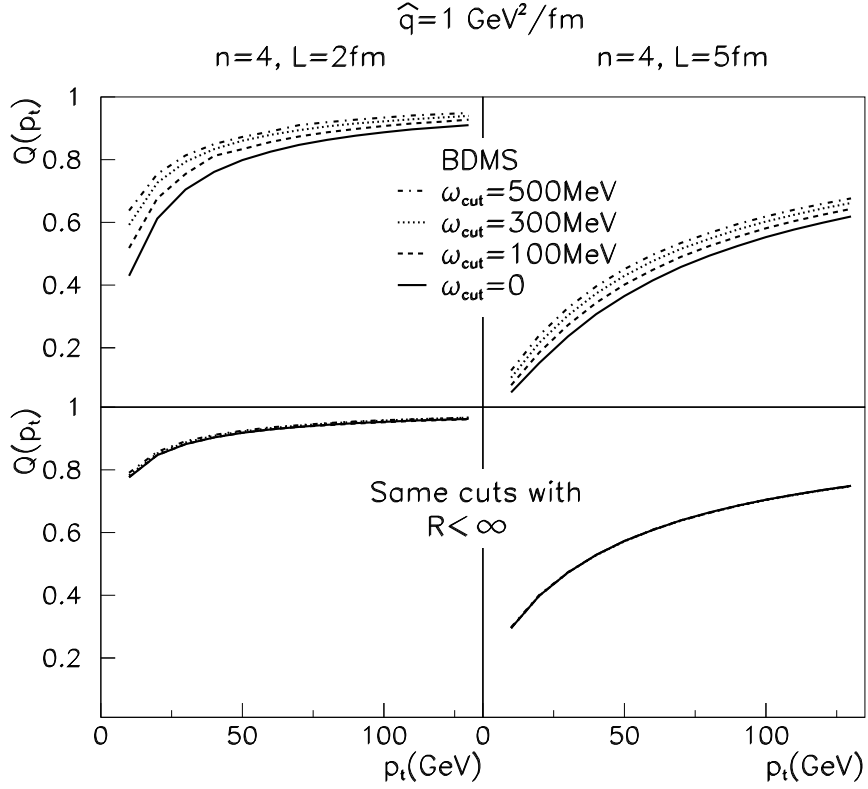}}
\caption{The quenching factor (\protect\ref{5.11}) calculated in the
multiple soft scattering limit. Upper row: calculation in the 
$R\to\infty$-limit but with a variable sharp cut-off on the 
infrared part of the gluon energy distribution. Lower row: the 
same calculation is insensitive to infrared contributions if 
the finite kinematic constraint $R=\omega_cL <\infty$ is included. 
Figure taken from \protect\cite{Salgado:2003gb}.
}\label{fig9}
\end{figure}
%
Here, the last line is obtained by assuming a powerlaw fall-off of 
the $p_T$-spectrum. The effective power $n$ depends in general
on energy and $p_T$. It is $n \simeq 7$ for the kinematic range 
relevant for RHIC, and it is smaller for LHC. Alternatively, instead 
of the quenching factor (\ref{5.11}),
the medium modification of hadronic transverse momentum spectra is often 
characterized by a shift factor $S(p_T)$,
\begin{equation}
  \frac{d\sigma^{{\rm med}}(p_T)}{dp^2_T}  \simeq
  \frac{d\sigma^{{\rm vac}}(p_T+S(p_T))}{dp^2_T} \, , 
  \label{5.12}
\end{equation} 
which is related to $Q(p_T)$ by
\begin{equation}
  Q(p_T) = \exp\left\{-\frac{n}{p_T}\cdot
  S(p_T)\right\} \, .
  \label{5.13}
\end{equation}
Most importantly, since the hadronic spectrum shows a strong
powerlaw decrease, what matters for the suppression is not
the average energy loss $\langle \Delta E \rangle$ but 
the least energy loss with which a hard parton is likely
to get away. One concludes that $S(p_T) < \langle \Delta E \rangle$
and depends on transverse momentum~\cite{Baier:2001yt}. 

Fig.~\ref{fig9} shows a calculation of the quenching factor
(\ref{5.11}) in the multiple soft scattering limit. A 
qualitatively similar result is obtained in the opacity
expansion. In general, quenching weights increase
monotonically with $p_T$ since the medium-induced
gluon radiation is independent of the total projectile energy
for sufficiently high energies. At very low transverse momenta,
the calculation based on (\ref{2.1}) is not reliable
and the interpretation of the medium modification of
hadronic spectra in nucleus-nucleus collisions will 
require additional input (e.g. modifications due to the
Cronin effect). Fig.~\ref{fig9} suggests, however, that
hadronic spectra at transverse momenta $p_T > 10$ GeV,
can be suppressed significantly due to partonic final state
rescattering. 

Finally, Fig.~\ref{fig9} allows to comment on the sensitivity
of the perturbative calculation of $\omega \frac{dI}{d\omega}$ 
on uncontrolled non-perturbative soft physics. The gluon energy
distribution in (\ref{2.1}) allows in principle for the emission
of arbitrarily soft gluons. It is clear, however, that the
calculation cannot be reliable in this soft regime. 
To quantify the sensitivity of the calculation to the low momentum 
region, Baier et al.\cite{Baier:2001yt} introduced a sharp 
cut-off on the $R\to \infty$ gluon energy distribution which was
varied between $\omega_{\rm cut} = 0$ and $\omega_{\rm cut} = 500$ MeV.
However, phase space constraints (i.e. finite $R$) deplete the
gluon radiation spectrum in the soft region, see 
Fig.~\ref{bdmpscomp}.
As seen in Fig.~\ref{fig9}, this decreases significantly the 
sensitivity of quenching factors to the uncontrolled infrared
properties of the radiation spectrum~\cite{Salgado:2003gb}.

\subsubsection{Medium-Modified Fragmentation Functions}
\label{sec342}
{\em Carlos Salgado}

In proton-proton collisions, the 
inclusive production of a hadron $h$  of high enough $p_T$ can be described
by the factorization (LO) formula (\ref{hcrossec}).
Both the parton distribution functions and the fragmentation functions
entering this expression are obtained from global fits to experimental 
data. The procedure is well know: an initial
condition containing all the non-perturbative information is
evolved, by DGLAP equations, to scales $Q^2$ and $\mu_F^2$ respectively
and then fitted, in a recursive procedure, to available data. 
A third scale, the renormalization scale $\mu_R^2$, is contained
in the perturbative cross section $d\sigma^{ij\to k}$ through the 
running of $\alpha_S(\mu_R^2)$. Eq.~(\ref{hcrossec})
leads to a fair description of the shape of high-$p_T$
hadronic spectra while the normalization has to be adjusted by
an energy-dependent $K$-factor (see e.g. \cite{Eskola:2002kv}).
Also to NLO, the disagreement between theory and experiment lies
essentially in an albeit reduced normalization factor~\cite{Aurenche:1999nz}.
However, the theoretical K-factor $\sigma^{NLO}/\sigma^{LO}$ shows, 
some $p_T$-dependence.

For proton-nucleus or nucleus-nucleus collisions, Eq.~(\ref{hcrossec})
is also expected to work, though, due to the enhanced power corrections
in the nucleus, the range of validity would be for $p_T$ larger than in 
proton-proton.
Apart from geometrical factors, the generalization to 
pA or AA collisions needs of nuclear PDF and medium-modified FF.
This {\it medium} that modifies the parton fragmentation could be 
the  nucleus itself ({\it cold nuclear matter}) 
in both pA and AA collisions and/or eventually the produced high-dense state
({\it hot and dense medium}) in AA collisions. The nuclear PDF have been
studied in several approaches and global fits similar to the ones for the
proton are available (see the pA section in this Yellow Report for 
the state-of-the-art in the field). 
The case for the FF is less clear, ideally one
should perform a new global fit for these medium-FF using modified evolution
equations. These new evolution equations would take care of the evolution, 
in the medium,
of a highly virtual parton to the final hadrons. Whether something like this
could be obtain for (factorized) 
leading twist FF, and how the evolution equations
would be modified by the medium, is still very unclear. 
Finite temperature 
modifications to DGLAP evolution have been calculated in \cite{Osborne:2002dx}
in the framework of a thermal field theory. The modified splitting functions
depend in this case on the temperature of the medium. However, multiple
scattering effects as induced radiation or interference (LPM) are not 
included.
There has
been an attempt of constructing medium-modified fragmentation functions
from a twist expansion in Refs. \cite{Guo:2000nz}\cite{Wang:2001if}
\cite{Wang:2002ri}\cite{Osborne:2002st}. In these references, the 
medium-modifications of the FF are given by one additional collision
of the parton with the medium (the first term in an opacity expansion). These
terms are of higher-twist nature. On the other hand, the medium induced 
radiation that could eventually lead to modified leading-twist
evolution equations including multiple scattering effects has been
computed in several approximations \cite{Baier:1996kr}-\cite{Gyulassy:2000fs}.
In summary, a full leading-twist DGLAP-like 
evolution containing all the relevant features of the problems
is still missing. Most of the present approaches using leading twist FF
rely on a model proposed in Ref. \cite{Wang:1996yh}. 
The medium-modified fragmentation functions are, in this model, given by 
\begin{eqnarray}
  D_{k\to h}^{(\rm med)}(z,\mu_F^2) = \int_0^{1-z} d\epsilon\, P_E(\epsilon)\,
  \frac{1}{1-\epsilon}\, D_{k\to h}(\frac{z}{1-\epsilon},\mu_F^2)\, .
  \label{salgado_eq2}
\end{eqnarray}
The picture is the following. The high energetic quark or gluon loses
some fraction $\epsilon$ of its energy when traveling through the medium and
then fragments in the vacuum with the normal vacuum FF, 
$D_{k\to h}(z/1-\epsilon,\mu_F^2)$, with the corresponding 
shifted momentum fraction. Any
modification of the virtuality dependence of the FF by the medium
is neglected. Also, the hadronized remnants
of the medium-induced soft radiation are neglected in the definition
of (\ref{salgado_eq2}). However, these remnants are expected to be
soft, and their inclusion would thus amount to an additional 
contribution to $D_{k\to h}^{(\rm med)}(z,\mu_F^2)$ for
$z < 0.1$ say. 
The only ingredient needed in Eq.~(\ref{salgado_eq2}) is
the probability distribution $P_E(\epsilon)$ for a parton of energy $E$ to 
lose a fraction $\epsilon$ of this energy. These quenching weights
are normally computed in the independent
gluon emission approximation (\ref{5.1}) with $P(\Delta E) =
P_E(\epsilon = \Delta E/E) / E$.

%
\begin{figure}[t]
\begin{center}
\includegraphics[width=10cm]{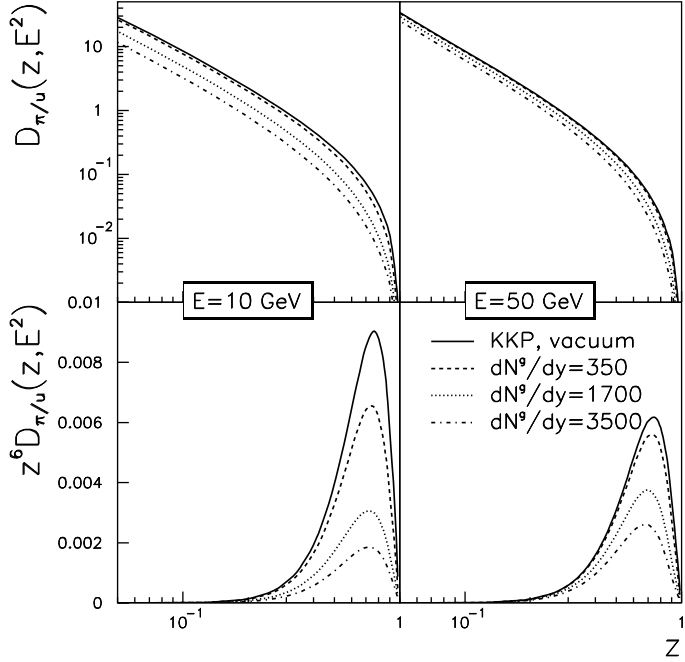}
\caption{
Medium-modified fragmentation functions for media of different
densities (upper two panels). Multiplying these FF by $z^6$ (lower two
panels) the
position of the maximum gives the relevant $z$ values in the integration
of Eq.~(\ref{hcrossec}). The vacuum fragmentation functions are from
Ref. \cite{Kniehl:2001}.
}
\label{salgado_fig1}
\end{center}
\end{figure}

Another approximation that has been taken for the quenching weight is just
\begin{equation}
P_E(\epsilon) = \delta\left(\epsilon-\frac{\Delta E}{E}\right)\ .
\label{salgado_eq4}
\end{equation}
 
\noindent
It has been argued that using Eq.~(\ref{salgado_eq4}) produces
a much stronger effect due to the rapid $p_T$-dependence of the
production spectrum (or equivalently to the rapid $z$-dependence of the FF).
This is clearly the case for the multiple soft scattering approximation
for which the quenching weight has a sharp maximum. In the 
hard scattering opacity expansion, the longer tails of the distributions
makes the difference between using 
Eqs. (\ref{5.1}) and (\ref{salgado_eq4})
smaller.

Medium-modified quark to pion 
fragmentation functions for different media are plotted in
Fig.~\ref{salgado_fig1} \cite{Salgado:2002cd}\cite{Salgado:2003gb}. 
They are calculated from Eq.~(\ref{salgado_eq2}) using 
the multiple scattering approximation for the quenching weights
and the LO KKP~\cite{Kniehl:2000fe}
parametrization of $D_{h/q}(z,Q^2)$. For this calculation, 
the virtuality $Q$ of $D_{h/q}(z,Q^2)$ is identified with
the (transverse) initial energy $E_q$ of the parton.
This is justified since $E_q$ and $Q$ are of the same order, and
$D_{h/q}(z,Q^2)$ has a weak logarithmic $Q$-dependence while
medium-induced effects change as a function of
$\epsilon = \frac{\Delta E}{Q} \approx O(\frac{1}{Q})$.
For a collision region expanding according to Bjorken scaling, 
the transport coefficient can be related to the initial gluon 
rapidity density~\cite{Baier:1996sk,Gyulassy:2000gk}.  
\begin{equation}
  R = \frac{1}{2}\hat{q}L^3 = \frac{L^2}{R_A^2}\, \frac{dN^g}{dy}\, ,
\label{5.15}
\end{equation}
That's what is done in Fig.~\ref{salgado_fig1}. Interestingly, Eq.~(\ref{5.15})
indicates how partonic energy loss changes with the particle 
multiplicity in nucleus-nucleus collisions. This allows to 
extrapolate parton energy loss effects from RHIC to LHC
energies~\cite{Salgado:2002cd}.

In principle, the medium modified fragmentation function
should be convoluted with the hard partonic cross section
and parton distribution functions in order to determine the
medium modified hadronic spectrum. For illustration, however,
one may exploit that hadronic cross sections 
weigh $D_{h/q}^{(\rm med)}(z,Q^2)$ by the partonic cross section
${d\sigma^q}/{dp_{\perp}^2} \sim {1}/{p_{\perp}^{n(\sqrt{s}, p_{\perp})}}$
and thus effectively test $z^{n(\sqrt{s}, p_{\perp})} 
D_{h/q}^{(\rm med)}(z,Q^2)$~\cite{Eskola:2002kv}. The value
$n = 6$ which characterizes~\cite{Eskola:2002kv} the power law
for typical values at RHIC ($\sqrt{s} = 200$ GeV and $p_T \sim 10$ GeV).
Then, the
position of the maximum $z_{\rm max}$ of  $z^6 D_{h/q}^{(\rm med)}(z,Q^2)$
corresponds to the most likely energy fraction $z_{\rm max} E_q$
of the leading hadron. And the suppression around its maximum translates 
into a corresponding relative suppression of this contribution to the
high-$p_T$ hadronic spectrum at $p_T \sim z_{\rm max} E_q$.
In general, the suppression of hadronic spectra extracted in this
way is in rough agreement with calculations of the quenching
factor (\ref{5.11}).

Medium-modified fragmentation functions have been applied to lepton-nucleus
and nucleus-nucleus collisions in this framework. The idea is to compare
the strength of the effect in both systems. In this way,  
 information about the relative densities of
the media could be obtained. 
In this section we summarize the results obtained up to now
for both lepton-nucleus and AA collisions.
%
\begin{figure}[htb]
\begin{minipage}[t]{80mm}
\vskip -9cm
\includegraphics[width=7cm]{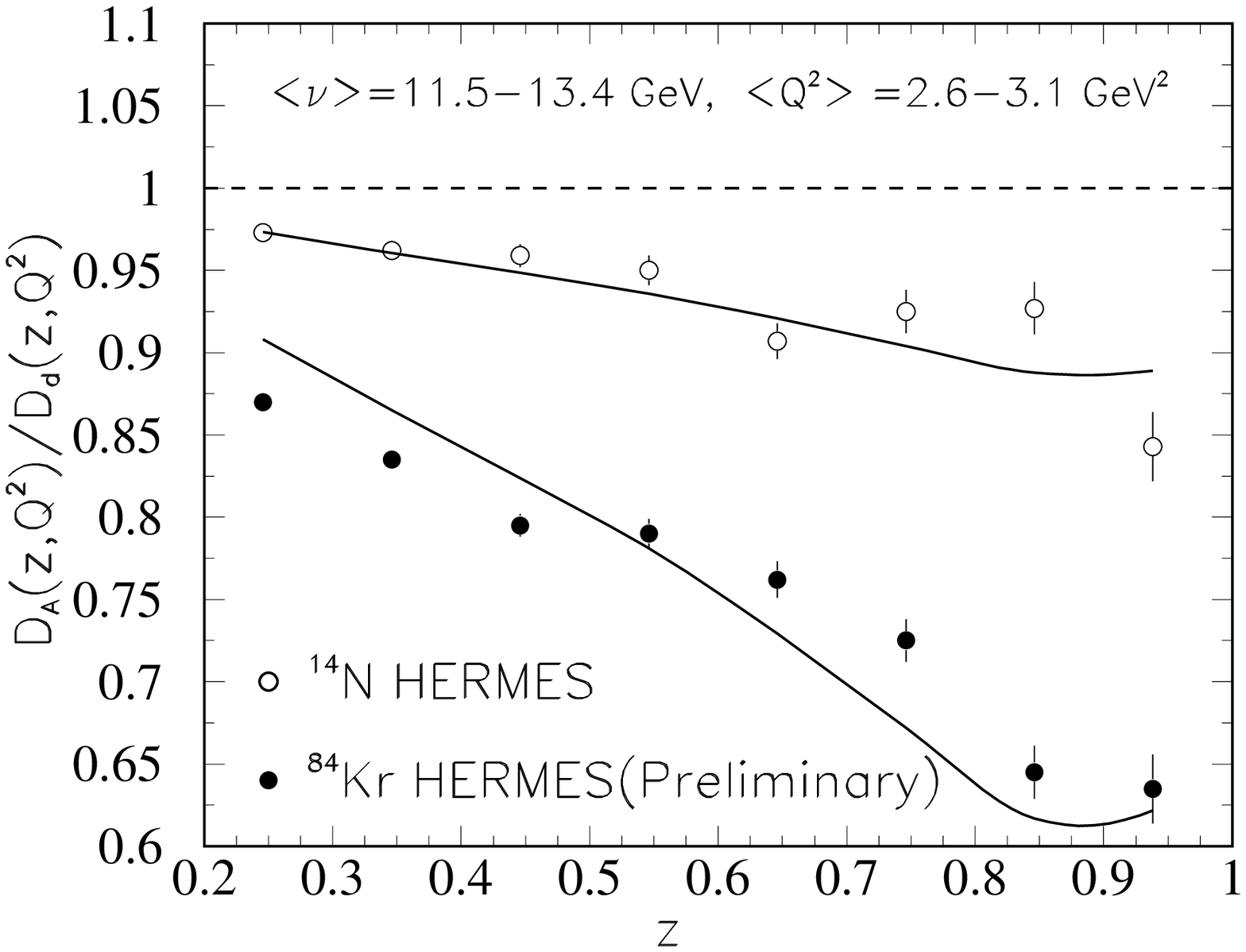}
\caption{Ratios of medium-induced FF over those for vacuum computed
in the twist expansion from Ref. \cite{Wang:2002ri}.}
\label{salgado_fig2}
\caption{
Theoretical calculations with (solid lines) and without (dashed lines) finite
formation times are computed \cite{Arleo:2003jz} in the
BDMPS formalism with $\hat q=0.72 $GeV/fm$^2$. This corresponds to 
$\Delta E/L\sim$ 0.6 GeV/fm. (See Ref. \cite{Arleo:2003jz} for details.)}
\label{salgado_fig3}
\end{minipage}
\hspace{\fill}
\begin{minipage}[t]{75mm}
\includegraphics[width=7cm]{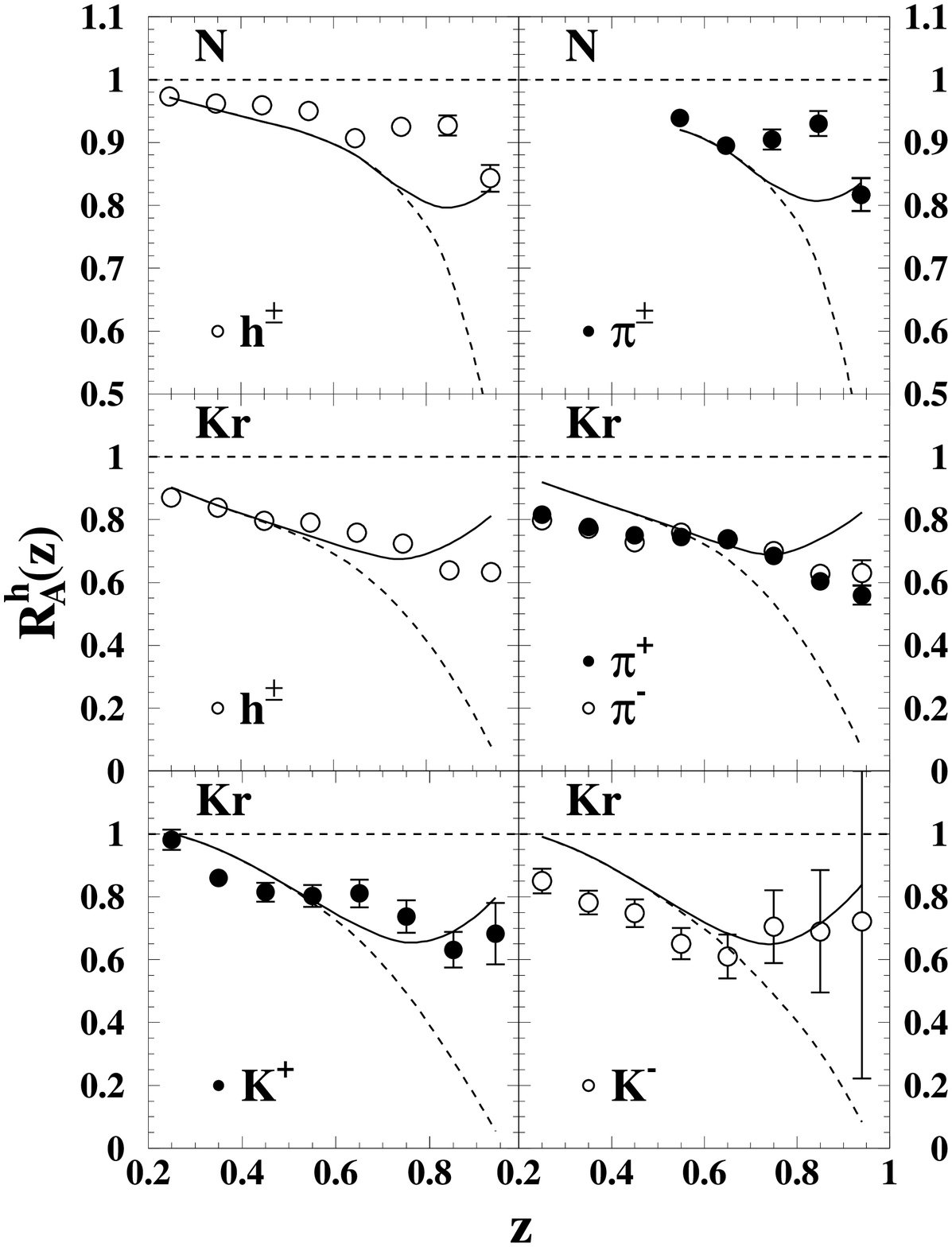}
\end{minipage}
\end{figure}
 
In the case of cold nuclear matter, experiments of lepton-nucleus scattering
measure the inclusive particle cross section for different nuclei. In the
kinematic regime of present experiments, the valence quarks give
the dominant contribution, so, the ratio of cross sections give direct
information about the ratio of FF:

\begin{equation}
R_A^{h}(z,\nu) = \frac{d\sigma_A^h(z,\nu)/d\nu\,dz}
{d\sigma_D^h(z,\nu)/d\nu\,dz}
\simeq \frac{D_q^h(z, Q^2, A)}{D_q^h(z, Q^2, D)}.
\label{salgado_eq5}
\end{equation}

\noindent
where $\nu$ is the energy of the virtual photon.  HERMES experiment
has measured the ratios (\ref{salgado_eq5}) for different nuclei 
\cite{Airapetian:2000ks}.
These data has been
studied in \cite{Wang:2002ri} 
in the the twist-expansion previously mentioned and
in \cite{Arleo:2003jz} using the BDMPS \cite{Baier:1996kr}\cite{Baier:1996sk}
 gluon radiation spectrum for
the energy loss. See Figs.~\ref{salgado_fig2} and~\ref{salgado_fig3} for
the comparison with data.
 Though the two approximations are rather different,
it is interesting that they both result in a similar average energy 
loss $dE/dL\sim $ 0.5 GeV/fm.

\begin{figure}
\begin{center}
\includegraphics[width=8cm]{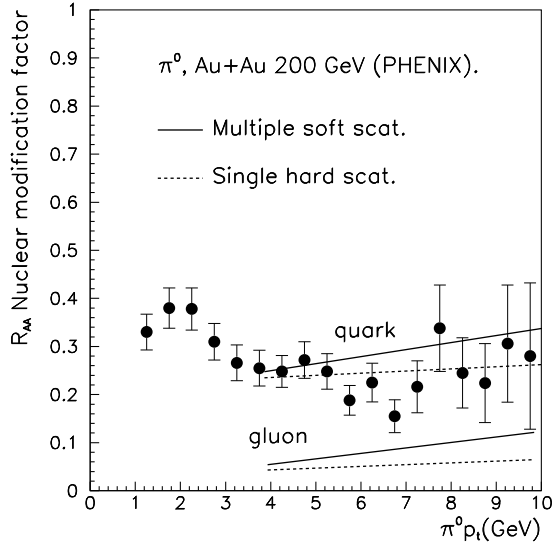}
\caption{Suppression of $\pi^0$ production at high-$p_T$ 
measured by PHENIX collaboration
\cite{Adler:2003qi}. Theoretical suppression is estimated 
\cite{Salgado:2003gb} from ratios of
$z^6 D^{med}_i(z,Q^2)$ at the maximum (see Fig.~\ref{salgado_fig1}). Solid
lines are for multiple soft scattering approximation and dashed for
single hard scattering in the opacity expansion of Ref.
\cite{Wiedemann:2000za}.} 
\label{salgado_fig4}
\end{center}
\end{figure}

In the case of nucleus-nucleus collisions, the produced medium is 
expected to be the main source of energy loss for particles produced 
in the central rapidity region. The new data from RHIC has been used 
to fix the amount of energy loss needed to reproduce the observed 
suppression of particles produced at large $p_T$. The size of the 
effect is compatible with this {\it jet quenching} explanation. 
The fact that the transport coefficient is proportional to the density 
of the medium allows to estimate the effect for the LHC. This can be 
read out from the lower two panels in Fig.~\ref{salgado_fig1}. For 
instance, the suppression for 10 GeV quarks in a medium of 350 gluons 
per unit rapidity is similar to the one for 50 GeV quarks in a medium 
of five time larger density. In Fig.~\ref{salgado_fig4} the suppression 
using medium-modified fragmentation functions is compared with the 
experimental data from PHENIX \cite{Adler:2003qi}.

\subsubsection{Nuclear Modification Factors} 
\label{sec343}
{\em I. Vitev}

\begin{figure}[htb!]
\begin{center} 
\hspace*{0.3in}
\includegraphics[height=2.7in,width=4.5in,angle=-90]{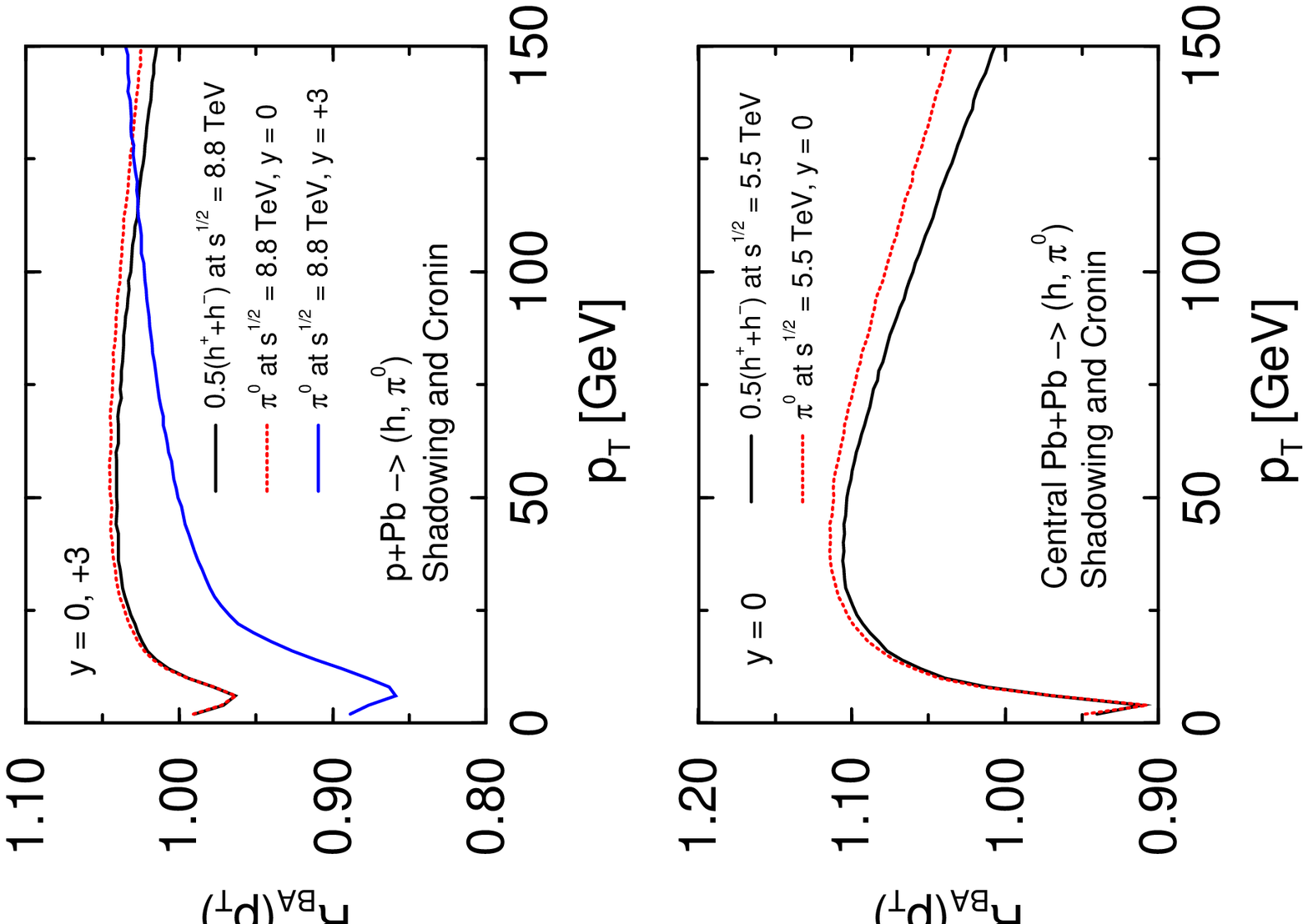}
\includegraphics[height=3.0in,width=4.5in,angle=-90]{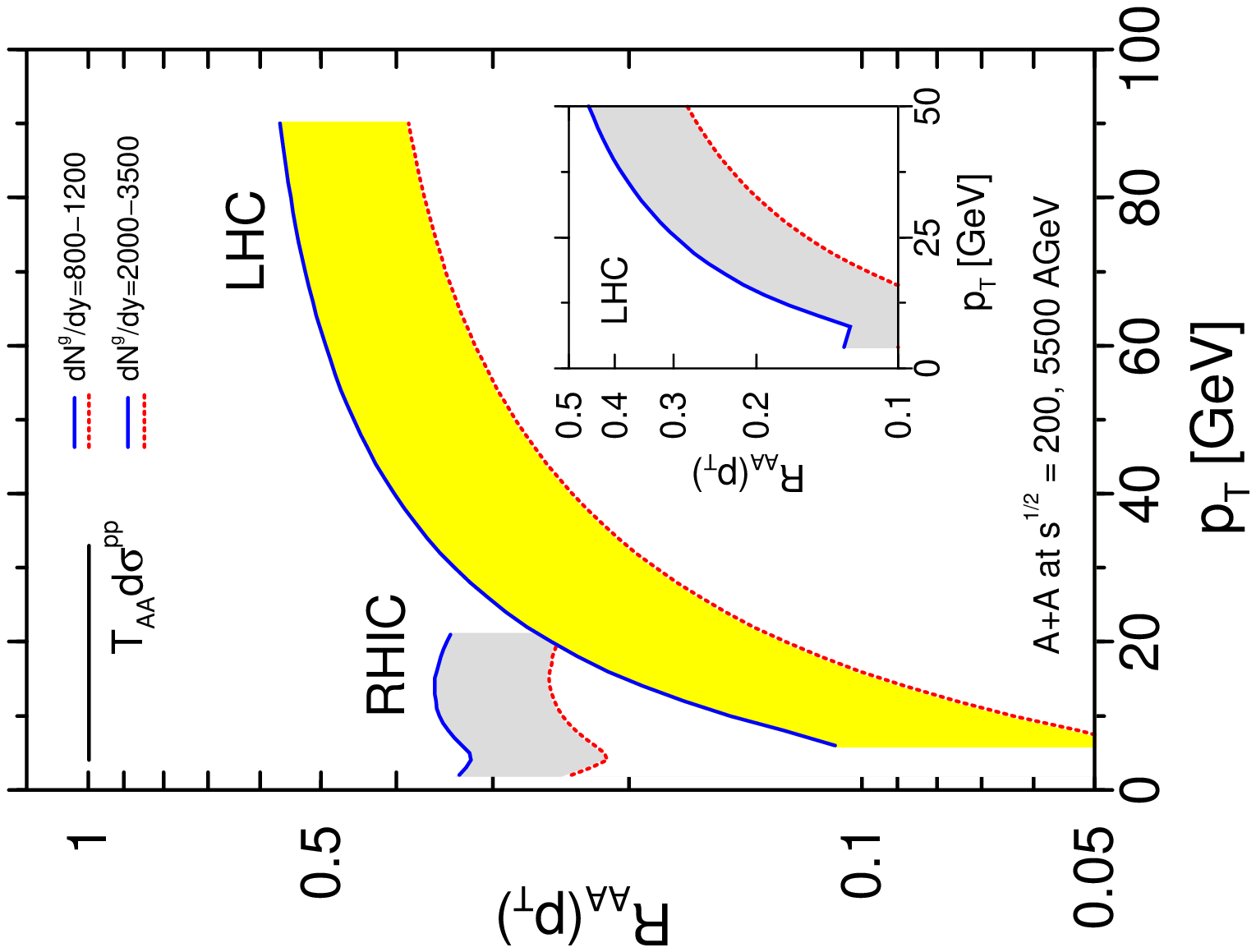}
\caption{\small The antishadowing and Cronin effects 
in $p+Pb$ and central $Pb+Pb$ without energy loss  at 
the LHC ($\sqrt{s}= 5.5$ and $8.8$~TeV)  are shown in the 
left top and bottom panels. The right panel demonstrates
the dominance of final state radiative energy loss effects 
at the LHC with  a much stronger $p_T$ dependence compared 
to RHIC. The possible restoration  of the participant scaling 
through hydrodynamic-like feedback  at $p_T\rightarrow 0$ 
is also shown~\cite{Vitev:2002pf}.  }
\label{lhc-h:fig3}
\end{center} 
\end{figure}

Dynamical  nuclear  effects  in $p+A$ and $A+A$ reactions     
are detectable through  the nuclear modification ratio
\begin{equation}         
R_{BA}(p_T)  = \left\{   
\begin{array}{ll}  \displaystyle 
\frac{d \sigma^{pA}}{dyd^2{p}_T} / 
\frac{ A \; d \sigma^{pp}} {dyd^2{p}_T} \;\; &  
{\rm in}  \; \;    p+A  \\[2.8ex]  \displaystyle 
\frac{d N^{AA}(b)}{dyd^2{p}_T} / 
\frac{ T_{AA}(b)\; d \sigma^{pp}}{dyd^2{p}_T}
   \;\;  &    {\rm in} \;  \; A+A  \;  \end{array} \right. ,  
\label{geomfact} 
\end{equation}          
where $A$  and $T_{AA}({b}) = 
\int  d^2{\bf r} \, T_A({\bf r})T_B({\bf r}-{\bf b})$ 
in terms  of nuclear thickness functions 
$T_A(r)=\int dz \,\rho_A({\bf r},z)$ 
are the corresponding Glauber scaling factors~\cite{Glauber:1970jm} 
of $d\sigma^{pp}$.  We note that in $R_{BA}(p_T)$ the uncertainty 
associated with the $K_{NLO}$ factors, discussed in the previous 
sections, drops out.  The reference calculations that follow  
include shadowing/antishadowing/EMC-effect (here referred to as 
``shadowing''), the Cronin effect, and the non-abelian energy 
loss of jets. The scale  dependent nuclear 
PDFs read: $f_{\alpha/A}(x,Q^2) = S_{\alpha/A} (x,Q^2) \,  ( Z/A \,  
 f_{\alpha/p}(x,Q^2) +  N/A \, f_{\alpha/n}(x,Q^2) )$,  where we 
take the isospin  effects on average and the EKS'98 
parameterization~\cite{Eskola:1998df} of the  shadowing  
functions  $S_{\alpha/A} (x,Q^2)$.  Initial   state 
multiple elastic  scatterings  have been discussed 
in~\cite{Accardi:2001ih,Qiu:2001hj,Gyulassy:2002yv}. 
From~\cite{Gyulassy:2002yv} the transverse momentum distribution 
of partons  that have undergone an average  $\chi = L/\lambda$  
incoherent interactions  in the medium can be evaluated 
exactly for any initial flux $dN^{i}({\bf p})$:
\begin{equation}
\frac{dN^f ({\bf p})}{d^2 {\bf p}} 
= \sum_{n=0}^\infty e^{-\chi} \frac{\chi^n}{n!}  
\int \prod_{i=1}^n d^2 {\bf q}_i \; \left[ \frac{1}{\sigma_{el}} 
 \frac{d \sigma_{el} }{d^2 {\bf q}_i}  \right] \;  
\frac{dN^{i}}{d^2{\bf p}}
({\bf p} -{\bf q}_1 - \cdots - {\bf q}_n)  \;. 
\label{glauber}
\end{equation} 
Numerical estimates of~(\ref{glauber}) 
show that for thin media with a few  semi-hard  
scatterings  the induced  transverse  momentum   broadening 
exhibits a weak logarithmic enhancement  with $p_T$  and  is
proportional to $L \propto A^{1/3}$. The transverse 
momentum transfer per unit length in cold nuclear matter 
is found to be  
$\mu^2 / \lambda_q \simeq 0.05$~GeV$^2$/fm~\cite{Vitev:2002pf}
from comparison to low energy $p+A$ 
data~\cite{Cronin:zm,Straub:xd,Antreasyan:cw}.  The left top and 
bottom panels of Fig.~\ref{lhc-h:fig3} show the predicted 
Cronin+shadowing effect in $p+Pb$ collisions at $\sqrt{s}=8.8$~TeV
and  central $Pb+Pb$ at $\sqrt{s}=5.5$~TeV without final 
state medium induced energy loss.  The 4\% (10\%) enhancement 
of $R_{BA}$ at $p_T \simeq 40$~GeV comes from antishadowing and in 
not related no multiple initial state scattering. 
The observed difference  between $\pi^0$ and $0.5 (h^+ + h^-)$ reflects 
the different  $S_\alpha(x,Q^2)$ for quarks and gluons. 
Cronin effect at  the LHC results  in  slowing 
down of the decrease of $R_{BA}$ at  small $x$ as seen 
in  the  $p_T \rightarrow 0$  limit. In contrast, at  RHIC  one  
finds  $\simeq 30\%$  enhancement in central $d+Au$ reactions  
at $\sqrt{s}=200$~GeV  and  $\simeq 60\%$  effect in central 
$Au+Au $  relative  to  the  {\em binary collision}  scaled  $p+p$  
result~\cite{Vitev:2002pf,Vitev:2003xu}. At CERN-SPS energies 
of $\sqrt{s}= 17$~GeV  the  results 
are most striking, with values reaching 250\% in $d+Au$ 
and 400\%  in central $Au+Au$ at $ p_T \simeq 4$~GeV.  
For a summary of results on midrapidity Cronin effect at the LHC 
see~\cite{pAwriteup,Accardi:2003jh}.     
       
The manifestation of multiple initial state scattering  
and nuclear shadowing at forward and backward rapidities 
$y=\pm 3$ in $p+Pb$ at the LHC (for CMS $\eta \leq 2.5$)   
and $d+Au$  at RHIC (for BRAHMS $ \eta \leq 3$ ) has also 
been studied in the framework of a fixed (or slowly varying)  
initial parton interaction strength~\cite{Vitev:2003xu}. At  LHC  
energies at  $y=+3$ (in the direction of the proton beam) the effect of the 
sequential projectile interactions is again small (due to the much 
flatter rapidity and transverse momentum distributions) and 
is overwhelmed  by  shadowing,  which is found  to  be a factor 
of  2-3  times larger than the $y=0$ result 
at small $p_T \sim$ few GeV and vanishes ($R_{BA}=1$) at 
$p_T \simeq 50$~GeV. As previously emphasized, initial state gluon 
showering can significantly change the low-$p_T$ behavior of 
the hadronic spectra at the LHC beyond  the  current  shadowing 
parameterization. At RHIC in $d+Au$ reactions  at $\sqrt{s} = 200$~GeV 
the nuclear modification ratio is qualitatively different. 
While near nucleus beam  (backward $y=-3$)  rapidity 
$R_{BA} \simeq 0.9 - 1$ at forward rapidities $y=+3$ the nuclear 
modification factor exhibits a  much more dramatic $p_T$ dependence. 
At  small transverse momenta  $p_T \sim 1$~GeV  hadron 
production is suppressed  relative to the
binary collision scaled $p+p$ result, $R_{BA} \leq 0.8$. The maximum 
Cronin enhancement $R_{BA}^{\max} \simeq 1.3$ (30\%) is essentially the
same as at midrapidity  but slightly  shifted 
to larger $p_T$. We emphasize  that {\em both} the suppression 
and enhancement regions are an integral 
part of the Cronin effect~\cite{Cronin:zm,Straub:xd,Antreasyan:cw} 
that is understood in terms of probability  conservation and momentum 
redistribution resulting from  multiple  initial state 
scattering~\cite{Vitev:2002pf,Zhang:2001ce,Glauber:1970jm,Accardi:2001ih,Qiu:2001hj,Gyulassy:2002yv,pAwriteup,Accardi:2003jh}.  
At forward (in the direction of the deuteron beam) rapidities a calculation as 
in~\cite{Vitev:2002pf,Vitev:2003xu} demonstrates a {\em  broader} 
Cronin enhancement  region with $R_{BA} \simeq 25 \%$  at 
$p_T=5$~GeV. This is understood in terms of the significantly 
steeper fall-off of the hadron spectra away from midrapidity that 
enhances the effect of the otherwise similar transverse momentum kicks.  
While the discussed moderate $p_T$ interval~\cite{Vitev:2003xu} lies at 
the very  edge of BRAHMS acceptance (at $y=+3$) the same qualitative  
picture holds at $y=+2$. 

The full solution for the medium induced gluon radiation 
off jets produced in a hard collision inside  the  nuclear  
medium  of length $L$ and computed {\em to all orders} 
in the correlations  between the  multiple  scattering  centers  
via the GLV  reaction operator approach~\cite{Gyulassy:2000er} 
has been discussed in section~\ref{sec313}.
At large jet energies the lowest order 
correlation between the jet production point one of the 
scatterings that follow has been shown to dominate and 
lead to a quadratic  mean energy loss  
dependence on the size of the plasma, $\Delta E \propto L^2$  
for {\em static} media~\cite{Gyulassy:2000fs}.  To improve 
the numerical accuracy for small parton energies  we 
include corrections to third order in opacity~\cite{Vitev:2002pf}  
from Eq.~(\ref{difdistro}). 
The dynamical expansion of the bulk soft matter is assumed to be 
of Bjorken type. In the Poisson approximation of independent gluon 
emission~\cite{Baier:2001yt,Gyulassy:2001nm,Wang:2002ri,Salgado:2002cd}   
the probability distribution $P(\epsilon,E)$ of the fractional 
energy loss  $\epsilon=\sum_i \omega_i/E$ can be obtained iteratively 
from the single inclusive gluon radiation  spectrum 
$dN(x,E)/dx$~\cite{Gyulassy:2001nm} as in Eq.~(\ref{poisiter}). 
If a fast parton looses  $\epsilon \, E$ of its initial energy prior 
to hadronization its momentum fraction $z_c$ is modified to  
$z_c^* = p_h/p_c(1-\epsilon) = z_c/(1-\epsilon)$. The observable 
suppressed  hadron  differential cross section can be 
computed  from  Eq.~(\ref{hcrossec}) with the substitution 
\begin{equation} 
D_{h/c}(z_c,Q^2) \longrightarrow  \int d \epsilon  \, P(\epsilon,p_c) 
\frac{z_c^*}{z_c} D_{h/c}(z^*_c,Q^2)    \; . 
\label{fragmod}
\end{equation}   
The nuclear  modification  factor $R_{AA}(p_T)$ at the LHC is 
shown on the right panel of Fig.~\ref{lhc-h:fig3} and is completely 
dominated by final state interactions (see left panel). It shows a 
{\em significantly stronger}  $p_T$  dependence as compared to RHIC,
where jet quenching was predicted to be {\em approximately 
constant} over the full measured 
moderate- to high-transverse momentum range~\cite{Vitev:2002pf} -    
the result of an interplay of shadowing, Cronin effect, and  
radiative energy loss.  The  variation of  $R_{AA}$ at the LHC 
is a  factor of 5: from  10-20 fold suppression 
at $p_T = 10$~GeV to only a factor  2-3 suppression 
at $p_T = 100$~GeV. The reason for such a prominent variation   
is the hardening of the particle transverse momentum spectra 
(see Fig.~\ref{lhc-h:fig2})
and the   insufficient  balancing action of multiple initial state 
scattering. In fact, the prediction from  Fig.~\ref{lhc-h:fig3}  
is  that the suppression  in central $Pb+Pb$ at $\sqrt{s}_{NN}=5.5$~TeV 
at $p_T \simeq 40$~GeV is comparable to the factor of 4-5 
suppression currently observed at RHIC.

The extrapolation of the LHC quenching calculations to small 
$p_T \rightarrow 0$ results into suppression below participant scaling. 
More careful examination of the mean energy loss of partons, 
in particular for  gluons radiating in  nuclear matter at
LHC densities, reveals sizable regions of phase  space 
with $\Delta E \geq E$. This indicates complete absorption of 
jets in nuclear matter. There is experimental evidence that this 
regime of extreme {\em final state} densities may have been 
achieved at RHIC~\cite{Mioduszewski:2002wt,Jacobs:2002pz,Kunde:2002pb}.
In this case Eq.~(\ref{fragmod}) has to be corrected to 
include the feedback of the radiated gluons into the system. 
This hydrodynamic-like feedback  is expected to recover
the $N_{part}$ scaling  in the soft $p_T$  region~\cite{Vitev:2002pf} - 
also illustrated  on the right panel of Fig.~\ref{lhc-h:fig3}. 
The effective initial gluon densities derived from the rapidity  
densities used in Fig.~\ref{lhc-h:fig3} are $\rho_g(RHIC) = 
30-50$/fm$^3$ and $\rho_g(LHC) = 130-275$/fm$^3$. These are 
one to two  orders of magnitude larger than  the density 
of cold nuclear matter and are suggestive of a deconfined 
QCD state  -  the quark-gluon plasma.  Interestingly, 
a recent study of  non-equilibrium  parton transport 
in central $Au+Au$ and $Pb+Pb$ at $\sqrt{s}_{NN}= 200$~GeV and 
$\sqrt{s}=5.5$~TeV  has found  initial  parton  densities  
corresponding to the lower bound of the intervals quoted 
above~\cite{Cooper:2002td}.

\begin{figure}[!t]
\begin{minipage}[t]{75mm}
\includegraphics[width=7cm]{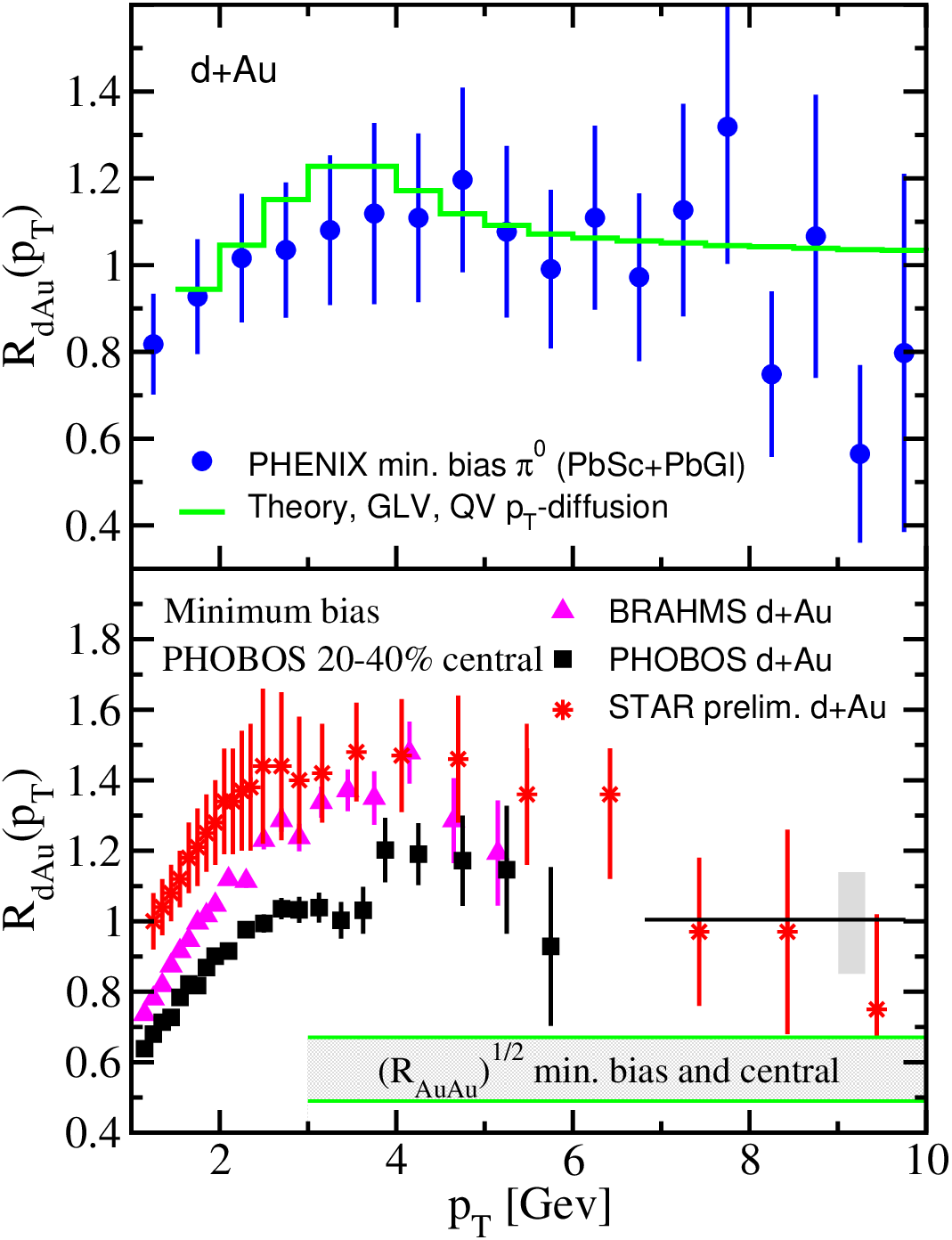}
\caption{Comparison of the prediced Cronin effect 
from~\cite{Vitev:2002pf,Vitev:2003xu} to the measured small 
enhancement of  single inclusive neutral pion production in $d+Au$ at 
$\sqrt{s}_{NN}=200$~GeV. Data is from PHENIX~\cite{Adler:2003ii}.  
Bottom panel: a test of a suggested  interpretation of high 
$p_T$-hadron suppression as a  result of initial state 
wavefunction modification, $R_{dAu} \approx \sqrt{R_{AuAu}}$. 
Data  from BRAHMS, PHOBOS  and 
STAR~\cite{Arsene:2003yk,Back:2003ns,Adams:2003im} exclude 
such possibility. Figure adapted from~\cite{Vitev:2003jg}. 
}
\label{cipanp:fig1b}
\end{minipage}
\hspace{\fill}
\begin{minipage}[t]{80mm}
\includegraphics[width=7cm]{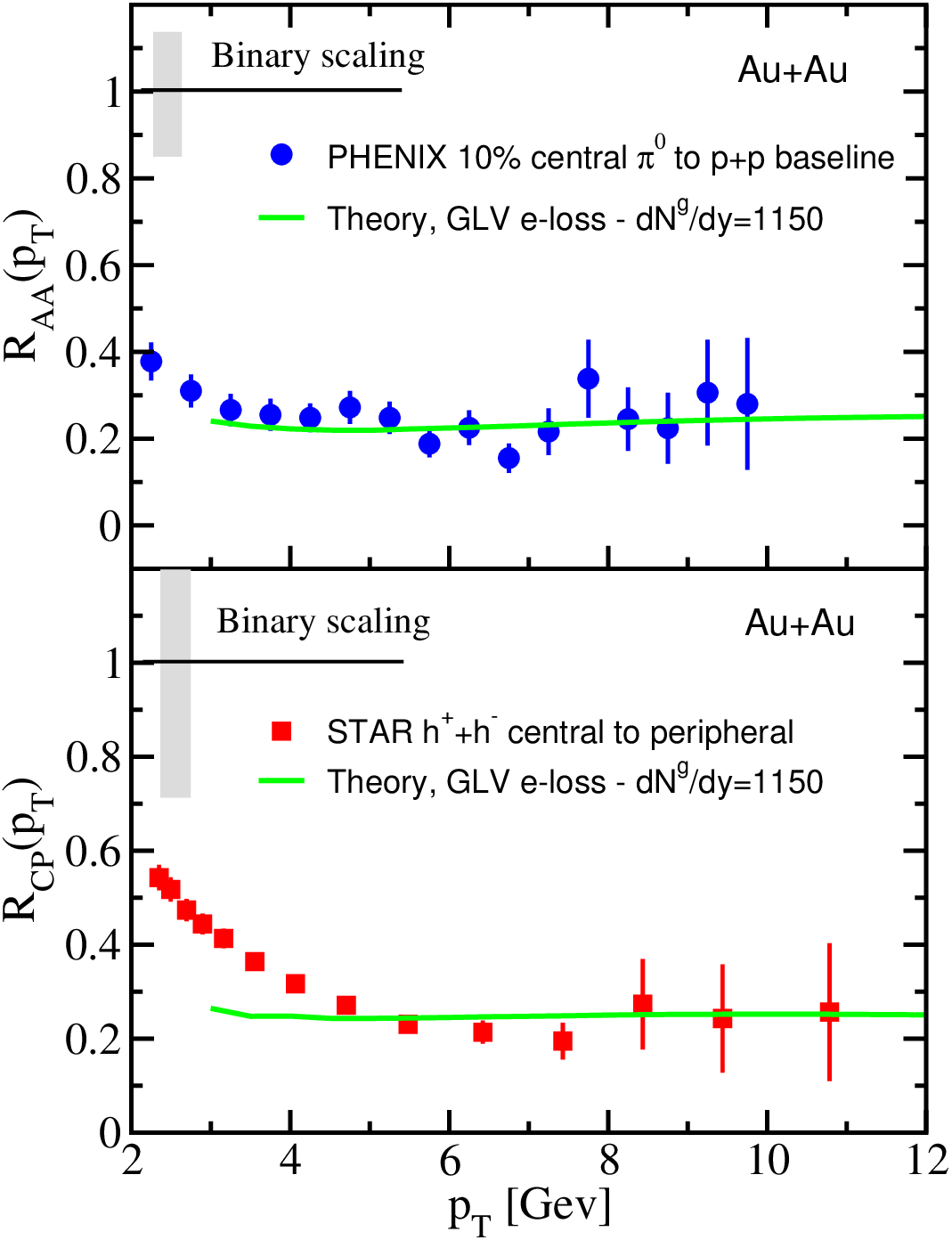}
\caption{Predicted~\cite{Vitev:2002pf} suppression for $\pi^0$ (top panel)
and $h^+ + h^-$ (bottom panel) in $Au+Au$ compared 
to PHENIX and STAR data~\cite{Adler:2003qi,Adams:2003kv}. Similar 
quenching is found by BRAHMS and PHOBOS~\cite{Back:2003qr,Arsene:2003yk}.
Figure adapted from~\cite{Vitev:2003jg}.
}
\label{cipanp:fig1a}
\end{minipage}
\end{figure}

At the time of the completion of the CERN  Yellow Report experimental
data on hadroproduction at RHIC $\sqrt{s}=200$~AGeV in 
$d+Au$ and $Au+Au$ reactions became available for comparison to 
theoretical predictions. In the top panel of Fig.~\ref{cipanp:fig1b} the 
Cronin enhancement, resulting from initial 
state parton broadening~\cite{Vitev:2002pf,Qiu:2003pm} is seen to  
compare qualitatively to the shape of the 
PHENIX $\pi^0$ measurement~\cite{Adler:2003ii} in minimum bias $d+Au$.
Larger enhancement of $h^++h^-$ production, consistent 
with results form low energy $p+A$ data, is also 
shown~\cite{Arsene:2003yk,Back:2003ns,Adams:2003im}.  The 
bottom  panel rules out the scenario for the initial wavefunction  
origin of moderate and high-$p_T$ hadron suppression, see 
Fig.~\ref{cipanp:fig1a}, since in this case 
$R_{dAu} \approx \sqrt{R_{AuAu}}$. 
Fig.~\ref{cipanp:fig1b} compares the predicted~\cite{Vitev:2002pf} 
approximately constant suppression of $\pi^0$ and $h^+ + h^-$ 
in $\sqrt{s} = 200$~AGeV  $Au+Au$ collisions at RHIC to 
PHENIX and STAR data~\cite{Adler:2003qi,Adams:2003kv}. 
The overall quenching magnitude and its centrality dependence are set 
by $(\langle L \rangle / A_T) dN^g/dy \propto N_{part}^{2/3}$, 
$dN^g/dy=1150$.  We again emphasize that the shape of $R_{AuAu}$ 
is a result of the interplay of all three nuclear effects: Cronin, 
shadowing, and jet quenching. The full numerical 
calculation takes into account the dynamical Bjorken 
expansion of the medium, finite kinematic bounds, higher
order opacity corrections and approximates multiple gluon 
emission by a Poisson distribution~\cite{Vitev:2002pf}. 
The remarkably good agreement between the {\em predicted} nuclear 
modification factors  and the experimental measurements give 
confidence in projecting the anticipated nuclear effects over
a much wider dynamical moderate- and high-$p_T$ range at the LHC.

{\bf Conclusions:} 
In central $A+A$ reactions  the nuclear 
modification factor $R_{AA}(p_T)$ at the LHC is shown to be  
completely dominated by final state multi-parton 
interactions~\cite{Vitev:2002pf}. For comparison, at RHIC 
Cronin effect and nuclear shadowing also play an important role, 
leading to an approximately constant suppression ratio. At the SPS  
initial state multiple elastic scatterings dominate, resulting 
in a net enhancement of hadron production.  At forward ($y=+3$) 
rapidities in $d+Au$ at RHIC the Cronin enhancement region is predicted 
to be broader in comparison to the $y=0$ case. In contrast in $p+Pb$ at 
the LHC nuclear shadowing dominates but in order to detect a sizable  
reduction relative to the binary collision scaled $p+p$ cross section 
measurements at close to proton rapidity ($y_{\max}=9.2$ for 
$\sqrt{s}=8.8$~TeV) are needed. 
    
The predicted decreasing $R_{AA}$ with  $p_T$ at the LHC, if confirmed, 
may have important experimental consequences. Comparative 
large-$E_T$ measurements of the difference in the 
{\em full structure} of the jet cone in $p+p$ and $A+A$ reactions 
may  prove difficult for weak signals and large
backgrounds. We emphasize that one of the easiest 
and most unambiguous approaches  for detecting  the non-abelian jet 
energy loss  and  performing 
jet-tomographic analysis of the properties of the hot and dense 
matter created  in ultra-relativistic heavy ion reactions is 
through  the suppression pattern of leading hadrons. Therefore these 
measurements should enter as an important part of the experimental
programs at the LHC.

\subsubsection{Heavy Quark Energy Loss Observables}
\label{sec344}
{\em R. Vogt}

Heavy quarks (HQ) are good probes of the QCD medium produced in heavy ion
collisions.  They are produced perturbatively in the initial hard
nucleon-nucleon collisions at timescales on the order
of $1/m_Q$.  Their production during other stages of the evolution of the
system is unlikely, except perhaps in the pre-equilibrium phase of the plasma, 
because $m_Q \gg T$.  (See 
Refs.~\cite{Shor:xc,Shor:yt,Muller:xn,Levai:1997bi} for some estimates of
thermal charm production.)  Thus the initially produced heavy quarks experience
the full collision history.

While the heavy quarks are in the medium, they can undergo energy loss by
two means: elastic collisions with light partons in the system (collisional)
and gluon bremsstrahlung (radiative).  We will briefly review some of the
predicted results for $-dE/dx$ of heavy quarks for both collisional and
radiative loss.  We then show the predicted effect on the charm and bottom
contributions to the dilepton continuum for both ALICE and CMS using a fixed
value of $-dE/dx$.

The collisional energy loss of heavy quarks through processes such as 
$Qg \rightarrow Qg$ and $Qq \rightarrow Qq$ depends logarithmically on
the heavy quark momentum, $-dE/dx \propto \ln(q_{\rm max}/q_{\rm min})$.
Treatments of the collisional loss vary with the values assumed or calculated
for the cutoffs.  These cutoffs are sensitive to the energy of the heavy quark
and the temperature and strong coupling constant in the medium.  Thus the
quoted value of the energy loss is usually for a certain energy and
temperature.  The calculation was first done by Bjorken
\cite{Bjorken:1982tu} who found $-dE/dx \approx 0.2$ GeV/fm for a 20 GeV quark
at $T = 250$ MeV.  Further work refined the calculations of the cutoffs
\cite{Thoma:1990fm,Thomas:1991ea,Mrowczynski:da}, with similar results.  
Braaten and Thoma
calculated the collisional loss in the limits $E \ll m_Q^2/T$ and $E \gg
m_Q^2/T$ in the hard thermal loop approximation, removing the cutoff
ambiguities.  They obtained $-dE/dx \approx 0.3$ GeV/fm for a 20 GeV charm
quark and 0.15 GeV/fm for a 20 GeV bottom quark at $T = 250$ MeV
\cite{Braaten:1991we}. 

Other models of heavy quark energy loss were presented in the context of
$J/\psi$ suppression:  Could a produced $c \overline c$ pair stay together in
the medium long enough to form a $J/\psi$?  Svetitsky \cite{Svetitsky:gq} 
calculated
the effects of diffusion and drag on the $c \overline c$ pair in the Boltzmann
approach and found a strong effect.  The drag\footnotetext{His drag coefficient
$A(p^2)$ is related to the energy loss per unit length through $A(p^2) =
(-dE/dx)/p^2$.} stopped the $c \overline c$ pair after traveling about 1 fm
but Brownian diffusion drove the $c$ and $\overline c$ apart quickly.  
The diffusion effect
increased at later times.  Essentially he predicted that the heavy quarks would
be stopped and then go with the flow.  His later calculations of $D$ meson
breakup and rehadronization \cite{Svetitsky:1996nj} 
while moving through plasma droplets
reached a similar conclusion.  Koike and Matsui calculated the energy loss of a
color dipole moving through a plasma using kinetic theory and found
$-dE/dx \sim 0.4-1.0$ GeV/fm for a 10 GeV $Q \overline Q$ \cite{Koike:xs}.

Thus the collisional loss was predicted to be rather small, less than 1 GeV/fm
for reasonable assumptions of the temperature.  The loss increases with the
energy and temperature.  Using the hard thermal loop approach, Mustafa
{\it et al.} found $-dE/dx \approx 1-2$ GeV/fm for a 20 GeV quark at
$T = 500$ MeV \cite{Mustafa:1997pm}. 

\begin{figure}[htb]
\setlength{\epsfxsize=0.95\textwidth}
\setlength{\epsfysize=0.25\textheight}
\centerline{\epsffile{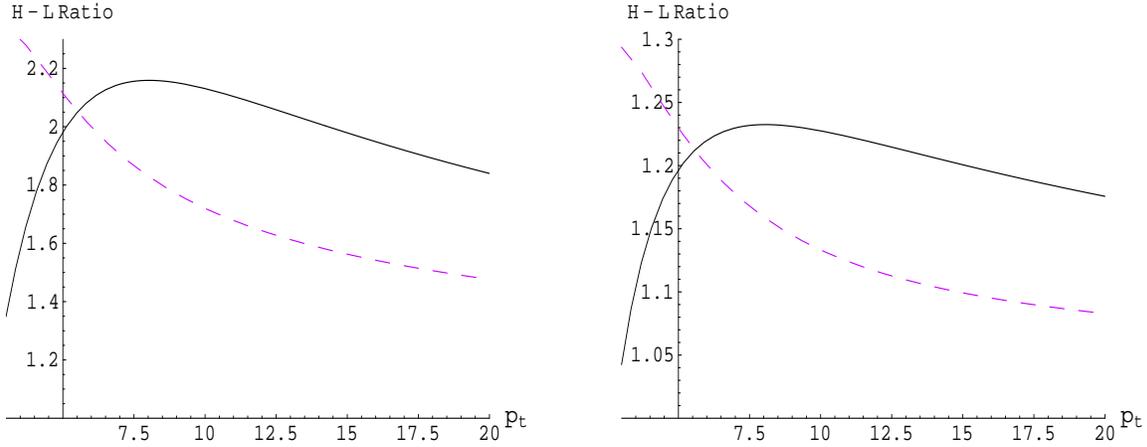}}
\caption{The ratio of quenching factors $Q_H(p_{\perp})/Q_L(p_{\perp})$ 
for charm and light quarks 
in hot matter with $\hat{q}=0.2$ GeV$^3$ ($L=5$ fm upper panel,
$L=2$ fm lower panel).  Solid lines correspond to unrestricted
gluon radiation, while the dashed lines are based on the calculation
with the cut $\omega>0.5$ GeV on gluon energies. From 
Ref.~\cite{Dokshitzer:2001zm}.}
\label{Fhotadj}
\end{figure}

Radiative energy loss of light quarks has been extensively studied and is
discussed in detail in the remainder of this chapter.  The first application
of radiative loss to heavy quarks was perhaps by Mustafa {\it et al.}
\cite{Mustafa:1997pm}.  
They included the effects of only a single scattering/gluon
emission, $Q q \rightarrow Qqg$ or $Q g \rightarrow Qgg$.  In this case,
the loss grows as the square of the logarithm $\ln(q_{\rm max}/q_{\rm min})$,
one power more than the collisional loss, but is of the same order in the
strong coupling constant~\cite{Braaten:1991we}.  
Thus the radiative loss is guaranteed to be higher
than the collisional in this approximation.  
The heavy quark mass enters their expressions only
in the definition of $q_{\rm max}$ so that the mass dependence of the energy 
loss is rather weak.  They found, for a 20 GeV quark at $T = 500$ MeV,
$-dE/dx \approx 12$ GeV/fm for charm and 10 GeV/fm for bottom.

These large values suggested that energy loss could be quite important for
heavy quarks.  If true, there would be a strong effect on the $Q \overline Q$
contribution to the dilepton continuum.  Shuryak \cite{Shuryak:1996gc} 
was the first
to consider this possibility for $AA$ collisions.  He assumed that low mass $Q
\overline Q$ pairs would be stopped in the medium, suppressing the dilepton
contribution from these decays substantially.  However, the stopped
heavy quarks should
at least expand with the medium rather than coming to rest, as discussed by
Svetitsky \cite{Svetitsky:gq}.  Lin {\it et al.} then calculated the effects of
energy loss at RHIC, including thermal fluctuations, for a constant $-dE/dx =
0.5-2$ GeV/fm \cite{Lin:1997cn}.  These results showed that the heavy quark
contributions to the dilepton continuum would be reduced albeit not completely
suppressed.  In any case, the energy loss does not affect the total cross
section.  The heavy quarks are thus piled up at low $p_T$ and at midrapidity if
stopped completely.  The work by Lin {\it et al.} was extended to the LHC for
$-dE/dx = 1$ GeV/fm \cite{Lin:1998bd}.  These results are shown here.

Other calculations of effects on the dilepton continuum have focused on 
higher mass lepton pairs.
Lokhtin and Snigirev have calculated the effect
of collisional and radiative energy loss on the 
correlated $b \overline b$ contribution to
the dilepton continuum in the CMS acceptance \cite{Lokhtin:2001nh} 
and find a large
depletion in the mass range $20<M<50$ GeV.
If the loss is large, the Drell-Yan and thermal dileptons could emerge from
under the reduced $b \overline b$ decay contribution at large masses.
Gallmeister {\it et al.} \cite{Gallmeister:2002tv} have recently considered
the amount of energy loss that the RHIC charm data can support.

Before presenting the results of the model calculations of 
Ref.~\cite{Lin:1998bd}, 
we note that Dokshitzer and Kharzeev recently pointed
out that soft gluon radiation from heavy quarks is suppressed at angles
smaller than $\theta_0 = m_Q/E$ \cite{Dokshitzer:2001zm}.  
Thus bremsstrahlung is suppressed for heavy
quarks relative to light quarks by the factor $(1 + \theta_0^2/\theta^2)^{-2}$,
the `dead cone' phenomenon.  The radiative energy loss of heavy quarks could
then be quite small.  In fact, PHENIX sees little to no energy loss in their
charm data \cite{Adcox:2002cg}.
These calculations and their implications are discussed
in detail in Chapter \cite{HQwriteup} of this report.  
We show how the ratio of quenching factors for heavy to light quarks,
$Q_H(p_T)/Q_L(p_T)$, depends on $L$, the path length through the medium, in
Fig.~\ref{Fhotadj}. These results are indicative of how the $D/\pi$ ratio 
might be modified in a heavy ion collision. 

We now turn to an illustration of how a constant 1 GeV/fm energy loss might
affect the heavy quark contribution to the dilepton continuum in 5.5 TeV
Pb+Pb collisions at the LHC~\cite{Lin:1998bd}. It is based on the
picture that a heavy quark with a transverse path, $l_T$, and mean-free path,
$\lambda$, undergoes on average $S = l_T/\lambda$ scatterings.
The main model assumption is that the actual number of scatterings, 
$n$, is generated from the Poisson distribution, $P(n,S)=e^{-S} S^n/n!$.  
\begin{figure}[htb]
\setlength{\epsfxsize=10.0cm}
\setlength{\epsfysize=7.0cm}
\centerline{\epsffile{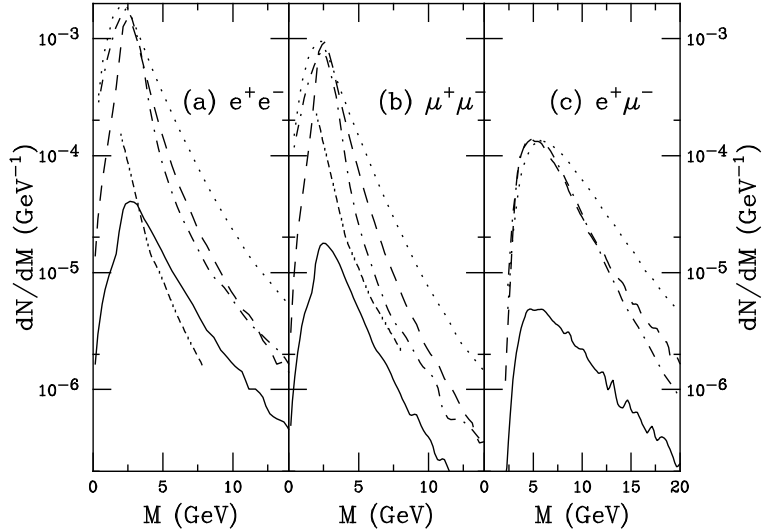}}
\caption{The dilepton invariant mass distributions in the ALICE acceptance.
The $e^+ e^-$ (a), $\mu^+ \mu^-$ (b) and $e \mu$ (c) channels are shown.
The dashed and dotted curves are the $D \overline D$ and summed single $B$ and
$B \overline B$ decays respectively without energy loss.  The solid and 
dot-dashed curves are the corresponding results with $-dE/dx = 1$ GeV/fm.  
The Drell-Yan rate is given by the dot-dot-dashed curve in (a) and (b).
From Ref.~\cite{Lin:1998bd}.}
\label{malice}
\end{figure}
In the rest frame of the medium, the heavy quark then experiences momentum loss
$\Delta p = n \lambda \; dE/dx$ so that its final momentum is $p_T' = p_T - 
\Delta p$.  The heavy quark will thermalize if 
$p^\prime_T$ is smaller than the average transverse 
momentum of thermalized heavy quarks with a temperature $T$.
These thermalized quarks are given a random thermal momentum in the
rest frame of the fluid and are then transformed back to the center-of-mass
frame of the collision. The calculation assumes 
$-dE/dx = 1$ GeV/fm, $\lambda=1$ fm and $T=150$ MeV.  Even a small energy loss
will suppress high $p_T$ and large invariant mass $Q \overline Q$ 
pairs as long as $|dE/dx| \geq \langle p_T 
\rangle/R_A \sim 0.4$ GeV/fm in Pb+Pb collisions at 5.5 TeV
where $\langle p_T \rangle$ is the average transverse momentum 
of the heavy quark and $R_A$ is the nuclear radius.

The results for $D \overline D$ and $B \overline B$ decay contributions to the
dilepton continuum in ALICE
are shown in Fig.~\ref{malice} for the $e^+e^-$, $\mu^+ 
\mu^-$ and $e \mu$ channels.  The pseudorapidity cuts are $| \eta|<0.9$ for
electrons and $2.5 \leq \eta \leq 4$ for muons.  A momentum cut of $p_T > 1$
GeV is used for both lepton types.  Full azimuthal coverage is assumed for both
the central barrel and forward muon arm.  Note that the $B \overline B$
contribution also includes opposite sign lepton pairs from chain decays of 
individual $B$
and $\overline B$ mesons.  In the calculation, 540 $c \overline c$ pairs and 7 
$b \overline b$ pairs were created in central
Pb+Pb collisions using the MRS D-' \cite{Martin:1992zi} parton densities
\cite{Lin:1998bd}. 
More recent parton densities such as MRST HO \cite{Martin:1998sq}
give smaller charm cross sections, $\sim 6$ mb, instead of the 17 mb
obtained with MRS D-. This
would reduce the charm rate relative to the bottom rate by 
nearly a factor of five.  The charm and bottom yields for low mass pairs are
similar without loss but energy loss suppresses the  charm yield much more
strongly than bottom.  The moderate loss assumed here still predicts a larger
$B \overline B$ contribution to the $e^+e^-$ and $\mu^+\mu^-$ continua than the
Drell-Yan yield.
\begin{figure}[htb]
\setlength{\epsfxsize=10.0cm}
\setlength{\epsfysize=7.0cm}
\centerline{\epsffile{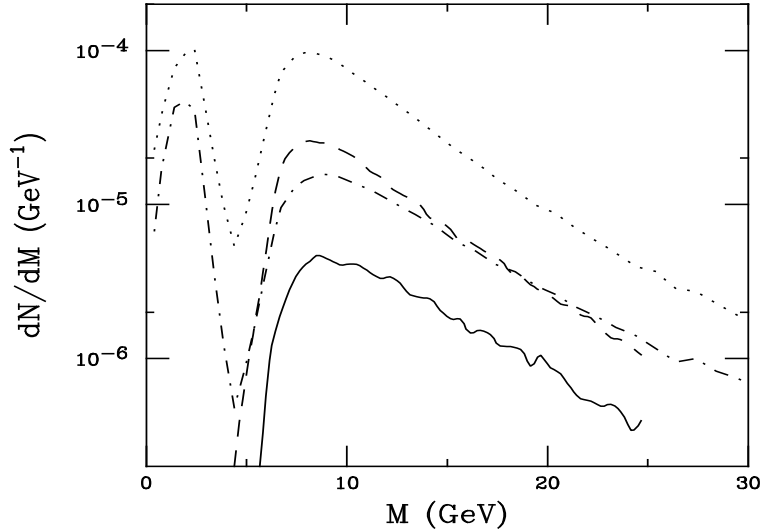}}
\caption{The dilepton invariant mass distributions in the CMS acceptance.
The dashed and dotted curves are the $D \overline D$ and summed single $B$ and
$B \overline B$ decays respectively without energy loss.  The solid and 
dot-dashed curves are the corresponding results with $-dE/dx = 1$ GeV/fm.
From Ref.~\cite{Lin:1998bd}.}
\label{mcms}
\end{figure}

The CMS 
muon acceptance is in the range $|\eta| \leq 2.4$ with a lepton
$p_T$ cut of 3 GeV.  After these simple cuts are applied, the results are
shown in Fig.~\ref{mcms} for both $D \overline D$ and $B \overline B$ decays.
Whereas for $M \leq 15$ GeV, the $D \overline D$ decays would dominate those of
$B \overline B$ before the cuts, the measured $B \overline B$ decays are 
everywhere larger
than those from charm mesons both before and after energy 
loss.  The generally larger momentum of muons from $B$ decays and the rather
high momentum cut result in larger acceptance for $B \overline B$ decays.
No $D \overline D$ decay pairs with $M \leq 5$ GeV survive the momentum cut.
A factor of 50 loss in rate at $M \sim 10$ GeV is found before energy loss.  
A loss in rate
by a factor of 100 is obtained when energy loss is included.  The corresponding
acceptance from $B \overline B$ decays is significantly larger, with a loss in
rate of a factor
of $\approx 8$ before energy loss and $\approx 15$ with energy loss.  
Interestingly, the leptons in the decay chain of a single $B$ meson
are energetic enough for both to pass the momentum cut, causing the peak at
$M \sim 2-3$ GeV.  These results suggest that rather than providing an indirect
measurement of the charm cross section, as postulated in Ref.~\cite{Gavin:ma},
the dilepton continuum above the $\Upsilon$ family could instead measure the $b
\overline b$ production cross section indirectly.  A comparison with the
spectrum from $pp$ interactions at the same energy would then suggest
the amount of energy loss, $-dE/dx$, of the medium.  For a calculation of the
effects of the dead cone on higher mass dileptons in CMS, see 
Ref.~\cite{Lokhtin:2002wu}.

We have so far shown results for the dilepton continuum.  However, the PHENIX
measurement was of single leptons~\cite{Adcox:2002cg}.
The single leptons are not as sensitive to the magnitude of the energy loss as
the dilepton continuum \cite{Lin:1998bd}.

Single leptons can be categorized as those from thermalized
heavy quarks and those from heavy quarks energetic enough to
escape after energy loss. 
The former mainly reflects the effective
thermalization temperature while the latter can provide  
us with information on the energy loss.  
Single leptons with energies greater than $1-2$ GeV 
are mainly from energetic heavy quarks and thus are more sensitive to the
energy loss.  Before energy loss, the single leptons from $D$ decays are larger
than those from $b$ hadron decays for $p_T < 2.5$ GeV.  After energy loss, the
$b$ hadron decays dominate the spectra over all $p_T$.

\begin{figure}[htb]
\begin{minipage}[t]{80mm}
\includegraphics[width=8.0cm]{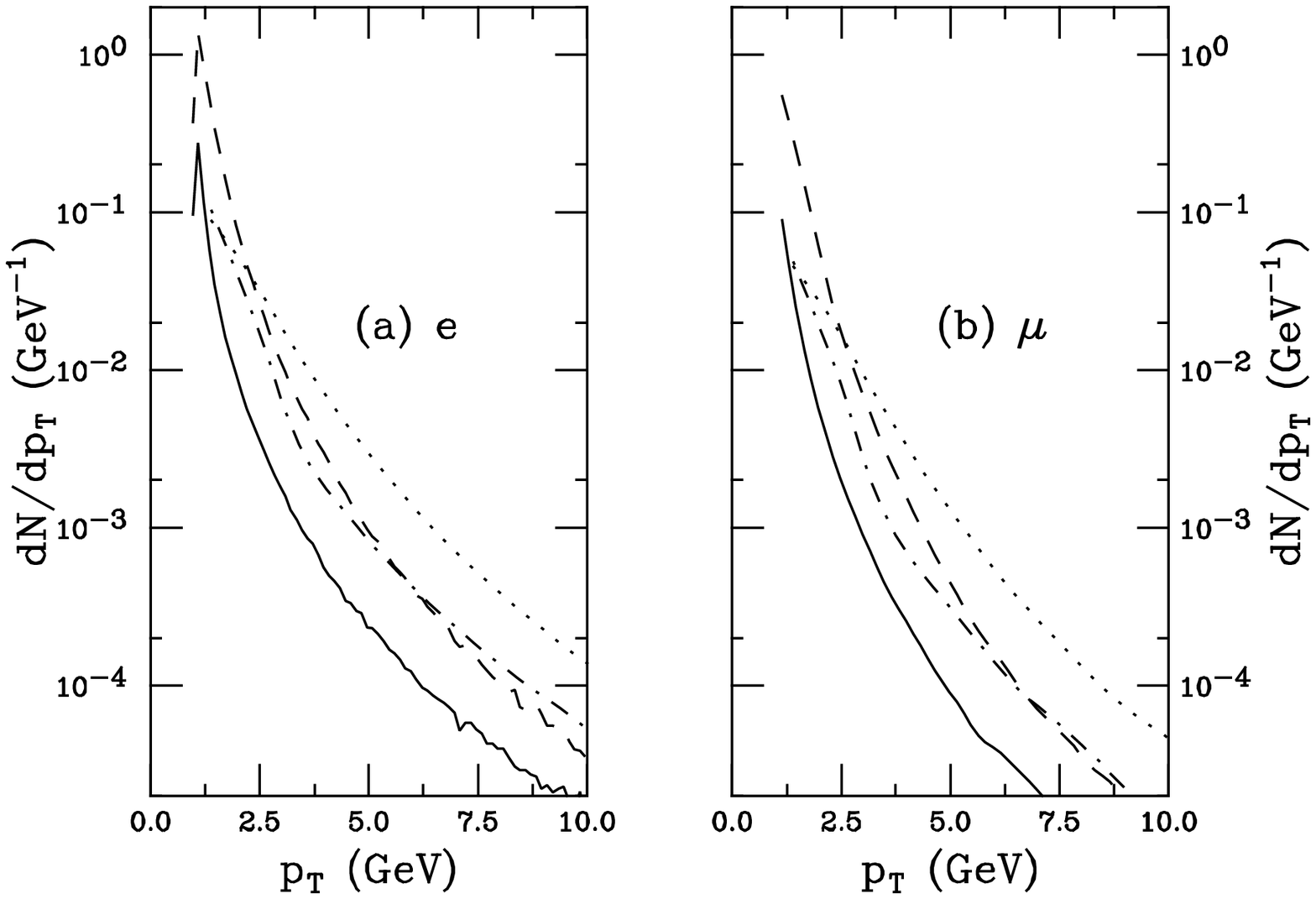}
\caption{The $p_T$ spectrum of single electrons (a) and muons (b) from 
charm and bottom decays within the ALICE acceptance.  
The dashed and dotted curves are the $D$ and $B$ meson 
decays respectively without energy loss.  The solid and 
dot-dashed curves are the corresponding results with $-dE/dx = 1$ GeV/fm. 
From Ref.~\cite{Lin:1998bd}.
}
\label{ptalice}
\end{minipage}
\hspace{\fill}
\begin{minipage}[t]{80mm}
\includegraphics[width=6.7cm]{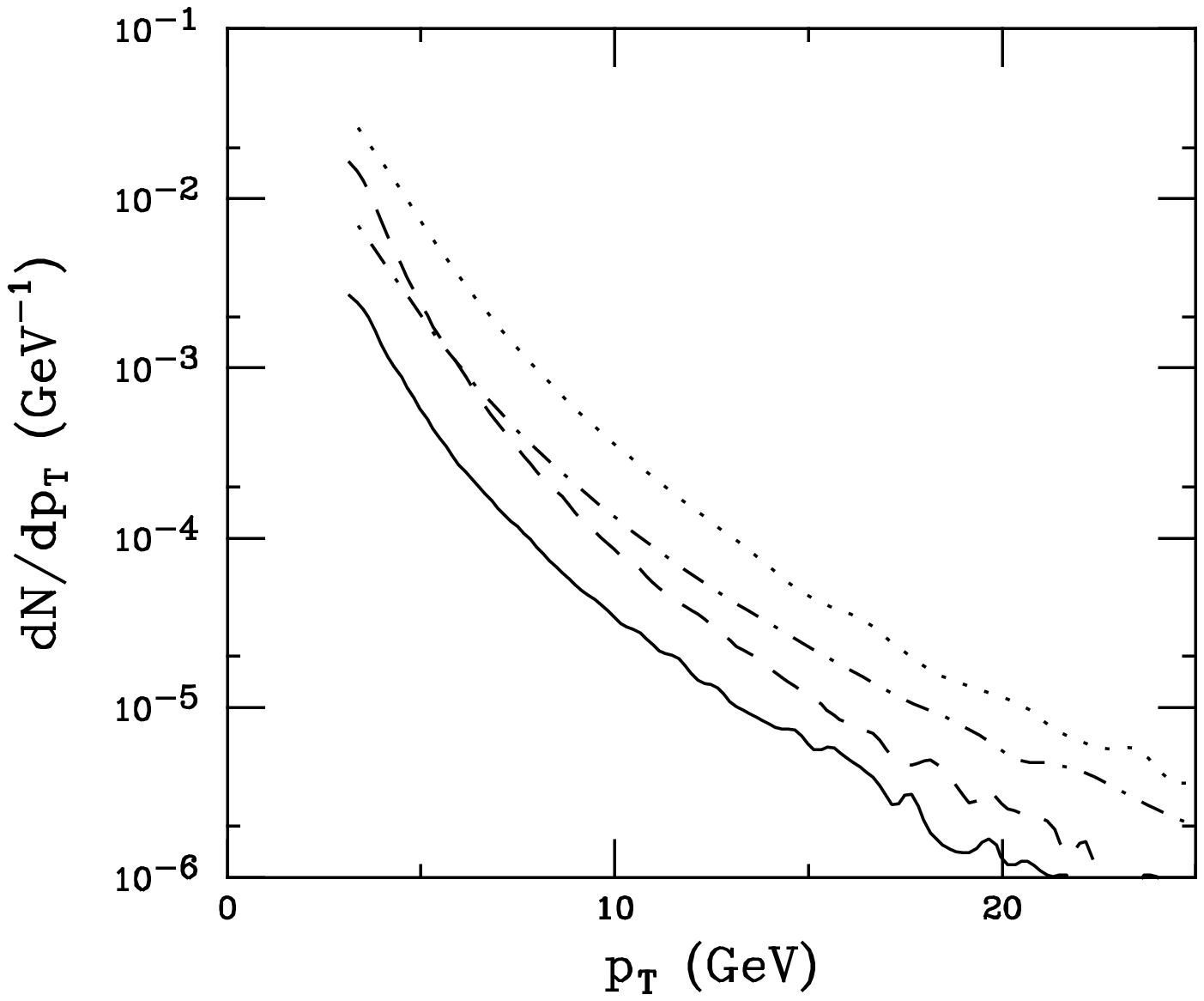}
\caption{The $p_T$ spectrum of single muons from charm and bottom
decays within the CMS acceptance. 
The dashed and dotted curves are the $D$ and $B$ meson 
decays respectively without energy loss.  The solid and 
dot-dashed curves are the corresponding results with $-dE/dx = 1$ GeV/fm. 
From Ref.~\cite{Lin:1998bd}.
}
\label{ptcms}
\end{minipage}
\end{figure}

We show the effect of energy loss on single electrons and muons within the
ALICE acceptance in Fig.~\ref{ptalice}.  
A comparison of the $p_T$ distributions of single muons in the CMS acceptance
from the decays of $D$ and $B$ mesons can also provide a measure of the $b$
cross section, shown in Fig.~\ref{ptcms}. The muon
$p_T$ distribution is clearly dominated by $B$ decays. 

%

\subsubsection{Medium-Modified Jet Shapes and Jet Multiplicities}
\label{sec345}
{\em C.A. Salgado, U.A. Wiedemann}

To discuss the medium-dependence of jet shape observables, 
one can start from the 
probability $P_{\rm tot}(\epsilon,\Theta)$ that a fraction
$\epsilon = \frac{\Delta E}{E}$ of the total jet energy $E$ is
emitted outside the cone angle $\Theta$. Under the assumption that
gluon emission follows an independent Poisson process
(see section~\ref{sec32}), this probability is given by 
\begin{eqnarray}
  P_{\rm tot}(\epsilon,\Theta) 
  = \int_C \frac{d\nu}{2\pi i}\, e^{\nu\, \epsilon}\, 
  \exp\left[ -\int_0^\infty d\omega\, 
    \left( \frac{dI^{>\Theta}_{\rm vac}}{d\omega} 
            + \frac{dI^{>\Theta}_{\rm med}}{d\omega} \right)
    \left( 1-e^{-\nu\omega}\right)\right]\, .
  \label{eqq3}
\end{eqnarray}
%
%
%
%
\begin{figure}[t!]\epsfxsize=8.7cm
\centerline{\epsfbox{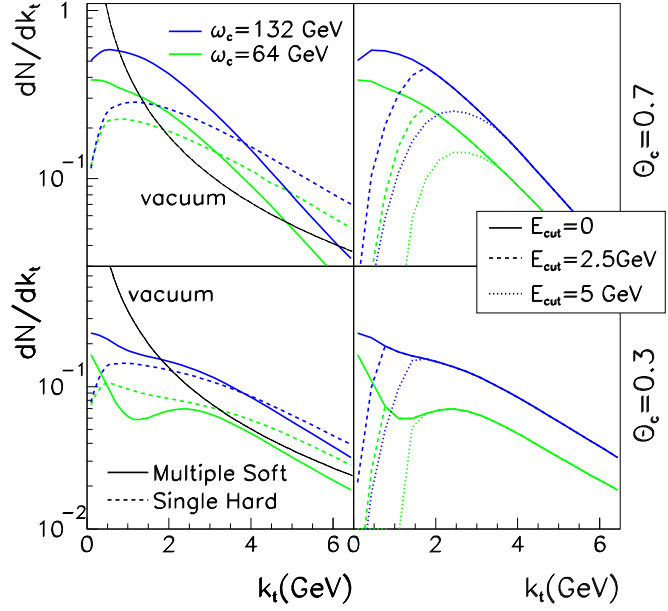}}
\caption{The gluon multiplicity distribution (\protect\ref{eq888})
inside a cone size $R=\Theta_c$, measured as a 
function of $k_t$ with respect to the jet axis. Removing gluons with
energy smaller than $E_{\rm cut}$ from the distribution (dashed and dotted
lines) does not affect the high-$k_t$ tails. Figure taken from 
Ref.~\cite{Salgado:2003rv}.
}\label{fig3multi}
\end{figure}

The expression 
(\ref{eqq3}) takes into account the angular energy distribution 
of the parton fragmentation process in the vacuum, 
$\frac{dI^{>\Theta}_{\rm vac}}{d\omega}$, as well as its
medium-modification 
$\frac{dI^{>\Theta}_{\rm med}}{d\omega} = \int_{\Theta}^\pi\, 
d\Theta\, \frac{dI_{\rm med}}{d\omega\, d\Theta}$. 
Since both contributions are additive, the total probability 
(\ref{eqq3}) can be written as a convolution of the vacuum
and the medium-induced probability,
\begin{equation}
  P_{\rm tot}(\epsilon,\Theta) = \int d\epsilon_1\, 
  P_{\rm vac}(\epsilon_1,\Theta)\,  
  P_{\rm med}(\epsilon - \epsilon_1,\Theta)\, .
  \label{eqq4}
\end{equation} 
We define 
the vacuum contribution $P_{\rm vac}(\epsilon,\Theta)$ in terms of
the jet shape $\rho(r)$. This jet shape is measured in elementary 
($pp$, $p\bar{p}$ or $e^+e^-$) collisions as the
average fraction of calorimeter cell $E_T$ in a jet subcone
of radius $r = \sqrt{(\Delta \eta)^2 + (\Delta \Phi)^2}$,
\begin{eqnarray}
  \rho(r) &=& \frac{1}{N_{\rm jets}} \sum_{\rm jets}
  \frac{E_T(r)}{E_T(r=1)}\, .
  \label{eqq5}
\end{eqnarray}
We use the Fermilab $D0$ parametrization~\cite{Abbott:1997fc} 
of $\rho(r)$ which is
based on jet shapes measured for average transverse energies of
$\approx 50 - 150$ GeV. We work in the dijet center of mass where
the jet width in pseudorapidity $\Delta \eta$ and azimuth $\Delta \Phi$
is related to the gluon emission angle $\Theta$ of our calculation
as $r = \Theta$. In general, 
$P_{\rm vac}(\epsilon,\Theta)$ is a probability distribution 
of some width in $\epsilon$ whose first moment determines the
jet shape. In the presence of medium-effects, however, the
vacuum part $\frac{dI_{\rm vac}}{d\omega}$ emits only a 
fraction $\frac{E - \Delta E}{E}$ of the total energy, and thus
we have
\begin{equation}
  \langle \epsilon \rangle_{\rm vac}(\Theta) =
  \int d\epsilon\, \epsilon\, P_{\rm vac}(\epsilon,\Theta)
  = \left[1-\rho(r=\Theta) \right]\, \frac{E - \Delta E}{E}\, .
  \label{eqq6}
\end{equation}
Since we have no experimental data about the width of
$P_{\rm vac}(\epsilon,\Theta)$, we choose
\begin{equation}
  P_{\rm vac}(\epsilon,\Theta) =
  \delta\left(\epsilon -  \frac{E - \Delta E}{E} 
  \left[1-\rho(r)\right]\right) 
  \vert_{r=\Theta}\, .
  \label{eqq7}
\end{equation}
The medium-modified jet shape 
$\rho_{\rm med}(r) = 1-\langle \epsilon \rangle_{\rm tot}(\Theta)$ 
is then defined in terms of the average jet energy fraction radiated 
outside an angle $\Theta$. One finds
\begin{eqnarray}
  \rho_{\rm med}(r) &\equiv&
  1 - \langle \epsilon \rangle_{\rm tot}(\Theta) 
  = 1 - \int d\epsilon\, \epsilon\, P_{\rm tot}(\epsilon,\Theta)
  \nonumber \\
  &=&  \rho(r)\, - \frac{\Delta E(\Theta)}{E}
       + \frac{\Delta E(\Theta=0)}{E} \left( 1 - \rho(r) \right)
      \label{eqq10}\, ,
\end{eqnarray}
where  $\frac{\Delta E}{E}(\Theta) = 
  \int d\epsilon\, \epsilon\, P_{\rm med}(\epsilon,\Theta)$.
For realistic parameters, one finds that this jet shape is modified
by a few percent only ~\cite{Salgado:2003rv}.

While the jet energy distribution is little affected by the medium,
the multiplicity distribution inside the jet cone is expected to
change significantly. This is seen from the medium-induced additional 
number of gluons 
with transverse momentum $k_\perp = \vert {\bf k}\vert$, produced 
within a subcone of opening angle $\theta_c$, 
\begin{eqnarray}
 \frac{dN^{\rm jet}}{dk_\perp} =  \int_{k_\perp/\sin\theta_c}^E d\omega\,
               \frac{dI_{\rm med}}{d\omega\, dk_\perp}\, .
  \label{eq888}
\end{eqnarray}
In Fig.~\ref{fig3multi}, this distribution is compared to the shape
of the corresponding perturbative component, 
$\frac{dN^{\rm vac}}{dk_\perp} \propto \frac{1}{k_\perp}\, 
\log(E\sin\theta_c/k_\perp)$.
The total partonic jet multiplicity is the sum of both components.
For realistic values of medium density and in-medium pathlength,
medium effects are seen to increase this multiplicity significantly 
(by a factor $> 2 - 5$) in particular in the high-$k_\perp$ tails. 
Also, the shape and width of the distribution (\ref{eq8}) changes
sensitively with the scattering properties of the medium.
Moreover, since gluons must have a minimal energy $\omega > 
k_\perp/\sin\Theta_c$ to be emitted inside the jet cone, this 
high-$k_\perp$ tail is unaffected by ``background'' cuts on the 
soft part of the spectrum, see Fig.~\ref{fig3multi}. 
This suggests that the measurement
of the transverse momentum distribution of hadrons with respect
to the jet axis is very sensitive to the transverse momentum
broadening of the underlying parton shower and should be detectable
above background. 

\subsubsection{Jet Quenching and High-$p_T$ Azimuthal Asymmetry}
{\em I.P.~Lokhtin, A.M.~Snigirev and I.~Vitev}
\label{sec346}

The azimuthal anisotropy of high-$p_T$ particle production in
non-central heavy ion collisions is among the most promising 
observables of partonic energy loss in an azimuthally non-symmetric 
volume  of quark-gluon plasma. We discuss the implications of nuclear 
geometry for the models of partonic energy loss in the context of 
recent RHIC data and consequences for observation of jet quenching 
at the LHC.

In order to interpret data on nuclear collisions from current  experiments at 
the Relativistic Heavy Ion Collider (RHIC) and  future  experiments at the 
Large Hadron Collider (LHC), it is  necessary to have knowledge of the 
{\em initial conditions}. There are large  uncertainties in  the estimates of 
the initial produced  gluon density,  $\rho_g (\tau_0) 
\sim 15 - 50 / {\rm fm}^3$  in central  $Au+Au$ at  $\sqrt{s}=130,200$~AGeV  
and $\rho_g (\tau_0) \sim 100  -  400/{\rm fm}^3$
in central $Pb+Pb$ reactions at $\sqrt{s}=5500$~AGeV,  since widely different 
models (e.g. see \cite{Wang:1991ht,Eskola:1999fc}) seem to 
be roughly consistent with data~\cite{Back:2000gw}. It is, therefore, 
essential to check the energy dependence of the density of the produced 
quark-gluon plasma (QGP) with observables complementary to  the  particle  
multiplicity $dN^{ch}/dy$ and  transverse energy  $dE_T/dy$ per unit rapidity. 
High-$p_T$ observables are  ideally suited for this task because they provide 
an estimate~\cite{Wang:xy} of the energy loss,  $\Delta E$,  of fast partons, 
resulting from medium induced non-abelian radiation along their 
path, as first discussed in~\cite{Gyulassy:1993hr,Wang:1994fx} in the
context of relativistic heavy ion reactions. The approximate linear dependence 
of  $\Delta E$  on  $\rho_g$  is the key that enables high-$p_T$ observables  
to convey information 
about the  initial conditions. However, 
$\Delta E$ also depends non-linearly on the size, $L$, 
of the medium~\cite{Baier:1998kq,Gyulassy:2000fs}   
and therefore differential observables which have well 
controlled geometric dependences are also highly desirable.

A new way to probe $\Delta E$ in variable geometries was recently  proposed  
in Refs.~\cite{Wang:2000fq,Gyulassy:2000gk}. The idea is to exploit the  
spatial  azimuthal asymmetry  of non-central  nuclear collisions. The 
dependence of $\Delta E$ on the path length $L(\phi)$ 
naturally results in a pattern of azimuthal asymmetry of high-$p_T$ hadrons 
which  can be measured  via the differential elliptic 
flow parameter (second Fourier coefficient),  
$v_2(p_T)$~\cite{Voloshin:1994mz,Poskanzer:1998yz,Ollitrault:bk}. 
Before we show the sensitivity of the high-$p_T$ $v_2(p_T > 2\;{\rm GeV})$ to 
different initial conditions we briefly discuss the various model 
calculations  for the ``elliptic flow'' coefficient $v_2$: 
\begin{enumerate}
\item The elliptic flow parameter $v_2$ was first introduced in the context of
{\em relativistic hydrodynamics}~\cite{Ollitrault:bk} and reflects the fact 
that due to the macroscopic sizes of large nuclei many aspects of $A+A$ 
collisions are  driven by nuclear geometry. In non-central 
collisions the interaction region has a  characteristic ``almond-shaped'' 
form as shown in Fig.~\ref{fig:ellips}. 
\vskip 0.7 cm
\begin{figure}[htb]
\begin{center}
\resizebox{75mm}{57mm}
{\includegraphics{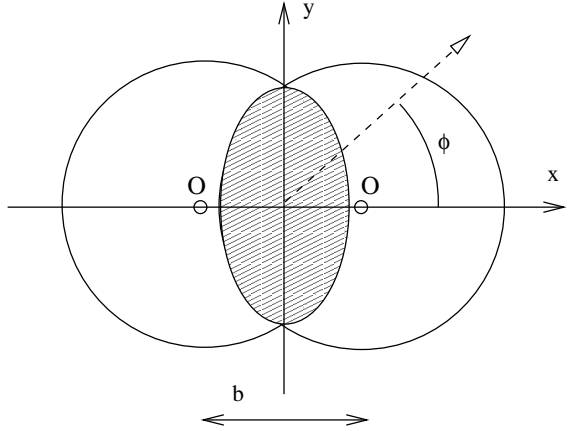}}
\vskip .2cm
\caption{\small The nuclear overlap region in non-central $A+A$ collisions
shows the importance of reaction geometry. Model calculation described here 
convert  the spatial anisotropy illustrated above into momentum 
anisotropy of  measured hadrons.}
\label{fig:ellips}
\end{center}
\end{figure}
Hydro calculations convert the ellipticity  of the reaction volume into 
momentum space azimuthal asymmetry  
\begin{equation}
\varepsilon=\frac{\langle x^2 \rangle - \langle y^2 \rangle}
{\langle x^2 \rangle + \langle y^2 \rangle }  \; \Longleftrightarrow \; 
\frac{\langle p_x^2 \rangle - \langle p_y^2 \rangle}
{\langle p_x^2 \rangle + \langle p_y^2 \rangle }  = 
\langle \cos 2 \phi \rangle =  
\frac{\int_{0}^{2 \pi}  d\phi\,\cos 2\phi\,
\frac{dN^h}{ dy\,p_{ T}\,dp_{T}\, d\phi }  }   
{ \int_{0}^{2 \pi}  d\phi\, \frac{dN^h}{ dy\,p_{ T}\,dp_{ T}\, d\phi  } }  
\; \; 
\label{anisrelat}  
\end{equation}
through the higher pressure gradient along the small axis. 
The elliptic flow is thus perfectly correlated to the 
reaction plane and can be used for its determination~\cite{Ollitrault:ba}. 
Hydrodynamic simulations~\cite{Kolb:2001qz,Teaney:2001av} typically describe 
well data from relativistic nucleus-nucleus collisions at 
$\sqrt{s}_{NN}=200$~GeV 
up to $p_T \simeq 1.5-2$~GeV and it is not unlikely that at LHC energies of  
$\sqrt{s}_{NN}=5.5$~TeV the region of validity of those calculations may extend
to $p_T \simeq 5$~GeV. 
 
\item  Initial conditions can also be mapped onto final state observable 
distributions  by solving covariant Boltzmann transport equations as in 
{\em cascade models} (partonic, hadronic, and multi-phase). Elliptic  flow in 
this approach is generated  via  multiple elastic scatterings. 
Calculations are 
sensitive to the choice of initial conditions~\cite{Molnar:2001ux} 
and are currently 
limited by statistics to $p_T \sim 6$~GeV. It is interesting to note that they 
can match the high-$p_T$ behavior of the $v_2$ but require extremely large 
initial gluon rapidity densities $dN^g/dy \simeq 16000$~\cite{Molnar:2001ux} 
and/or string melting~\cite{Lin:2001zk}.
    
\item  Memory of the initial parton density, reaction geometry, and 
the consequent dynamical evolution is  also retained by large transverse 
momentum partons (and fragmented hadrons) 
through their {\em jet quenching} pattern~\cite{Wang:2000fq,Gyulassy:2000gk}.  
While this approach is discussed in more detail below, it is important to 
emphasize here that at the single inclusive jet (or hadron) 
level the resulting 
high-$p_T$ azimuthal asymmetry is also perfectly coupled to the 
reaction plane.  
It has been suggested that in the limit of very large energy loss the momentum
asymmetry is driven by jet production from the boundary of the interaction 
volume~\cite{Shuryak:2001me}.   

\item Recently, a classical computation of the elliptic flow  at transverse 
momenta $k_T^2 > Q_s^2$ in the  framework of {\em gluon saturation}  models 
has been performed~\cite{Teaney:2002kn}. It was found 
that the azimuthal asymmetry is generated already at proper 
time $\tau=0$, i.e. it is built in the coherent initial conditions. 
The resulting elliptic flow  coefficient was found to vanish quickly 
$v_2(k_T) \propto k_T^{-2} (R_x^{-2} - R_y^{-2})$ above $Q_s$ 
($\sim 1$~GeV for RHIC  and $\sim 1.4-2$ for LHC energies) which is  not 
supported by the current data.     

\item An approach that {\em does not} associate azimuthal asymmetry with the 
reaction plane has also been presented~\cite{Kovchegov:2002nf}. Both  
high-$p_T$ and low-$p_T$ $v_2$ emerge as a {\em back-to-back jet correlation 
bias}  (with arbitrary direction relative to the reaction geometry for every 
$p_T$ bin). For large transverse momenta $v_2 \propto \ln p_T/\mu$ 
suggest an easily detectable factor of 3 increase in going 
from  $p_T=5$~GeV to $p_T = 100$~GeV at LHC. The $p_T$-integrated 
$v_2 \propto 1/Q_s$ at LHC exhibits $\sim 50\%$ reduction relative to RHIC. 
(It can also be deduced  that $v_2$ is larger at the SPS in comparison 
to RHIC in this model.)   

\end{enumerate}

The methods for $v_2$ analysis can be broadly divided in two  categories:
two-particle methods discussed, e.g., in~\cite{Wang:qh} and multi-particle 
methods~\cite{Borghini:2001vi,Borghini:2001zr}. 
In two-particle methods the error on the determined $v_2$ from non-flow 
(non-geometric) correlations is ${\cal O}(1/(v_2 M))$, where $M$ is the 
measured multiplicity.  With multi-particle methods this error goes down 
typically to ${\cal O} ( 1/(v_2 M^2))$, i.e., smaller by a factor of order 
$M$. Although it is not possible to {\em completely} eliminate the non-flow 
components to $v_2$, experimental techniques based on higher oder cumulant 
analysis~\cite{Borghini:2001vi,Borghini:2001zr} will be able in many 
cases to  {\em clearly distinguish} between between reaction geometry 
generated  azimuthal asymmetry and back-to-back jet bias.

\noindent
{\bf Parton energy loss and  nuclear geometry}
For nucleus-nucleus collisions  the co-moving  plasma  produced in an $A+B$ 
reaction at impact parameter $b$ at formation time 
$\tau_0$ has a transverse coordinate distribution at mid-rapidity  
$\rho_g({\bf r},z=0,\tau_0)$.  In studying jet production and 
propagation in nuclear environment it is not always  technically possible 
to perform the Monte-Carlo averaging over the 
jet production points coincidentally with the simulation of 
parton fragmentation.  It is therefore useful to separate the medium 
dependence of the mean jet  energy loss as a function of the extent 
of the nuclear matter traversed and the
azimuthal angle $\phi$ relative to the reaction plane. 
The total energy loss is proportional to 
a line integral along the jet trajectory
${\bf r}(\tau,\phi)={\bf r}+\hat{v}(\phi)(\tau-\tau_0)$,
averaged over the distribution  of the jet production points 
\begin{eqnarray}
F(b,\phi) &=&  \int d^2{\bf r} \; \frac{T_A(r)T_B(|{\bf r}-{\bf b}|)}
{T_{AB}(b)}  \int_{\tau_0}^\infty
 d \tau \; \tau\; \left(\frac{\tau_0}{\tau}\right)^\alpha 
\rho_0({\bf r}+\hat{v}(\phi)(\tau-\tau_0))\;.
\label{linint}
\end{eqnarray}

\vskip 0.5cm
\begin{figure}[htb]
\begin{center} 
\hspace*{-.4in}\includegraphics[height=4.8in,width=3.6in,
bbllx=90,bblly=0,bburx=600,bbury=700, clip=,angle=-90]{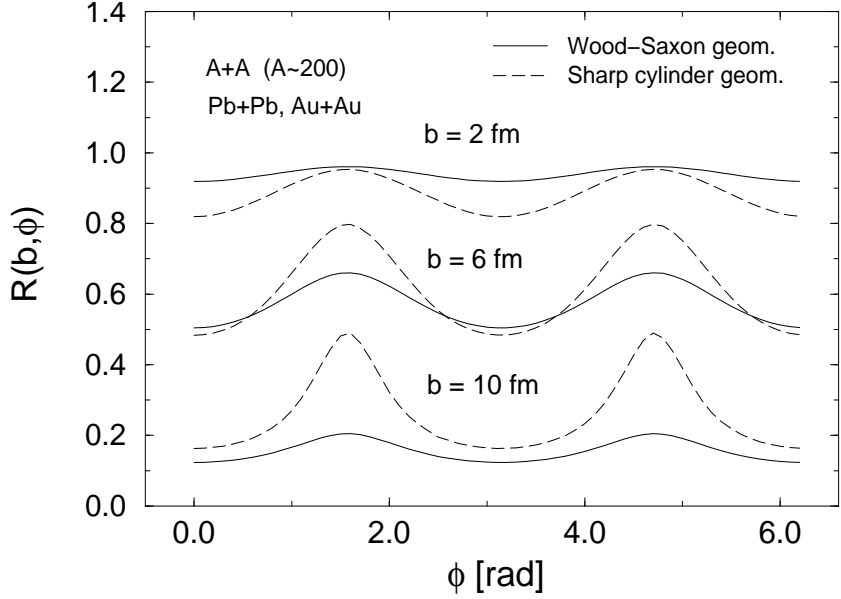}
\vspace*{-.4in}
\caption{\small The modulation function  $R(b,\phi)$   
is plotted versus $\phi$ for several impact parameters $b=2, 6, 10$~fm
from Ref.~\protect{\cite{Gyulassy:2000gk}}.
Diffuse Wood-Saxon versus uniform sharp 
cylinder geometries are compared. 
The most drastic difference between these geometries
occurs at  high impact parameters.}
\label{ng:fig2-vv}
\end{center} 
\end{figure}
\noindent 
$T_A(r)=\int dz\, \rho_A({\bf r},z)$ and  $T_{AB}(b) = \int d^2{\bf r}  \, 
T_A({\bf r})T_B({\bf r- b})$ depend on the geometry.
In particular, for a sharp uniform cylinder of radius $R_{\rm eff}$
 one readily gets ${T_A}(r)=(A/\pi R^2_{\rm eff})
\theta (R_{\rm eff}-|{\bf r}|)$ and ${T_{AB}}(0)=A^2/\pi R^2_{\rm eff}$.
We can therefore define the  effective radius of the sharp cylinder 
equivalent to a diffuse Wood-Saxon geometry via
$ F (0,\phi)_{\rm Wood-Saxon} =  F (0,\phi)_{\rm Sharp \;\; cylinder}$.
For $Au+Au$ collisions and $\alpha=1$
the above constraint gives $R_{\rm eff}\approx 6$~fm. 

For a non-vanishing  impact parameter $b$ 
and jet direction ${\hat{v}(\phi)}$,  we calculate the energy loss as
\begin{equation}
\frac{\Delta E(b,\phi )}{E}
= \frac{F(b,\phi )}{F(0,\phi)} \, \frac{\Delta E(0)}{E} 
\equiv R(b,\phi ) \, \frac{\Delta E(0)}{E} \;,
\label{separat}
\end{equation}
where the modulation function $ R(b,\phi)$ captures in the 
{\em linearized} approximation the $b$ and $\phi$ 
dependence of the jet energy loss and also provides a rough estimate 
of the maximum ellipticity generated via correlations to the 
reaction plane.
Fig.~\ref{ng:fig2-vv} shows the  $R(b,\phi)$  modulation factor
plotted against the azimuthal angle $\phi$ for impact parameters 
$b=2, 6, 10$~fm. Note that $R(b,\phi)$ reflects not only 
the dimensions of the characteristic ``almond-shaped'' cross section 
of the interaction volume but also the rapidly decreasing  initial plasma 
density  as a function of the impact parameter.

In order to compare to data at $p_T < 2$~GeV at RHIC and  $p_T < 5$~GeV  
at LHC, one must also take into account the  soft non-perturbative component
that cannot be computed with the eikonal jet quenching formalism.
The hydrodynamic elliptic flow~\cite{Ollitrault:bk} was found 
in~\cite{Kolb:2001qz} to have the monotonically growing form
$v_{2s}(p_T) \approx {\rm tanh}(p_T/(10\pm 2\;{\rm GeV}))$ 
at $\sqrt{s}=200$~AGeV and to be less sensitive to the initial conditions
than the high-$p_T$  jet quenching   studied  here. The interpolation between 
the low-$p_T$ relativistic hydrodynamics region and the high-$p_T$ 
pQCD-computable region can be evaluated as in~\cite{Gyulassy:2000gk}. 

\vskip 0.5cm
\begin{figure}[htb]
\begin{center} 
\hspace*{-.4in}\includegraphics[height=4.8in,width=3.6in,
bbllx=90,bblly=0,bburx=600,bbury=700, clip=,angle=-90]{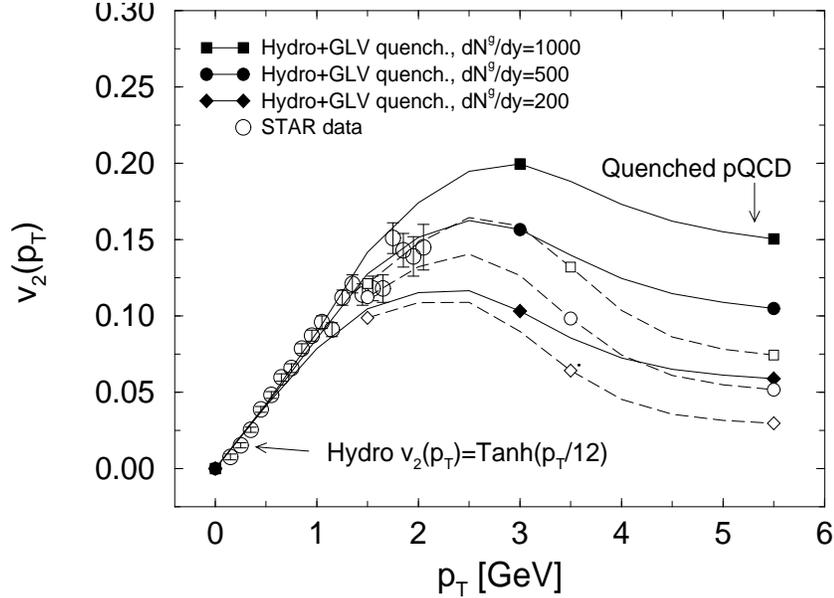}
\vspace*{-.4in}
\caption{\small The interpolation of $v_2(p_T)$
between the soft hydrodynamic~\protect{\cite{Kolb:2001qz}} 
and hard pQCD regimes is shown for $b=7$ fm adapted from 
Ref.~\protect{\cite{Gyulassy:2000gk}}.
Solid (dashed) curves  correspond to sharp cylindrical (diffuse 
Wood-Saxon) geometries presented in Fig.~\ref{ng:fig2-vv}.}
\label{ng:fig4-vv}
\end{center} 
\end{figure}

Fig.~\ref{ng:fig4-vv} shows the predicted pattern of high-$p_T$ anisotropy.
Note the difference between sharp cylinder and diffuse Wood-Saxon geometries
at $b=7$~fm  approximating roughly  20-30\% central events.  
While  the central ($b = 0$)  inclusive quenching 
is insensitive to the density profile, non-central events clearly exhibit  
large sensitivity to the actual distribution. 
We conclude that $v_2(p_T>2\;{\rm GeV},b)$ 
provides essential complementary information about the geometry and impact
parameter dependence of the initial conditions in $A+A$. In particular, the 
rate at which the $v_2$  coefficient  decreases at high $p_T$ is an  
indicator of the diffuseness of that geometry. Minimum bias STAR data at
RHIC~\cite{Filimonov:2002xk,Jacobs:2002pz} for $p_T \geq 6$~GeV  now seem to 
support the predicted~\cite{Wang:2000fq,Gyulassy:2000gk} 
slow decrease of $v_2$ at large transverse momenta.      
Recently in~\cite{Vitev:2002pf} hadron suppression in $Au+Au$ ($Pb+Pb$) 
relative to the binary scaled $p+p$ result at $p_T \simeq 5$~GeV for RHIC 
conditions ($\sqrt{s}_{NN}, dN^g/dy$) was found to be approximately equal 
to the quenching factor at LHC at a much larger transverse momentum scale 
$p_T\simeq 50$~GeV. One may thus anticipate proportionally large 
($\sim 10-15\%$) azimuthal asymmetry for high $p_T$ at the LHC.

\noindent
{\bf Energy loss of jets in transversely expanding medium} 
We next address the question of the effect of possibly large 
transverse expansion in relativistic heavy ion reactions on $v_2$.
In non-central collisions, the azimuthal asymmetry of the  mean 
energy loss can be expanded in a Fourier series and characterized as 
\begin{eqnarray}
\Delta E^{(1)}_{3D}(\phi)&=&\Delta E(1+ 2 
                               \delta_2(E)\cos 2\phi +\cdots) \; .   
\label{debjv2}  
\end{eqnarray}
It is correlated to the final measured elliptic ``flow'' of jets and hadrons 
and  has been evaluated by using a full
hydrodynamic calculation from  Ref.~\cite{Kolb:2001qz}. In this case we use the
parameterization eBC of~\cite{Kolb:2001qz} to initialize the system and treat
gluon number as conserved current to calculate the density evolution needed
in the line integral Eq.~(\ref{linint}), where it replaces the naive 
Bjorken $(\tau_0/\tau)^\alpha$ expansion. We average over the jet
formation points the density of which is given by
the number of binary collisions per unit area as in the Woods-Saxon geometry
used in  Ref.~\cite{Gyulassy:2000gk}. We find that the  azimuthal asymmetry of 
the energy loss  is strongly  reduced for realistic hydrodynamic flow 
velocities. This implies a much smaller $v_2$ at high $p_T$  than obtained in 
Ref.~\cite{Gyulassy:2000gk} where transverse  
expansion was not considered and poses questions about the observability of the
effect at LHC. 

{\bf LHC-specific remarks}
\label{sbsec:remarks}

There are several important aspects in which LHC and RHIC will differ 
significantly. We briefly discuss the implications of those differences 
for high-$p_T$ $v_2$ measurements:   

\begin{enumerate}

\item    Currently at RHIC at $\sqrt{s}= 200$~AGeV the $p_T \geq 2-3$~GeV 
regime is perturbatively  computable~\cite{Vitev:2002pf} (modulo 
uncertainties in the baryonic sector~\cite{Vitev:2001zn}).  
At LHC the $p_T$ region which is not accessible through the 
pQCD approach may extend to transverse momenta as high as 5-10~GeV. This
would imply the validity of the relativistic hydrodynamics in this
domain, the extent of which  can be tested by looking 
for marked deviations in the growth of $v_2(p_T)$, saturation, and turnover.
 
\item   Estimates of the initial gluon rapidity density at LHC vary from 
$dN^g/dy=2500$  to  $dN^g/dy=8000$. This would imply very large parton energy 
loss, at least in some regions of phase space. In this case jet production 
for moderate transverse momenta may be limited to a small shell on 
the surface of the interaction region,  leading to a constant $v_2(p_T)$ 
purely determined by geometry~\cite{Shuryak:2001me}. 

\item    Since mean transverse expansion velocities at RHIC have been  
estimated to be on the order of $v_T \simeq 0.5$ 
through relativistic hydrodynamics fits,
it is natural to expect even larger  values at LHC. This may lead to a 
significant reduction of the observed azimuthal asymmetry as discussed above. 
An important prediction of the approach put forth in~\cite{Gyulassy:2000gk} is 
that  $v_2(p_T)$ exhibits a slow decrease with increasing transverse momentum. 
This can be  used to distinguish  azimuthal anisotropy generated 
through energy loss  from alternative mechanisms.  
\end{enumerate}

\noindent
{\bf Jet impact parameter dependence at the LHC} 
In light of the discussion in Sec.~\ref{sbsec:remarks}  it is important to 
asses the feasibility of  azimuthal asymmetry measurements for 
large-$E_T$ jets via  detailed simulations.
The impact parameter dependence of jet rates in $Pb+Pb$ collisions at the 
LHC was analyzed in~\cite{Lokhtin:2000wm}. The initial jet spectra at 
$\sqrt{s} = 5.5$~TeV were generated with PYTHIA$5.7$~\cite{Sjostrand:1993yb}. 
The initial distribution of jet pairs over impact
parameter $b$ of $A+A$ collisions (without  collective nuclear effects) was 
obtained by multiplying the corresponding nucleon-nucleon jet cross section, 
$\sigma _{NN}^{\rm jet}$, by the number of binary 
nucleon-nucleon sub-collisions~\cite{Vogt:1999jp}: 
\begin{equation} 
\label{jet_prob}
\frac{d^2 \sigma^0_{\rm jet}}{d^2b}(b,\sqrt{s})=T_{AA} (b)
\sigma _{NN}^{\rm jet} (\sqrt{s})
\left[ 1 - \left( 1- \frac{1}  
{A^2}T_{AA}(b) \sigma^{\rm in}_{NN} (\sqrt{s}) \right) ^{A^2} \right]   
\end{equation} 
with the total inelastic non-diffractive nucleon-nucleon cross section  
$\sigma^{\rm in}_{NN} \simeq 60$ mb. 

The rescattering and energy loss of jets in a gluon-dominated plasma, created 
initially in the nuclear overlap zone in $Pb+Pb$ collisions at different 
impact  parameters, were simulated. For details of this model one 
can refer to~\cite{Lokhtin:2000wm,Lokhtin:2001kb}. To be specific, we treated 
the medium as a boost-invariant longitudinally expanding fluid according to 
Bjorken's solution~\cite{Bjorken:1982qr} and used the initial conditions 
expected for central $Pb+Pb$ collisions at
LHC~\cite{Eskola:1994vm,Eskola:1997ki,Eskola:1997au}: 
formation time $\tau_0 \simeq 0.1$ fm/c, initial temperature 
$T_0 \simeq 1$ GeV, gluon plasma density $\rho_g \approx 1.95T^3$. For our 
calculations we have used the collisional part of the energy loss and 
the differential scattering  cross section from~\cite{Lokhtin:2000wm}; 
the energy spectrum of coherent  medium-induced gluon radiation was 
estimated using the BDMS  formalism~\cite{Baier:1998kq}. 

\begin{figure}[htb]
\begin{minipage}[t]{80mm}
\includegraphics[width=8.0cm]{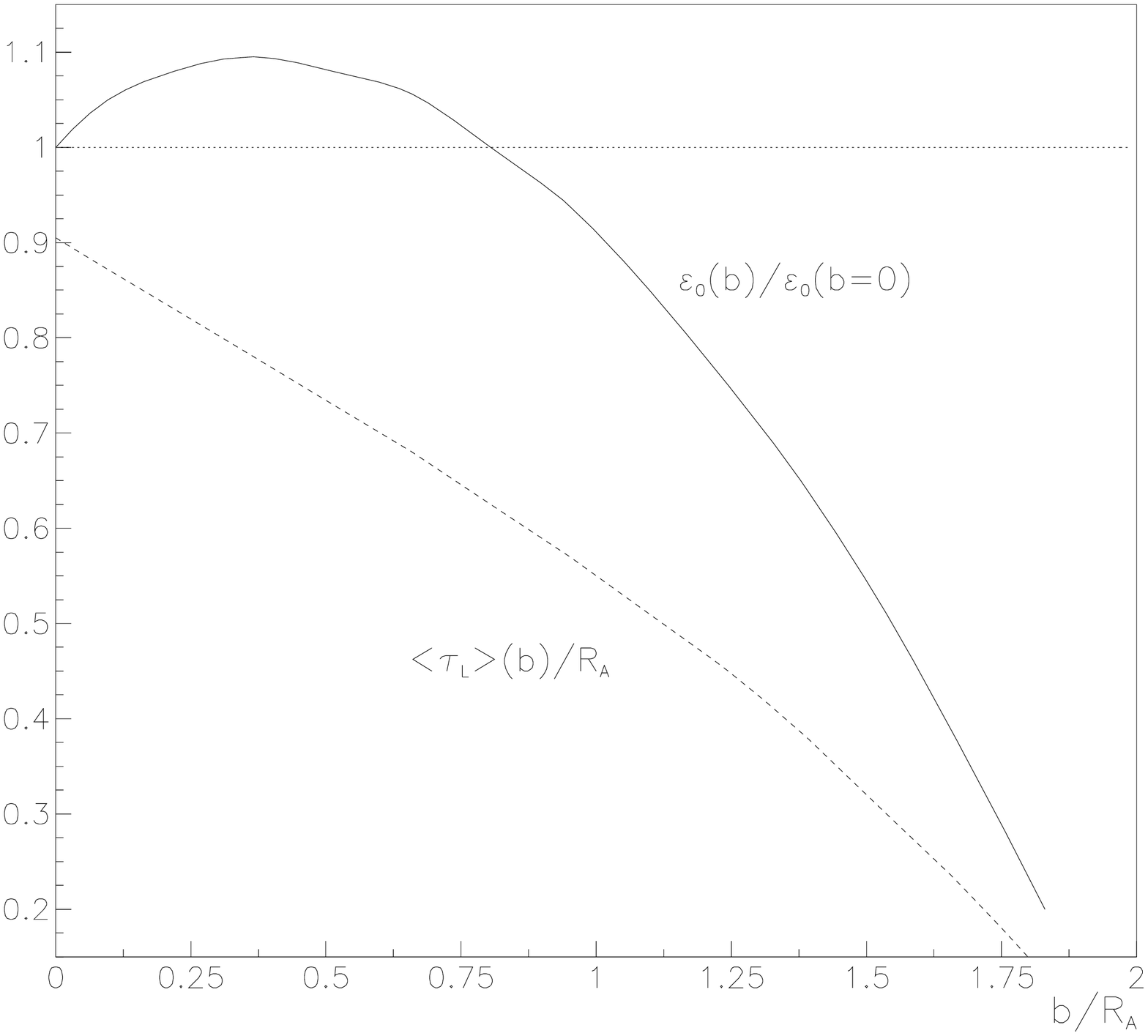}
\caption{\small The impact parameter dependence of the initial energy density 
$\varepsilon_0 (b) / \varepsilon_0 (b=0)$ in nuclear overlap zone 
(solid curve), and  the average proper jet escape  time 
$\left< \tau_L \right> / R_A$ of from the dense matter 
(dashed curve)~\cite{Lokhtin:2000wm}.
}
\label{ng:fig3}
\end{minipage}
\hspace{\fill}
\begin{minipage}[t]{80mm}
\includegraphics[width=8.0cm]{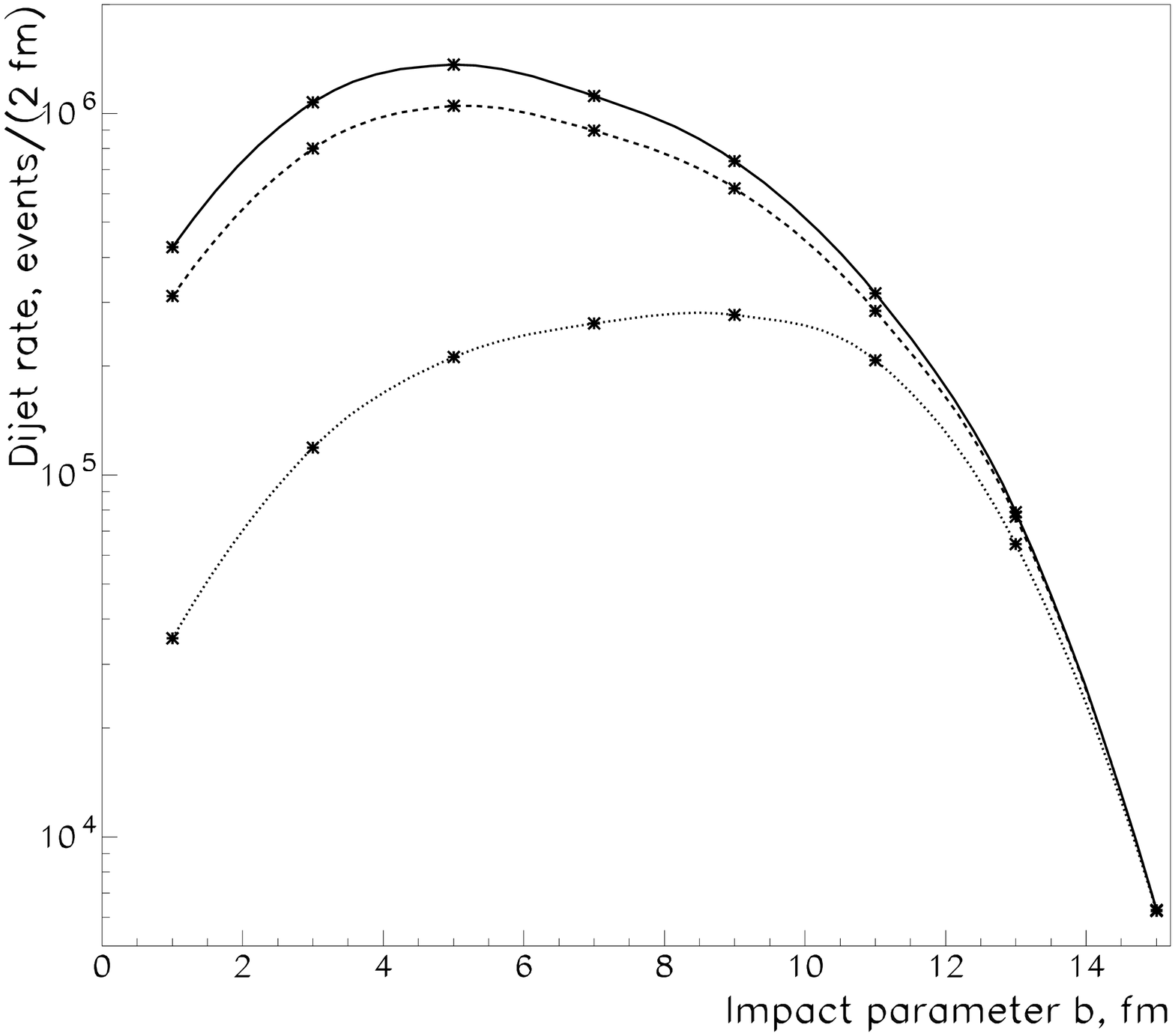}
\caption{\small The jet$+$jet rates for $E_T^{\rm jet} > 100$ GeV and 
$|\eta^{\rm jet}| < 2.5$ in 
different impact parameter bins: without energy loss (solid curve), with 
collisional loss (dashed curve), with collisional and radiative loss (dotted
curve)~\cite{Lokhtin:2000wm}.
}
\label{ng:fig4}
\end{minipage}
\end{figure}

The impact parameter dependences of the initial energy density $\varepsilon_0$ 
and  the averaged over $\varphi$  jet escape  time $\left< \tau_L \right>$  
from the dense zone are shown in Fig.~\ref{ng:fig3}~\cite{Lokhtin:2000wm}.   
$\left< \tau_L \right>$  goes down almost linearly with increasing impact 
parameter $b$. On the other hand,  $\varepsilon_0$ is very weakly dependent 
of $b$ ($\delta \varepsilon_0 /\varepsilon_0  
 \mathrel{\lower.9ex \hbox{$\stackrel{\displaystyle  <}{\sim}$}}  10 \%$) 
up to $b$ on the order of nucleus radius $R_A$, and decreases rapidly only 
at $b  \mathrel{\lower.9ex \hbox{$\stackrel{\displaystyle  >}{\sim}$}}  R_A$. 
This suggests that for impact parameters $b < R_A$, 
where $\approx 60 \%$ of jet   pairs are produced, the difference in 
rescattering intensity and   energy loss is determined mainly by the 
different path lengths rather than the  initial energy density.  

Fig.~\ref{ng:fig4} shows dijet rates in different impact parameter bins for 
$E_T^{\rm jet} > 100$ GeV and the pseudorapidity acceptance of central part 
of the CMS  calorimeters, $|\eta^{\rm jet}| < 2.5$, for three cases: 
$(i)$ without energy loss, $(ii)$ with collisional loss only, $(iii)$ with 
collisional and radiative loss. The total impact parameter integrated rates 
are normalized to the expected number of $Pb+Pb$ events during a two week LHC 
run,  $R= 1.2 \times 10^6$ s, assuming luminosity 
$L = 5 \times 10^{26}~$cm$^{-2}$s$^{-1}$. The maximum and mean values of 
$dN^{\rm dijet}/db$ distribution get shifted towards the larger $b$, because 
jet quenching is much stronger in central collisions than in peripheral ones. 
Since the coherent Landau-Pomeranchuk-Migdal radiation induces a strong 
dependence of the radiative energy loss of a jet on the angular cone 
size~\cite{Lokhtin:1998ya,Baier:1998yf}, the corresponding result for jets with 
non-zero cone size $\theta_0$ is expected to be somewhere between $(iii)$ 
($\theta_0 \rightarrow 0$) and $(ii)$ cases. Thus the observation of a 
dramatic change in the $b$-dependence of dijet rates in heavy ion collisions 
as compared  to what is expected from independent nucleon-nucleon reaction 
pattern, would indicate the existence of medium-induced parton rescattering. 

Of course, such kind of measurements require the adequate determination of 
impact parameter in nuclear collision with high enough accuracy. 
It has been shown in~\cite{Damgov:et} that for the CMS experiment the 
very forward pseudorapidity region  $3 \le |\eta| \le 5$ can provide a 
measurement of impact parameter via  the  energy flow in the very forward (HF) 
CMS calorimeters with resolution  $\sigma_b \sim 0.5$ fm for central and 
semi-central $Pb+Pb$  collisions (see details in  the section on jet 
detection at CMS). 
%

\noindent
{\bf Jet azimuthal anisotropy at the LHC} 
While at RHIC the {\em hadron}  azimuthal asymmetry at high-$p_T$  is 
being analyzed,  at LHC energies one can hope to observe  similar effects 
for the hadronic jet itself~\cite{Lokhtin:2001kb} due to the large inclusive 
cross section  for  hard jet  production  on  a  scale  $E_T \sim 100$ GeV. 
 
\begin{figure}[htb]
\begin{center} 
\resizebox{95mm}{95mm} 
{\includegraphics{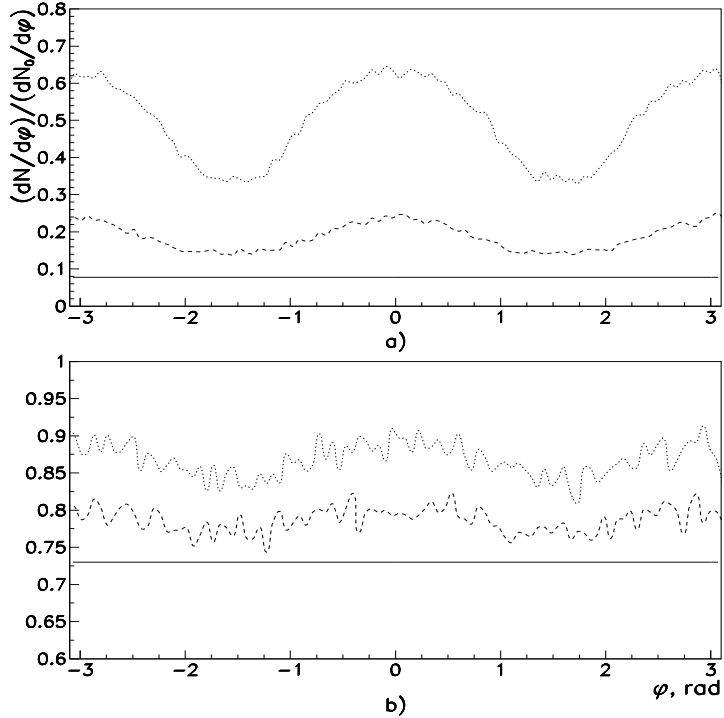}} 
\caption{\small The jet distribution over azimuthal angle for the cases with 
collisional and radiative loss (a) and collisional loss only (b), 
$E_T^{\rm jet} > 100$ GeV and $|\eta^{\rm jet}| < 
2.5$~\cite{Lokhtin:2001kb}. The histograms (from bottom to top) correspond 
to the impact parameter values $b = 0$, $6$ and $10$ fm.} 
\label{ng:fig5}
\end{center} 
\end{figure}

The  anisotropy of the energy loss ($\Delta E$)  goes up with increasing $b$, 
because the azimuthal  asymmetry of the  interaction volume gets stronger. 
However, the absolute  value of the energy loss goes down with increasing $b$ 
due to the reduced  path length $L$ (and $\varepsilon _0$ at 
$b \mathrel{\lower.9ex \hbox{$\stackrel{\displaystyle  >}{\sim}$}}  R_A$, 
see Fig.~\ref{ng:fig3}). The non-uniform dependence of $\Delta E$ on 
the azimuthal angle $\varphi$ is then mapped onto the jet spectra in  
semi-central collisions. Fig.~\ref{ng:fig5} from~\cite{Lokhtin:2001kb} shows 
the distribution of jets over $\varphi$ for the cases 
with collisional and radiative  loss (a) and collisional loss only (b) for 
$b = 0$, $6$ and $10$ fm. The same conditions and kinematical acceptance as
in Fig.~\ref{ng:fig4} were fulfilled.  
The distributions are normalized by the distributions of jets as a function of
$\varphi$ in $Pb+Pb$ collisions without energy loss. 
The azimuthal anisotropy becomes stronger in going from central to semi-central
reactions, but the absolute suppression factor is reduced with increasing $b$. 
For jets with finite cone size one can expect the intermediate result between
cases (a) and (b), because, as we have mentioned before, radiative loss
dominates at relatively small angular sizes of the jet cone 
$\theta_0 (\rightarrow 0)$,  while the relative contribution of collisional 
loss grows with increasing $\theta_0$. 

In non-central collisions the jet distribution over $\varphi$ is approximated 
well by the form $A (1+B\cos{2 \varphi})~$, where 
$A=0.5(N_{\rm max}+N_{\rm min})$ 
and $B=(N_{\rm max}-N_{\rm min})/(N_{\rm max}+N_{\rm min})= 
2\left< \cos{2\varphi}\right> $. 
In the model~\cite{Lokhtin:2001kb} the coefficient of jet azimuthal anisotropy, 
$v^{\rm jet}_{2} \equiv 
\langle \cos{2\varphi ^{\rm jet}} \rangle_{\rm event}$,  
increases almost linearly with the impact parameter $b$ and 
becomes maximum at $b \sim 1.2 R_A$. After that 
$v_2^{\rm jet}$ drops rapidly with increasing $b$: this is the domain of 
impact parameter  values  where the effect of decreasing energy loss due to 
the reduction of the  effective transverse  size of the dense zone 
and the initial energy  density of the medium is crucial and cannot be
compensated  by the stronger volume ellipticity. Anther important feature 
is  that  the jet azimuthal anisotropy decreases with increasing jet 
energy, because the energy dependence of medium-induced loss is 
rather weak (absent in the BDMS formalism and $\sim \ln E$ in the 
GLV formalism for the radiative part at high $E_T$).  

The advantage of azimuthal jet observables is that one needs to 
reconstruct only the direction of the jet, not its total energy. 
It can be  done with high accuracy, while reconstruction of the jet energy 
is more ambiguous. However, analysis of jet production as a function 
of the azimuthal angle  requires event-by-event measurement of the angular 
orientation of the reaction plane. The methods summarized in 
Ref.~\cite{Voloshin:1994mz,Poskanzer:1998yz,Ollitrault:1997di,Poskanzer:1998yz} present ways 
for reaction plane  determination. 
They are applicable for studying anisotropic particle flow in current 
heavy ion dedicated experiments at  the SPS and RHIC, and may be also used 
at the LHC~\cite{Lokhtin:2001kb}. Recently a method for measuring  jet azimuthal 
anisotropy coefficients without event-by-event reconstruction of the 
reaction plane was proposed~\cite{Lokhtin:2002vq}. 
This technique is based on the correlations between 
the azimuthal position of jet axis and the angles of hadrons not 
incorporated in the jet. The method has been generalized by taking as 
weights the particle momenta or the energy deposition in the calorimetric 
sectors. It was shown that the accuracy of the method improves with 
increasing multiplicity and particle (energy) flow azimuthal anisotropy, 
and is  practically independent of the absolute values of azimuthal 
anisotropy of the jet itself. 

\noindent
{\bf Conclusions} 
The azimuthal anisotropy of high-$p_T$ hadron production in
non-central heavy ion collisions is shown to  provide a 
valuable experimental tool  for studying  both gluon bremsstrahlung
in non-abelian media and the properties of the reaction volume such 
as its size, shape, initial parton (number and energy) rapidity  densities, 
and their subsequent dynamical evolution. The {\em saturation} and  
the {\em gradual decrease} at large transverse momentum  of the reaction 
geometry generated $v_2$, predicted as a signature complementary 
to jet quenching of strong  radiative energy loss in a dense QCD 
plasma~\cite{Gyulassy:2000gk}, seem now supported by preliminary 
data extending up to $p_T \simeq 10$~GeV at RHIC.             

The initial gluon densities in $Pb+Pb$ reactions at $\sqrt{s}_{NN}=5.5$~TeV   
at the Large Hadron Collider are expected to be significantly higher
than at RHIC, implying even stronger  partonic 
energy loss. This may result in interesting novel features of jet 
quenching, such as modification of the jet distribution 
over impact parameter~\cite{Lokhtin:2000wm} in addition to the  
azimuthal anisotropy of the jet spectrum. The predicted large cross 
section for hard jet production on a scale of $E_T \sim 100$ GeV will 
allow for a systematic study of the differential  nuclear geometry related 
aspects of jet physics at the LHC.   

\subsubsection{Rapidity Distribution and Jets}
\label{sec347}
{\em I.P.~Lokhtin, S.V.~Shmatov, P.I.~Zarubin}
%
\begin{figure}[h]\epsfxsize=12.7cm 
\centerline{\epsfbox{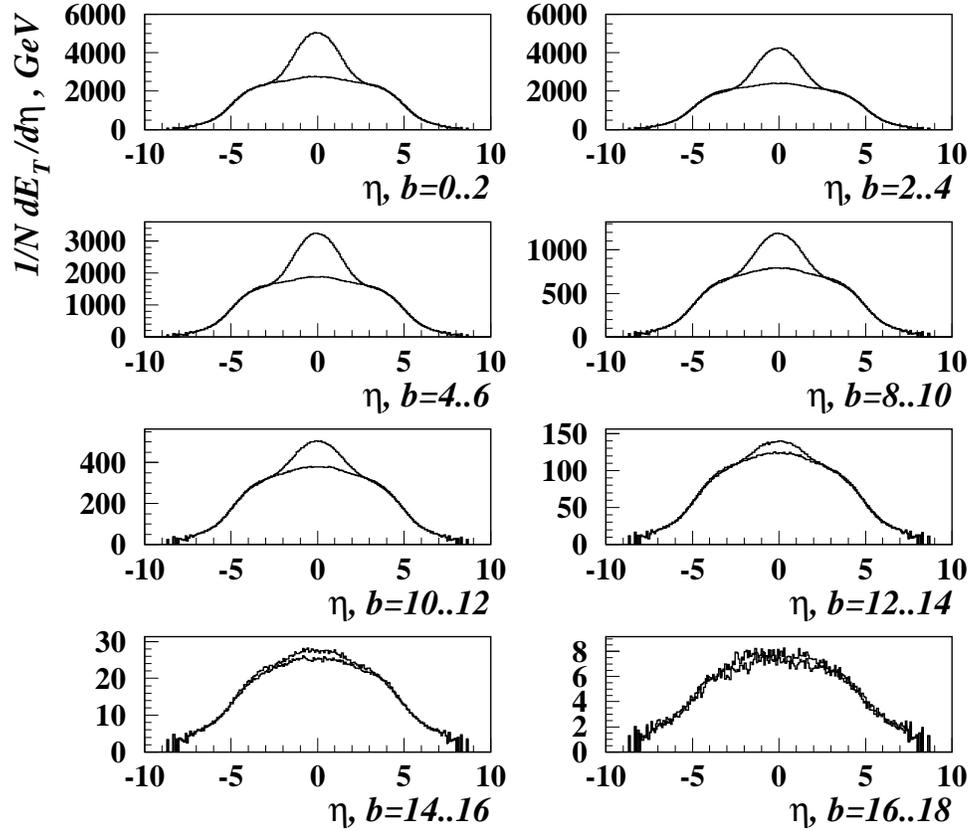}}
\vspace{0.5cm}
\caption{Normalized differential distribution of the total transverse energy 
$dE_T/d\eta$ over pseudo-rapidity $\eta$ for $10.000$ minimum bias Pb$-$Pb 
collisions 
at $\sqrt{s}= 5.5 A$ TeV for various impact parameters~\cite{Damgov:et}. 
The two cases are included: with jet quenching (the top histogram) and 
without jet quenching (the lower histogram).
}\label{figdamgov}
\end{figure}

Medium-induced parton energy loss may result in observable modifications
in the rapidity distributions of the transverse energy flow and charged 
multiplicity, $dE_T/d \eta$, 
$dE_T^{\gamma}/d \eta$, and $dn_{ch}/d \eta$~\cite{Savina:2000dy,Damgov:et}. 

Indeed, in several Monte Carlo simulations of ultrarelativistic heavy
ion collisions, one observes for example the appearance of a wide bump 
in the pseudo-rapidity interval $-2 \la \eta \la 2$, which is due to 
jet quenching. 
Fig.~\ref{figdamgov} from~\cite{Damgov:et} demonstrates the evolution of 
the effect in Pb$-$Pb collisions as a function of impact parameter 
(HIJING~\cite{Wang:1991ht,Gyulassy:ew} prediction for $\sqrt{s}= 5.5 A$ TeV). 
One can see that even peripheral Pb$-$Pb collisions show the effects of 
energy loss with the central enhancement still evident at impact parameters 
up to $12$ fm. Since jet quenching due to final state re-interactions is 
effective only for the mid-rapidity region (where the initial energy 
density of minijet plasma is high enough), the very forward rapidity region, 
$|\eta| \ga 3$, remains practically unchanged. A scan of collisions of 
different nuclear systems provides an additional test of jet quenching. 
Because smaller nuclei require a shorter transverse distance for the 
partons to traverse before escaping the system, the central enhancement 
due to energy loss decreases with system size as obvious from the comparison 
with and without energy loss. Although the effect has only 
been shown here for the global $E_T$ distributions $dE_T/d \eta$, 
qualitatively the same picture is seen when neutral or charged particle 
production is studied instead of $E_T$.   

The greater the medium-induced energy loss, the more transverse energy 
is piled up at central $\eta$ values. This leads to an increase in energy 
density or "stopping" in the mid-rapidity region, in contradiction to 
the assumption of nuclear transparency. Results qualitatively similar
to those shown in Fig.~\ref{figdamgov} can be obtained using the VENUS 
generator~\cite{Werner:1993uh} or the Parton Cascade Model 
VNI~\cite{Geiger:1994he,Geiger:1997pf}. However, the physics of the 
VENUS nucleon rescattering or VNI parton rescattering modes is very 
different from that of the radiative energy loss mechanism in HIJING. 
This may be due to the fact that various different nuclear collective 
effects provide effective forms of "nuclear 
stopping". 

One can consider rapidity spectrum of jets itself or high-p$_T$ products 
of jet fragmentation (the latter case has been discussed for RHIC energy 
in a paper~\cite{Polleri:2001cf}). One expects an anti-correlation between
the rapidity distribution of the hard jets and the global $dE_T/d \eta$ 
spectrum: in the region in which jets are suppressed most, the multiplicity
should be the highest. Since jet quenching 
due to in-medium parton energy loss is strongest in mid-rapidity, 
the maximal suppression of jet rates as compared to what is expected from 
independent nucleon-nucleon interactions extrapolation can be observed at 
central rapidity, while the very forward rapidity domain remains again 
almost unchanged. 
Thus analyzing the correlation between rapidity distributions of global 
energy (particle) flow and hard jets, by scanning the wide rapidity region 
(up to $|\eta|\le 5$ under acceptance of CMS experiment~\cite{Baur:2000}), 
might provide the important information about the pseudo-rapidity size of 
dense QCD-matter area. 

\subsection{Medium Enhanced Higher Twist Effects}
\label{sec35}

\subsubsection{Formalism of Medium Enhanced Higher Twist Effects}
{\em R.J.~Fries}
\label{sec351}

In perturbative QCD the so-called leading twist approximation is widely 
used to describe a large class of phenomena. There are quantities of 
non-perturbative nature which cannot be described by perturbation theory, 
e.g.\ the bound states of QCD. Nevertheless it is possible to separate 
perturbative (short range) and non-perturbative (long range) physics in a 
scattering reaction. Factorization theorems (see e.g.\ \cite{Collins:gx}) 
enable us to shift non-perturbative physics into a set of well-defined, 
gauge-invariant (i.e.\ observable) and universal (i.e.\ process independent) 
quantities. These quantities can be expressed by matrix elements of parton 
operators between hadron states.

It is possible to establish a hierarchy between the matrix elements in 
terms of an expansion in inverse powers of the momentum transfer. The 
expansion parameter is $\lambda^2 / Q^2$, where $Q$ is the perturbative 
hard scale and $\lambda$ (for massless QCD) has to be some non-perturbative 
scale. The leading contribution in this expansion is called leading twist 
or (in the cases relevant here) twist-2. Factorization theorems can strictly 
be proved only for certain processes and only up to a certain level of higher
twist (see section 2 of Ref.\cite{pAwriteup}).

The leading twist contribution always consists of one hard 
scattering on the parton level. In the simple example of deep inelastic 
scattering, the showcase for pQCD, the hard scattering takes place between
the virtual photon and a quark from the target.
The non-perturbative part is described by a matrix 
element which encodes the process of taking one quark out of the nucleon and 
putting it back (in the complex conjugated graph). These matrix elements
define the well known parton distributions $f_q \sim \langle \bar q q
\rangle$, $f_g \sim \langle FF \rangle$ for quarks and gluons respectively.

In a nuclear environment, more precisely in $A+A$ collisions, the 
factorization theorems are still valid, but
obviously the picture of a dominant single hard scattering process is
doubtful. From the point of view of the twist expansion,
the matrix elements which are factors in front of the expansion 
parameter $\lambda^2/Q^2$, can be numerically larger compared to the case
of the same observable in $p+p$ collisions. This is clear since the 
matrix elements encode the non-perturbative long-range behaviour and will
be sensitive to the size of the system. In fact parton distributions are
expected to scale roughly with the mass number $A$ of the nucleus, when we
neglect shadowing corrections for the moment. However it can happen that 
some higher twist matrix elements scale more strongly with the nuclear size. 
They have to contain more operators of parton fields, corresponding to 
more partons that enter the hard scattering. On the level of twist-4 e.g.\ 
one has a matrix element of the form $T_{qg} \sim \langle \bar q q FF 
\rangle$ which is a correlator of a quark and a gluon. When the indices of the
parton fields are contracted in the right way, this matrix element scales with
$A^{4/3}$. Generally, on the
level of twist-$(n+2)$, there exists a set of matrix elements that scale with
$A^{1+n/3}$. These matrix elements are called nuclear enhanced. 
The reason for the additional factors of $A^{1/3}$ is, that the different 
partons can come from different nucleons in the nucleus.

These matrix elements and their nuclear enhancement explain the trivial 
fact that multiple scatterings are important in collisions of large nuclei.
Luo, Qiu and Sterman pointed out some time ago \cite{Luo:fz,Luo:ui,Luo:np}, 
that for large nuclei with $A \gg 1$ one can replace the twist expansion by 
an effective expansion in $\lambda^2 A^{1/3} / Q^2$, keeping only the nuclear 
enhanced nuclear effects. These correspond to multiple scatterings on the 
parton level in the nuclear collision \cite{Fries:2002dn}.

On the level of twist-4 there exist calculations for jet production in 
lepton or photon induced reactions on nuclei. They deal with the transverse
momentum broadening in jet production \cite{Guo:1998rd} and the cross section
for dijet production \cite{Luo:np}. The twist-4 contributions in these cases
correspond to an additional final state interaction of the jets, more
precisely a rescattering of the outgoing jet in the nuclear medium. No 
attempt was made so far to calculate these medium corrections for hadron 
induced jet production or nucleus-nucleus collisions. For jet production 
in $p+A$ or $A+A$ a complete twist-4 calculation would contain both initial 
and final state interactions of the partons. For an overview of calculations
available in the case of $p+A$, see Ref.\cite{pAwriteup}.
%
\begin{figure}[htb]
\begin{minipage}[t]{80mm}
\hskip 1cm
\includegraphics[width=5cm]{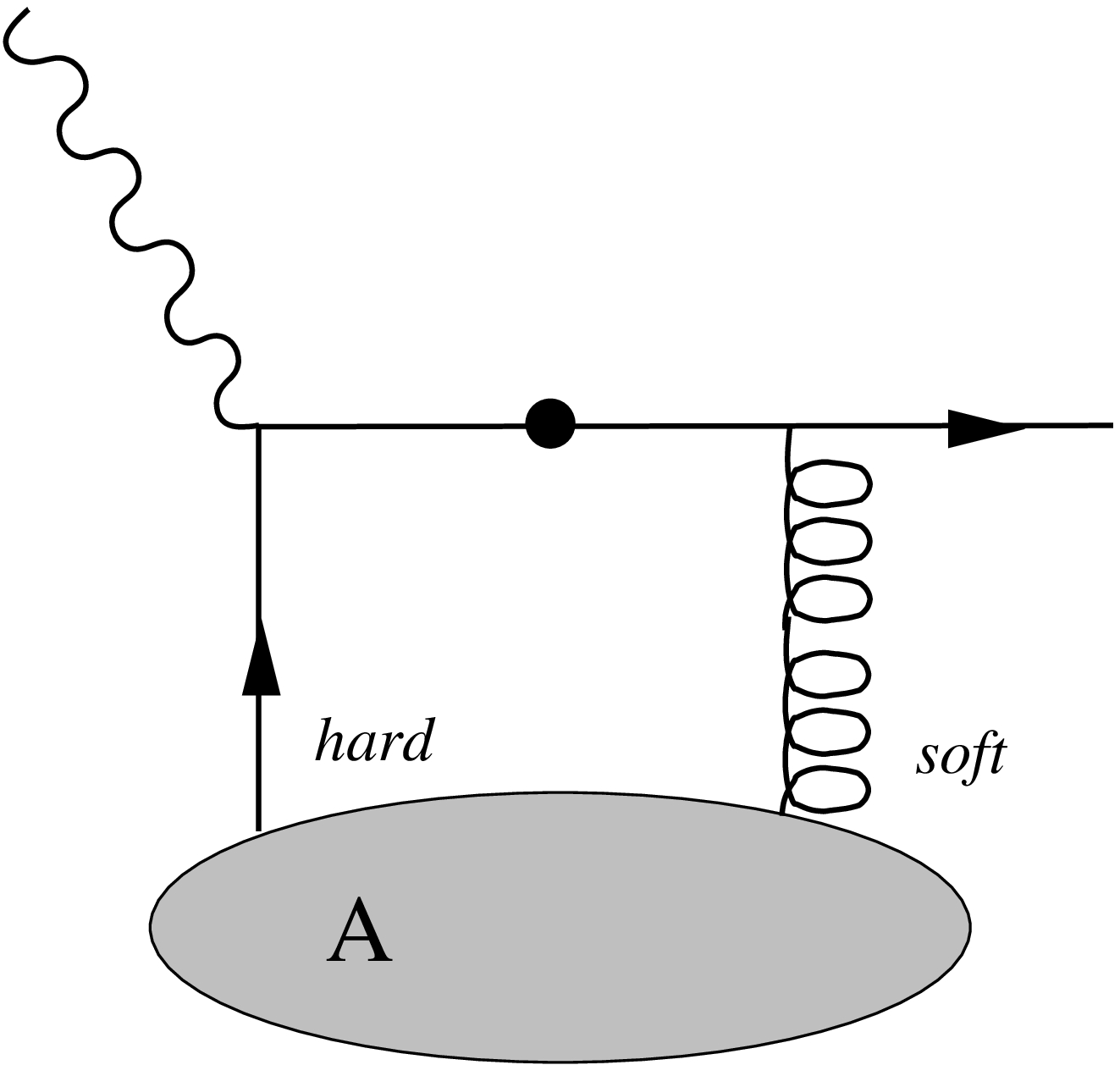}
\caption{Soft hard scattering: the parton from the primary hard scattering
is on the mass shell (indicated by the blob) and scatters off a soft 
gluon. The soft gluon 
together with the parton that enters the primary hard scattering from 
the nucleus is described by a so called soft hard matrix element $T^{SH}$
of twist-4.}
\label{fig:softrescat}
\end{minipage}
\hspace{\fill}
\begin{minipage}[t]{80mm}
\hskip 1cm
\includegraphics[width=5cm]{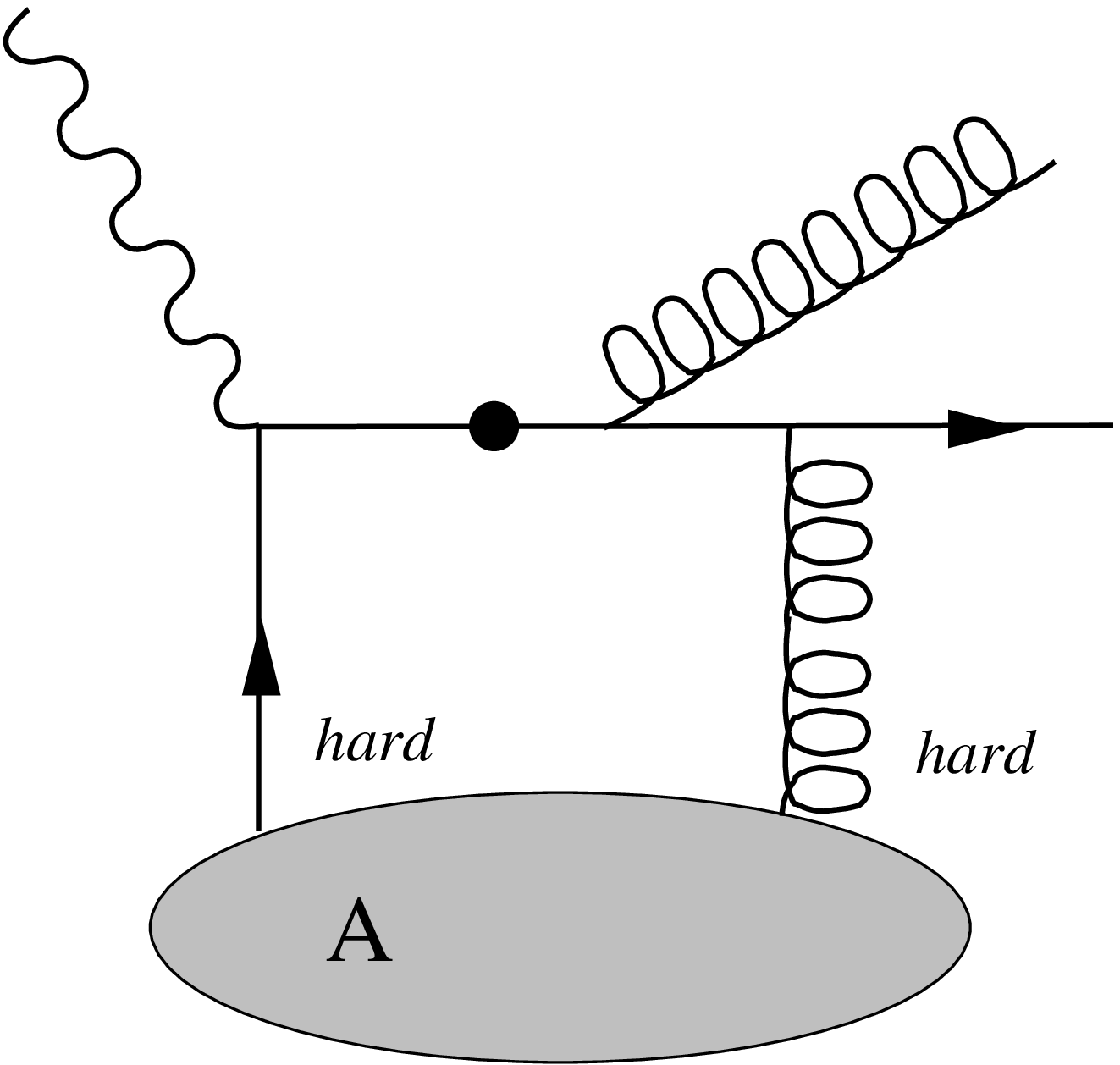}
\caption{Double hard scattering: the parton from the primary hard scattering
     is on the mass shell and undergoes an interaction with a second hard 
     parton. It has to radiate a gluon to come back to the mass shell.
     The two hard partons from the nucleus are
    described by a so called double hard matrix element $T^{DH}$ of twist-4.
}
\label{fig:hardrescat}
\end{minipage}
\end{figure}

In the case of pure final state interactions there are two important
mechanisms at the level of twist-4. A parton that already underwent one hard
scattering and is on the mass shell afterwards can interact with the soft 
gluon field of the nucleus, see Fig.~\ref{fig:softrescat}.
The second important case is that the parton leaving the primary hard 
scattering interacts with a hard parton from the nucleus and has to radiate
a gluon in order to fall back onto the mass shell. This medium induced 
radiation is shown in Fig.~\ref{fig:hardrescat}.

The factorization formula for the cross section for the twist-4 contribution
of hadron induced double scattering takes the form
\begin{equation}
  \sigma \sim f_A \otimes H_{AB} \otimes T_B + T_A \otimes H_{BA}
  \otimes f_B
\end{equation}
where $f_A$, $f_B$ are parton distributions for nucleus $A$, $B$ respectively,
describing one parton entering the parton cross sections $H_{AB}$, $H_{BA}$.
$T_A$, $T_B$ are nuclear enhanced twist-4 matrix elements, describing 
two partons from the respective nucleus. The generalization of double 
scattering (twist-4) to arbitrary nuclear enhanced twist is possible in 
situations where factorization theorems allow, see Ref.\cite{pAwriteup}.

For quantitative estimates models for the twist-4 matrix elements have to be
introduced. For soft hard matrix elements the effect of the soft gluon 
amounts to the appearance of an additional soft energy scale $\lambda$. The
dependence on the parton momentum fraction of the hard parton is taken to
be the same as in the parton distribution of this parton. One therefore
sets $T^{SH} = \lambda^2 A^{4/3} f$ where $f$ is the parton distribution
of the hard parton normalized to one nucleon. 
Similarly double hard matrix elements are approximated by the product
of the two parton distributions for both partons $T^{DH} = C A^{4/3} f_1 f_2 $.
$C$ is normalization constant.
 
\subsubsection{Parton Energy Loss and Modified Fragmentation Functions}
{\em E. Wang, X.N. Wang, B. Zhang}
\label{sec352}

The formalism of medium-induced higher twist effects was extended
recently to the calculation of
medium effects on fragmentation functions \cite{Guo:2000nz,Wang:2001if}. 
This is of importance for jet physics at LHC since the energy loss
of a parton cannot be observed directly. One has to resort to 
particle distributions within a jet and study the effect of parton energy loss 
by measuring the modification of the fragmentation function of 
the produced parton, $D_{a\rightarrow h}(z,\mu^2)$ which can be measured
directly. This modification can be directly translated into the 
energy loss of the leading parton.

Here we give an account of this approach which so far includes 
applications to $eA$ DIS and Au+Au collisions at RHIC. The main
results will be seen to be consistent with calculations described
in sections~\ref{sec31}.

\paragraph{Parton energy loss in a nuclear medium}

As a first example, we consider the process $eA$ 
DIS \cite{Wang:2001if,Guo:2000nz,Zhang:2003yn}. Here,
we consider the semi-inclusive processes,
$e(L_1) + A(p) \longrightarrow e(L_2) + h (\ell_h) +X$,
where $L_1$ and $L_2$ are the four-momenta of the incoming and the
outgoing leptons, and $\ell_h$ is the observed hadron momentum.
The differential
cross section for the semi-inclusive process can be expressed as
\begin{equation}
E_{L_2}E_{\ell_h}\frac{d\sigma_{\rm DIS}^h}{d^3L_2d^3\ell_h}
=\frac{\alpha^2_{\rm EM}}{2\pi s}\frac{1}{Q^4} L_{\mu\nu}
E_{\ell_h}\frac{dW^{\mu\nu}}{d^3\ell_h} \; ,
\label{sigma}
\end{equation}
where $p = [p^+,0,{\bf 0}_T] \label{eq:frame}$
is the momentum per nucleon in the nucleus,
$q =L_2-L_1 = [-Q^2/2q^-, q^-, {\bf 0}_T]$ the momentum transfer,
$s=(p+L_1)^2$ and $\alpha_{\rm EM}$ is the electromagnetic (EM)
coupling constant. $L_{\mu\nu}$ is the leptonic tensor
while $W_{\mu\nu}$ is the semi-inclusive hadronic tensor.

In the parton model with the collinear factorization approximation,
the leading-twist contribution to the semi-inclusive cross section
can be factorized into a product of parton distributions,
parton fragmentation functions and the partonic cross section.
Including all leading log radiative corrections, the lowest order
contribution (${\cal O}(\alpha_s^0)$) from a single
hard $\gamma^*+ q$ scattering can be written as
\begin{eqnarray}
& &\frac{dW^S_{\mu\nu}}{dz_h}
= \sum_q e_q^2 \int dx f_q^A(x,\mu_I^2) H^{(0)}_{\mu\nu}(x,p,q)
D_{q\rightarrow h}(z_h,\mu^2)\, . \label{Dq} 
\end{eqnarray}
Here, $H^{(0)}_{\mu\nu}(x,p,q)$ is the LO hard matrix element.
The momentum fraction carried by the hadron is defined as
$z_h=\ell_h^-/q^-$ and $x_B=Q^2/2p^+q^-$ is the Bjorken variable.
$\mu_I^2$ and $\mu^2$ are the factorization scales for the initial
quark distributions $f_q^A(x,\mu_I^2)$ in a nucleus and the fragmentation
functions $D_{q\rightarrow h}(z_h,\mu^2)$, respectively.

In a nuclear medium, the propagating quark in DIS will experience additional
scatterings with other partons from the nucleus. The rescatterings may
induce additional gluon radiation and cause the leading quark to lose
energy. Such induced gluon radiations will effectively give rise to
additional terms in the evolution equation leading to the modification of the
fragmentation functions in a medium. These are the so-called higher-twist
corrections since they involve higher-twist parton matrix elements and
are power-suppressed. We will consider those contributions that
involve two-parton correlations from two different nucleons inside
the nucleus.

\begin{figure}
\centerline{\includegraphics[width=3.in,height=2.in]{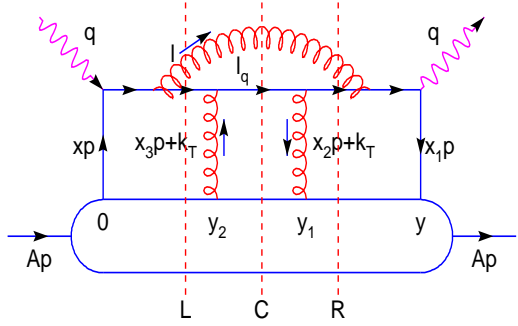}}
\caption{A typical diagram for quark-gluon re-scattering processes with three
possible cuts, central(C), left(L) and right(R).}
\label{fig:tw4-1}
\end{figure}

The generalized factorization is usually applied to these multiple 
scattering processes\cite{Luo:fz,Luo:ui,Luo:np}. In this approximation, 
the double scattering contribution to radiative correction from processes 
like the one illustrated in Fig.~\ref{fig:tw4-1} can be written in the 
following form,
%
%
%
\begin{eqnarray}
 \frac{W_{\mu\nu}^{D,q}}{dz_h} 
 &=&\sum_q \,\int dx H^{(0)}_{\mu\nu}(xp,q)
 \int_{z_h}^1\frac{dz}{z}D_{q\rightarrow h}(z_h/z)
 \frac{\alpha_s}{2\pi} C_A \frac{1+z^2}{1-z}
 \int \frac{d\ell_T^2}{\ell_T^4} \frac{2\pi\alpha_s}{N_c}
 T^{A}_{qg}(x,x_L).
 \;\; , \label{wd1}
\end{eqnarray}
%
%
%
Here, $T^{A}_{qg}(x,x_L)$ twist-four parton matrix elements of the nucleus
which can be expressed in terms of 
$\langle A | \bar{\psi}_q(0)\,
\gamma^+\, F_{\sigma}^{\ +}(y_{2}^{-})\, F^{+\sigma}(y_1^{-})\,\psi_q(y^{-})
| A\rangle$.
The fractional momentum is defined as
$x_L =\ell_T^2/2p^+q^-z(1-z)$ and $x=x_B=Q^2/2p^+q^-$ is 
the Bjorken variable. 

Using the factorization approximation
~\cite{Wang:2001if,Guo:2000nz,Luo:fz,Luo:ui,Luo:np,Osborne:2002st}, 
they can be related to the twist-two parton distributions of nucleons 
and the nucleus,
\begin{eqnarray}
T^A_{qg}(x,x_L)&=&\frac{C}{x_A}
(1-e^{-x_L^2/x_A^2}) [f_q^A(x+x_L)\, x_Tf_g^N(x_T)
+f_q^A(x)(x_L+x_T)f_g^N(x_L+x_T)] \, ,
\end{eqnarray}
where C is a constant, $x_A=1/MR_A$, $f_q^A(x)$ is the quark
distribution inside a nucleus, and $f_g^N(x)$ is the gluon
distribution inside a nucleon. A Gaussian distribution in the
light-cone coordinates was assumed for the nuclear distribution,
$\rho(y^-)=\rho_0 \exp({y^-}^2/2{R^-_A}^2)$, where
$R^-_A=\sqrt{2}R_AM/p^+$ and $M$ is the nucleon mass. We should
emphasize that the twist-four matrix element is proportional to
$1/x_A=R_AM$, or the nuclear size~\cite{Osborne:2002st}.

Including the virtual corrections and the single scattering
contribution, we can rewrite the semi-inclusive tensor in
a factorized form with a nuclear modified fragmentation function,
\begin{eqnarray}
\widetilde{D}_{q\rightarrow h}(z_h,\mu^2)&\equiv&
D_{q\rightarrow h}(z_h,\mu^2)
+\int_0^{\mu^2} \frac{d\ell_T^2}{\ell_T^2}
\frac{\alpha_s}{2\pi} \int_{z_h}^1 \frac{dz}{z}
\left[ \Delta\gamma_{q\rightarrow qg}(z,x,x_L,\ell_T^2)
D_{q\rightarrow h}(z_h/z) \right. \nonumber \\
&+& \left. \Delta\gamma_{q\rightarrow gq}(z,x,x_L,\ell_T^2)
D_{g\rightarrow h}(z_h/z)\right] \, , \label{eq:MDq}
\end{eqnarray}
where $D_{q\rightarrow h}(z_h,\mu^2)$ and
$D_{g\rightarrow h}(z_h,\mu^2)$ are the leading-twist
fragmentation functions. The modified splitting functions are
given as
\begin{eqnarray}
\Delta\gamma_{q\rightarrow qg}(z,x,x_L,\ell_T^2)&=&
\left[\frac{1+z^2}{(1-z)_+}T^{A}_{qg}(x,x_L) +
\delta(1-z)\Delta T^{A}_{qg}(x,\ell_T^2) \right]
\frac{2\pi\alpha_s C_A}
{\ell_T^2 N_c\widetilde{f}_q^A(x,\mu_I^2)}\, ,
\label{eq:r1}\\
\Delta\gamma_{q\rightarrow gq}(z,x,x_L,\ell_T^2)
&=& \Delta\gamma_{q\rightarrow qg}(1-z,x,x_L,\ell_T^2). \label{eq:r2}
\end{eqnarray}

To further simplify the calculation, we assume
$x_T\ll x_L \ll x$. The modified parton matrix elements can be
approximated by
\begin{equation}
T^{A}_{qg}(x,x_L)\approx \frac{\widetilde{C}}{x_A}
(1-e^{-x_L^2/x_A^2}) f_q^A(x),
\label{modT2}
\end{equation}
where $\widetilde{C}\equiv 2C x_Tf^N_g(x_T)$ is a coefficient which
should in principle depend on $Q^2$ and $x_T$. Here we will simply take
it as a constant.

In the above matrix element, one can identify $1/x_Lp^+=2q^-z(1-z)/\mu^2$ 
as the formation time of the emitted gluons. For large formation time 
as compared to the nuclear size, the above matrix element vanishes,
demonstrating a typical LPM interference effect.
Additional scattering will not induce more gluon radiation, 
thus limiting the energy loss of the leading quark. 


Since the LPM interference suppresses 
gluon radiation whose formation time ($\tau_f \sim Q^2/\ell_T^2p^+$)
is larger than the nuclear 
size $MR_A/p^+$ in our chosen frame, $\ell_T^2$ should then have a 
minimum value of $\ell_T^2\sim Q^2/MR_A\sim Q^2/A^{1/3}$. 
Here $M$ is the nucleon mass.
Therefore, the leading higher-twist
contribution proportional to $\alpha_s R_A/\ell_T^2 \sim \alpha_s R_A^2/Q^2$
due to double scattering depends quadratically on the nuclear size $R_A$.

With the assumption of the factorized form 
of the twist-4 nuclear parton matrices, there is only one free 
parameter $\widetilde{C}(Q^2)$
which represents quark-gluon correlation strength inside nuclei.
Once it is fixed, one can predict the $z$, energy and
nuclear dependence of the medium modification of the fragmentation
function. Shown in Fig.~\ref{salgado_fig2} are the
nuclear modification factor of the fragmentation functions for $^{14}N$ 
and $^{84}Kr$ targets as compared to the recent 
HERMES data\cite{Airapetian:2000ks,Muccifora:2001zn}.
The predicted shape of the $z$- and $\nu$-dependence agrees 
well~\cite{Wang:2002ri} 
with the experimental data.  A remarkable feature of the prediction
is the quadratic $A^{2/3}$ nuclear size dependence, which is verified 
for the first time by an experiment.
By fitting the overall suppression for one nuclear target, 
we obtain the only parameter in our calculation,
$\widetilde{C}(Q^2)=0.0060$ GeV$^2$ 
with $\alpha_{\rm s}(Q^2)=0.33$ at $Q^2\approx 3$ GeV$^2$.

%
%
%
%

We can quantify the modification of the fragmentation
by the quark energy loss which is defined as the momentum fraction
carried by the radiated gluon,
\begin{eqnarray}
\langle\Delta z_g\rangle(x_B,\mu^2)
&=& \int_0^{\mu^2}\frac{d\ell_T^2}{\ell_T^2}
\int_0^1 dz \frac{\alpha_s}{2\pi}
 z\,\Delta\gamma_{q\rightarrow gq}(z,x_B,x_L,\ell_T^2) \nonumber \\
&=&\widetilde{C}\frac{C_A\alpha_s^2}{N_c}
\frac{x_B}{x_AQ^2} \int_0^1 dz \frac{1+(1-z)^2}{z(1-z)}
\int_0^{x_\mu} \frac{dx_L}{x_L^2}(1-e^{-x_L^2/x_A^2}),
\label{eq:heli-loss}
\end{eqnarray}
where $x_\mu=\mu^2/2p^+q^-z(1-z)=x_B/z(1-z)$ if we choose the
factorization scale as $\mu^2=Q^2$.
When $x_A\ll x_B\ll 1$ we can estimate the leading quark energy
loss roughly as
\begin{eqnarray}
\langle \Delta z_g\rangle(x_B,\mu^2)& \approx &
\widetilde{C}\frac{C_A\alpha_s^2}{N_c}\frac{x_B}{Q^2
x_A^2}6\sqrt{\pi}\ln\frac{1}{2x_B}\, .
\label{eq:appr1-loss}
\end{eqnarray}
Since $x_A=1/MR_A$, the energy loss $\langle \Delta
z_g\rangle$ thus depends quadratically on the nuclear size.

In the rest frame of the nucleus, $p^+=m_N$, $q^-=\nu$, and
$x_B\equiv Q^2/2p^+q^-=Q^2/2m_N\nu$. One can
get the averaged total energy loss as
$ \Delta E=\nu\langle\Delta z_g\rangle
\approx  \widetilde{C}(Q^2)\alpha_{\rm s}^2(Q^2)
m_NR_A^2(C_A/N_c) 3\ln(1/2x_B)$.
With the determined value of $\widetilde{C}$, 
$\langle x_B\rangle \approx 0.124$ in the HERMES 
experiment\cite{Airapetian:2000ks,Muccifora:2001zn}
and the average distance $\langle L_A\rangle=R_A\sqrt{2/\pi}$
for the assumed Gaussian nuclear distribution,
one gets the quark energy 
loss $dE/dL\approx 0.5$ GeV/fm inside a $Au$ nucleus
(see section~\ref{sec314} for a comparison to other cold
nuclear matter estimates).

\paragraph{Energy Loss and Jet Quenching in Hot Medium at RHIC}

To extend our study of modified fragmentation functions to 
jets in heavy-ion collisions and to relate to results obtained
in the opacity expansion approach, we can
assume $\langle k_T^2\rangle\approx \mu^2$ (the Debye screening mass)
and a gluon density profile
$\rho(y)=(\tau_0/\tau)\theta(R_A-y)\rho_0$ for a 1-dimensional 
expanding system. Since the initial jet production 
rate is independent of the final gluon density which can be 
related to the parton-gluon scattering cross 
section\cite{Baier:1996sk} 
[$\alpha_s x_TG(x_T)\sim \mu^2\sigma_g$], one has then
\begin{equation}
\frac{\alpha_s T_{qg}^A(x_B,x_L)}{f_q^A(x_B)} \sim
\mu^2\int dy \sigma _g \rho(y)
[1-\cos(y/\tau_f)],
\end{equation}
where $\tau_f=2Ez(1-z)/\ell_T^2$ is the gluon formation time. One
can recover the form of energy loss in a thin plasma obtained 
in the opacity expansion approach\cite{Gyulassy:2000gk},
\begin{eqnarray}
\langle\Delta z_g\rangle &=&\frac{C_A\alpha_s}{\pi}
\int_0^1 dz \int_0^{\frac{Q^2}{\mu^2}}du \frac{1+(1-z)^2}{u(1+u)}
\int_{\tau_0}^{R_A} d\tau\sigma_g\rho(\tau) 
\left[1-\cos\left(\frac{(\tau-\tau_0)\,u\,\mu^2}{2Ez(1-z)}\right)\right].
\end{eqnarray}
Keeping only the dominant contribution and assuming 
$\sigma_g\approx C_a 2\pi\alpha_s^2/\mu^2$ ($C_a$=1 for $qg$ and 9/4 for
$gg$ scattering), one obtains the averaged energy loss,
\begin{equation}
\langle \frac{dE}{dL}\rangle \approx \frac{\pi C_aC_A\alpha_s^3}{R_A}
\int_{\tau_0}^{R_A} d\tau \rho(\tau) (\tau-\tau_0)\ln\frac{2E}{\tau\mu^2}.
\label{effloss}
\end{equation}
Neglecting the logarithmic dependence on $\tau$, the averaged energy loss
in a 1-dimensional expanding system can be expressed as
$\langle\frac{dE}{dL}\rangle_{1d} \approx (dE_0/dL) (2\tau_0/R_A)$,
where $dE_0/dL\propto \rho_0R_A$
is the energy loss in a static medium with the same gluon density $\rho_0$ 
as in a 1-d expanding system at time $\tau_0$.
Because of the expansion, the averaged energy loss $\langle dE/dL\rangle_{1d}$
is suppressed as compared to the static case and does not depend linearly
on the system size.

In order to calculate the effects of parton energy loss on the
attenuation pattern of high $p_T$ partons in nuclear collisions, 
we use a simpler effective modified fragmentation
function\cite{Wang:1996yh,Wang:1996pe},
\begin{eqnarray}
D_{h/c}^\prime(z_c,Q^2,\Delta E_c) 
&=&(1-e^{-\langle \frac{\Delta L}{\lambda}\rangle})
\left[ \frac{z_c^\prime}{z_c} D^0_{h/c}(z_c^\prime,Q^2)
+\langle \frac{\Delta L}{\lambda}\rangle
\frac{z_g^\prime}{z_c} D^0_{h/g}(z_g^\prime,Q^2)\right]
\nonumber \\
+ e^{-\langle\frac{\Delta L}{\lambda}\rangle} D^0_{h/c}(z_c,Q^2),
\label{modfrag} 
\end{eqnarray}
where $z_c^\prime,z_g$ are the rescaled momentum fractions.
The first term is the fragmentation function of 
the jet $c$ after losing  energy $\Delta E_c(p_c,\phi)$ 
due to {\em medium induced} gluon radiation. 
The second term is the  feedback due to the fragmentation  
of the $N_g(p_c,\phi)=\langle \Delta L/\lambda\rangle$ 
radiated gluons.  
This effective model is found to reproduce the pQCD result 
from Eq.~(\ref{eq:MDq}) very well, but only when
$\Delta z=\Delta E_c/E$ is set to
 be $\Delta z\approx 0.6 \langle z_g\rangle$.
Therefore the actual averaged parton energy loss should be
$\Delta E/E=1.6\Delta z$ with $\Delta z$ extracted from the 
effective model. The factor 1.6 is mainly
caused by the unitarity correction effect in 
the pQCD calculation.

Since gluons are bosons, there should also
be stimulated gluon emission and absorption by the propagating parton
because of the presence of thermal gluons in the hot medium.  Such detailed
balance is crucial for parton thermalization and should also be important
for calculating the energy loss of an energetic parton in a hot 
medium\cite{Wang:2001cs}. Taking into account such detailed balance in
gluon emission, one can then get the
asymptotic behavior of the effective energy loss
in the opacity expansion framework \cite{Wang:2001cs},
\begin{eqnarray}
   {\Delta E\over E}\approx &&
   {{\alpha_s C_F \mu^2 L^2}\over 4\lambda_gE}
   \left[\ln{2E\over \mu^2L} -0.048\right] -
   {{\pi\alpha_s C_F}\over 3} {{LT^2}\over {\lambda_g E^2}}
   \left[
   \ln{{\mu^2L}\over T} -1+\gamma_{\rm E}-{{6\zeta^\prime(2)}\over\pi^2}
\right],
 \end{eqnarray}
where the first term is from the induced bremsstralung and the second
term is due to gluon absorption in detailed balance which effectively
reduce the total parton energy loss in the medium.

\begin{figure}
\centerline{\includegraphics[width=4.in,height=4.in]{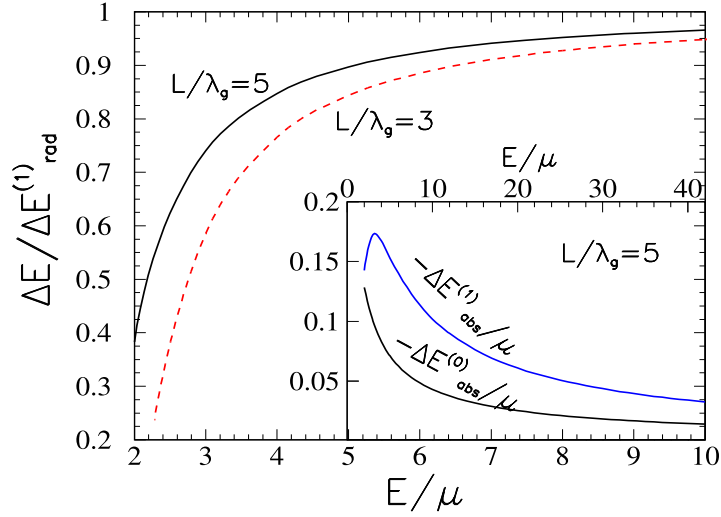}}
\vspace*{-3cm}
\caption{The ratio of effective parton energy loss
with ($\Delta E=\Delta E^{(0)}_{abs}+\Delta E^{(1)}_{abs} +\Delta
E^{(1)}_{rad}$) and without ($\Delta E^{(1)}_{rad}$) absorption as
a function of $E/\mu$. Inserted box: energy gain via absorption
with ($\Delta E^{(1)}_{abs}$) and without ($\Delta E^{(0)}_{abs}$)
rescattering.}
\label{fig:balance}
\end{figure}

Shown in Fig.~\ref{fig:balance} are numerical results of the
ratios of the calculated radiative
energy loss with and without stimulated emission and thermal
absorption as functions of $E/\mu$ for $L/\lambda_g=3$,5 and
$\alpha_s=0.3$. Shown in the inserted
box are the energy gain via gluon absorption with ($\Delta
E^{(1)}_{abs}$) and without ($\Delta E^{(0)}_{abs}$) rescattering.
For partons with very high energy the effect of the gluon
absorption is small and can be neglected. 
However, the thermal absorption reduces the effective
parton energy loss by about 30-10\% for intermediate values of
parton energy. This will increase the energy dependence of the
effective parton energy loss in the intermediate energy region.
One can parameterize such energy dependence as,
\begin{equation}
 \langle\frac{dE}{dL}\rangle_{1d}=\epsilon_0 (E/\mu-1.6)^{1.2}
 /(7.5+E/\mu),
\label{eq:loss}
\end{equation}
The threshold is the consequence of gluon absorption that competes
with radiation that effectively shuts off the energy loss. The
parameter $\mu$ is set to be 1 GeV in the calculation.

To calculate the modified high $p_T$ spectra in $A+A$ collisions,
we use a LO pQCD model \cite{Wang:1998ww,Wang:2003mm},
\begin{eqnarray}
  \frac{d\sigma^h_{AA}}{dyd^2p_T}&=&K\sum_{abcd} 
  \int d^2b d^2r dx_a dx_b d^2k_{aT} d^2k_{bT}  t_A(r)t_A(|{\bf b}-{\bf r}|) 
  g_A(k_{aT},r)  g_A(k_{bT},|{\bf b}-{\bf r}|) 
  \nonumber \\
  &\times& f_{a/A}(x_a,Q^2,r)f_{b/A}(x_b,Q^2,|{\bf b}-{\bf r}|)
  \frac{D_{h/c}^\prime (z_c,Q^2,\Delta E_c)}{\pi z_c}  
  \frac{d\sigma}{d\hat{t}}(ab\rightarrow cd), \label{eq:nch_AA}
\end{eqnarray}
with medium modified fragmentation funcitons $D_{h/c}^\prime$
given by Eq.~\ref{modfrag} and the fragmentation functions in 
free space $D^0_{h/c}(z_c,Q^2)$ are given by the BBK 
parameterization \cite{Binnewies:1994ju}.
Here, $z_c=p_T/p_{Tc}$, $y=y_c$, $\sigma(ab\rightarrow cd)$ are 
elementary parton scattering cross sections and $t_A(b)$ is the 
nuclear thickness function normalized to $\int d^2b t_A(b)=A$. 
We will use a hard-sphere model of nuclear distribution in this paper.
The $K\approx 1.5-2$ factor is used to account for higher order pQCD 
corrections.
The parton distributions per nucleon $f_{a/A}(x_a,Q^2,r)$
inside the nucleus are assumed to be factorizable into the parton 
distributions in a free nucleon given by the MRS D$-^{\prime}$  
parameterization and the impact-parameter dependent 
nuclear modification factor which will given by the new 
HIJING parameterization \cite{Li:2001xa}. The initial transverse momentum
distribution $g_A(k_T,Q^2,b)$ is assumed to have a Gaussian form
with a width that includes both an intrinsic part in a nucleon and 
nuclear broadening. This model has been fitted to the nuclear
modification of the $p_T$ spectra in $p+A$ collisions at up
to the Fermilab energy $\sqrt{s}=40$ GeV~\cite{Wang:1998ww}.
The initial multiple
scattering in nuclei can give some moderate Cronin enhancement
of the high $p_T$ spectra. Therefore, any suppression of the
high $p_T$ spectra in $Au+Au$ collisions has to be caused by
jet quenching.


We assume a 1-dimensional expanding medium with a gluon 
density $\rho_g(\tau,r)$ that is proportional to the 
transverse profile of participant nucleons.
According to Eq.~\ref{effloss}, we will calculate impact-parameter
dependence of the energy loss as
\begin{equation}
\Delta E(b,r,\phi)\approx \langle\frac{dE}{dL}\rangle_{1d}
\int_{\tau_0}^{\Delta L} d\tau\frac{\tau-\tau_0}{\tau_0\rho_0}
\rho_g(\tau,b,\vec{r}+\vec{n}\tau),
\end{equation}
where $\Delta L(b,\vec{r},\phi)$ is the distance a jet, produced at
$\vec{r}$, has to travel along $\vec{n}$ at an azimuthal 
angle $\phi$ relative to the reaction plane in a collision 
with impact-parameter $b$. Here, $\rho_0$ is the averaged 
initial gluon density at $\tau_0$ 
in a central collision and $\langle dE/dL\rangle_{1d}$ 
is the average parton energy loss over a distance $R_A$
in a 1-d expanding medium with an initial uniform gluon 
density $\rho_0$. The corresponding energy loss 
in a static medium with a uniform gluon density 
$\rho_0$ over a distance $R_A$ is \cite{Wang:2002ri}
$dE_0/dL=(R_A/2\tau_0)\langle dE/dL\rangle_{1d}$.
We will use the parameterization in Eq.~(\ref{eq:loss})
for the effective energy dependence of the parton
quark energy loss.

\begin{figure}[htb]
\begin{minipage}[t]{3.0in}
\includegraphics[width=3.0in]{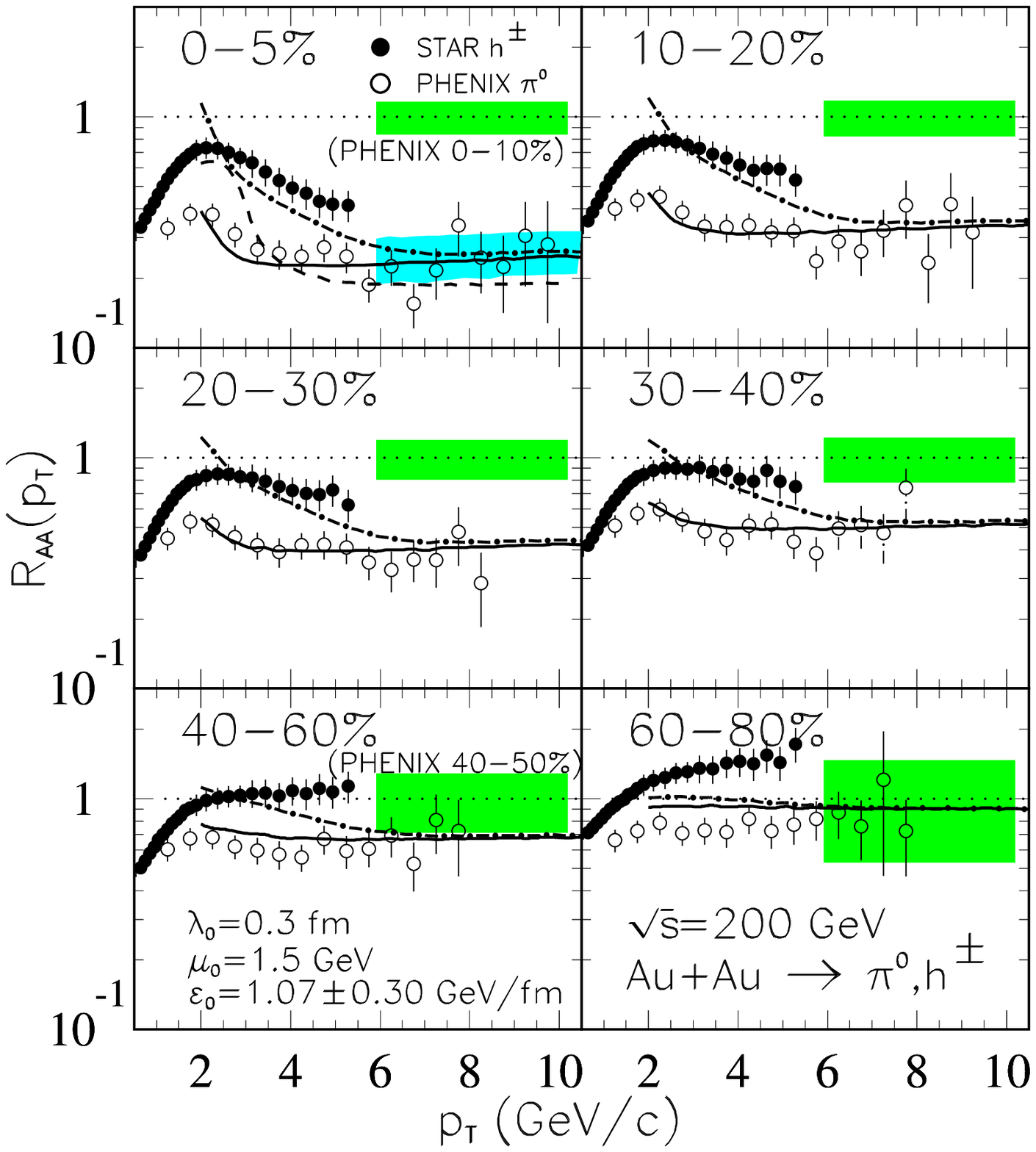}
\caption{Hadron suppression factors in $Au+Au$ collisions
as compared to data from STAR\protect\cite{Adler:2002xw,Klay:2002xj} and 
PHENIX \protect\cite{Adcox:2001jp,Adler:2003qi}. See text for a detailed explanation.}
\label{wangfig1}
\end{minipage}
\hspace{\fill}
\begin{minipage}[t]{3.0in}
\includegraphics[width=3.0in,height=3.3in]{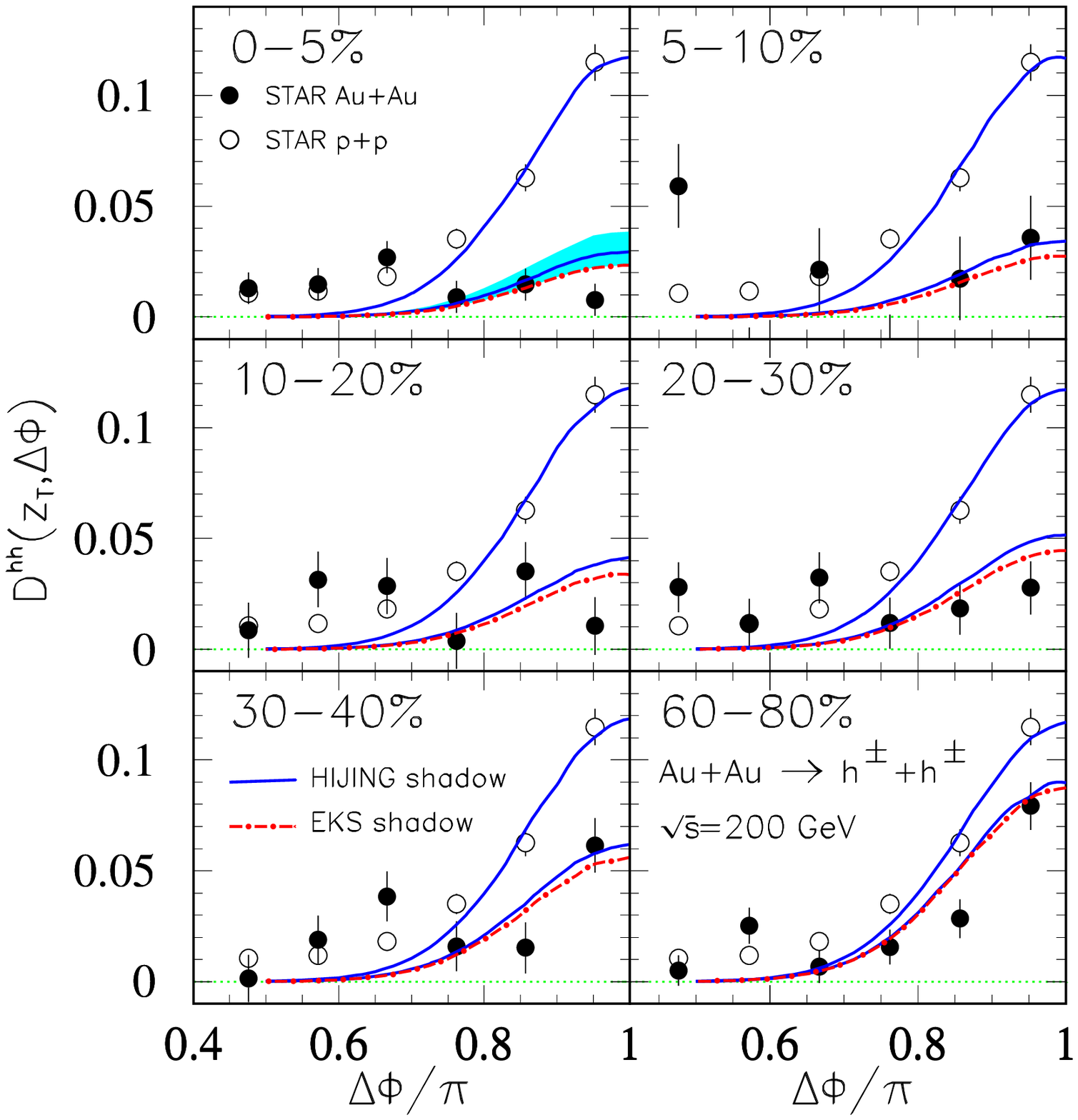}
\caption{Back-to-back correlations for charged hadrons 
with $p^{\rm trig}_T>p_T>2$ GeV/$c$, 
$p^{\rm trig}_T=4-6$ GeV/$c$ and $|y|<0.7$ in $Au+Au$ (lower curves) 
and $p+p$ (upper curves)
collisions as compared to the STAR\protect\cite{Adler:2002tq} data.}
\label{wangfig3}
\end{minipage}
\end{figure}

Shown in Fig.~\ref{wangfig1} are the calculated nuclear 
modification factors
$R_{AB}(p_T)=d\sigma^h_{AB}/\langle N_{\rm binary}\rangle d\sigma^h_{pp}$
for hadron spectra ($|y|<0.5$) in $Au+Au$ collisions 
at $\sqrt{s}=200$ GeV, as compared to experimental 
data \cite{Adler:2002xw,Klay:2002xj,Adcox:2001jp,Adler:2003qi}. Here,
$\langle N_{\rm binary}\rangle=\int d^2bd^2r t_A(r)t_A(|\vec{b}-\vec{r}|)$.
To fit the observed $\pi^0$ suppression (solid lines) in the most 
central collisions, we have used $\mu=1.5$ GeV,
$\epsilon_0=1.07$ GeV/fm and $\lambda_0=1/(\sigma\rho_0)=0.3$ fm.
The hatched area (also in other figures in this paper) indicates 
a variation of $\epsilon_0=\pm 0.3$ GeV/fm.
The hatched boxes around $R_{AB}=1$ represent experimental
errors in overall normalization.
Nuclear $k_T$ broadening and parton shadowing together give a slight 
enhancement of hadron spectra at intermediate $p_T=2-4$ GeV/$c$ 
without parton energy loss.

The flat $p_T$ dependence of the $\pi^0$ suppression is 
a consequence of the strong energy dependence of the
parton energy loss. The slight rise of $R_{AB}$
at $p_T<4$ GeV/$c$ in the calculation is due to the detailed
balance effect in the effective parton energy loss. In this
region, one expects the fragmentation picture to gradually 
lose its validity and is taken over by other non-perturbative 
effects, especially for kaons and baryons.
As a consequence, the $(K+p)/\pi$ ratio in central $Au+Au$
collisions is significant larger than in peripheral $Au+Au$ or
$p+p$ collisions. To take into account this effect, 
we add a nuclear dependent (proportional to
$\langle N_{\rm binary}\rangle$) soft
component to kaon and baryon fragmentation functions so that
$(K+p)/\pi\approx 2$ at $p_T\sim 3$ GeV/$c$ in the most 
central $Au+Au$ collisions and approaches its $p+p$ value 
at $p_T>5$ GeV/$c$. The resultant suppression for
total charged hadrons (dot-dashed) and the centrality dependence 
agree well with the STAR data. One can directly relate $h^{\pm}$ 
and $\pi^0$ suppression via the $(K+p)/\pi$ ratio: 
$R_{AA}^{h^{\pm}}=R_{AA}^{\pi^0}[1+(K+p)/\pi]_{AA}/[1+(K+p)/\pi]_{pp}$.
It is clear from the data that $(K+p)/\pi$ becomes the same for
$Au+Au$ and $p+p$ collisions at $p_T>5$ GeV/$c$.
To demonstrate the sensitivity to the parameterized
parton energy loss in the intermediate $p_T$ region, 
we also show $R_{AA}^{h^{\pm}}$ in 0-5\% centrality (dashed line)
for $\mu=2.0$ GeV and $\epsilon_0=2.04$ GeV/fm without the 
soft component.

%

In the same LO pQCD parton model, one can also calculate di-hadron
spectra,
\begin{eqnarray}
  E_1E_2\frac{d\sigma^{h_1h_2}_{AA}}{d^3p_1d^3p_2}&=&\frac{K}{2}\sum_{abcd} 
  \int d^2b d^2r dx_a dx_b d^2k_{aT} d^2k_{bT} 
  t_A(r)t_A(|{\bf b}-{\bf r}|) g_A(k_{aT},r)  g_A(k_{bT},|{\bf b}-{\bf r}|)
  \nonumber \\
  && \times f_{a/A}(x_a,Q^2,r)
  f_{b/A}(x_b,Q^2,|{\bf b}-{\bf r}|) D_{h/c}(z_c,Q^2,\Delta E_c)
  \nonumber \\
  & & \times 
  D_{h/d}(z_d,Q^2,\Delta E_d) 
 \frac{\hat{s}}{2\pi z_c^2 z_d^2} \frac{d\sigma}{d\hat{t}}(ab\rightarrow cd)
 \delta^4(p_a+p_b-p_c-p_d),
 \label{eq:dih}
\end{eqnarray}
for two back-to-back hadrons from independent fragmentation
of the back-to-back jets. 
Let us assume hadron $h_1$ is a triggered hadron 
with $p_{T1}=p_T^{\rm trig}$. One can define a hadron-triggered FF 
as the back-to-back correlation with respect to the triggered hadron:
\begin{equation}
  D^{h_1h_2}(z_T,\phi,p^{\rm trig}_T)=
  \frac{d\sigma^{h_1h_2}_{AA}/d^2p^{\rm trig}_T dp_Td\phi}
  {d\sigma^{h_1}_{AA}/d^2p^{\rm trig}_T},
\end{equation}
similarly to the direct-photon triggered FF \cite{Wang:1996yh,Wang:1996pe} 
in $\gamma$-jet events. Here, $z_T=p_T/p^{\rm trig}_T$ and 
integration over $|y_{1,2}|<\Delta y$ is implied. 
In a simple parton model, the two jets should be
exactly back-to-back. The initial parton transverse momentum distribution
in our model will give rise to a Gaussian-like angular distribution.
In addition, we also take into account transverse momentum smearing
within a jet using a Gaussian distribution with a width of
$\langle k_T\rangle=0.6$ GeV/$c$. Hadrons from 
the soft component are assumed to be uncorrelated.

Shown in Fig.~\ref{wangfig3} are the calculated back-to-back correlations 
for charged hadrons in $Au+Au$ collisions as compared to the STAR 
data \cite{Adler:2002tq}. The same energy loss that is used to calculate 
single hadron suppression and azimuthal anisotropy can also describe
well the observed away-side hadron suppression and its centrality
dependence. In the data, a background 
$B(p_T)[1+2v_2^2(p_T)\cos(2\Delta\phi)]$ from uncorrelated hadrons
and azimuthal anisotropy has been subtracted. The value of $v_2(p_T)$
is measured independently while
$B(p_T)$ is determined by fitting the observed correlation in the
region $0.75<|\phi|<2.24$ rad \cite{Adler:2002tq}.

With both the single spectra and dihadron spectra, 
the extracted average energy loss in this model calculation
for a 10 GeV quark in the expanding medium is 
$\langle dE/dL\rangle_{1d}\approx 0.85 \pm  0.24$ GeV/fm, which
is equivalent to $dE_0/dL\approx 13.8 \pm 3.9$ GeV/fm in a static and
uniform medium over a distance $R_A=6.5$ fm. This value 
is about a factor of 2 larger than a previous estimate \cite{Wang:2002ri}
because of the variation of gluon density along the propagation
path and the more precise RHIC data considered .

%
\begin{figure}[htb]
\begin{minipage}[t]{3.0in}
\includegraphics[width=3.0in,height=2.25in]{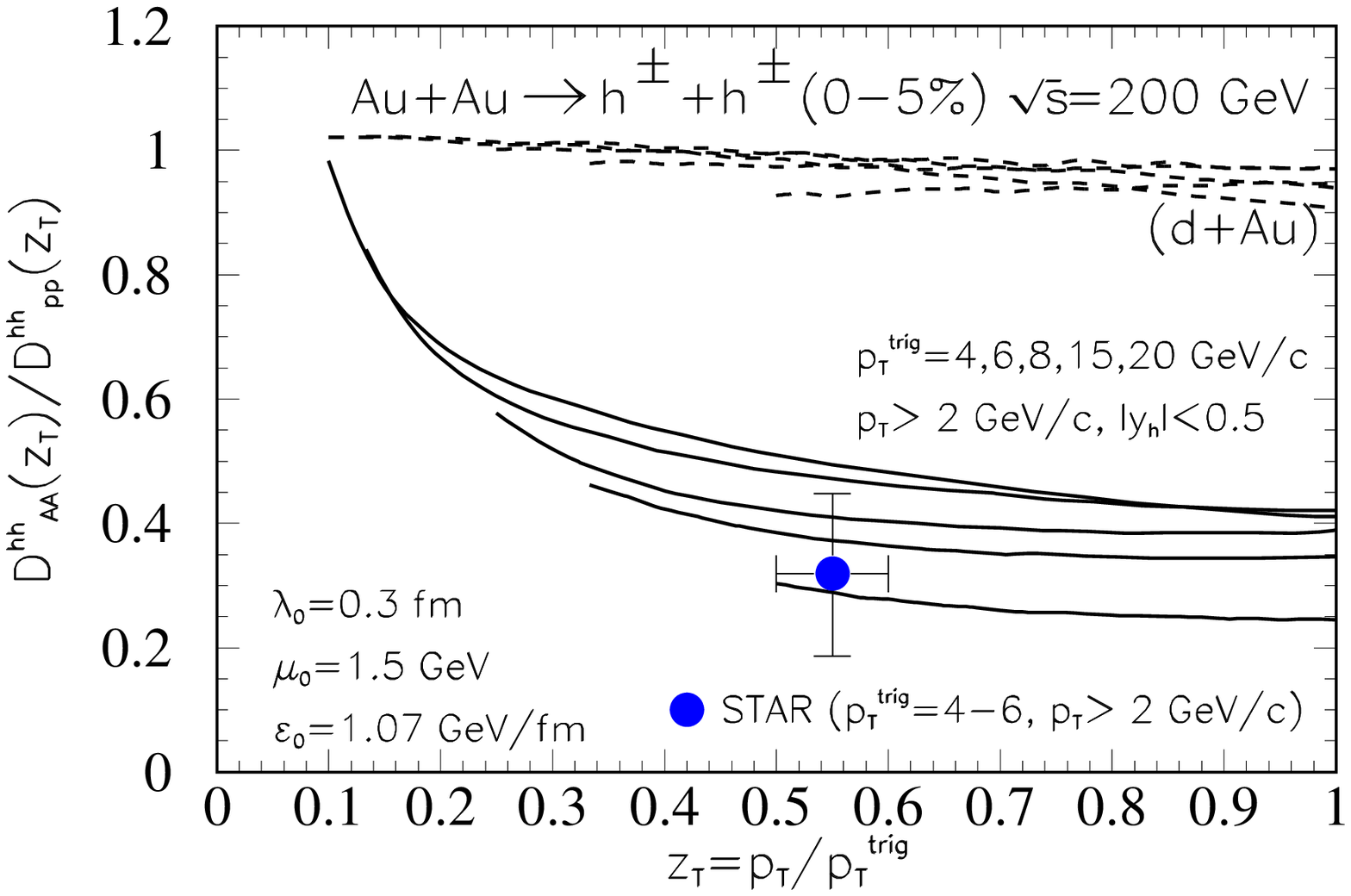}
\caption{The suppression factor for hadron-triggered fragmentation
functions in central (0-5\%) $Au+Au$ (d+Au) collisions as compared to
the STAR data \protect\cite{Adler:2002tq}.
}
\label{wangfig4}
\end{minipage}
\hspace{\fill}
\begin{minipage}[t]{3.0in}
\includegraphics[width=3.0in,height=2.25in]{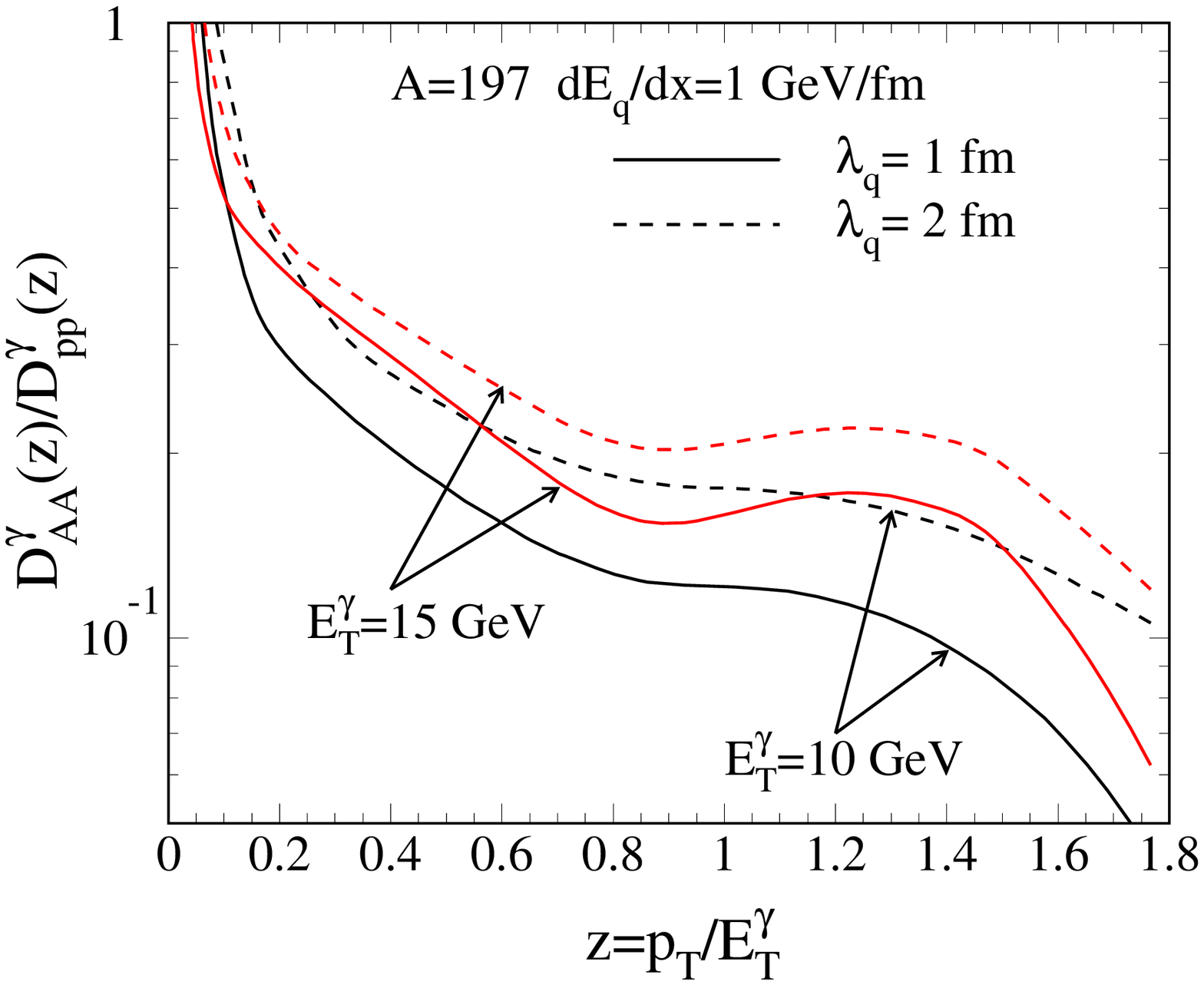}
\caption{The modification factor of the photon-tagged inclusive jet
  fragmentation function in central $Au+Au$ collisions at 
  $\protect\sqrt{s}=200$ GeV for a fixed $dE_q/dx=1$ GeV/fm.}
\label{wangfig6}
\end{minipage}
\end{figure}

Integrating over $\phi$, one obtains a hadron-triggered FF,
$D^{h_1h_2}(z_T,p_T^{\rm trig})=\int_{\pi/2}^{\pi} 
d\phi D^{h_1h_2}(z_T,\phi,p^{\rm trig}_T)$. Shown in Fig.~\ref{wangfig4} are
the suppression factors of the hadron-triggered FF's for different 
values of $p^{\rm trig}_T$ in central $Au+Au$ collisions
as compared to a STAR data point that is obtained by integrating
the observed correlation over $\pi/2<|\Delta\phi|<\pi$.
The dashed lines illustrate
the small suppression of back-to-back correlations
due to the initial nuclear $k_T$ broadening in $d+A$ collisions.
The strong QCD scale dependence 
on $p^{\rm trig}_T$ of FF's is mostly canceled in 
the suppression factor. The approximately universal shape
reflects the weak $p_T$ dependence of the hadron spectra 
suppression factor in Fig.~\ref{wangfig1}, due to a unique
energy dependence of parton energy loss. 
Shown in Fig.~\ref{wangfig6} are the suppresion factors for the
direct-photon-tagged jet fragmentation function. They are very
similar to the direct-triggered fragmentation function, except
that the photon's energy is more closely related to the
original jet energy.

\subsection{Other Possible Medium-Modifications of High-$p_T$ 
Hadronic Spectra}
\label{sec36}
%

\subsubsection{Recombination Models at the LHC}
\label{sec361}
{\em R.J.~Fries}

Recent results from RHIC show interesting phenomena in hadron production at 
intermediate transverse momenta of 2 to 5 GeV/$c$. This is a region where 
perturbative QCD starts to be a valid description of hadron dynamics, but
non-perturbative effects can still be expected to play a crucial role.
The key observations at RHIC are the anomalous enhancement of baryon
production, seen e.g.\ in a p/$\pi^0$ ratio of about one and the lack of 
nuclear suppression for baryons between 1.5 and 4 GeV/$c$ \cite{Adler:2003kg} 
and the different behavior of elliptic flow for mesons and baryons \cite{Adler:2003kt,Snellings:2003mh}.

These observations have lead to the hypothesis that hadron production
at intermediate $p_T$ is dominated by recombination from a hot and dense
parton phase instead of fragmentation of fast partons from hard scatterings
\cite{Fries:2003vb,Greco:2003xt,Hwa:2002tu,Voloshin:2002wa}. 
In the recombination picture a quark-antiquark pair close
in phase space can form a meson at hadronization, while three (anti)quarks can 
find together to be a (anti)baryon. The spectrum of mesons from recombination
can be written as \cite{Fries:2003rf,Fries:2003kq}
\begin{equation}
  \label{eq:mesreco}
  E \frac{N_M}{d^3 P} = C_M \int\limits_\Sigma d\sigma 
  \frac{P\cdot u(\sigma)}{(2\pi)^3} \int\limits_0^1  {d x} \>  
  w_a\big( {\sigma} ; x {P^+} \big)
   \> \left| \phi_M (x) \right|^2 \>
  w_b\big( {\sigma}; (1-x) {P^+} \big)\, ,
\end{equation}
if the energy $E$ of the meson is large compared to $\Lambda_{\rm QCD}$
and the constituent quark masses.
Here $\phi_M (x)$ is the meson wave function in light cone coordinates,
$x$ is the momentum fraction of one of the quarks, $C_M$ is a degeneracy factor
and the $w( {\sigma} ; p)$ are classical phase space distributions of the 
partons before hadronization. Transverse momenta relative to the hadron 
momentum have been integrated out in this equation.
A similar expression can be found for baryons.

One can easily show that recombination is more effective than fragmentation 
for an exponential parton spectrum. On the other hand, fragmentation will
win over recombination at high $P_T$ if the parton spectrum follows a
power law.
It has been shown that the shape of the wave function plays only little
role if the input parton spectrum is exponential \cite{Fries:2003kq}. 
It is therefore
a good approximation to assume that the momentum is equally shared by
the valence quarks (1/2 for mesons, 1/3 for baryons). Note that the thermal
parton phase at hadronization is assumed to have effective degrees of freedom 
with constituent quarks and no dynamical gluons.

It turns out that all spectra, the nuclear suppression factors and the 
anisotropic flow coefficient $v_2$ for hadrons in Au+Au collisions at RHIC 
for $p_T> 1.5$ GeV/$c$ can be explained by the competition between
recombination from a thermalized parton phase with temperature $T=175$ MeV 
and radial flow velocity $v_T=0.55 c$ (for central collisions) and pQCD
fragmentation of hard partons including energy loss \cite{Fries:2003kq}. 
It is worthwhile to note that it is the strong energy loss of partons in
the medium that allows recombination to dominate for $p_T<4$ GeV/$c$ for 
mesons and for $p_T<6$ GeV/$c$ for baryons.

\paragraph{Numerical estimates for LHC}

In above calculations for RHIC the parameters for the parton phase were 
determined to match existing data on hadron production. To present estimates 
for LHC, we fix the temperature of the parton phase at hadronization again at
175 MeV as predicted by lattice QCD \cite{Karsch:2001vs}.
The average radial flow will be increased at LHC compared to RHIC. 
We choose $v_T=0.75 c$ as the radial flow velocity in accordance with 
\cite{Kolb:2000sd}. 
The geometric assumptions about the fireball remain the same as those for 
RHIC \cite{Fries:2003kq}. This is certainly a lower bound for LHC.

The contribution from fragmentation is calculated in leading order (LO) pQCD
using the parton spectrum given in \cite{Srivastava:2002ic}
and KKP fragmentation functions \cite{Kniehl:2000fe}. 
The partonic energy loss is taken into account as in \cite{Fries:2003kq}. 
Its magnitude is fixed to match the mean nuclear suppression factor of about 
0.1 for a 10 GeV pion at LHC estimated in \cite{Vitev:2002pf}.

\begin{figure}[t]
  \includegraphics[width=7.5cm]{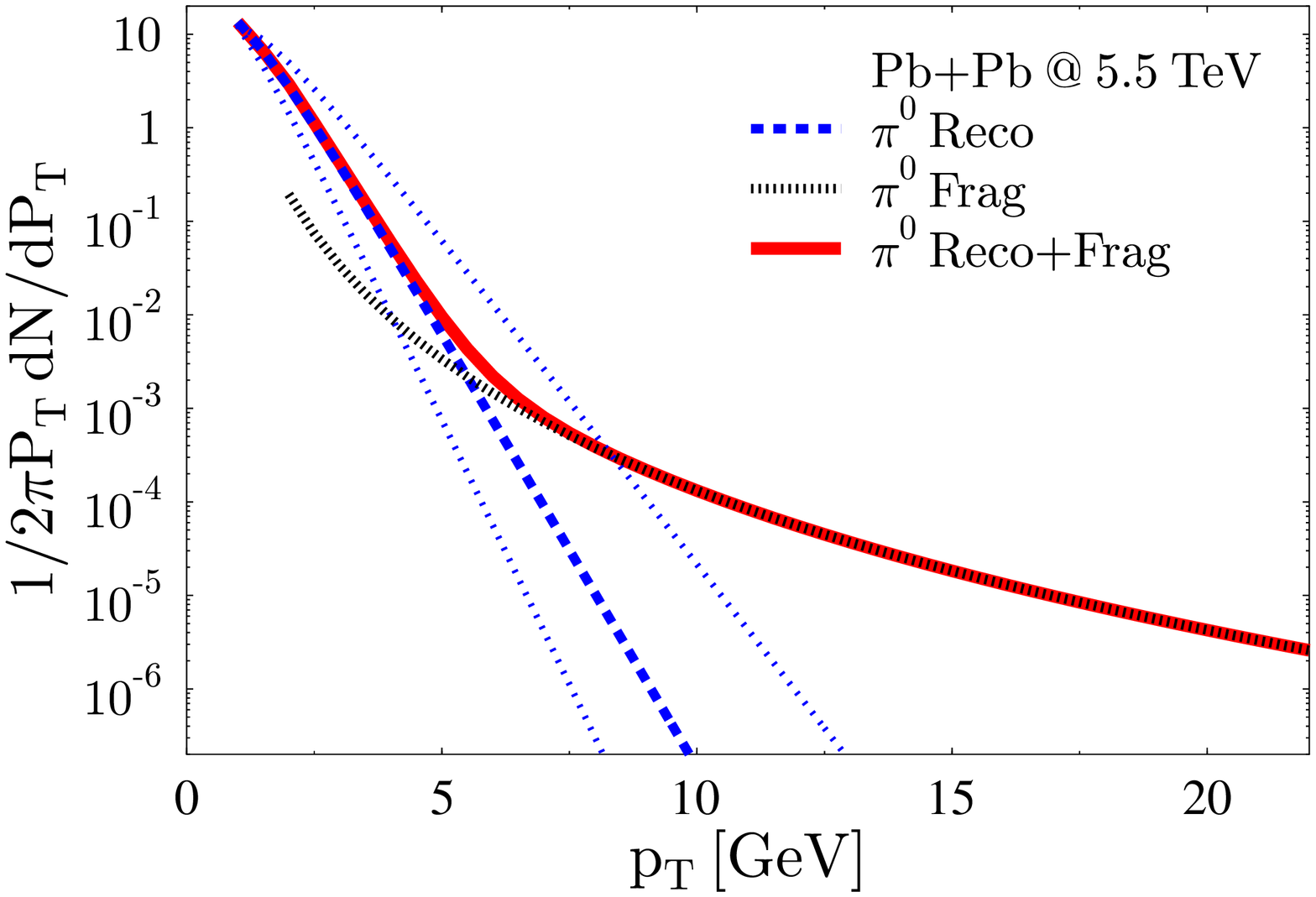}  
  \includegraphics[width=7.5cm]{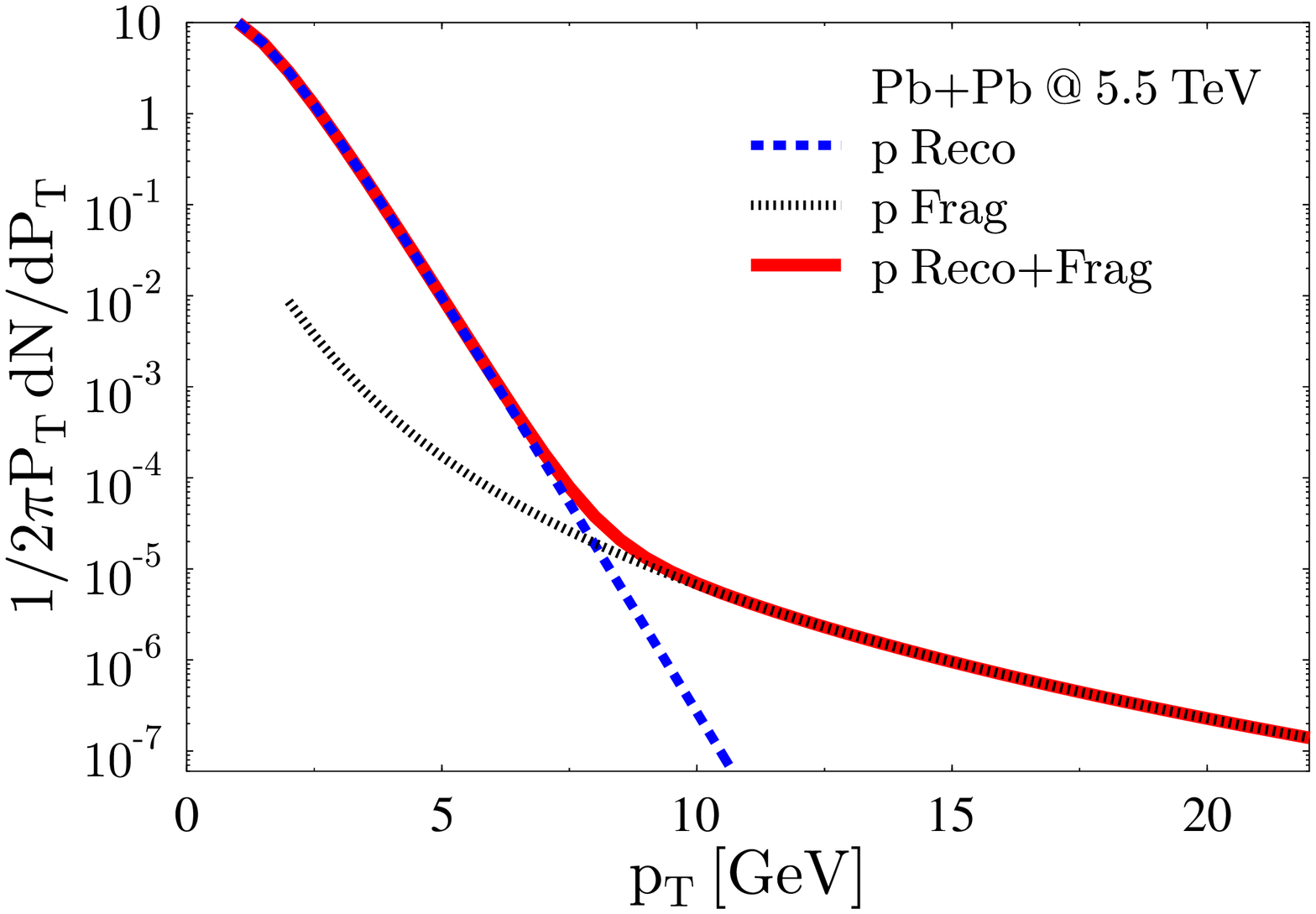}
  \caption{Spectra of $\pi^0$ (left) and $p$ (right) as a function of 
   transverse momentum $p_T$ at midrapidity for central Pb+Pb collisions at 
   $\sqrt{s}=5.5$ TeV. The recombination from the thermal parton phase 
   (long dashed line), fragmentation with energy loss from LO pQCD (dotted 
   line) and the sum of both (solid line) are shown.
   For $\pi^0$ we also give the recombination contribution for different 
   values of the the radial flow $v_T=0.65 c$ (lower short dashed line) and
   $v_T=0.85 c$ (upper short dashed line).}
  \label{fig:pi0r+f}
\end{figure}

In Fig.~\ref{fig:pi0r+f} we show the spectra for
neutral pions and protons for central Pb+Pb collisions at
$\sqrt{s}=5.5$ TeV. The recombination part for $\pi^0$ is also given for 
two other values of the radial flow to estimate the theoretical uncertainty
that is inherent in our ansatz for the parton phase. Larger emission volumes
could shift the recombination curve trivially up without changing the slope 
while leaving the fragmentation contribution nearly unchanged.

\begin{figure}[t]
\hskip 4cm
  \includegraphics[width=7.5cm]{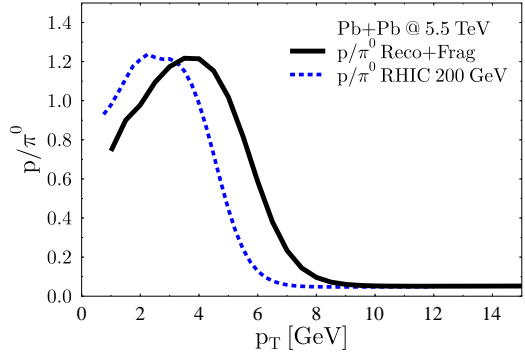}
  \caption{The ratio $p/\pi^0$ in central Pb+Pb collisions at LHC,
   $\sqrt{s}=5.5$ TeV, (solid line) and for Au+Au collisions at RHIC,
   $\sqrt{s}=200$ GeV, (dashed line) \cite{Fries:2003kq}.}
  \label{fig:protpi0ratio}
\end{figure}

The cross over between the fragmentation domain and the recombination domain
is at about 6 GeV for pions (4 GeV at RHIC) and 8 GeV for protons 
(6 GeV at RHIC) using $v_T=0.75 c$. A larger hadronization surface, as likely,
will shift these values to even higher $p_T$. In Fig.~\ref{fig:protpi0ratio}
we show the ratio $\pi^0/p$ from our calculation in comparison with the same
quantity calculated for RHIC \cite{Fries:2003kq}. We note that the surprising 
baryon enhancement is shifted to even higher transverse momenta at LHC.

\subsubsection{Transverse Momentum Diffusion and the Broadening of the 
back-to-back Di-Hadron Correlation Function}
\label{sec362}
{\em J.W.~Qiu and I.~Vitev}

Multiple parton interactions in relativistic heavy-ion reactions 
result in transverse momentum diffusion and medium  induced non-abelian 
energy loss of the hard probes traversing cold and hot nuclear matter. 
The corresponding modification of the single inclusive hadron spectra 
carries information about the dynamical properties of the medium created
in such reactions and constitutes the basis for ``jet 
tomography''~\cite{Vitev:2002pf,Levai:2001dc,Gyulassy:2001nm,Arleo:2002kh,Wang:2002ri,Salgado:2002cd,Vitev:2003xu}.  
It has been demonstrated  that the competition between nuclear shadowing, 
multiple scattering and jet quenching may lead to distinctly different 
enhancement/suppression pattern of moderate and high-$p_T$ hadron production 
in $p+A$ and $A+A$ collisions at SPS, RHIC and the LHC~\cite{Vitev:2002pf}. 
Additional experimental tools that can complement the single inclusive 
measurements, however, are highly desirable. A natural extension of the
jet-tomographic technique, first quantitatively discussed
in~\cite{Gyulassy:2001zv}, 
is ``di-jet tomography''. In this case  the medium response to the propagation
of hard partons leads to an associated increase of di-jet  
acoplanarity~\cite{Appel:dq,Blaizot:1986ma,Bodwin:1988fs},
measured  via the broadening  of the back-to-back di-hadron  correlation 
function~\cite{Qiu:2003pm,Vitev:2003jg}, as well as to a quenching of the
away-side Gaussian~\cite{Vitev:2003jg,Hirano:2003hq,Wang:2003mm}.  
These experimental observables are potentially free of the uncertainties 
in Glauber  scaling of the baseline $p+p$ result that are present in the
comparison of single inclusive spectra.

Particle production from a single hard scattering with 
momentum exchange much larger than 1/fm is localized in space-time. 
It is multiple parton scattering before 
or after the hard collision that is sensitive to the properties of the 
nuclear matter. By comparing the high-$p_T$ observables
in $p+p$, $p+A$ and $A+A$ reactions, we are able to study the strong 
interaction dynamics of QCD in the vacuum, cold nuclear matter and hot
dense medium of quarks and gluons, respectively.  
We here address the elastic (no-radiation) scattering of jets 
($\sqrt{p^2}/p^0 \simeq 0$) in nuclear 
matter~\cite{Qiu:2003pm,Vitev:2003jg,Gyulassy:2002yv,Luo:ui,Luo:np,Qiu:2001hj}
that is sensitive to the zeroth line integral moment, 
$\int dz \; z^0 \rho(z) \propto \langle L \rangle /\lambda  = \chi$, 
of the matter density. A closed form solution can be obtained via the GLV 
reaction operator approach~\cite{Gyulassy:2002yv}. Recently, 
we computed the  power corrections due to the recoil of  
the medium~\cite{Qiu:2003pm} and related the momentum distribution of partons 
that have traversed nuclear matter to their initial distribution as follows:   
\begin{eqnarray}
\frac{d^3N^{f}(p^+,{\bf p}_T)}{d p^+ d^2 {\bf p}_T }
|_{ p^- =\frac{ {\bf p}_T^2}{2 p^+ } }  
& = &  \sum_{n=0}^{\infty} \frac{\chi^n}{n!} 
\int  \prod_{i=1}^n   d^2 {\bf q}_{i\,\perp} 
 \left[  \frac{1}{\sigma_{el}} 
\frac{d\sigma_{el}(R,T)}{d^2{\bf q}_{i\,\perp}} \,
 \left(   e^{-{\bf q}_{i\, \perp} \cdot 
\stackrel{\rightarrow}{\nabla}_{{\bf p}_T} } \, 
 e^{\frac{1}{{2}} ({\bf q}^2_{i\,\perp} / (\sqrt{2}P) )\, 
\partial_{p^+} }  - 1  \right) \, \right]  \;   \nonumber \\[1ex]
&&  \times  
\frac{d^3N^{i}(p^+,{\bf p}_T)}{d p^+ d^2 {\bf p}_T }
|_{p^- =  \frac{ {\bf p}_T^2}{2 p^+ } }    \; .
\label{ropit}
\end{eqnarray}
For any initial jet flux  the opacity series 
in Eq.~(\ref{ropit}) is most easily resummed in the impact parameter 
space $(b^-,{\bf b}_T)$ conjugate to $( p^+,{\bf p}_T )$. 
For the case of a normalized forward monochromatic beam 
$p^\mu=(P,{\bf 0}_T,P)$
in the small angle scattering limit we find:
\begin{equation}
\frac{d^3N^{f}(p^+,{\bf p}_T)}{d p^+ d^2 {\bf p}_T }
|_{p^- =  \frac{ {\bf p}_T^2}{2 p^+ } } 
= \frac{1}{2\pi} \frac{e^{-\frac{{\bf p}^2_T}{2\,\chi \,\mu^2 \xi}}}
 {\chi\, \mu^2\, \xi } \,  \delta\left[p^+ - 
\left( \sqrt{2}P - \frac{1}{{2}} \frac{2 \chi \mu^2 \xi}{\sqrt{2}P}  
 \right) \right]  \;\;.
\label{gauss}  
\end{equation}
The medium induced $p_T$ broadening  and the corresponding longitudinal 
momentum reduction can be evaluated from Eq.~(\ref{gauss}): 
\begin{equation}
\langle \Delta {\bf p}_T^2 \rangle   \approx
  2\xi \int dz \, \frac{\mu^2}{\lambda_{q,g}} 
= 2 \xi \int dz \,  \frac{3  C_R \pi \alpha_s^2}{2} \rho^g(z) 
= \left\{ \begin{array}{ll}   2 \xi \,  
\frac{3  C_R \pi \alpha_s^2}{2} \, \rho^g \langle L \rangle \, , 
&  {\rm static} \\[1ex]
 2\xi \, \frac{3  C_R \pi \alpha_s^2}{2} 
\frac{1}{A_T} \frac{dN^{g}}{dy}
\,   \ln \frac{  \langle L  \rangle }{\tau_0}\, . & 1+1D  
\end{array}  \right. 
\label{broad}
\end{equation}

\begin{equation}
- \frac{dp_\parallel}{dz}  \approx - 
\frac{\Delta p_\parallel}{ \langle L \rangle } 
= \frac{  \mu^2 \, (2\xi) }{\lambda_{q,g}} \, \frac{1}{2 p_\parallel}  
= \left(\frac{\mu^2}{\lambda_{q,g}}\right)_{eff} 
\frac{1}{2 p_\parallel}  \;\; . 
\label{parshift}
\end{equation}
In Eqs.(\ref{broad},\ref{parshift}) the factor 2 comes from 2D diffusion,  
$\xi \simeq {\cal O}(1)$ and $\rho^g$ is the  effective gluon density. 
For the 1+1D Bjorken expansion scenario $A_T$ is the 
transverse area of the interaction region, $\tau_0$ is the initial 
equilibration time and  $dN^g/dy$ is the effective gluon rapidity
density. 
We note that  $-\Delta p_\parallel$  may mimic 
small elastic  energy  loss if the full structure  of  
${d^3{N}^{f}}/{d p^+ d^2 {\bf p}_T }$  is not observed.    

\begin{figure}[!t]
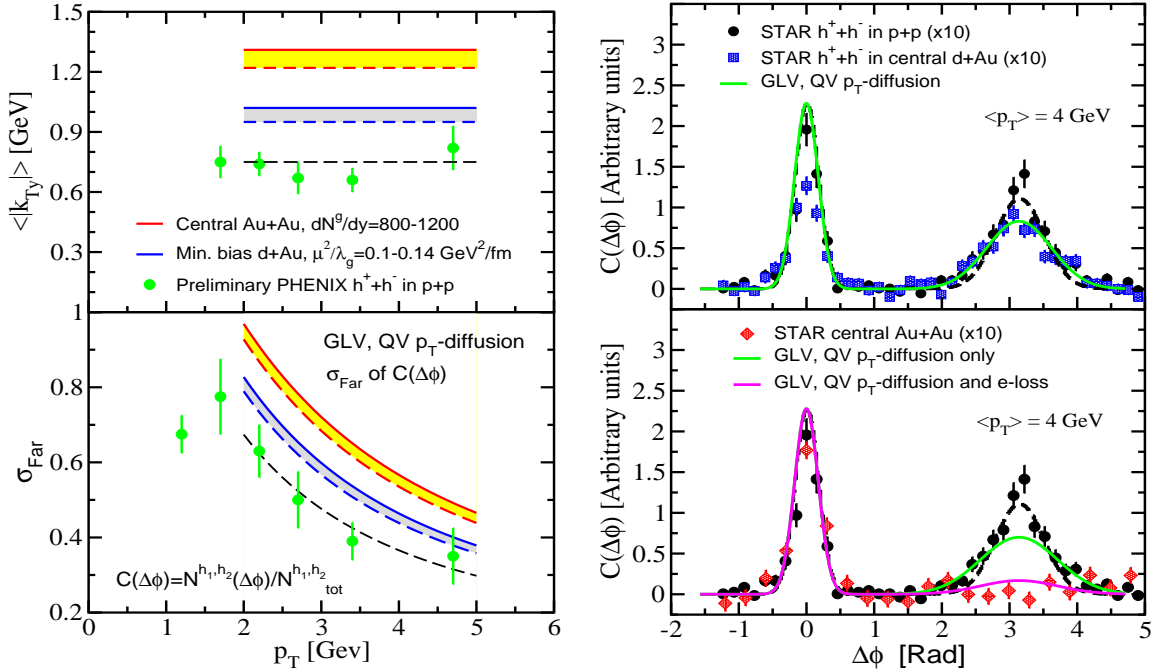

\hspace*{0.2cm}
  \includegraphics[width=2.9in,height=3.5in]{fig3-cipanp.eps}
\hspace*{0.3cm}
  \includegraphics[width=2.9in,height=3.5in]{fig4-cipanp.eps}
\vspace*{0.3cm}
  \caption{Left panel: predicted enhancement of 
$\langle |{\bf k}_{T\;y}| \rangle$  and  $\sigma_{Far}$ in minimum bias
$d+Au$ and central $Au+Au$ reactions at RHIC from 
$p_T$-diffusion~\cite{Qiu:2003pm}. Preliminary $p+p$ data is from 
PHENIX~\cite{Rak:2003ay}.  Right panel:  the broadening of the 
far-side di-hadron correlation function in central $d+Au$ and $Au+Au$ 
compared to scaled (x10) STAR data~\cite{Adler:2002tq,Adams:2003im}. In the 
bottom right panel the broadening with and without suppression, 
approximately given by $R_{AA}$, are shown. We have used the 
predicted quenching factor~\cite{Vitev:2002pf}, confirmed
by experimental data~\cite{Adler:2003qi,Back:2003qr,Adams:2003kv}. }
\label{fig2}
\end{figure}

As an application of the multiple initial and  final state elastic 
scattering formalism elaborated in~\cite{Qiu:2003pm,Gyulassy:2002yv}
 we consider the nuclear induced  
broadening of the  back-to-back jet correlations associated with
hard QCD  $ab \rightarrow cd$ partonic subprocesses.  We  will limit 
the  discussion  to the Gaussian 2D random walk approximation, 
Eq.~(\ref{gauss}), to make use of its additive dispersion property. 
Measurements of intra-jet correlations find an approximately Gaussian 
jet cone shape. If one defines $\langle |{\bf j}_{T\,y}| \rangle$ 
to be the average particle transverse momentum  relative to  the hard 
scattered parent parton in the plane normal to the collision axis,  
it  can be  related  to the width $\sigma_{Near}$ of the near-side 
$(\Delta \phi < \pi/2)$ di-hadron correlation function  
$C(\Delta \phi) = N^{h_1,h_2}(\Delta \phi) / N_{tot}^{h_1,h_2}$ 
as follows:  $\langle | {\bf j}_{T\,y} | \rangle = 
\langle | {\bf p}_T | \rangle  \sin ( \sigma_{Near} / \sqrt{\pi}) $. 
It is the away-side ($\Delta \phi > \pi/2$) correlation function, 
however, that measures the di-jet acoplanarity.  
The total vacuum+nuclear induced broadening for the two partons
in a  plane perpendicular to the collision axis in $p+A$ ($A+A$) 
reads~\cite{Qiu:2003pm}:
\begin{equation}
 \langle  {\bf k}_T^2 \rangle  = 
\langle  {\bf k}_T^2 \rangle_{vac} +   
\begin{array}{c} 1_{jet} \\ 
(2_{jets})
\end{array} 
\, \left( \frac{\mu^2}{\lambda} \right)_{eff} 
\langle L \rangle_{IS}  
+  2_{jets}\,  \left(\frac{1}{2} \right)_{projection}  \,
\left( \frac{\mu^2 }{\lambda} \right)_{eff} 
\langle L \rangle_{FS} \;\;.
\label{eq:netbr}
\end{equation}
A typical range for the cold nuclear matter transport coefficient 
for gluons $({\mu^2}/{\lambda}_g)_{eff,\;IS \approx FS} = 
2\times 0.1$~GeV$^2$/fm - $2\times 0.15$~GeV$^2$/fm is 
extracted from the analysis of low energy $p+A$ 
data~\cite{Vitev:2002pf,Vitev:2003xu}. This can be tested via the predicted  
Cronin enhancement  in $d+Au$ collisions  at RHIC 
$\sqrt{s}=200$~AGeV~\cite{Vitev:2002pf,Vitev:2003xu}, which compares well 
to BRAHMS, PHENIX, PHOBOS and STAR 
measurements~\cite{Vitev:2003jg,Adams:2003im,Arsene:2003yk,Adler:2003ii,Back:2003ns}. 
For the case  of FS scattering in a 1+1D Bjorken expanding quark-gluon plasma 
the final state broadening can  be evaluated from Eq.~(\ref{broad}).
The relation between $\langle |{\bf k}_{T\;y}| \rangle = 
\sqrt{\langle  {\bf k}_T^2 \rangle_{1\, parton}/\pi}$, 
$\langle {\bf k}_T^2 \rangle_{1\, parton} = 
\langle {\bf k}_T^2 \rangle/2$, the near-side and away-side widths 
$\sigma_{Near},\sigma_{Far}$ and $\langle |{\bf p}_T| \rangle $
in the hard fragmentation  $z=p_h/p_z \rightarrow 1$ limit 
is approximately given by:
\begin{equation} 
\langle | {\bf k}_{T\,y} | \rangle = \langle |{\bf p}_{T}| 
\rangle \cos \left( \frac{\sigma_{Near}}{\sqrt{\pi}} \right)
\sqrt{ \frac{1}{2}\tan^2 \left(  \sqrt{\frac{2}{\pi}}\, \sigma_{Far} \right) 
- \tan^2 \left( \frac{\sigma_{Near}}{\sqrt{\pi}}   \right) } \;\; .
\label{eq:relat}
\end{equation}

The left panel in Fig.~\ref{fig2} shows two measures of the predicted 
increase in di-jet acoplanarity for minimum bias $d+Au$ and central $Au+Au$ 
reactions~\cite{Qiu:2003pm}: $\langle |{\bf k}_{T\;y}| \rangle$
and the away-side width $\sigma_{Far}$  
of the di-hadron correlation function  $C(\Delta \phi) = 
N^{h_1,h_2}(\Delta \phi)/N_{tot}^{h_1,h_2}$. $C(\Delta \phi)$ is 
approximated here by near-side and far-side Gaussians for a symmetric 
$p_T^{h1} \approx p_T^{h2}$ case and the vacuum widths are  
taken from PHENIX~\cite{Rak:2003ay}. 
In the right panel of Fig.~\ref{fig2} di-hadron correlations in 
$d+Au$ are shown to be qualitatively similar to the $p+p$ case 
and in agreement with STAR measurements~\cite{Adler:2002tq,Adams:2003im}.
In $Au+Au$ reactions at RHIC di-jet acoplanarity is 
noticeably larger, but this effect alone does not lead to the 
reported  disappearance of the back-to-back 
correlations~\cite{Adler:2002tq}. To first approximation the 
coefficient of the away-side Gaussian
(the area under $C(\Delta \phi)$, $\Delta \phi > \pi/2$),  
is determined by jet energy loss and given by $R_{AA} \propto 
N_{part}^{2/3}$ in the GLV 
approach~\cite{Gyulassy:2000er,Gyulassy:2000fs,Gyulassy:2003mc}. 
Broadening with and  without away-side quenching 
is shown the bottom right panel of Fig.~\ref{fig2}. 
Combined $d+A$u and $Au+Au$ experimental data in Fig.~\ref{fig2} 
also rule out the existence of monojets at RHIC. For further 
discussion on di-hadron correlations 
see~\cite{Qiu:2003pm,Hirano:2003hq,Wang:2003mm}.
  
\begin{figure}
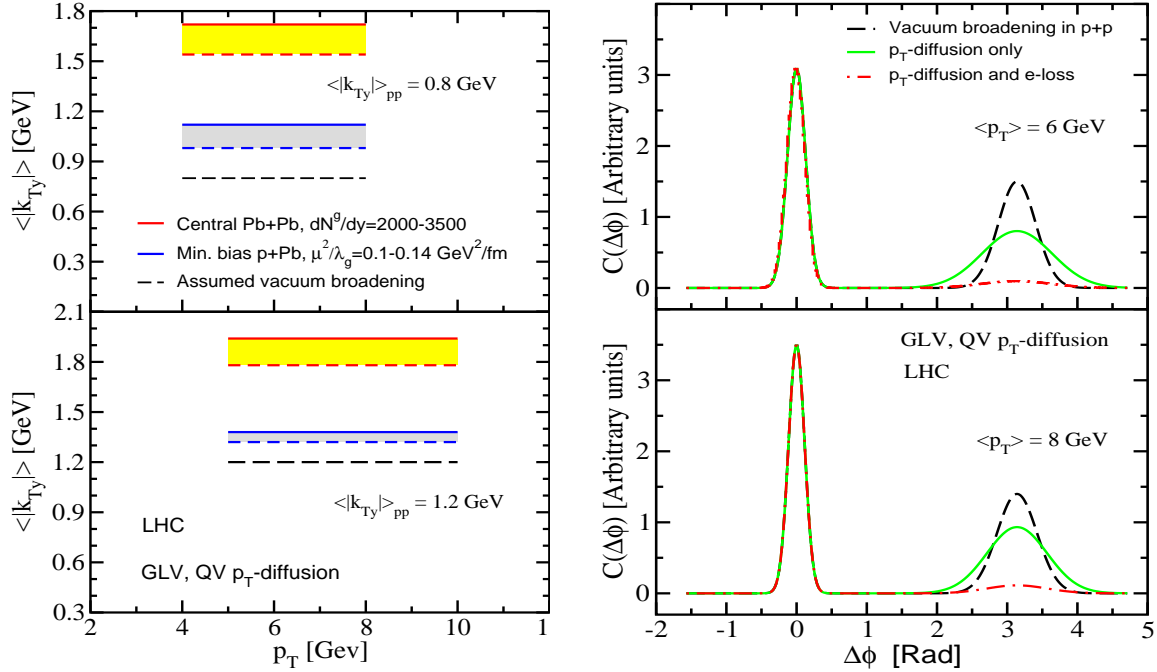

\hspace*{0.2cm}
  \includegraphics[width=2.9in,height=3.5in]{cern5-cor.eps}
\hspace*{0.3cm}
  \includegraphics[width=2.9in,height=3.5in]{cern6-cor.eps}
\vspace*{0.3cm}
  \caption{Left panel: di-jet acoplanarity quantified via the enhancement 
of  $\langle |{\bf k}_{T\;y}| \rangle$  in min. bias $p+Pb$ and 
central $Pb+Pb$ reactions at the LHC.  Two assumed baseline values 
for  $p+p$ vacuum broadening,  $\langle |{\bf k}_{T\;y}| \rangle_{vac} = 
0.8$~GeV, $1.2$~GeV.  Right panel:  the broadening of the 
far-side di-hadron correlation function in central $Pb+Pb$ collisions
with and without suppression. The quenching factor $R_{AA}$ is taken  
from~\cite{Vitev:2002pf}.}
\label{fig3}
\end{figure}

The broadening of the away-side di-hadron correlation function in $p+Pb$ and 
$Pb+Pb$ reactions at the LHC is shown in Fig.~\ref{fig3}. 
The near side  width  $\sigma_{Near}$  at 
$\langle |{\bf p}_T| \rangle = 6$~GeV, $8$~GeV 
is extrapolated  from the PHENIX~\cite{Rak:2003ay} and 
STAR~\cite{Adler:2002tq,Adams:2003im} measurements.
We use two baseline values for the vacuum radiation induced di-jet 
acoplanarity in $p+p$ reactions to account for its possible 
growth with $\sqrt{s}$ and $p_T$ relative to the PHENIX measurement,
$\langle |{\bf k}_{T\;y}| \rangle_{vac}=0.8$~GeV, $1.2$~GeV. In $p+A$ 
reactions the broadening of  $C(\Delta \phi)$, $\Delta \phi > \pi/2$      
comes from transverse momentum diffusion in cold nuclear matter. The 
band reflects a range of transport coefficients $\mu^2/\lambda = 
0.1$~GeV$^2$/fm - $0.14$~GeV$^2$/fm  as in Fig.~\ref{fig2} with 
$\sim 30\%$ increase of $\langle |{\bf k}_{T\;y}| \rangle$ 
relative to the vacuum case. In central $Pb+Pb$ reactions, where the 
hot and dense quark-gluon plasma is expected to be formed, 
the away-side width $\sigma_{Far}$ grows by approximately a factor of 
two relative to the $p+p$ case. 
The final state scattering strength in proportional to the 
gluon rapidity density of the medium and the band represents 
values in the range $dN^g/dy = 2000-3500$. 
In the right panel of Fig.~\ref{fig3} the broadening of the di-hadron 
correlation function in central $Pb+Pb$ with and without the 
corresponding  suppression  factor is shown. We note that a direct 
calculation that does not include the hydrodynamic feedback  at the
LHC energy and number densities  will result in suppression 
factors $R_{AA} < (N_{part}/2)/N_{bin}$~\cite{Vitev:2002pf}. 
In this case the $R_{AA}$ has been set to  $(N_{part}/2)/N_{bin} 
\approx  0.12 $ for $Pb+Pb$ at $\sqrt{s} = 5.5$~TeV. 
Because of its significantly larger  dynamical $p_T$ range,  
LHC may offer the best possibility to explore  the relation between 
single inclusive hadron suppression and the broadening and 
disappearance of the 
back-to-back jet correlations.

\subsubsection{High-$p_T$ Particle Production in Saturation Models}
\label{sec363}
{\em R. Baier, U.A. Wiedemann}

At RHIC, the production of high $p_T$ hadrons in central 
Au-Au collisions shows substantial differences compared to elementary
p-p collisions, see section~\ref{sec4}. As discussed in previous
sections, the observed depletion/suppression may be explained due 
to induced multiple gluon radiation off the large $p_T$ parton 
(``jet quenching''). Here, we consider an alternative
possibility due to initial state gluon radiation effects, especially
saturation effects, put forward first in~\cite{Kharzeev:2002pc}.

Instead of a detailed prediction for hadron production based on saturation
models \cite{pAwriteup}, we concentrate in the following on the problem
of suppression vs. enhancement of gluon production in A--A collisions
with shortly mentioning the relevant comparison in p--A scattering. 
This way, we do not include the fragmentation functions of gluons into 
hadrons and their possible medium dependence discussed in section~\ref{sec342}.
To proceed, we use the $k_T$ factorised formalism for calculating
gluon production which is expected to give a qualitatively reasonable
description of this process  \cite{Gribov:tu,Kharzeev:2002pc,Baier:2003hr}.

The basic factorised formula for the gluon yield at central
rapidity in a collision of identical nuclei is
\begin{equation}
   E{dN\over d^3p\, d^2b}
   ={4\pi^2\alpha_s S_{AA}(b)N_c\over N_c^2-1}{1\over p_T^2}
    \int d^2k_t\, \phi_A(y,k_t)\, \phi_A(y,p_T-k_t)\, .
 \label{eq2}
\end{equation}
Here $S_{AA}(b)$ is the overlap area in the transverse plane
between the nuclei at fixed impact parameter $b$, and $y$ is the
rapidity difference between the central rapidity and the
fragmentation region.
$\phi_A(y,k_t)$ is the intrinsic
momentum dependent nuclear gluon distribution function, related to
the standard gluon distribution by
\begin{equation}
  \phi_A(y,k_t)={d(xG_A(x,k^2_t))\over d^2k_t\, d^2b}\, .
  \label{eq3}
\end{equation}
In the following, we will also use the modified gluon
distribution 
\begin{equation}
  h_A(k_t) = k_t^2\, \nabla^2_{k_t}\, \phi_A(k_t)\, ,
  \label{hhh}
\end{equation}
which enters some calculations of the gluon yield instead of
$\phi_A$ \cite{Braun:2000bh,Kharzeev:2003wz,Albacete:2003iq}.
In general, and especially at low momenta, the distributions
$h_A$ and $\phi_A$ are different. However,
they coincide for the leading order perturbative 
distribution which at impact parameter $b$ has the shape
\begin{equation}
   \phi(k_t,b)
   \simeq {\alpha_s(N_c^2-1)\over 2\pi^2}{\rho_{\rm part}(b)\over 2}
    {1\over k^2_t}\, .
   \label{eq4}
\end{equation}
Here, $\rho_{\rm part}(b)$ is the density of participants, i.e.
for central collisions $\rho_{\rm part}(b) \propto A^{1/3}$
and $S_{AA} \propto A^{2/3}$. 

Since models for the gluon distribution are reviewed in 
chapter \cite{pAwriteup}, we limit the present discussion
to shortly reviewing their main features.

\paragraph{McLerran--Venugopalan gluon}
The McLerran-Venugopalan 
model~\cite{McLerran:1993ni,McLerran:1993ka} achieves saturation by
taking into account the Glauber-Mueller multiple scattering
effects. The intrinsic gluon distribution in this model was
calculated in \cite{Kovchegov:1996ty,Jalilian-Marian:1996xn}:
\begin{equation}
  \phi^{MV}_A(k_t)
  ={N^2_c-1\over 4\pi^4 \alpha_s N_c}
  \int {d^2{\bf x} \over {\bf x}^2}
  \left(1-e^{-{\bf x}^2Q_s^2({\bf x}^2)/4}\right)\,
  e^{i\, k_t \cdot {\bf x}}\, .
  \label{eq5}
\end{equation}
We will
take the saturation momentum to be ${\bf x}$-dependent
\begin{equation}
    Q_s^2({\bf x}^2,b)
    ={4\pi^2\alpha_s N_c\over N_c^2-1}\,
     xG(x,1/{\bf x}^2) {\rho_{\rm part}(b)\over 2}\, ,
  \label{eq6}
\end{equation}
with $G(x,k_t^2=1/{\bf x}^2)$ being the {\it nucleon} gluon
distribution:
\begin{equation}
   xG(x,1/{\bf x}^2)
   \simeq \frac{\alpha_s(N_c^2-1)}{2\pi}
    \ln\left( \frac{1}{ {\bf x}^2\Lambda^2_{\rm QCD}}
              \right)\, .
   \label{eq7}
\end{equation}

\paragraph{Evolved gluons}

The MV gluon distribution does not contain any evolution in 
Bjorken $x$ which is necessary to explore the energy 
dependence of the gluon spectrum. Also, small-$x$ evolution
leaves a distinct imprint in the $k_t$-dependence of the produced gluon.
It is argued~\cite{Iancu:2002tr,Mueller:2002zm} that in the wide region 
of momenta $Q_s<|k_t|<{Q_s^2\over Q_0}$ (typically, $Q_0 \sim O(1)$ GeV),
the evolved distribution behaves as (``geometric scaling'')
\begin{equation}
   \phi_A(k_t)
   \propto \left[{Q_s^2\over k_t^2}\right]^\gamma
   \label{eq8}
\end{equation}
with the anomalous dimension $\gamma=0.64$.
However, it is important to know how the distribution behaves
outside this scaling window since this decides about suppression
vs. enhancement of large $k_t$ gluon production. This is illustrated
by two models in \cite{Baier:2003hr}:

A gluon distribution with {\it fast} (F) crossover is
\begin{eqnarray}
  \phi^{\rm F}_A(k_t) = {N^2_c-1\over 4\pi^3 \alpha_s N_c}
   \left( {\frac{{\hat Q}^2_s}{k_t^2 + {\hat Q}_s^2}}
          \right)^{\gamma(k_t)} \, ,
\label{eq10}
\end{eqnarray}
where ${\hat Q}_s^2
= {2\pi\alpha_s^2 N_c} \, {\rho_{\rm part}(b)\over 2}$,
and $\gamma (k_t)$ is chosen to approach 1 rapidly 
(like a power law) for $k_t \gg \hat{Q}_s^2/Q_0$
\cite{Baier:2003hr}.

A {\it slow} (S) crossover can be modelled by a function
with fixed anomalous dimension:
\begin{eqnarray}
  \phi^{\rm S}_A(k_t) = {N^2_c-1\over 4\pi^3 \alpha_s N_c}
   \left( {\frac{{\hat Q}^2_s}{k_t^2 + {\hat Q}_s^2}}
          \right)^{0.64} \, .
\label{slow}
\end{eqnarray}
%
\begin{figure}[h]\epsfxsize=8.7cm
\vspace*{-0.6cm}
\centerline{\epsfbox{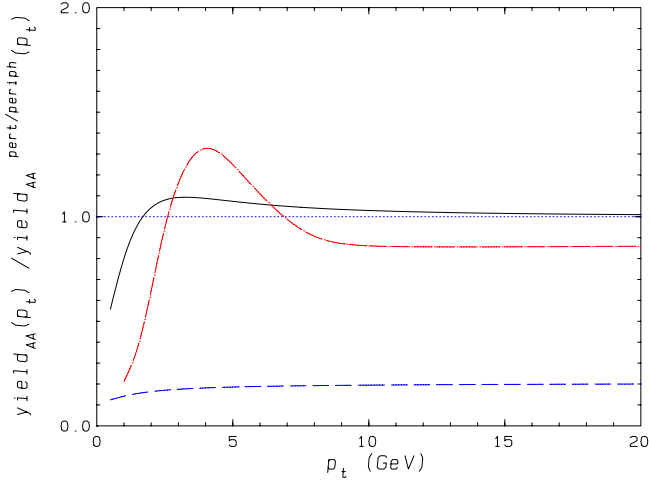}}
\caption{Cronin effect in the $p_{t}$-dependence of gluon production 
yields for head-on A-A collisions for $Q_s^2 = 2$ GeV$^2$.
The solid curve is for the MV-gluon distribution normalised to the 
perturbative yield, the dot-dashed curve is for the evolved gluon 
distribution (\ref{eq10}), and the dashed line is for the evolved 
gluon distribution (\ref{slow}).}
\label{Fig2cronin}
\end{figure}
%

The non-linear evolution of the nuclear gluon distributions has 
recently been calculated numerically \cite{Albacete:2003iq}, using 
the Balitsky-Kovchegov (BK) evolution equation 
\cite{Balitsky:1995ub,Kovchegov:1999yj}. The solutions for
$h(k_t)$ plotted versus the scaled variable $\rho = k_t/Q_s(x)$ 
approach a universal soliton-like shape independent of initial 
conditions if evolved sufficiently far in rapidity. This numerical
solution indicates that Eq.~(\ref{slow}) provides a more realistic
parametrisation of the evolved gluon distribution than Eq.~(\ref{eq10}) 
with a fast crossover.

Coming to gluon production in A-A collisions, one finds
with Eq.~(\ref{eq2}) 
for the perturbative gluon distribution Eq.~(\ref{eq4}), 
\begin{equation}
  \frac{dN^{\rm pert}(b)}{dy\, d^2p_T}\Bigg\vert_{y=0}
  \simeq 2 S_{AA}(b){N_c^2-1\over 4\pi^3 \alpha_s N_c}
  {{\hat Q}_s^4 (b)\over p^4_t}~
 \left( \ln {p_T^2\over 4 \Lambda^2_{\rm QCD}} \right) \, .
 \label{eq15}
\end{equation}
This ``reference'' spectrum
scales for all transverse momenta with $N_{\rm coll} \sim S_{AA}(b)\,
\rho_{\rm part}^2(b)$ ($\sim A^{4/3}$ at $b=0$), 
as expected perturbatively.

In contrast, using the saturated gluon distribution in the MV model,
the gluon yield Eq.~(\ref{eq2}) is suppressed at small momenta 
compared to the perturbative one and scales with the 
number of participants, $N_{\rm part}(b) = S_{AA}(b)~\rho_{\rm part}(b)$
($\sim A$ for $b=0$). It approaches the perturbative yield (\ref{eq15})
from above at large $p_T$. 

In Fig.~\ref{Fig2cronin}, we summarise the results for the normalised
ratio of central over peripheral (perturbative) yields, corresponding to
the nuclear modification factor
\begin{eqnarray}
   R_{AA} = { {dN_{AA}\over dyd^2p\, d^2b}\over A^{4/3} 
            {dN_{pp}\over dyd^2p\, d^2b}}\, ,
          \nonumber
\end{eqnarray}
here quoted at $b=0$. For the MV model,
one sees a small but clear Cronin enhancement
for momenta just above the saturation scale.
The distribution $\phi^{\rm F}_A$ displays a clear 
Cronin effect similar to the MV gluon, while $\phi_A^{\rm S}$ shows 
uniform suppression for the central/peripheral ratio for all momenta.
This illustrates, indeed, that the ratio $R_{AA}$ is very sensitive to
the way in which the distribution behaves outside the scaling window.

\begin{figure}
\hspace*{0.2cm}
  \includegraphics[width=3.1in,height=2.9in]{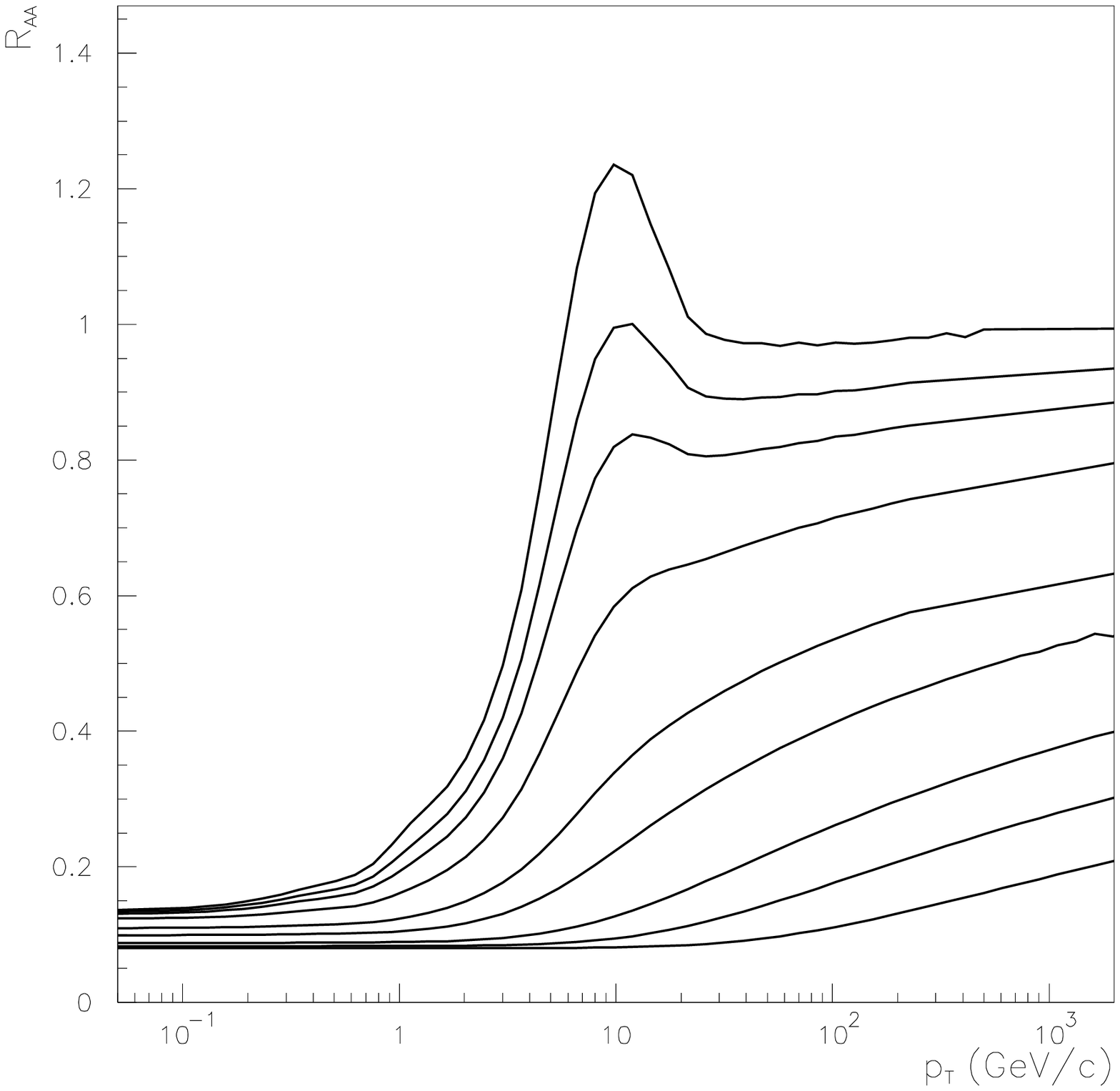}
\hspace*{0.3cm}
  \includegraphics[width=3.1in,height=2.9in]{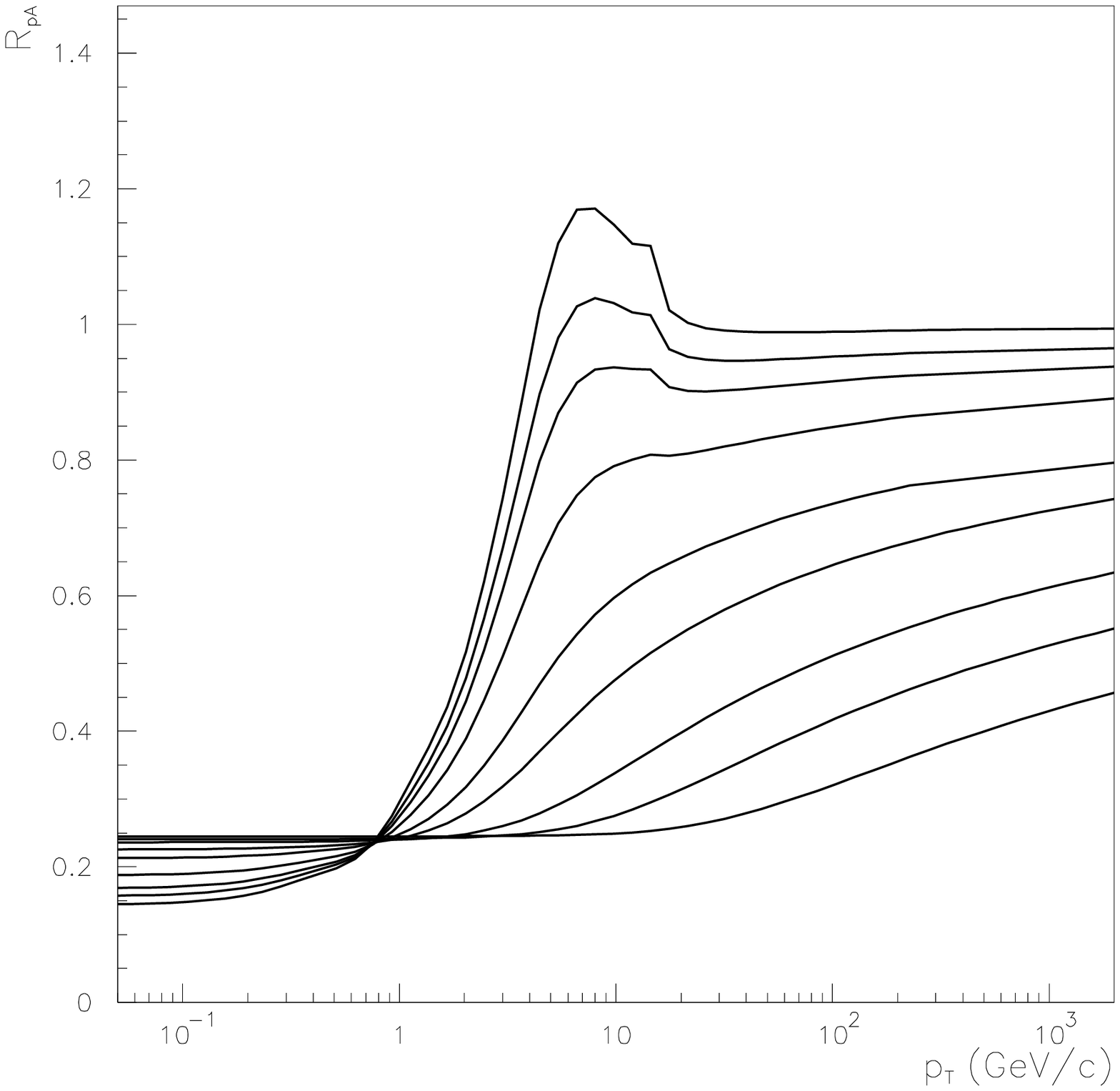}
\vspace*{0.3cm}
  \caption{Ratios $R_{AA}$ and $R_{pA}$ of gluon yields in 
A--A (LHS) and p--A (RHS) for BK evolution, with MV
as initial condition with $Q^2_s=0.1$ GeV$^2$ for p and 2 GeV$^2$ for A.
Lines from top to bottom correspond to 
$y= (\alpha_s\, N_c/\pi) \ln 1/x = 0$, 0.05, 0.1, 0.2, 0.4, 0.6, 1,
1.4 and 2. 
}
\label{fig3anticronin}
\end{figure}

For the numerically evolved gluon distribution $h$, the factor 
$R_{AA}$ is shown in Fig.~\ref{fig3anticronin}.
%

%
The non-linear BK evolution quickly
wipes out any initial Cronin enhancement not only on the level
of single parton distribution functions but also on the level of 
particle spectra. Several checks establish that this behaviour is 
generic \cite{Albacete:2003iq}.
For 'realistic' initial conditions this 
disappearance occurs within half a unit of rapidity. We note that in 
our units the evolution from $130$ GeV to $200$ GeV corresponds to 
$\delta y\simeq 0.1$ for $\alpha_s = 0.2$, and thus is not sufficient to 
completely eliminate an initial enhancement at central rapidity. For 
forward rapidity, $\delta y$ is greater.
The evolution to the LHC energy corresponds to 
$\delta y \sim 1$. Thus the BK evolution suggests the reduction
of the Cronin effect in d--Au for forward rapidities at RHIC and
predicts its disappearance for p--A 
collisions at LHC.

This numerical study 
~\cite{Albacete:2003iq} strongly indicates that crossover from the scaling
regime to the perturbative one is very slow and gradual, and that the Cronin
effect which is present in the MV gluon is wiped out by the quantum evolution
at high energies. Thus, $\phi_A^{\rm S}$ in Eq.~(\ref{slow}) seems
to provide a more realistic parametrisation of the evolved gluon
distribution than $\phi_A^{\rm F}$ in Eq.~(\ref{eq10}).

On the qualitative level, however, we observe that the gluon distributions 
which lead to the Cronin effect in d-Au collisions also lead to the Cronin
enhancement in the Au-Au collisions (see Fig.~\ref{fig3anticronin}). 
And vice versa, if no Cronin effect appears in Au-Au, none is seen in 
d-Au collisions.
Given the recent experimental observation of the Cronin enhancement
in d-Au collisions at RHIC (see section~\ref{sec4}), this supports 
the view that
significant final state (``quenching'') effects are needed in
order to account for the Au-Au data.

At LHC, due to higher energies, quantum 
evolution according to the BK equation will suppress gluon production
in p--A as well as in A--A collisions (see Fig.~\ref{fig3anticronin}).

\section{EXPERIMENTAL STATUS AT RHIC}
\label{sec4}
{\em D.~d'Enterria}

We summarize the main results on hard scattering processes in Au+Au, 
p+p, and d+Au collisions at $\sqrt{s_{NN}}$ = 200 GeV obtained after 
3 years of operation at the BNL Relativistic Heavy-Ion Collider (RHIC).
The main observations so far at RHIC are the following:

\begin{itemize}
\item  The high $p_{T}$ yields of inclusive charged hadrons and $\pi^0$ 
in central Au+Au at 
$\sqrt{s_{_{NN}}}$ = 130~\cite{Adcox:2001jp,Adler:2002xw,Adcox:2002pe} 
and 200 GeV~\cite{Adler:2003qi,Adams:2003kv,Adler:2003au,Back:2003qr,Arsene:2003yk},
are suppressed by as much as a factor 4 -- 5 compared to p+p and 
peripheral Au+Au yields scaled by $T_{AB}$ (or $N_{coll}$).

\item At intermediate $p_{T}$'s ($p_{T}\approx$ 2. -- 4. GeV/$c$) in 
central Au+Au, at variance with mesons ($\pi^0$~\cite{Adler:2003qi} 
and $K$'s~\cite{Adams:2003am}) no suppression is seen for baryons 
($p,\bar{p}$~\cite{Adcox:2001mf,Adler:2003kg} and 
$\Lambda,\bar{\Lambda}$~\cite{Adams:2003am}), yielding 
an ``anomalous'' baryon over meson ratio $p/\pi\sim$ 1 much larger 
than the ``perturbative'' $p/\pi\sim$ 0.1 -- 0.3 ratio observed in 
p+p collisions~\cite{Alper:1975jm,Angelis:fk} and in $e^+e^-$ 
jet fragmentation~\cite{Abreu:2000nw}. 

\item The near-side azimuthal correlations of high $p_{T}$ (leading) 
particles emitted in central and peripheral Au+Au 
reactions~\cite{Adler:2002tq,Chiu:2002ma} are, on the one hand, clearly 
reminiscent of jet-like parton fragmentation as found in p+p collisions. 
On the other, away-side azimuthal correlations (from back-to-back jets) 
in central Au+Au collisions are found to be significantly suppressed 
\cite{Adler:2002tq}.

\item At low $p_{T}$ the strength of the azimuthal anisotropy parameter 
$v_{2}$ is found to be large and consistent with hydrodynamical 
expectations for elliptic flow. Above $p_{T}\sim$ 2 GeV/$c$ where the 
contribution from collective behaviour is negligible, $v_{2}$ has still 
a sizeable value with a flat (or slightly decreasing) behaviour as a 
function of $p_{T}$~\cite{Adams:2003am,Adler:2002ct,Adler:2003kt}.

\item High $p_{T}$ production in ``cold nuclear matter'' as probed in 
d+Au reactions~\cite{Adler:2003qs,Adams:2003im,Back:2003ns,Arsene:2003yk} 
not only is not suppressed 
{\it enhanced} compared to p+p collisions, in a way very much reminiscent 
of the ``Cronin enhancement'' observed in p+A collisions at lower 
center-of-mass energies~\cite{Antreasyan:cw}.
\end{itemize}

All these results point to strong medium effects at work in central 
Au+Au collisions, and have triggered extensive theoretical discussions
based on perturbative 
or ``classical''-field QCD. Most of the studies on the high $p_{T}$ 
suppression are based on the 
prediction~\cite{Gyulassy:1990ye,Gyulassy:1993hr,Baier:1994bd} that 
a deconfined and dense medium would induce multiple gluon radiation 
off the scattered partons, effectively leading to a depletion 
of the high-$p_{T}$ hadronic fragmentation products (``jet quenching''), 
though alternative interpretations have been also put forward 
based on initial-state gluon saturation effects (``Color Glass 
Condensate'', CGC) ~\cite{Kharzeev:2002pc}, 
or final-state hadronic reinteractions~\cite{Gallmeister:2002us}. 
The different behaviour of baryons and mesons at moderately high 
$p_{T}$'s has been interpreted, among others, in terms of 
``quark recombination'' (or coalescence) effects in a thermalized 
partonic (QGP-like) medium 
~\cite{Hwa:2003bn,Greco:2003xt,Fries:2003vb}, whereas the 
disappearance of the back-to-back azimuthal correlations 
can be explained in both QGP energy loss and CGC monojet scenarios. 
Finally, the large value of $v_{2}$ above 2 GeV/$c$ has been 
addressed by jet energy loss~\cite{Gyulassy:2000gk}, gluon 
saturation~\cite{Kovchegov:2002nf}, and quark 
recombination~\cite{Molnar:2003ff} models.

This summary report presents the $p_{T}$, $\sqrt{s_{NN}}$, 
centrality, particle-species, and rapidity dependence of the 
inclusive high $p_{T}$ particle production, plus the
characteristics of the produced jets and collective elliptic flow signals
extracted from the azimuthal correlations at large $p_{T}$, 
as measured by the four experiments at RHIC (BRAHMS, PHENIX, PHOBOS and STAR) 
in Au+Au, p+p, and d+Au collisions. The whole set of experimental data puts
strong constraints on the different proposed physical explanations for 
the underlying QCD medium produced in heavy-ion collisions at RHIC and 
at LHC energies.\\


\subsection{High-$p_{T}$ Hadron Production in p+p Collisions at 
$\sqrt{s}$ = 200 GeV}

\paragraph{p+p inclusive cross-sections:}

Proton-proton collisions are the baseline ``vacuum'' reference to which 
one compares the Au+Au results in order to extract information about 
the QCD medium properties. At $\sqrt{s}$ = 200 GeV, there currently 
exist three published measurements of high $p_{T}$ hadron cross-sections 
in $p+p(\bar{p})$ collisions: UA1 $p+\bar{p}\rightarrow h^\pm$ 
($|\eta|<2.5$, $p_{T} <$ 7 GeV/$c$)~\cite{Albajar:1989an}, PHENIX 
$p+p\rightarrow \pi^0$ ($|\eta|<0.35$, $p_{T}<$ 14 GeV/$c$)
~\cite{Adler:2003pb}, and STAR $p+p\rightarrow  h^\pm$ ($|\eta|<0.5$, 
$p_{T}<$ 10 GeV/$c$)~\cite{Adams:2003kv}. At $\sqrt{s}$ = 130 GeV, an 
interpolation between the ISR inclusive charged hadron cross-section 
and UA1 and FERMILAB data, has been also used as a reference for Au+Au 
at this value of $\sqrt{s}$. Globally the spectra can be reasonably 
well parametrized by a power-law form  $A \cdot (1+p_{T} /p_0)^{-n}$ with 
the parameters reported in Table 4. 
We note that
the fit parameters $p_0$ and $n$ are actually strongly correlated 
via the mean $p_{T}$ of the collision: 
$\langle p_{T} \rangle = 2p_{0}/(n-3)$.

\begin{table}[htb]
\begin{center}
\begin{tabular}{l|c|c|c|c|c}
\hline\hline
\hspace{1mm} 
system & $\sqrt{s}$ (GeV) &  $p_{T}^{min}$ ($\frac{\rm GeV}{\rm c}$) 
& $A$ ($\frac{\rm mb\, c^3}{\rm GeV^2}$) &  $p_0$ ($\frac{\rm GeV}{\rm c}$) 
& $n$ \\\hline
 $p+p \rightarrow h^\pm$ (\small{inel., interpolation}~\cite{Adcox:2001jp}) &  130  & 0.4 & 330 & 1.72  &  12.4\\
 $p+\bar{p} \rightarrow h^\pm$ (NSD, UA1)~\cite{Albajar:1989an}                    &  200  & 0.25  & 286   & 1.8   &  12.14\\
 $p+p \rightarrow h^\pm$ (NSD, STAR)~\cite{Adams:2003kv}               &  200  & 0.4   & 286   & 1.43  &  10.35\\
 $p+p \rightarrow \pi^0$ (inel., PHENIX)~\cite{Adler:2003pb}       &  200  & 1.0   & 386   & 1.22  &   9.99\\
\hline\hline
\end{tabular}
\label{tab:powerlaw_fits}
\end{center}
{\small Table 4: Parameters of the fit $ Ed^{3}\sigma/dp^{3} = A 
\cdot (1+p_{T} /p_0)^{-n}$ to the inclusive $p_{T}$ distributions of all 
existing $p+p(\bar{p})$ hadron (inelastic or non-singly-diffractive)
cross-section measurements at 
$\sqrt{s}$ = 200 GeV.}
\end{table}

\begin{figure}[htbp]
\begin{center}
\begin{minipage}[t]{75mm}
\includegraphics[height=9.cm]{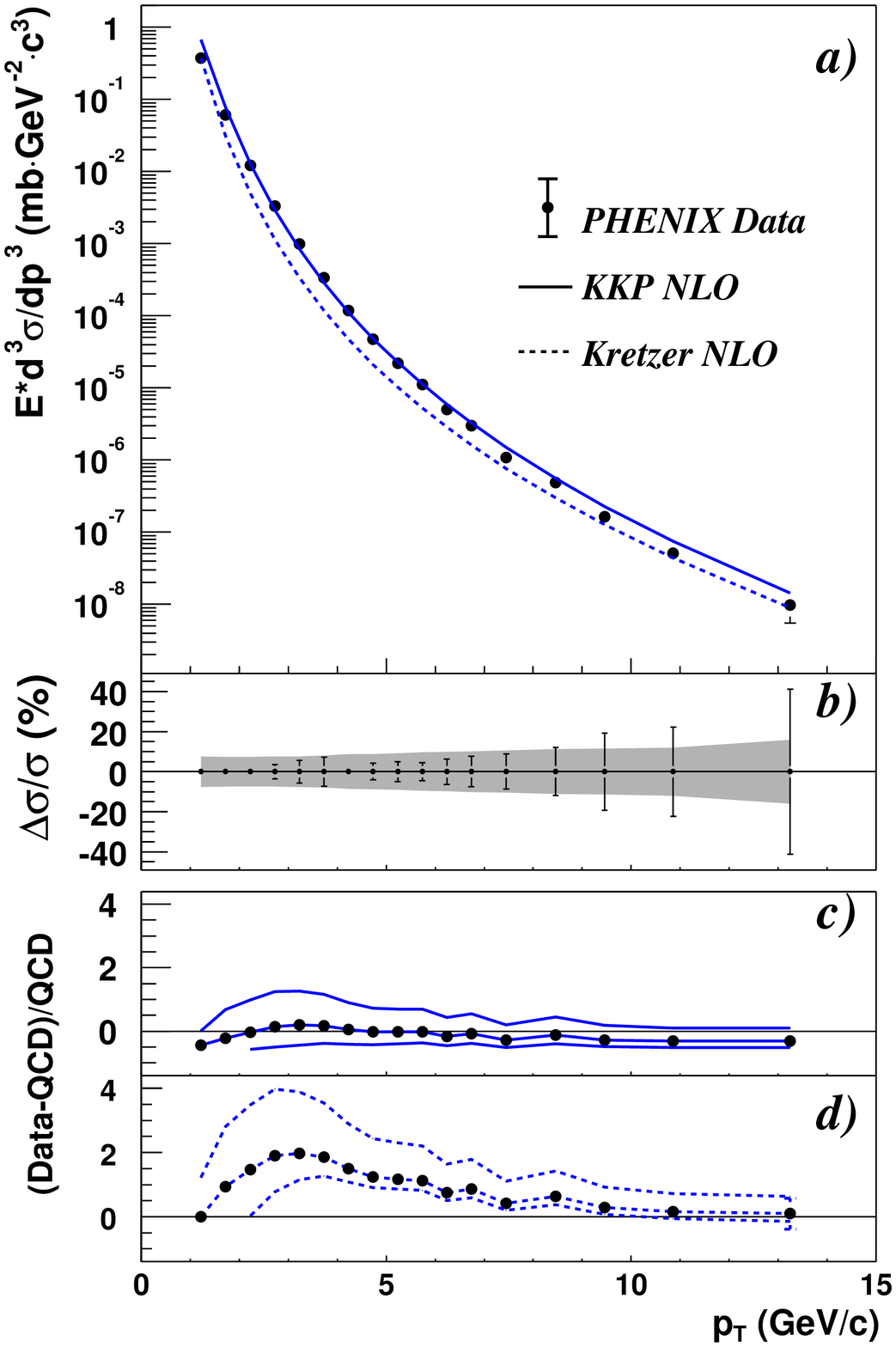}
\end{minipage}
\begin{minipage}[t]{75mm}
\includegraphics[height=9.6cm]{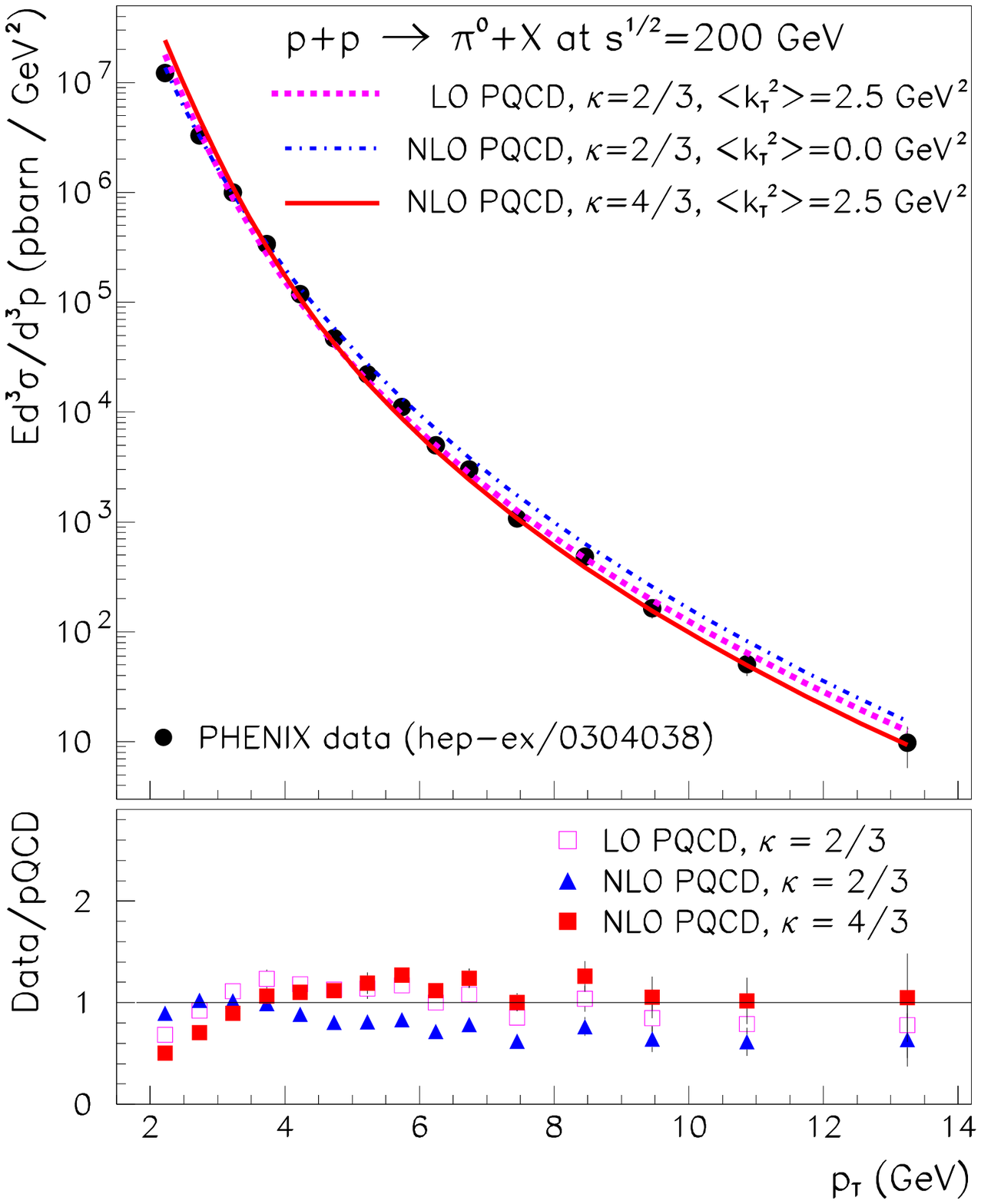}
\end{minipage}
\end{center}
\vspace*{-0.5cm}
\caption{High $p_{T}$ $\pi^0$ cross-section in p+p collisions at 
$\sqrt{s}$ = 200 GeV (PHENIX) compared to the results of two different 
NLO pQCD calculations: \cite{Adler:2003pb} ({\it left}),
~\cite{Barnafoldi:2002xp} ({\it right}).}
\label{fig:pp_pi0_vs_pQCD}
\end{figure}

In general, all experimental results are consistent within each other, 
although it is claimed~\cite{Adams:2003kv} STAR p+p inclusive charged 
yield is smaller by a factor of 0.79 $\pm$ 0.18 compared to UA1 
$p+\bar{p}$ results (approximately independent of $p_{T}$), the 
difference due in large part to differing non-singly-diffractive  
(NSD) cross section measured (35 $\pm$ 1 mb~\cite{Albajar:1989an} 
in the first and 30.0 $\pm$ 3.5 mb~\cite{Adams:2003kv} in the later). 
(The PHENIX high $p_{T}$ $\pi^0$ cross-section is inclusive and contains, 
in principle, all inelastic (including diffractive) channels.)
Standard next-to-leading-order (NLO) perturbative QCD calculations 
describe well the available high $p_{T}$ p+p data at $\sqrt{s}$ = 200 GeV 
(see Fig.~\ref{fig:pp_pi0_vs_pQCD} for $\pi^0$).

\vspace*{-0.5cm}

\paragraph{p+p azimuthal correlations:}

PHENIX~\cite{Rak:2003ay} has studied the azimuthal correlations at high 
$p_{T}$ in p+p collisions at $\sqrt{s}$ = 200 GeV extracting several 
parameters characterizing the produced jets:

\begin{itemize}
\item Mean jet fragmentation transverse momentum: 
$\langle |j_{\perp y}|\rangle$ = 373 $\pm$ 16 MeV/$c$, 
in agreement with previous measurements at ISR~\cite{Angelis:1980bs}
and showing no significant trend with increasing $\sqrt{s}$.
\item Average parton transverse momentum (fitted to a constant above 
1.5 GeV/$c$): $\langle |k_{\perp y}|\rangle$ = 725 $\pm$ 34 MeV/$c$. 
The momentum of the pair $p_{\perp}$ is related to the individual 
parton $\langle |k_{\perp y}|\rangle$ via 
$\sqrt{\langle |p_{\perp}^{2}|\rangle_{pair}}\, = 
\, \sqrt{2\pi} \,\langle |k_{\perp y}|\rangle$.
The extracted $\sqrt{\langle |p_{\perp}^{2}|\rangle_{pair}}$ = 
1.82 $\pm$ 0.85 GeV/$c$ is in agreement with the existing systematics 
of dimuon, diphoton and dijet data in hadronic 
collisions~\cite{Apanasevich:1998ki}.
\end{itemize}


\subsection{High-$p_{T}$ Hadron Production in Au+Au Collisions}

There is a significant amount of high $p_{T}$ Au+Au experimental 
spectra ($p_{T}>$ 2 GeV/$c$) measured by the 4 experiments at RHIC: 
inclusive charged hadrons at 130~\cite{Adcox:2001jp,Adler:2002xw,Adcox:2002pe} 
and 200 GeV~\cite{Adams:2003kv,Adler:2003au,Back:2003qr,Arsene:2003yk}, 
neutral pions at 130~\cite{Adcox:2001jp} and 200 GeV~\cite{Adler:2003qi}, 
protons and antiprotons at 130~\cite{Adcox:2001mf} and 200 
GeV~\cite{Adler:2003kg}, $K^0_s$ at 200 GeV~\cite{Adams:2003am}, and
$\Lambda,\bar{\Lambda}$ at 200 GeV~\cite{Adams:2003am}. 
Moreover, all these 
spectra are measured for different centrality bins and permit to address
the impact parameter dependence of high $p_{T}$ production.

\begin{figure}[htbp]
\begin{center}
\begin{minipage}[t]{75mm}
\includegraphics[height=6.0cm]{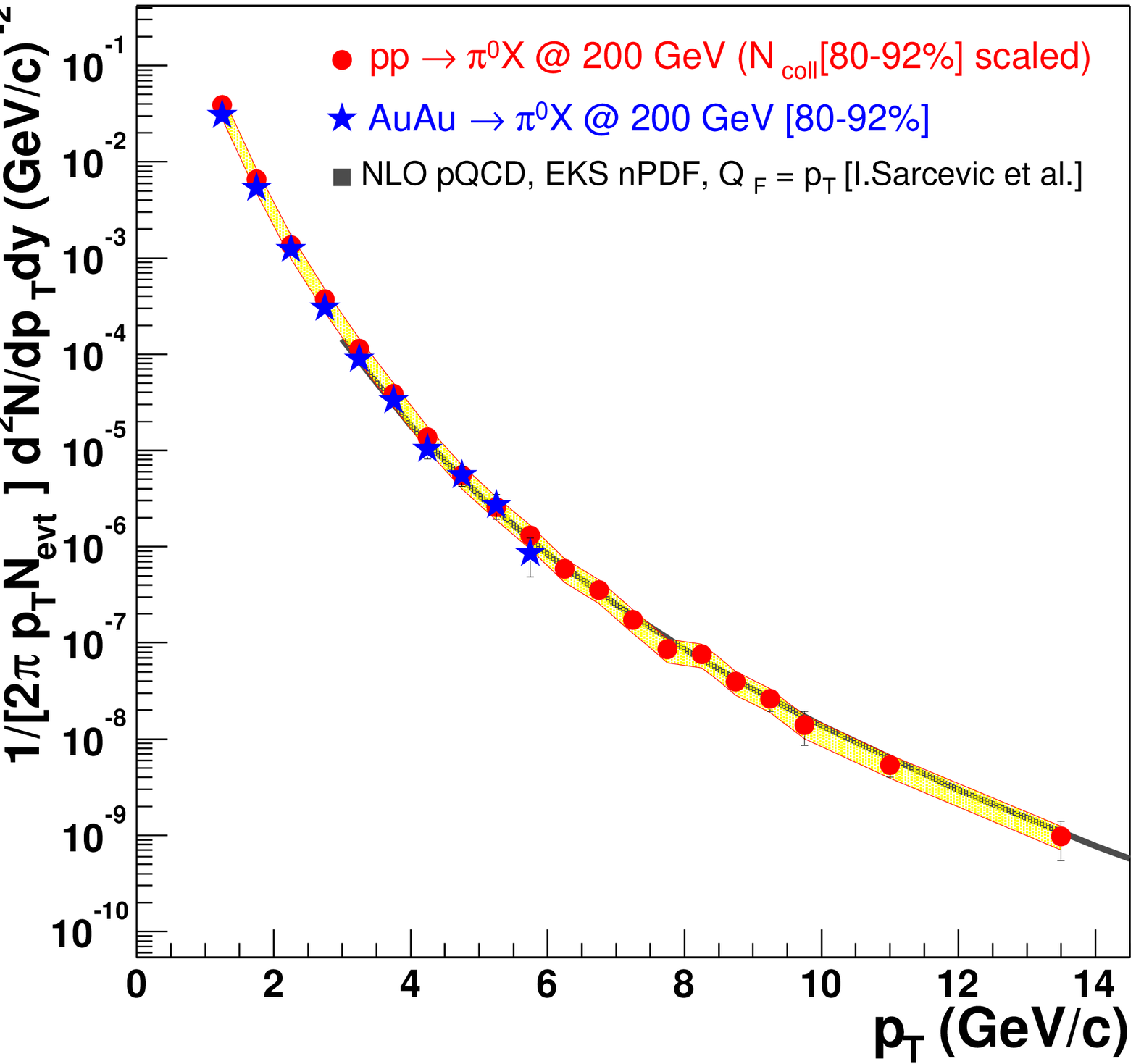}
\end{minipage}
\begin{minipage}[t]{75mm}
\includegraphics[height=6.0cm]{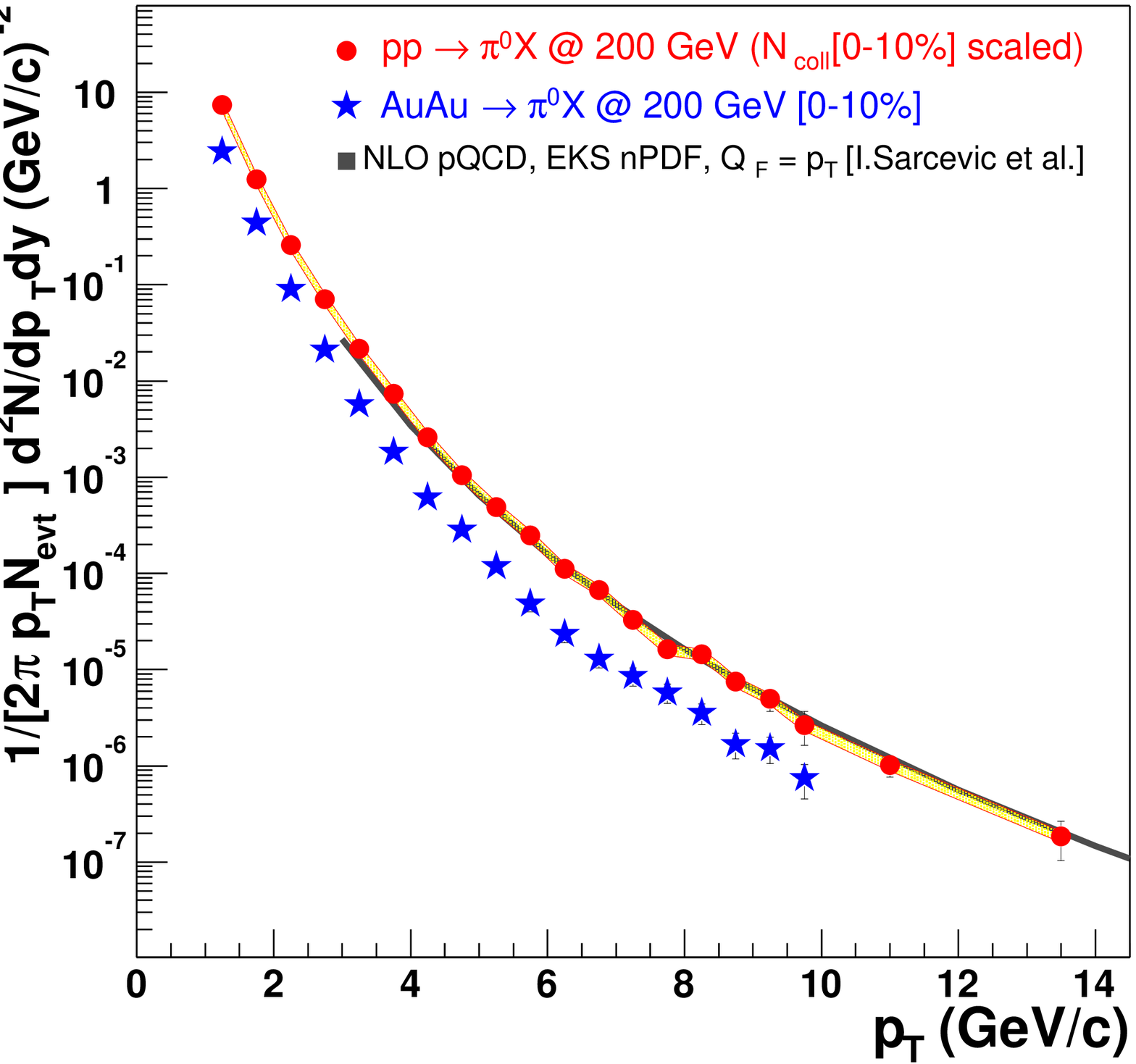}
\end{minipage}
\end{center}
\vspace*{-0.3cm}
\caption{Invariant $\pi^0$ yields measured by PHENIX in peripheral 
({\it left}) and in central ({\it right}) Au+Au collisions (stars), 
compared to the $N_{coll}$ scaled p+p $\pi^0$ yields (circles) and to
a NLO pQCD calculation~\cite{Jeon:2002dv} (gray line). The yellow band 
around the scaled p+p points includes in quadrature the absolute 
normalization errors in the p+p and Au+Au spectra as well as the 
uncertainties in $T_{AB}$. Updated version of Fig. 1 of 
\protect\cite{d'Enterria:2002bw} with final published data 
\protect\cite{Adler:2003qi,Adler:2003pb}.}
\label{fig:phenix_pi0_pp_AuAu}
\end{figure}

Details on hadron production mechanisms in $AA$ are usually studied
via their scaling behavior with respect to p+p collisions. On the one hand, 
soft processes ($p_{T}<$ 1 GeV/$c$) are expected to scale with 
$N_{part}$~\cite{Bialas:1976ed} (and they actually approximately 
do~\cite{Adcox:2000sp,Back:2002uc,d'Enterria:2003qs}). On the other hand, 
in the framework of collinear factorization, hard processes are 
incoherent and thus expected 
to scale with $N_{coll}$. 
result at RHIC in the high $p_{T}$ sector is the {\it breakdown} of this 
$N_{coll}$ scaling for central Au+Au collisions. Fig. 
\ref{fig:phenix_pi0_pp_AuAu} shows the comparison of the p+p $\pi^0$ 
spectrum to peripheral (left) and central (right) Au+Au spectra, and 
to pQCD calculations. Whereas peripheral data is consistent with a simple 
superposition of individual $NN$ collisions, central data shows a 
suppression factor of 4 -- 5 with respect to this expectation.

It is customary to quantify the medium effects at high $p_{T}$ using the 
{\it nuclear modification factor} given by the ratio of the $AA$ to the 
p+p invariant yields scaled by the nuclear geometry ($T_{AB}$):
\begin{equation} 
R_{AA}(p_{T})\,=\,\frac{d^2N^{\pi^0}_{AA}/dy dp_{T}}{\langle 
T_{AB}\rangle\,\times\, d^2\sigma^{\pi^0}_{pp}/dy dp_{T}}.
\label{eq:R_AA}
\end{equation}
$R_{AA}(p_{T})$ measures the deviation of $AA$ from an incoherent 
superposition of $NN$ collisions in terms of suppression ($R_{AA}<$1) 
or enhancement ($R_{AA}>$1). 


\paragraph{High $p_{T}$ suppression: magnitude and $p_{T}$ dependence}

\begin{figure}[htbp]
\begin{minipage}[t]{75mm}
\includegraphics[height=7.0cm]{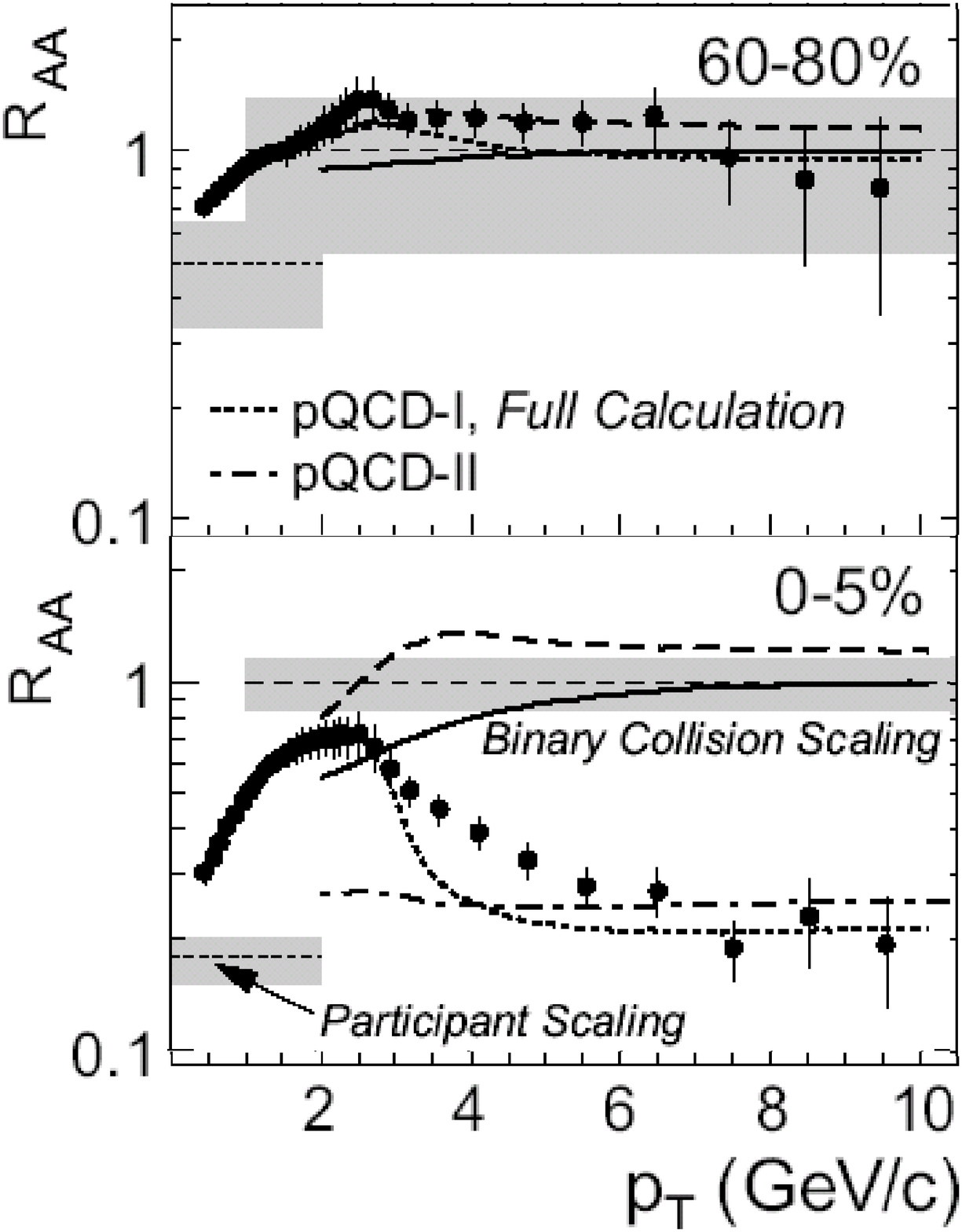}
\end{minipage}
\begin{minipage}[t]{75mm}
\includegraphics[height=6.8cm]{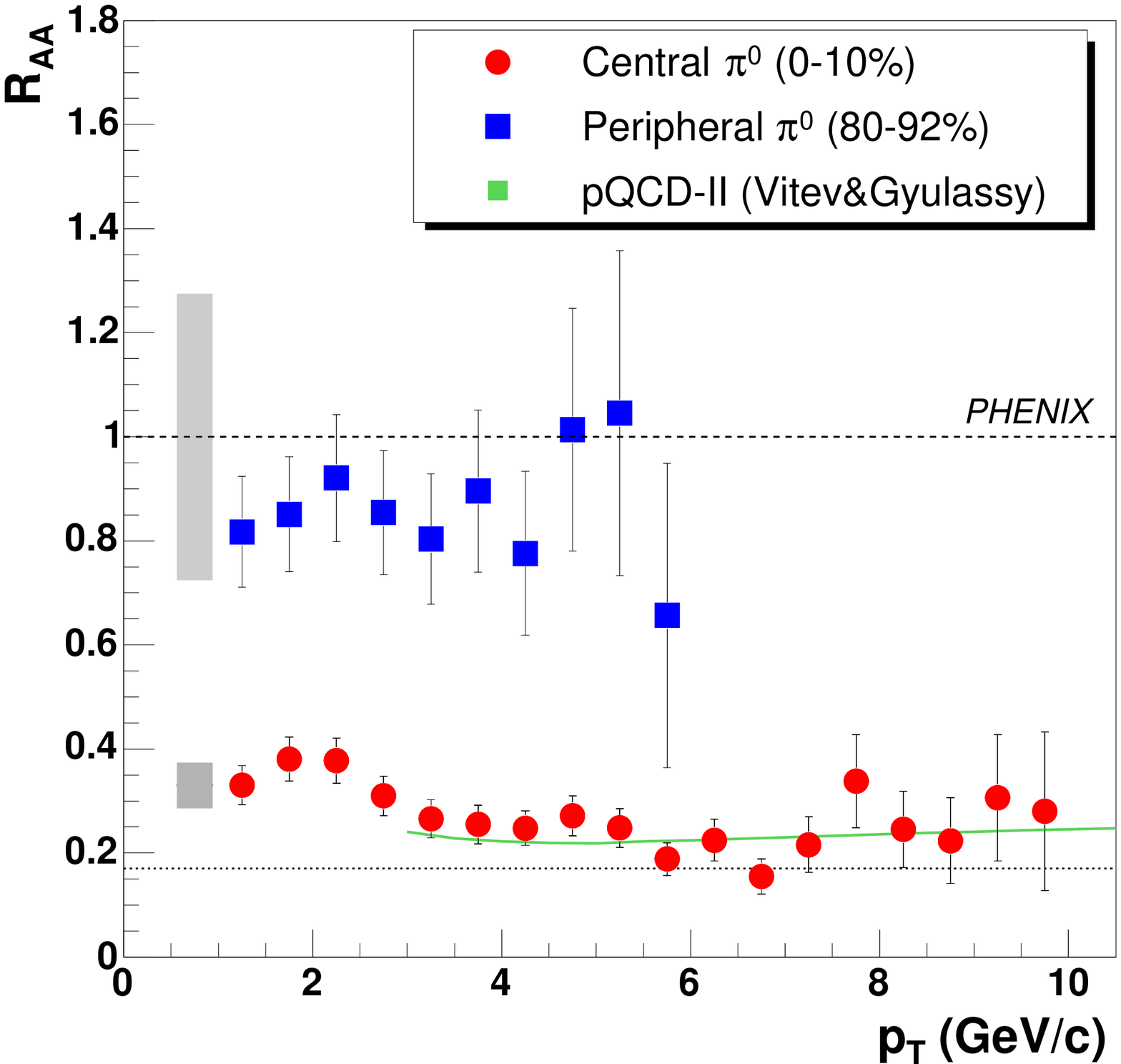}
\end{minipage}
\vspace*{-0.3cm}
\caption[]{Nuclear modification factor, $R_{AA}(p_{T})$, in peripheral 
and central Au+Au reactions for charged hadrons ({\it left}) and $\pi^0$ 
({\it right}) measured at $\sqrt{s_{NN}}$ = 200 GeV by STAR and PHENIX 
respectively. A comparison to theoretical curves: pQCD-I~\cite{Vitev:2002pf}, 
pQCD-II~\cite{Wang:2003mm}, is also shown.}
\label{fig:R_AA_phenix_star}
\end{figure}

Fig.~\ref{fig:R_AA_phenix_star} shows $R_{AA}(p_{T})$ for 
$h^\pm$ (STAR~\cite{Adams:2003kv}, left) and $\pi^0$ 
(PHENIX~\cite{Adler:2003qi}, right) measured in peripheral 
(upper points) and central (lower points) Au+Au reactions at 
$\sqrt{s_{NN}}$ = 200 GeV. As seen in Fig.~\ref{fig:phenix_pi0_pp_AuAu}, 
peripheral collisions are consistent with p+p collisions plus 
$N_{coll}$ scaling as well as with standard pQCD 
calculations~\cite{Vitev:2002pf,Wang:2003mm}, while 
central Au+Au are clearly suppressed by a factor $\sim$ 4 -- 5.
(Although peripheral STAR charged hadron data seems to be slightly 
above $R_{AA}$ = 1 and PHENIX $\pi^0$ data seems to be below, within errors 
both measurements are consistent with ``collision scaling''.)

The high $p_{T}$ suppression in central collisions for both $\pi^0$ and 
$h^\pm$ is smallest at $p_{T}$ = 2 GeV/$c$ and increases to an approximately 
constant suppression factor of 1/$R_{AA}\approx$~4 -- 5 over $p_{T}$~=
~5 -- 10~GeV/$c$. {\it Above} 5 GeV/$c$ the data are consistent within 
errors with ``participant scaling'' given by the dotted line at 
$R_{AA}\approx$ 0.17 in both plots (actually both STAR and PHENIX data 
are systematically slightly above this scaling). The magnitude and 
$p_{T}$ dependence of $R_{AA}$ in the range $p_{T}$ = 1 -- 10 GeV/$c$ 
(corresponding to parton fractional momenta $x_{1,2}=p_{T}/\sqrt{s}(
e^{\pm y_{1}}+e^{\pm y_{2}})\approx 2p_{T}/\sqrt{s}\sim$ 
0.02 -- 0.1 at midrapidity), is alone inconsistent with ``conventional'' 
nuclear effects like leading-twist shadowing of the nuclear parton 
distribution functions (PDFs)~\cite{Eskola:1998df,Klein:2002pi}.
Different pQCD-based jet quenching calculations~\cite{Vitev:2002pf,Wang:2003mm,Arleo:2002kh,Barnafoldi:2002xp,Salgado:2002cd,Salgado:2003gb} based 
on medium-induced radiative energy loss, can reproduce the {\it magnitude} 
of the $\pi^0$ suppression assuming the formation of a hot and dense 
partonic system characterized by different, but related, properties: i) 
large initial gluon densities $dN^{g}/dy\sim$ 1000 ~\cite{Vitev:2002pf}, 
ii) large ``transport coefficients'' $\hat{q}_{0}\sim$ 3.5  
GeV/fm$^2$~\cite{Arleo:2002kh}, iii) high opacities $L/\lambda\sim$ 
3 -- 4~\cite{Barnafoldi:2002xp}, or iv) effective parton 
energy losses of the order of $dE/dx\sim$ 14 GeV/fm~\cite{Wang:2003mm}. 

The {\it $p_{T}$ dependence} of the quenching predicted by all models 
that include the QCD version of the Landau-Pomeranchuck-Migdal (LPM) 
interference effect (BDMPS~\cite{Baier:2000mf} and 
GLV~\cite{Gyulassy:2000fs} approaches) is a slowly (logarithmic) 
increasing function of $p_{T}$, a trend not compatible with the data 
over the entire measured $p_{T}$ range. Other approaches, such 
as constant energy loss per parton scattering, are also not supported 
as discussed in~\cite{Jeon:2002dv}. Analyses which combine LPM jet 
quenching together with shadowing and initial-state $p_{T}$ broadening 
(``pQCD-II''~\cite{Vitev:2002pf} in Fig.~\ref{fig:R_AA_phenix_star}) 
globally reproduce the observed flat $p_{T}$ dependence of $R_{AA}$, 
as do recent approaches that take into account detailed balance between 
parton emission and absorption (``pQCD-I''~\cite{Wang:2003mm} in Fig. 
\ref{fig:R_AA_phenix_star}, left). 

At variance with parton energy loss descriptions, a gluon saturation 
calculation~\cite{Kharzeev:2002pc} is able to predict the magnitude 
of the observed suppression, although it fails to reproduce exactly 
the flat $p_{T}$ dependence of the quenching~\cite{Adams:2003kv}. 
Similarly, {\it semi-quantitative} estimates of final-state interactions 
in a dense {\it hadronic} medium~\cite{Gallmeister:2002us} yield the 
same amount of quenching as models based on partonic energy loss, 
however it is not yet clear whether the $p_{T}$ evolution of the hadronic 
quenching factor is consistent with the data or not~\cite{Adams:2003kv}.

The amount of suppression for $\pi^0$~\cite{Adler:2003qi} and 
$h^\pm$~\cite{Adams:2003kv,Adler:2003au} is the same above 
$p_{T}\approx$ 4 -- 5 GeV/$c$ for all centrality classes~\cite{Adler:2003au}. 
However, below $p_{T}\sim$ 5 GeV/$c$, $\pi^0$'s are more suppressed than 
inclusive charged hadrons in central collisions (as can be seen by 
comparing the right and left plots of Fig.~\ref{fig:R_AA_phenix_star}). 
This is due to the enhanced baryon production contributing to the total 
charged hadron yield in the intermediate $p_{T}$ region ($p_{T}\approx$ 
1 -- 4 GeV/$c$) in Au+Au collisions~\cite{Adler:2003kg,Adams:2003am} 
(see section~\ref{hadron_composition} below).

\paragraph{High $p_{T}$ suppression: $\sqrt{s_{NN}}$ dependence}

\begin{figure}[htbp]
\begin{center}
\includegraphics[height=6.2cm]{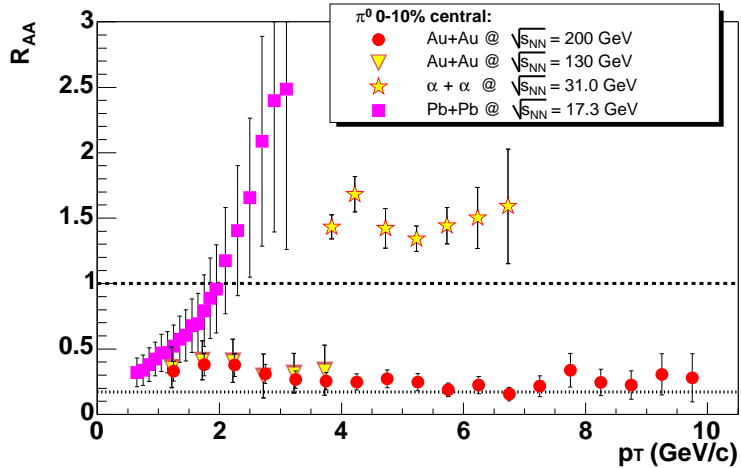}
\end{center}
\vspace*{-.5cm}
\caption[]{Nuclear modification factor, $R_{AA}(p_{T})$, for $\pi^0$ 
measured in central ion-ion reactions at 
CERN-SPS~\protect\cite{Aggarwal:2001gn} (squares), 
CERN-ISR~\protect\cite{Angelis:fk} (stars), 
and BNL-RHIC (triangles~\protect\cite{Adcox:2001jp}, 
circles~\protect\cite{Adler:2003qi}) energies.}
\label{fig:R_AA_pi0_syst}
\end{figure}

Fig.~\ref{fig:R_AA_pi0_syst} shows $R_{AA}(p_{T})$ for several 
$\pi^0$ measurements in high-energy $AA$ collisions at different 
center-of-mass energies~\cite{D'Enterria:2003rr}. The PHENIX 
$R_{AA}(p_{T})$ values for central Au+Au collisions at 200 GeV (circles) 
and 130 GeV (triangles) are noticeably below unity in contrast to the 
enhanced production ($R_{AA}>$1) observed at CERN-ISR (min. bias 
$\alpha+\alpha$~\cite{Angelis:fk}, stars) and CERN-SPS energies 
(central Pb+Pb~\protect\cite{Aggarwal:2001gn}, squares) and 
interpreted in terms of initial-state $p_{T}$ broadening 
(``Cronin effect''~\cite{Antreasyan:cw}).

\begin{figure}[htbp]
\begin{center}
\includegraphics[height=6.0cm]{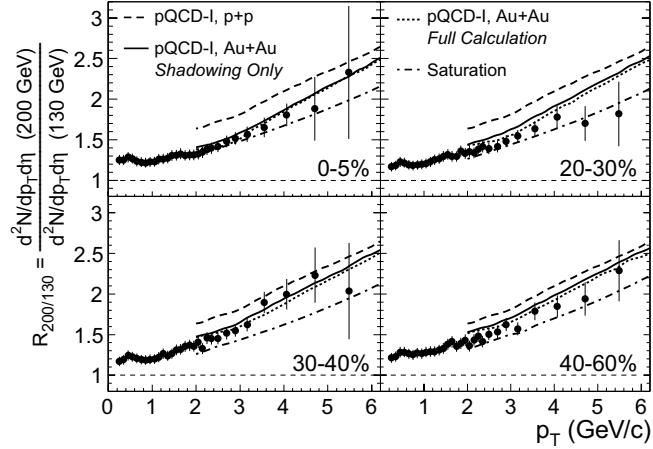}
\end{center}
\vspace*{-.8cm}
\caption{$R_{200/130}(p_{T})$ vs. $p_{T}$ for $(h^++h^-)/2$ for four 
different centrality bins measured by STAR compared to pQCD and gluon 
saturation model predictions~\protect\cite{Adams:2003kv}.}
\label{fig:star_hipt_200_130}
\end{figure}

Fig.~\ref{fig:star_hipt_200_130} shows $R_{200/130}(p_{T})$, 
the ratio of Au+Au charged hadron yields at 
$\sqrt{s_\mathrm{_{NN}}}$ = 130 and 200 GeV in 4 centrality classes, 
compared to pQCD and gluon saturation model 
predictions~\cite{Adams:2003kv}. The increase in high $p_{T}$ 
yields between the two center-of-mass energies is a factor $\sim$2 
at the highest $p_{T}$'s, whereas at low $p_{T}$, the increase is 
much moderate, of the order of 15\%. The large increment of the hard 
cross sections is naturally consistent with pQCD expectations due to 
the increased jet contributions at high transverse momenta. In the 
saturation model~\cite{Kharzeev:2002pc} the increase at high $p_{T}$ 
is accounted for by the enhanced gluon densities at 
$\sqrt{s_\mathrm{_{NN}}}$ = 200 GeV compared to 130 GeV in the 
``anomalous dimension'' $x_{T}$ region of the Au parton distribution 
function.

\begin{figure}[htbp]
\begin{center}
\includegraphics[height=5.8cm]{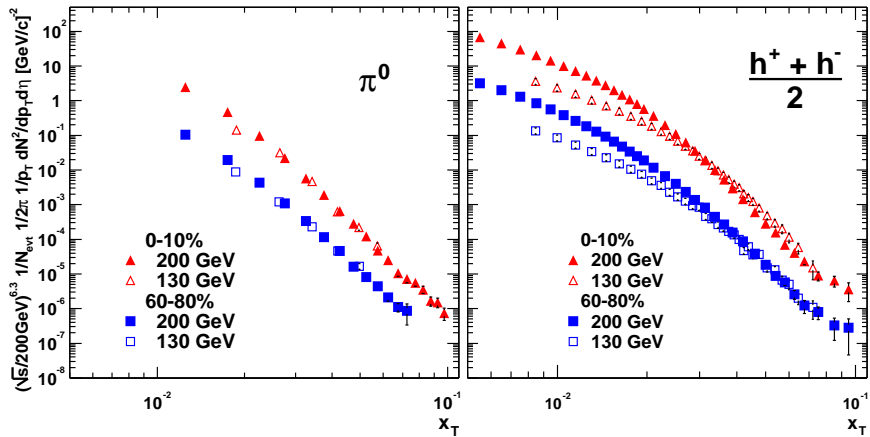}
\end{center}
\vspace*{-0.7cm}
\caption{$x_{T}$ scaled spectra for $\pi^0$ ({\it left}) and 
$(h^+ + h^-)/2$ ({\it right}) measured in central and peripheral 
collisions at $\sqrt{s_{_{NN}}}$ = 130 and 200 GeV by 
PHENIX~\cite{Adler:2003au}. Central (Peripheral) $x_{T}$ spectra 
are represented by triangles (squares), and solid (open) symbols 
represent $x_{T}$ spectra from $\sqrt{s_{_{NN}}}$ = 200 GeV 
($\sqrt{s_{_{NN}}}$ = 130 GeV scaled by a factor of $[130/200]^{6.3}$).}
\label{fig:xT_scaling_phenix}
\end{figure}

PHENIX~\cite{Adler:2003au} has addressed the $\sqrt{s}$ dependence of 
high $p_{T}$ production by testing the validity of ``$x_{T}$ scaling'' 
in Au+Au, i.e. verifying the parton model prediction that hard scattering 
cross sections can be factorized in 2 terms depending on $\sqrt{s}$ and 
$x_{T} = 2p_{T}/\sqrt{s}$ respectively: 
\begin{equation}
 E \frac{d^3\sigma}{dp^3}=\frac{1}{p_{T}^{n(\sqrt{s})}} \: F(x_{T}) \,\,
 \Longrightarrow E \frac{d^3\sigma}{dp^3}={1\over 
{\sqrt{s}^{{\,n(x_{T},\sqrt{s})}} }} \: G({x_{T}})\,.
\label{eq:x_{T}}
\end{equation}
In (\ref{eq:x_{T}}), $F(x_{T})$ embodies all the $x_{T}$ dependence coming 
from the parton distribution (PDF) and fragmentation (FF) functions
(PDFs and FFs, to first order, scale as the ratio of $p_{T}$ at 
different $\sqrt{s}$.), while the exponent $n$, related to the underlying 
parton-parton scattering, is measured to be $n\approx$ 4 -- 8 in a wide 
range of $p+p,\bar{p}$ collisions~\cite{Adler:2003au}.
Fig.~\ref{fig:xT_scaling_phenix} compares the $x_{T}$-scaled hadron yields 
in $\sqrt{s_{NN}}$ = 130 GeV and 200 GeV Au+Au central and peripheral 
collisions. According to Eq.~(\ref{eq:x_{T}}), the ratio of inclusive 
cross-sections at fixed $x_{T}$ should equal $(200/130)^{n}$.
On the one hand, $x_{T}$ scaling holds (in the kinematical region, 
$x_{T}>0.03$, where pQCD is expected to hold) in Au+Au with the same 
scaling power $n= 6.3\pm0.6$ for neutral pions (in central and peripheral 
collisions) and charged hadrons (in peripheral collisions) as measured 
in p+p~\cite{Adler:2003au}. 
This is consistent with equal (pQCD-like) production dynamics 
in p+p and Au+Au, and disfavours final-state effects described with 
medium-modified FF's that violate $x_{T}$ scaling (e.g. constant parton 
energy losses independent of the parton $p_{T}$). Equivalently, models 
that predict strong initial-state effects (e.g. gluon saturation) 
respect $x_{T}$ scaling as long as their predicted modified 
nuclear PDFs are depleted, independently of $\sqrt{s}$, by the same 
amount at a given $x_{T}$ (and centrality). On the other hand, 
Fig.~\ref{fig:xT_scaling_phenix} (right) shows that charged hadrons 
in central collisions (triangles) break $x_{T}$ scaling which is 
indicative of a non perturbative modification of particle composition
spectra from that of p+p at intermediate $p_{T}$'s (see section 
\ref{hadron_composition} below).

\paragraph{High $p_{T}$ suppression: centrality dependence}

\begin{figure}[htbp]
\begin{center}
\begin{minipage}[t]{75mm}
\includegraphics[height=5.5cm]{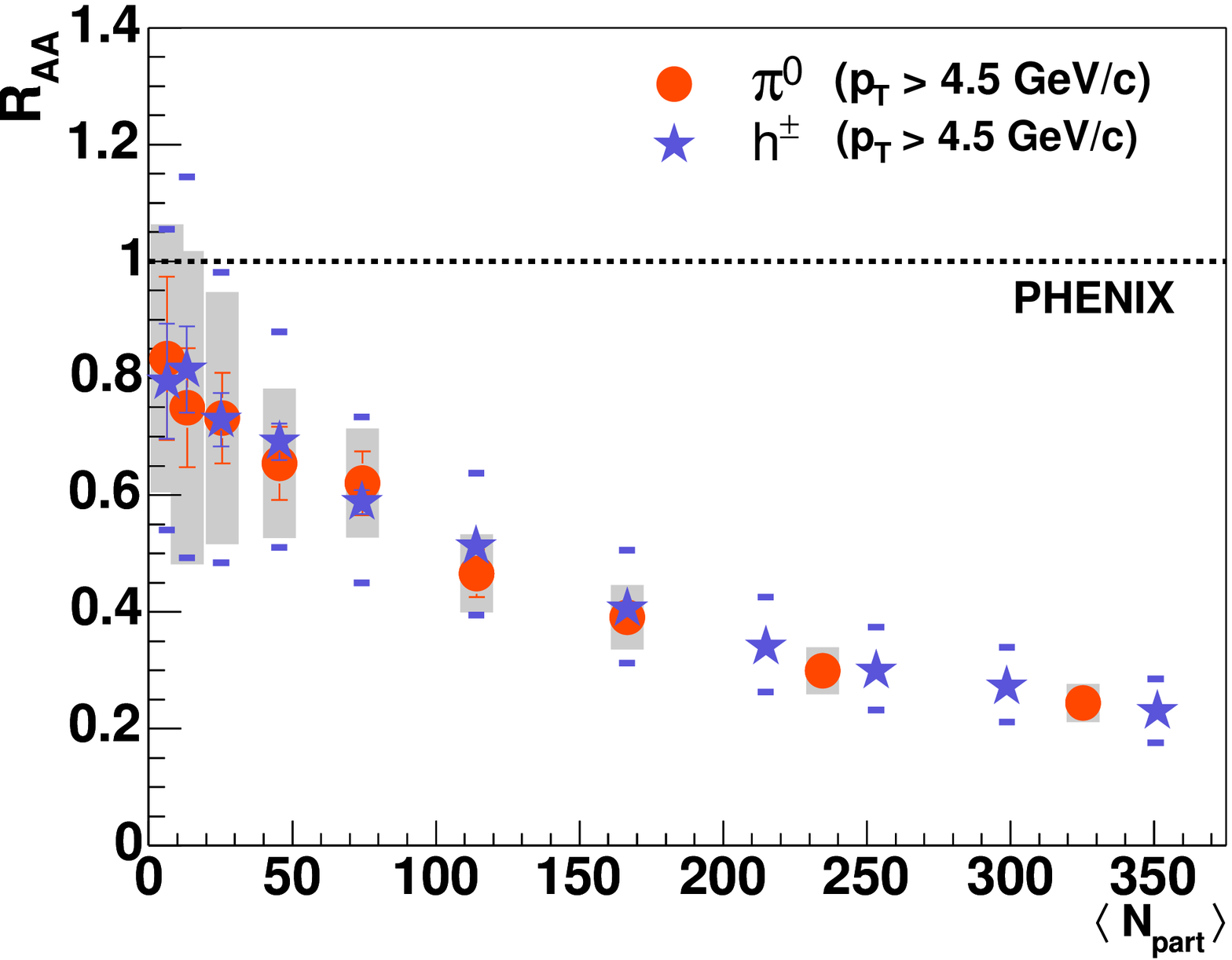}
\end{minipage}
\hspace*{.1cm}
\begin{minipage}[t]{75mm}
\includegraphics[height=5.3cm]{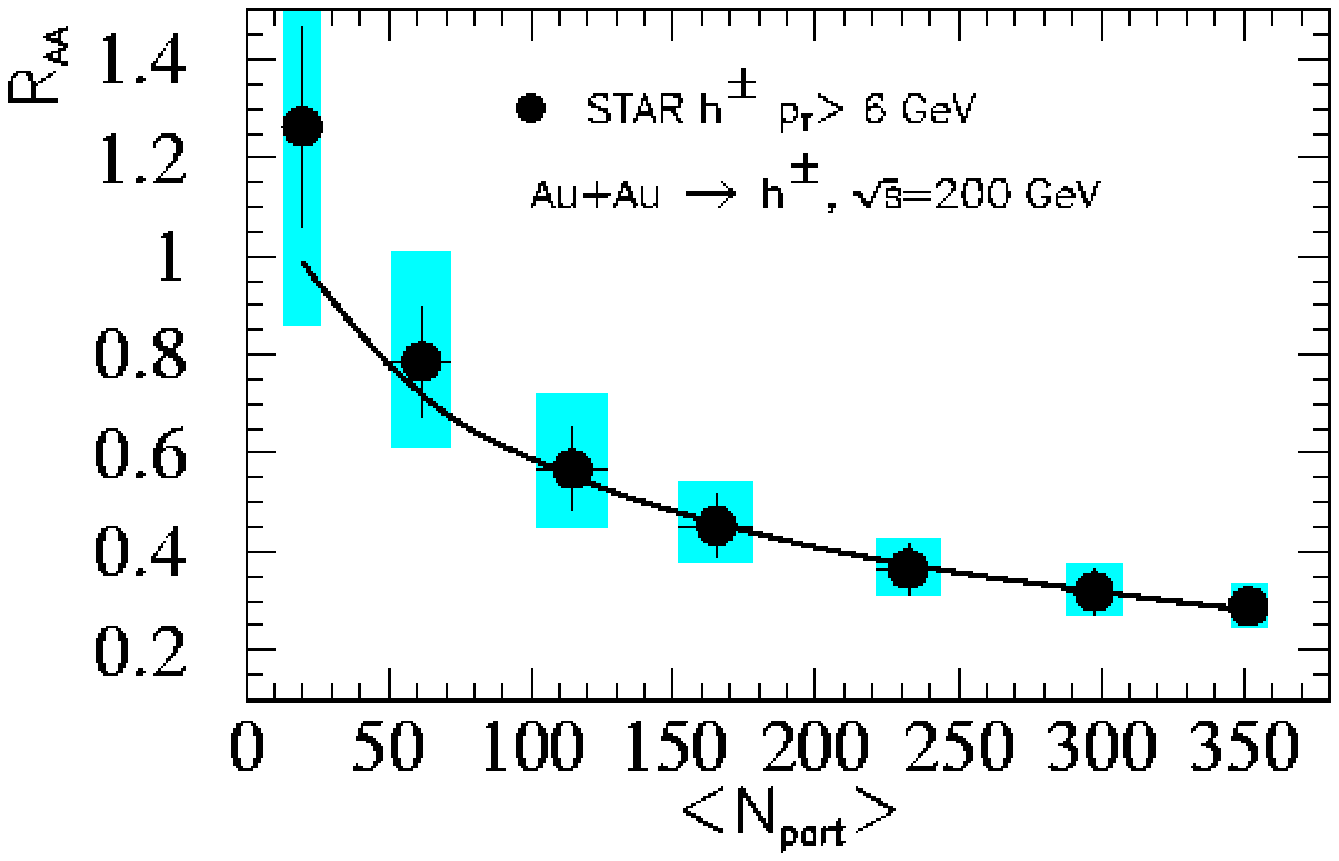}
\end{minipage}
\end{center}
\vspace*{-0.7cm}
\caption[]{Evolution of the high $p_{T}$ $\pi^0$ and $h^\pm$ suppression, 
$R_{AA}(p_{T}>$ 4.5 GeV/$c$), as a function of centrality given by 
$\langle N_{part} \rangle$ (PHENIX, {\it left}). Same evolution shown 
as $R_{AA}(p_{T}>$6.0 GeV/$c$) for STAR $h^\pm$ data~\cite{Wang:2003aw}
({\it right}).}
\label{fig:R_AA_vs_cent}
\end{figure}

In each centrality bin, the value of the high $p_{T}$ suppression can 
be quantified by the ratio of Au+Au over $N_{coll}$-scaled p+p yields 
integrated above a given (large enough) $p_{T}$. The centrality 
dependence of the high $p_{T}$ suppression for $\pi^0$ and charged 
hadrons, given by $R_{AA}(p_{T}>$ 4.5 GeV/$c$), is shown in 
Fig.~\ref{fig:R_AA_vs_cent} (left) as a function of $\langle N_{part} 
\rangle$ for PHENIX data. The transition from the $N_{coll}$ scaling 
behaviour ($R_{AA}\sim$~1) apparent in the most peripheral region, 
$\langle N_{part} \rangle \lesssim$ 40, to the strong suppression 
seen in central reactions ($R_{AA}\sim$~0.2) is smooth. Whether there is 
an abrupt or gradual departure from $N_{coll}$ scaling in the peripheral 
range cannot be ascertained within the present experimental 
uncertainties~\cite{D'Enterria:2003rr}. The data, however, is inconsistent 
with $N_{coll}$ scaling (at a $2\sigma$ level) for the 40--60\% centrality 
corresponding to $\langle N_{part} \rangle\approx$ 40 -- 80 
~\cite{Adcox:2002pe,D'Enterria:2003rr}, whose estimated ``Bjorken'' energy 
density ($\epsilon_{Bj}\approx$ 1 GeV/fm$^3$)~\cite{D'Enterria:2003rr} is
in the ball-park of the expected ``critical'' QCD energy density. A 
similar centrality dependence of the high $p_{T}$ suppression is seen 
in STAR $h^\pm$ data (Fig.~\ref{fig:R_AA_vs_cent}, right)

\begin{figure}[h!]
\hspace*{-2.8cm}
\begin{center}
\begin{minipage}[t]{75mm}
\includegraphics[height=6.6cm]{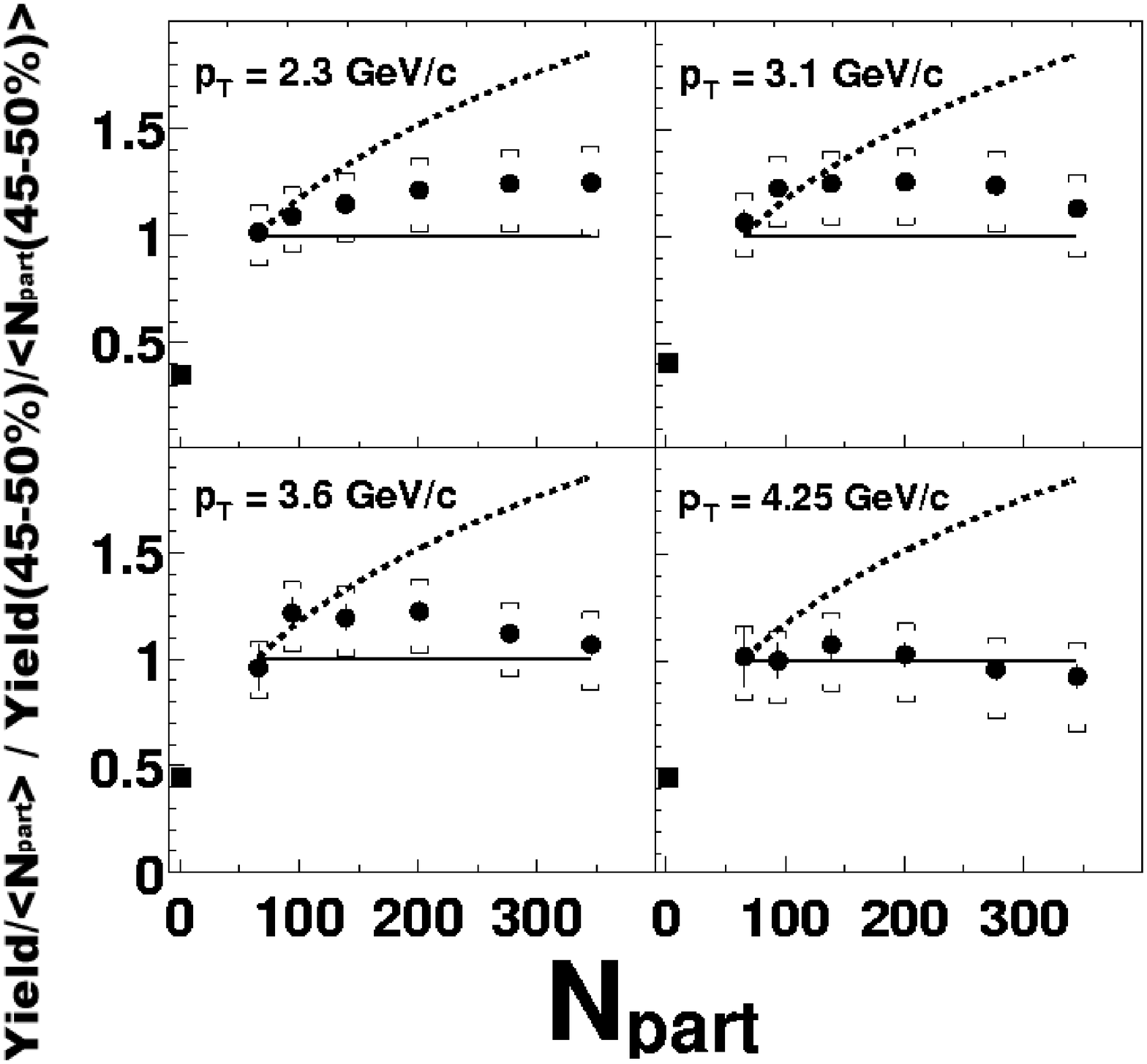}
\end{minipage}
\hspace*{.3cm}
\begin{minipage}[t]{75mm}
\includegraphics[height=6.4cm]{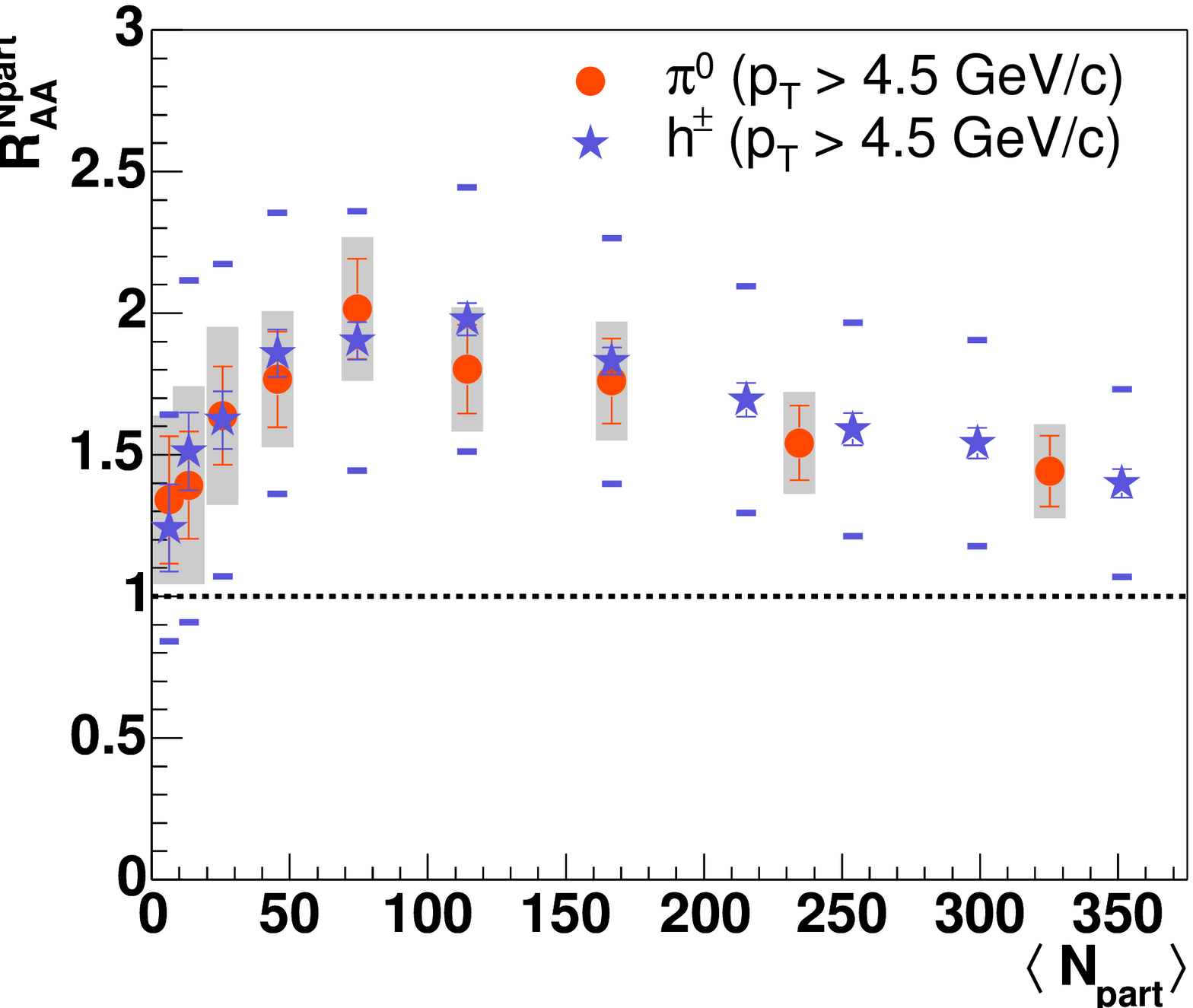}
\end{minipage}
\end{center}
\vspace*{-0.5cm}
\caption[]{{\it Left:} Ratio of central to semi-central $h^\pm$ yields 
normalized by $N_{part}$ in 4 different $p_{T}$ bins, as a function of 
$N_{part}$ measured by PHOBOS~\cite{Back:2003qr}. The dashed (solid) line 
shows the expectation for $N_{coll}$ ($N_{part}$) scaling. {\it Right:} 
Ratio of Au+Au over p+p $\pi^0$ and $h^\pm$ yields above 4.5 GeV/$c$
normalized by $N_{part}$ $(R_{AA}^{N_{part}})$ as a function of centrality 
given by $\langle N_{part} \rangle$ as measured by 
PHENIX ~\cite{Adler:2003qi,Adler:2003au}. The dashed line indicates the 
expectation for $N_{part}$ scaling.}
\label{fig:R_AA_Npart_vs_cent}
\end{figure}

$N_{part}$ (instead of $N_{coll}$) scaling at high $p_{T}$ is 
$N_{part}^{pp}$ = 2) expected in scenarios dominated either by gluon 
saturation~\cite{Kharzeev:2002pc} or by surface emission of the quenched 
jets~\cite{Muller:2002fa}. ``Approximate'' $N_{part}$ scaling has been 
claimed by PHOBOS~\cite{Back:2003qr}: the ratio of central to a {\it fit} 
to {\it mid-central} yields in the range $p_{T}\approx$ 2. -- 4. GeV/$c$ 
stays flat as a function of centrality 
(Fig.~\ref{fig:R_AA_Npart_vs_cent}, left). 
However, at higher $p_{T}$ values, where the suppression is seen to saturate 
at its maximum value, the centrality dependence of the ratio of 
$N_{part}$-scaled Au+Au over p+p yields for $\pi^0$ and $h^\pm$ measured 
by PHENIX (Fig.~\ref{fig:R_AA_Npart_vs_cent}, right) does not show a true 
participant scaling ($R_{AA}^{part}>$ 1 for all centralities). Nonetheless, 
the fact that the production per participant pair above 4.5~GeV/{\it c} is, 
within errors, approximately constant over a wide range of intermediate 
centralities, is in qualitative agreement with a gluon saturation model 
prediction~\cite{Kharzeev:2002pc}.

\paragraph{High $p_{T}$ suppression: particle species dependence} 
\label{hadron_composition}

\begin{figure}[!]
\hspace*{-.8cm}
\begin{center}
\includegraphics[height=6.8cm]{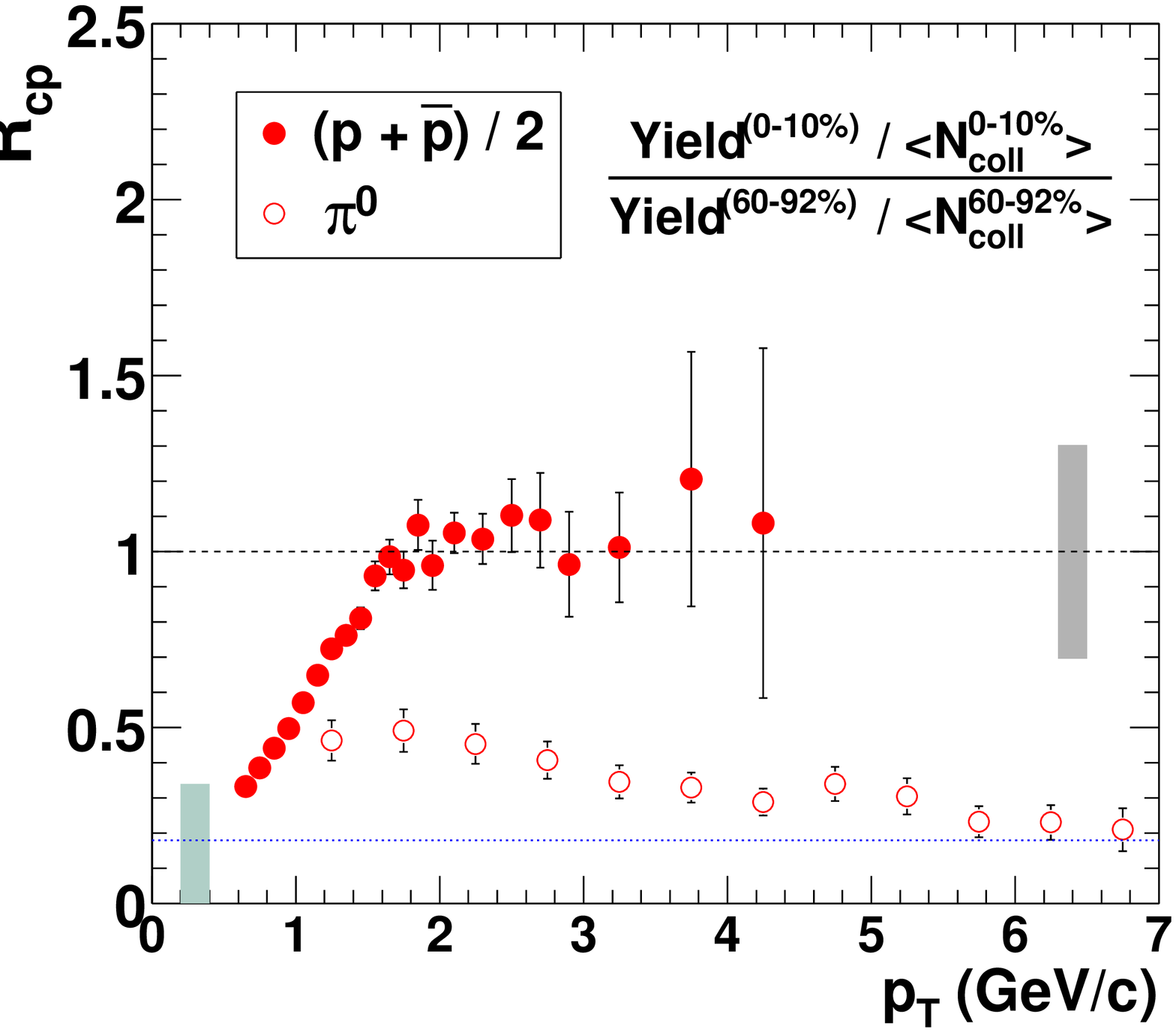}
\hspace*{4mm}
\includegraphics[height=7.cm]{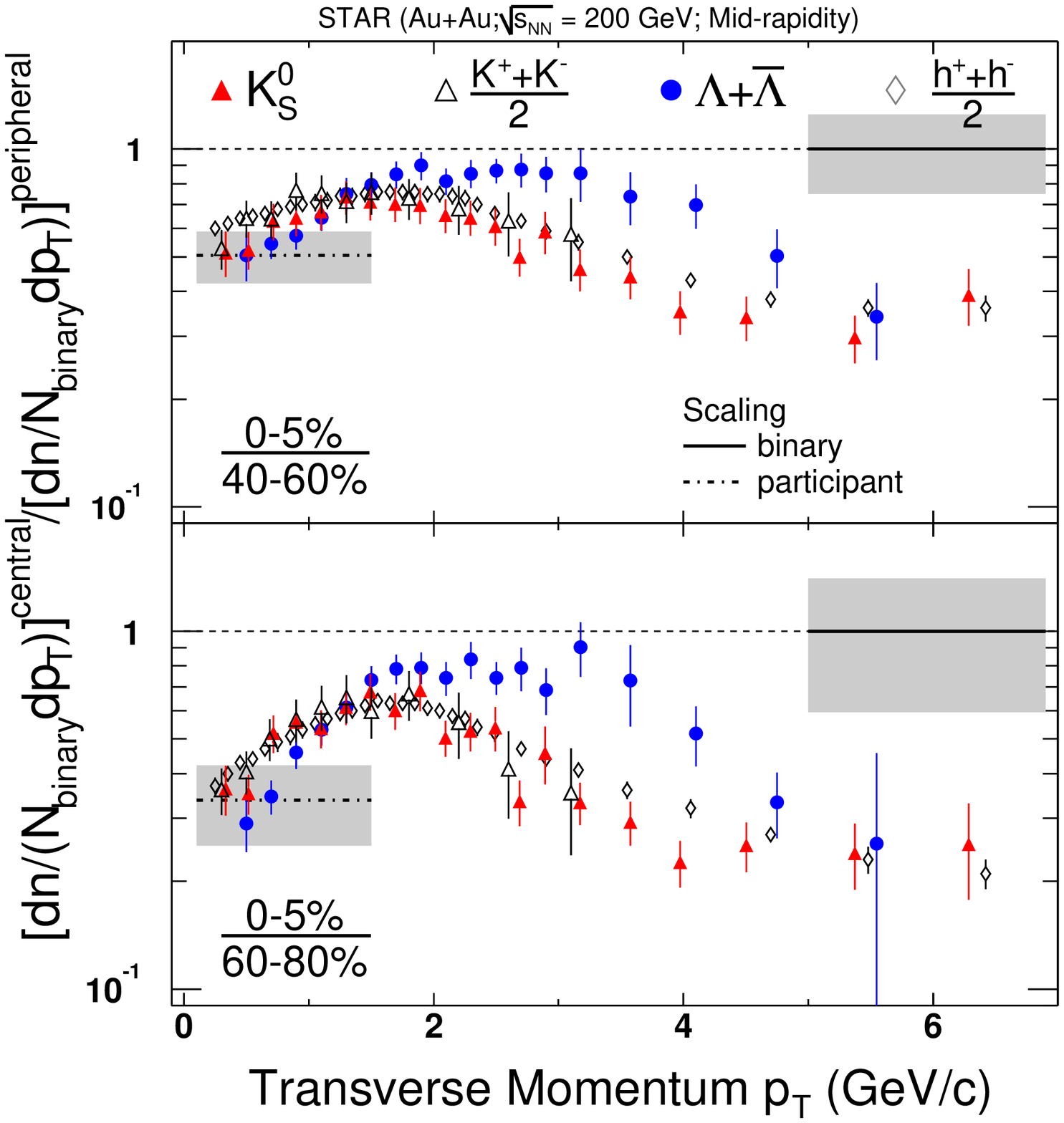}
\end{center}
\vspace*{-0.7cm}
\caption[]{Ratio of central over peripheral $N_{coll}$ scaled yields, 
$R_{cp}(p_{T})$, as a function of $p_{T}$ for different species measured 
in Au+Au collisions: $(p+\bar{p})/2$ (dots) and $\pi^0$ (circles) by 
PHENIX~\protect\cite{Adler:2003kg} ({\it left}), and $\Lambda,\bar{\Lambda}$ 
(circles) and $K^0_s,K^\pm$ (triangles) by STAR~\protect\cite{Adams:2003am} 
({\it right}).} 
\label{fig:flavor_dep}
\end{figure}

One of the most intriguing results of the RHIC program so far is the 
different suppression pattern of baryons and mesons at moderately high
$p_{T}$'s. Fig.~\ref{fig:flavor_dep} (left) compares the $N_{coll}$ 
scaled central to peripheral yield ratios \footnote{Since the 
60--92\% peripheral Au+Au (inclusive and identified) spectra scale 
with $N_{coll}$ when compared to the p+p 
yields~\cite{Adler:2003qi,Adams:2003kv,Adler:2003kg}, $R_{cp}$ carries 
basically the same information as $R_{AA}$.} 
for $(p+\bar{p})/2$ and $\pi^0$: $R_{cp} = (yield^{(0-10\%)} / 
N_{coll}^{0-10\%}) / (yield^{(60-92\%)} /  N_{coll}^{60-92\%})$. 
From 1.5 to 4.5 GeV/$c$ the (anti)protons are not suppressed 
($R_{cp}\sim$ 1) at variance with the pions which are reduced by a 
factor of 2 -- 3 in this $p_{T}$ range. 
If both $\pi^0$  and  $p,\bar{p}$ originate from the fragmentation 
of hard-scattered partons that lose energy in the medium, the nuclear 
modification factor $R_{ cp}$ should be independent of particle species 
contrary to the experimental result. The same discussion applies for
strange mesons and baryons as can be seen from the right plot of 
Fig.~\ref{fig:flavor_dep}. Whereas the kaon yields in central collisions 
are suppressed with respect to ``$N_{coll}$ scaling'' for 
all measured $p_{T}$, the yield of $\Lambda+\bar{\Lambda}$ is close to 
expectations from collision scaling in the $p_{T}$ range 1.8 -- 3.5 GeV/$c$. 
Interestingly, above $p_{T} \sim$ 5.0 GeV/$c$, the $K_S^0K^\pm$, 
$\Lambda+\bar{\Lambda}$, and charged hadron yields are suppressed from 
binary scaling by a similar factor. 

\begin{figure}[htbp]
\vspace*{-.3cm}
\hspace*{-.8cm}
\begin{center}
\includegraphics[height=5.8cm]{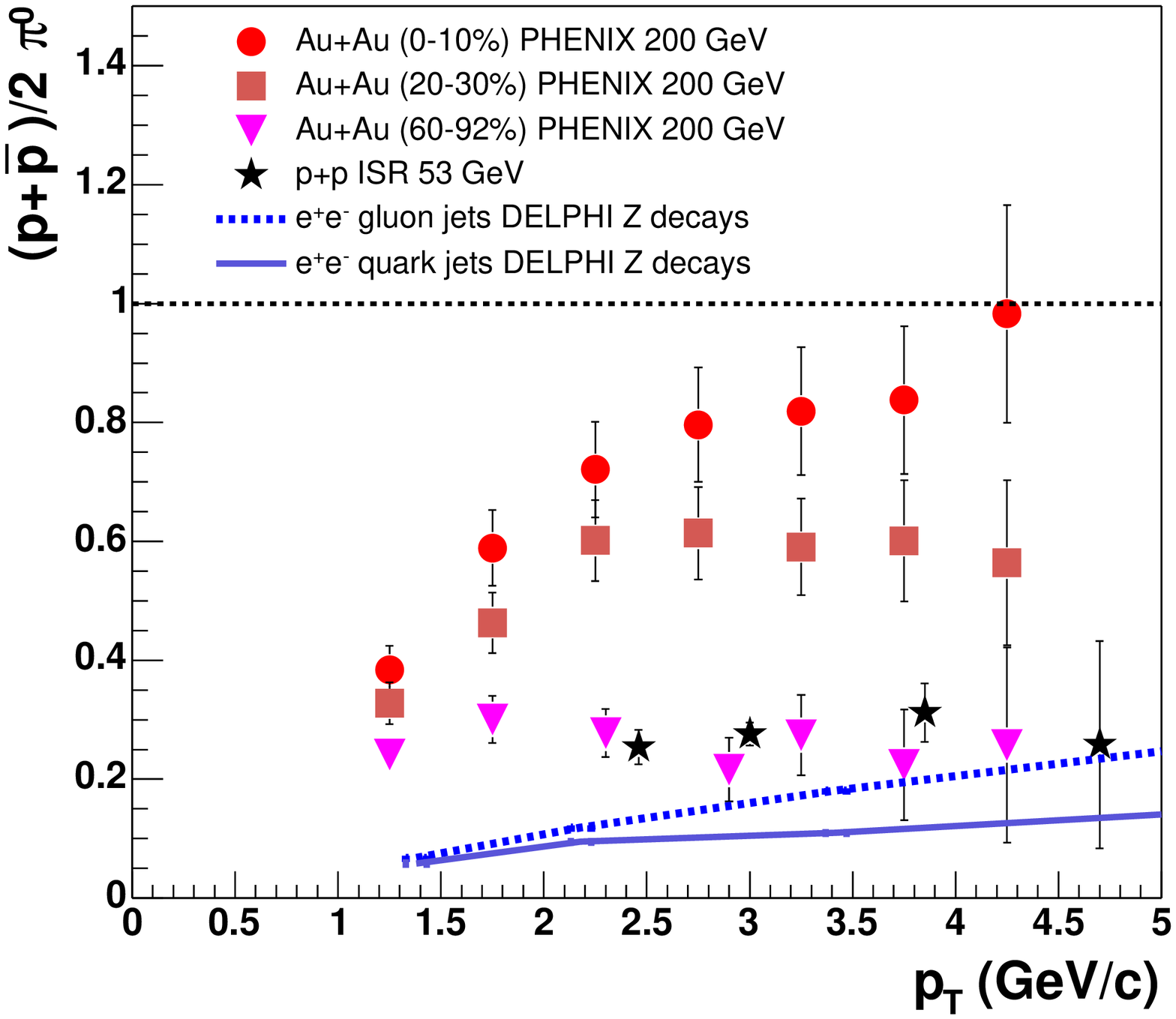}
\hspace*{4mm}
\includegraphics[height=6.0cm]{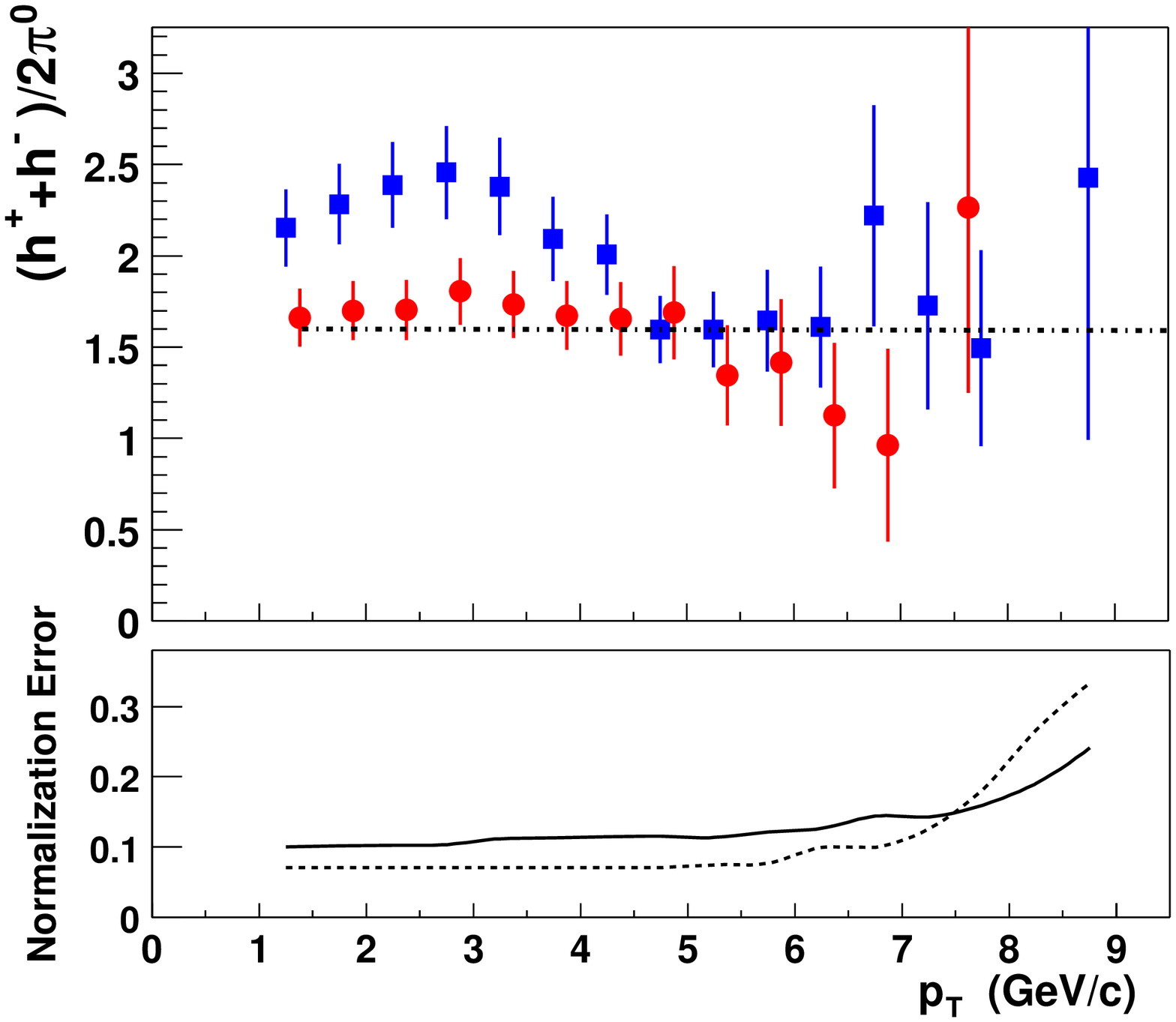}
\end{center}
\vspace*{-0.7cm}
\caption[]{{\it Left}: Ratios of  $(p+\bar{p})/2$ over $\pi^0$ versus 
$p_{T}$ in central (dots), mid-central (squares), and peripheral (triangles) 
Au+Au, and in p+p (crosses), and $e^{+}e^{-}$ (dashed and solid lines) 
collisions~\cite{Adler:2003kg}. {\it Right}: Ratio of charged hadron to 
$\pi^0$ in central (0--10\% -  squares) and peripheral (60--92\% - circles) 
Au+Au collisions compared to the $h/\pi\sim$ 1.6 ratio (dashed-dotted line) 
measured in p+p collisions~\cite{Adler:2003kg}.}
\label{fig:flavor_dep2}
\end{figure}

Fig.~\ref{fig:flavor_dep2} (left) shows the ratios of $(p+\bar{p})/2$ 
over $\pi^0$ as a function of $p_{T}$ measured by PHENIX in central 
(0--10\%, circles), mid-central (20--30\%, squares), and peripheral 
(60--92\%, triangles) Au$+$Au collisions~\cite{Adler:2003kg}, 
together with the corresponding ratios measured in p+p collisions at 
CERN-ISR energies~\cite{Alper:1975jm,Angelis:fk} (crosses) and in gluon 
and quark jet fragmentation from $e^{+}e^{-}$ collisions~\cite{Abreu:2000nw} 
(dashed and solid lines resp.). Within errors, peripheral Au+Au results 
are compatible with the p+p and $e^{+}e^{-}$ ratios, but central Au+Au 
collisions have a $p/\pi$ ratio $\sim$ 4 -- 5 times larger. Such a result 
is at odds with standard perturbative production mechanisms, since in 
this case the particle ratios $\bar{p}/\pi$ and $p/\pi$ should be described 
by a universal fragmentation function independent of the colliding system, 
which favors the production of the lightest particle. Beyond $p_{T} 
\approx 4.5$~GeV/$c$, the identification of charged particles is not yet
possible with the current PHENIX configuration, however the measured 
$h/\pi^0\sim$~1.6 ratio above $p_{T}\sim$ 5 GeV/$c$ in central and 
peripheral Au+Au is consistent with that measured in p+p collisions 
(Fig.~\ref{fig:flavor_dep2}, right). This result together with STAR 
$R_{cp}$ result on strange hadrons (Fig.~\ref{fig:flavor_dep}, right)
supports the fact that for large $p_{T}$ values the properties of the 
baryon production mechanisms approach the (suppressed) meson scaling, 
thus limiting the observed baryon enhancement in central Au$+$Au collisions 
to the intermediate transverse momenta $p_{T} \lesssim$ 5 GeV/$c$.

Several theoretical explanations (see refs. in 
~\cite{Adcox:2001mf,Adler:2003kg}) have been proposed to justify the 
different behaviour of mesons and baryons at intermediate $p_{T}$'s based 
on: (i) quark recombination (or coalescence), (ii) medium-induced 
difference in the formation time of baryons and mesons, (iii) different 
``Cronin enhancement'' for protons and pions, or (iv) ``baryon junctions''.
In the recombination picture~\cite{Hwa:2003bn,Greco:2003xt,Fries:2003vb} 
the partons from a thermalized system coalesce and with the addition of 
quark momenta, the soft production of baryons extends to much larger 
values of $p_{T}$ than that for mesons.  In this scenario, the effect is
limited to $p_{T} < 5$\,GeV, beyond which fragmentation becomes the
dominant production mechanism for all species.

\paragraph{High $p_{T}$ suppression: pseudorapidity dependence}

\begin{figure}[htbp]
\begin{center}
\begin{minipage}[t]{75mm}
\includegraphics[height=5.2cm]{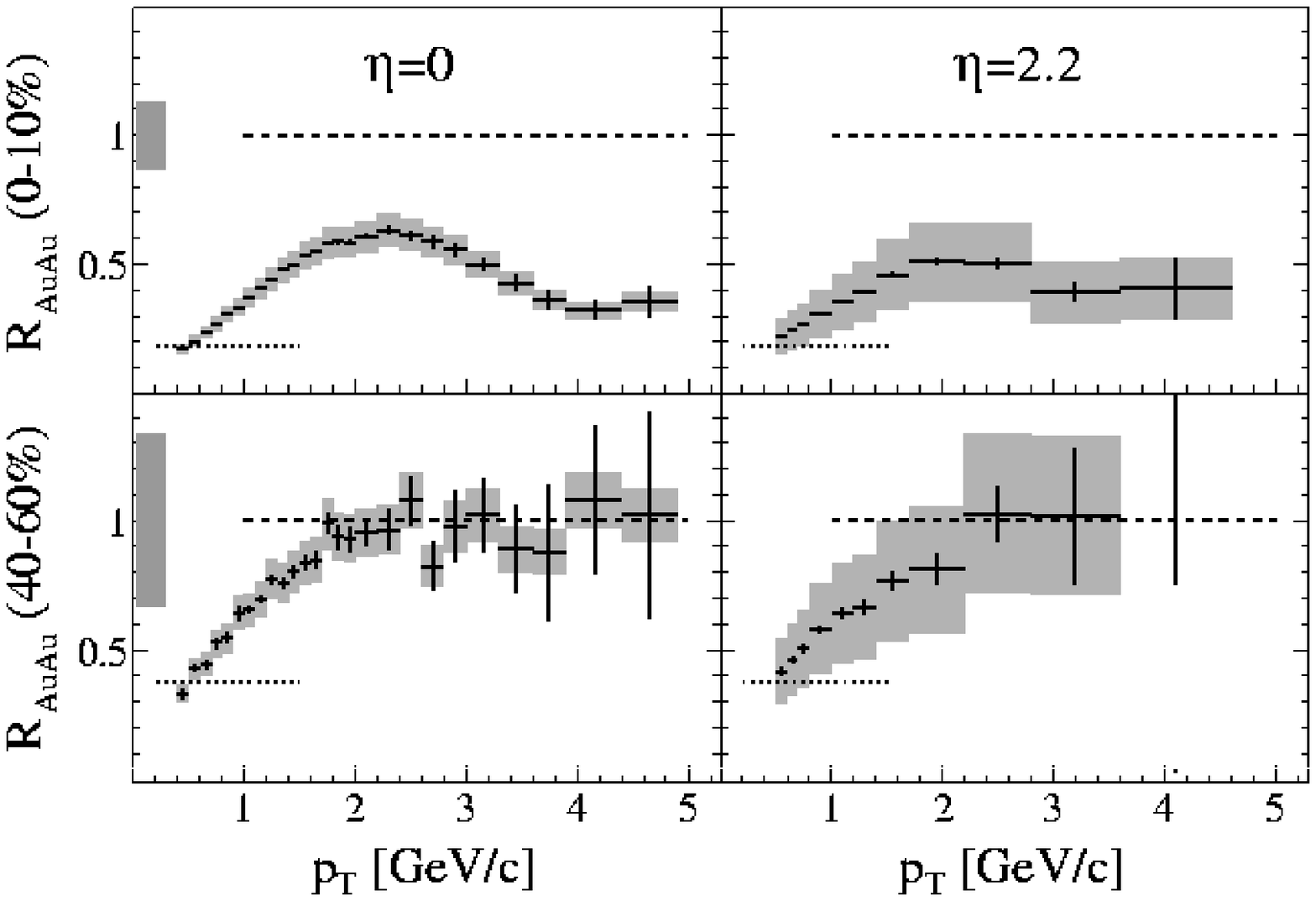}
\end{minipage}
\hspace*{0.5cm}
\begin{minipage}[t]{75mm}
\includegraphics[height=5.4cm]{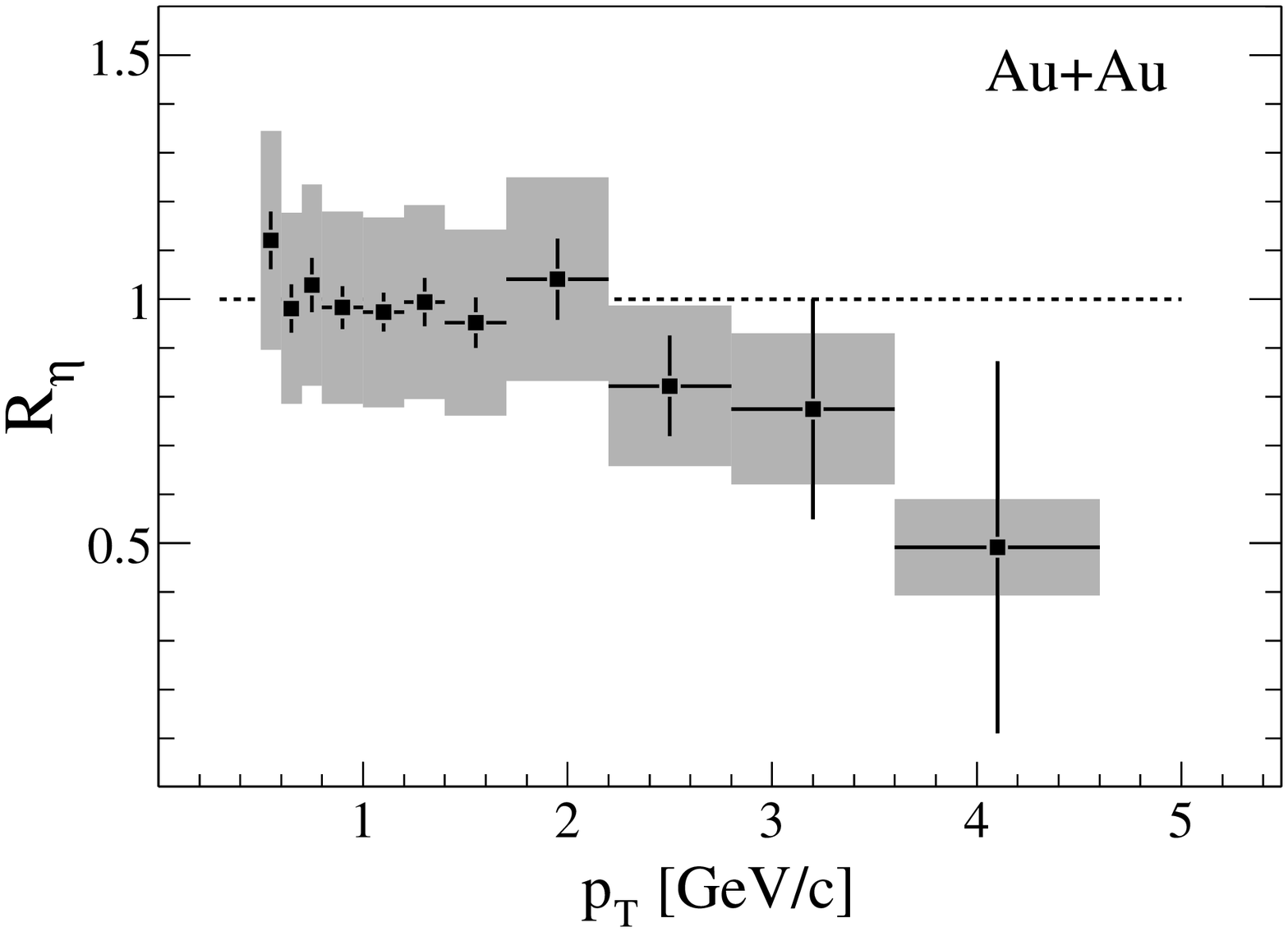}
\end{minipage}
\end{center}
\vspace*{-0.3cm}
\caption{{\it Left}: $R_{AA}(p_{T})$  measured by BRAHMS at $\eta$ = 0 
and $\eta$ = 2.2 for 0--10\% most central and for semi-peripheral 
(40-60\%) Au+Au collisions. {\it Right}: Ratio $R_\eta$ of $R_{cp}$ 
distributions at $\eta=2.2$ and $\eta=0$. Figs. from 
~\protect\cite{Arsene:2003yk}.}
\label{fig:brahms_RAA}
\end{figure}

BRAHMS is, so far, the only experiment at RHIC that has measured high 
$p_{T}$ inclusive charged hadron spectra off mid-rapidity. 
Fig.~\ref{fig:brahms_RAA} (left) shows the nuclear modification factors 
$R_{AA}(p_{T})$ for central and semi-peripheral Au+Au measurements 
at mid-pseudorapidity ($\eta$ = 0) and at $\eta$ = 2.2~\cite{Arsene:2003yk}.
The high $p_{T}$ suppression is not limited to central rapidities but it 
is clearly apparent at forward $\eta$'s too. Fig.~\ref{fig:brahms_RAA} 
(right) shows the ratio of suppressions at the two pseudorapidity values, 
$R_\eta = R_{cp}(\eta=2.2)/R_{cp}(\eta=0)$. The high $p_{T}$ deficit at 
$\eta$ = 2.2 is similar to, or even larger than, at $\eta$ = 0 indicating 
that the volume causing the suppression extends also in the longitudinal 
direction. 

\paragraph{High $p_{T}$ azimuthal correlations in Au+Au collisions}

There are two main sources of azimuthal correlations at high $p_{T}$ in 
heavy-ion collisions:

\begin{itemize}
\item The {\it fragmentation} of hard-scattered {\it partons} results in 
jets of high $p_{T}$ hadrons correlated in both rapidity and azimuthal 
angle. Such correlations are short range ($\Delta\eta\lesssim$ 0.7, 
$\Delta\phi\lesssim$ 0.75), involve comparatively large transverse 
momentum particles ($p_{T}>$ 2 GeV/$c$), and are unrelated 
(in principle) to the orientation of the $AA$ reaction plane.

\item {\it Collective (elliptic) flow}: 
The combination of (i) the geometrical asymmetry in non-central $AA$ 
reactions (``almond''-like region  of the overlapping nuclei), and (ii) 
multiple reinteractions between the produced particles in the overlap 
region; results in pressure gradients in the collision ellipsoid which 
transform the original coordinate-space asymmetry into a momentum-space 
anisotropy. The amount of elliptic flow (a true collective effect absent
in p+p collisions) is measured by the second harmonic coefficient, 
$v_{2}\equiv \langle cos(2\phi)\rangle$, of the Fourier expansion 
of the particles azimuthal distribution with respect to the reaction plane.
\end{itemize}

Additionally, there are other second-order sources of angular correlations
like resonance decays, final state (particularly Coulomb) interactions, 
momentum conservation, or other experimental effects like photon conversions,
which have to be subtracted out in order to extract the interesting 
``jet-like'' or ``flow-like'' signals.

\subsection{High-$p_{T}$ Azimuthal Correlations: Jet Signals}

\begin{figure}[htbp]
\includegraphics[width=.45\textwidth]{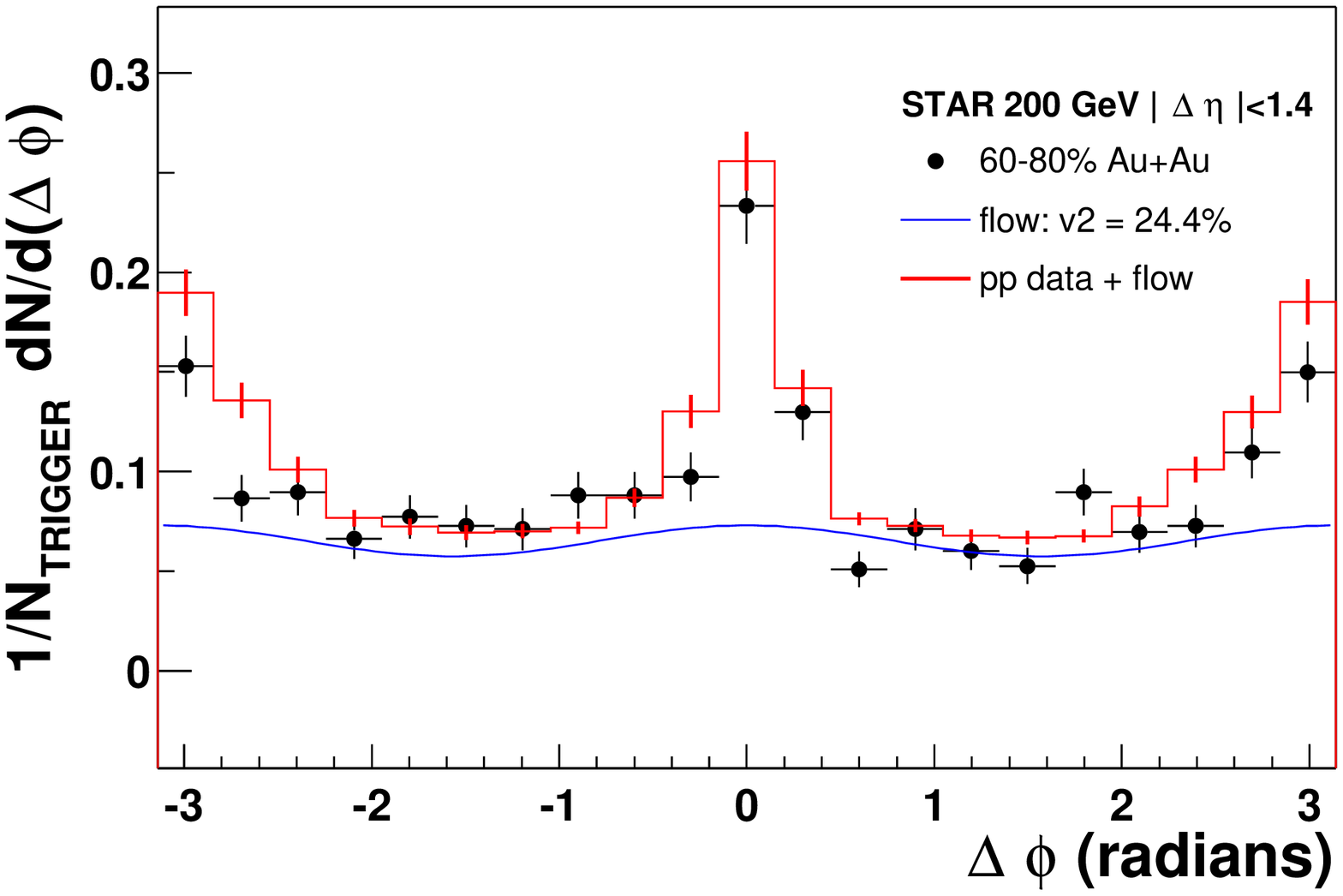}
\hspace{.1\textwidth}
\includegraphics[width=.45\textwidth]{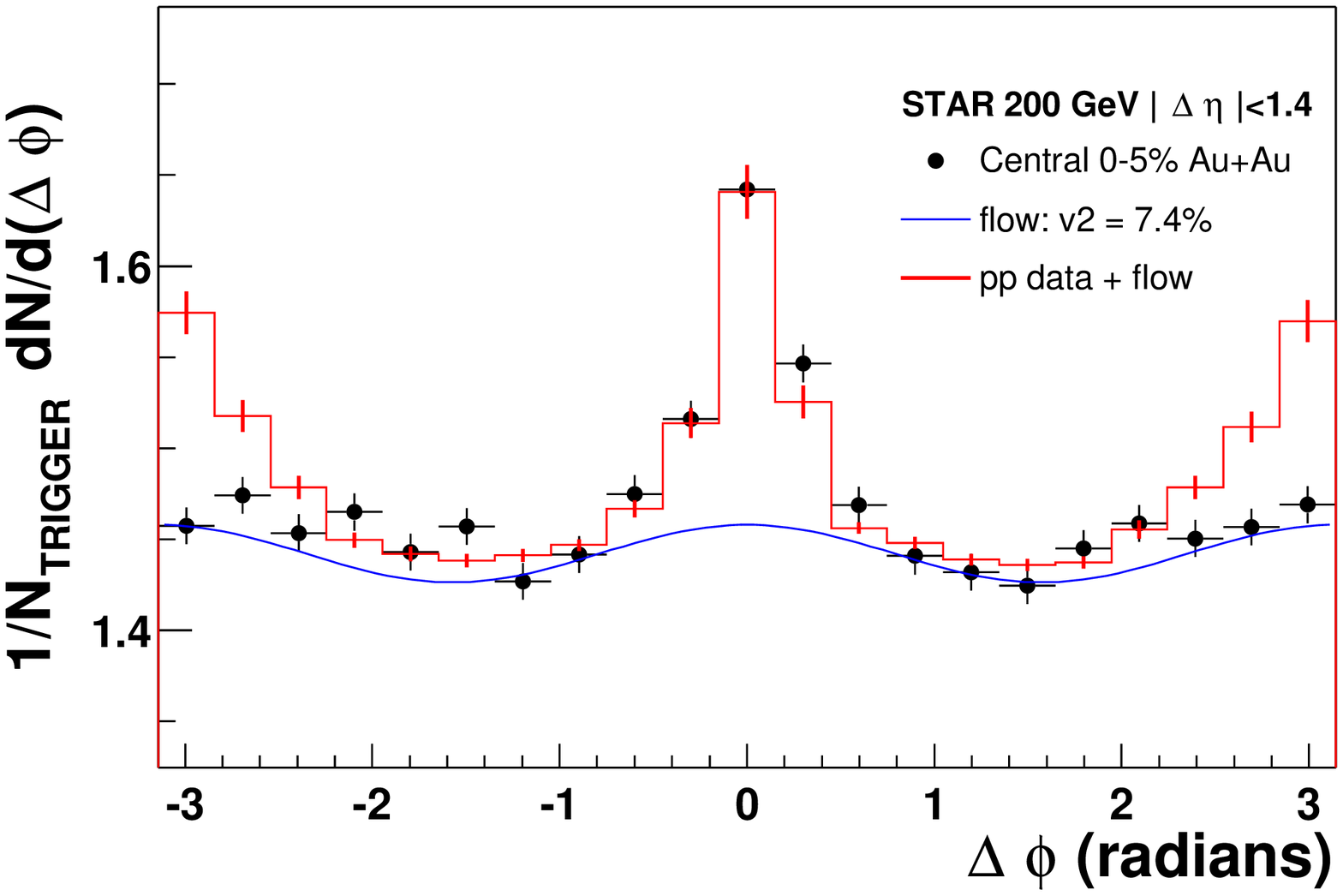}
\caption{Azimuthal correlations for peripheral ({\it left}) and central 
({\it right}) Au+Au collisions compared to the pedestal and flow-scaled 
correlations in p+p collisions. Fig. from ~\cite{Jacobs:2003bx}.}
\label{fig:jets_azim_corr_star}
\end{figure}

Although, standard jet reconstruction algorithms fail below $p_{T}\approx$ 
40 GeV/$c$ when applied to the soft-background dominated environment of 
heavy-ion collisions, angular correlations of pairs of high $p_{T}$ 
particles~\cite{Adler:2002tq,Chiu:2002ma} have been very successfully used 
to study on a statistical basis the properties of the produced jets. For 
each event with ``trigger'' particle(s) with $p_{T}$ = 4 -- 6 GeV/$c$ and 
``associated'' particle(s) with $p_{T}$ = 2 -- 4 GeV/$c$ and $|\eta|<$ 0.7, 
STAR~\cite{Adler:2002tq} determines the two-particle azimuthal distribution
\begin{equation}
D(\Delta \phi) \propto \frac{1}{N_{trigger}}
\frac{dN}{d(\Delta\phi)}\;.
\end{equation}
Fig.~\ref{fig:jets_azim_corr_star} shows $D(\Delta \phi)$ for peripheral 
(left) and central (right) Au+Au collisions (dots) compared to 
$D(\Delta \phi)$ from p+p collisions (histogram), and to a superposed 
$cos(2\Delta\phi)$ flow-like term (blue curve). On the one hand, the 
correlation strength at small relative angles ($\Delta \phi\sim$ 0) 
in peripheral and central Au+Au as well as at back-to-back angles 
($\Delta \phi\sim \pi$) in peripheral Au+Au are very similar to the 
scaled correlations in p+p collisions. The near-side peaks in all three 
collision systems are characteristic of jet fragmentation~\cite{Adler:2002tq} 
(a result also observed by PHENIX using {\it neutral} trigger 
particles~\cite{Chiu:2002ma}). On the other hand, the away-side peak 
($\Delta \phi\sim \pi$) in central collisions is strongly suppressed.

\begin{figure}[htbp]
\begin{center}
\includegraphics[height=6.cm]{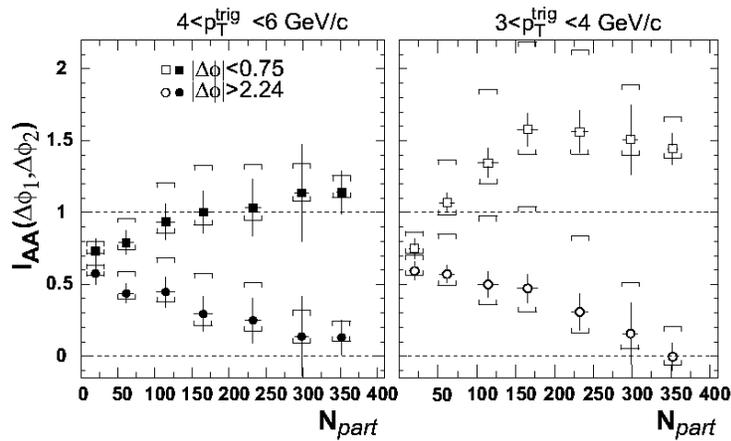}
\end{center}
\vspace*{-0.1cm}
\caption[]{Ratio of Au+Au over p+p integrated azimuthal correlations, 
Eq.~(\protect{\ref{eq:I_AA}}), for small-angle (squares, $|\Delta \phi|<0.75$ 
radians) and back-to-back (circles, $|\Delta \phi|>2.24$ radians) azimuthal 
regions versus number of participating nucleons for trigger particle 
intervals $4<p_{T}^{trig}<6$ GeV/$c$ (solid) and $3<p_{T}^{trig}<4$ GeV/$c$ 
(hollow)~\cite{Adler:2002tq}.}
\label{fig:I_AA_star}
\end{figure}

In order to study the evolution as a function of centrality of the
the near-side, $D^{AuAu}(\Delta \phi<0.75)$, and away-side, 
$D^{AuAu}(\Delta \phi>2.24)$, angular correlations in Au+Au compared to 
p+p, $D^{pp}$, STAR has constructed the quantity
\begin{eqnarray}
I_{AA}(\Delta \phi_1,\Delta \phi_2) = \frac{\int_{\Delta 
\phi_1}^{\Delta \phi_2} d(\Delta \phi) [D^{\mathrm{AuAu}}- B(1+2v_{2}^2 
\cos(2 \Delta \phi))]}{\int_{\Delta \phi_1}^{\Delta \phi_2} d(\Delta \phi) 
D^{\mathrm{pp}}},
\label{eq:I_AA}
\end{eqnarray}
where $B$ accounts for overall background and $v_{2}$ the azimuthal 
correlations due to elliptic flow. Fig.~\ref{fig:I_AA_star} shows $I_{AA}$ 
for the near-side (squares) and away-side (circles) correlations as a 
function of the number of participating nucleons ($N_{part}$). On the one 
hand, the near-side correlation function is relatively suppressed compared 
to the expectation from Eq.~(\ref{eq:I_AA}) in the most peripheral region 
(a result not completely understood so far) and increases slowly with 
$N_{part}$. On the other hand, the back-to-back correlation strength 
above the background from elliptic flow, decreases with increasing $N_{part}$ 
and is consistent with zero for the most central collisions.
The disappearance of back-to-back jet-like correlations is consistent 
with large energy loss effects in a system that is opaque to the propagation 
of high-momentum partons or their fragmentation products.   

\paragraph{High $p_{T}$ azimuthal correlations: Collective elliptic flow}

\begin{figure}[htbp]
\begin{center}
\includegraphics[height=4.1cm]{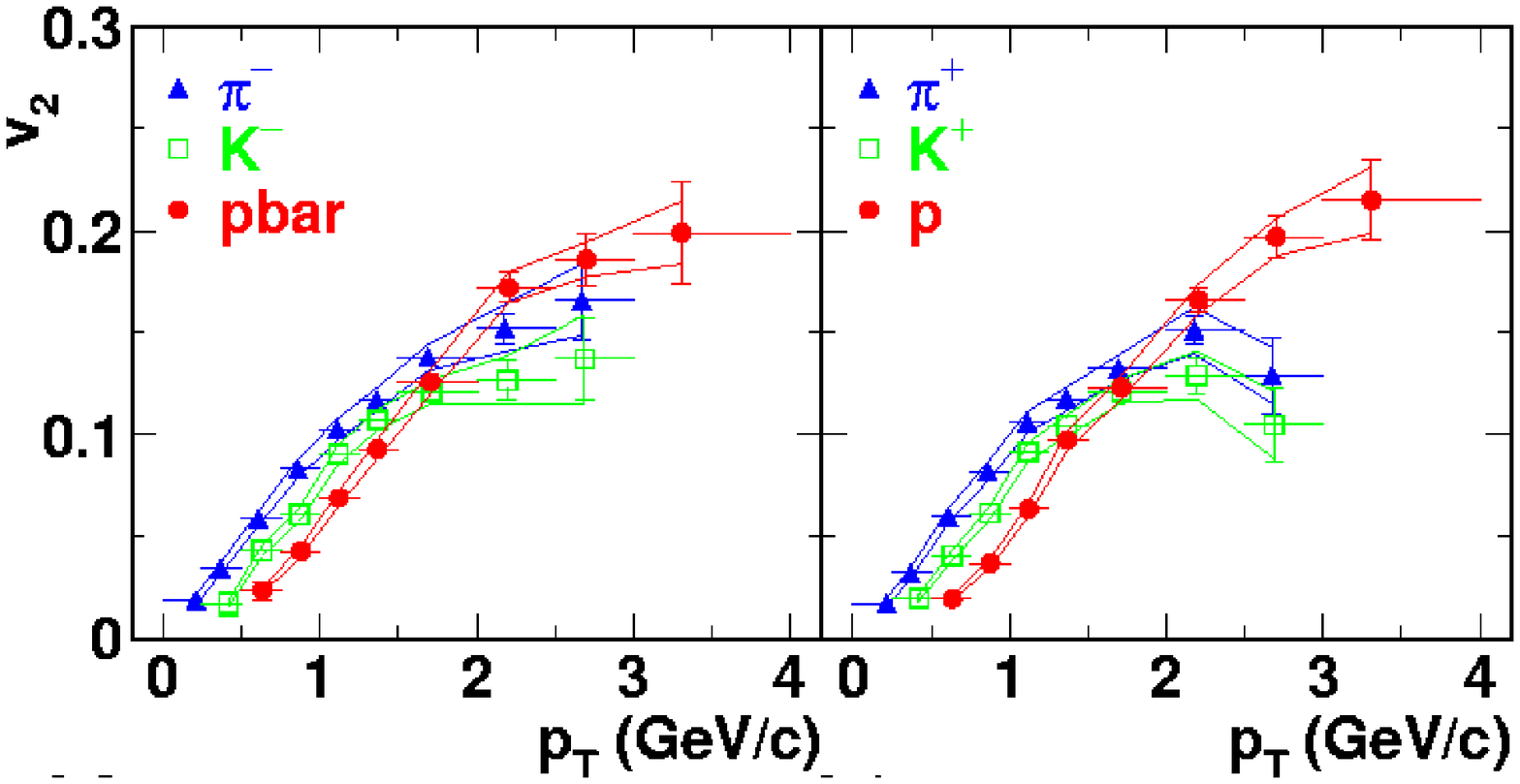}
\hspace*{0.4cm}
\includegraphics[height=4.4cm]{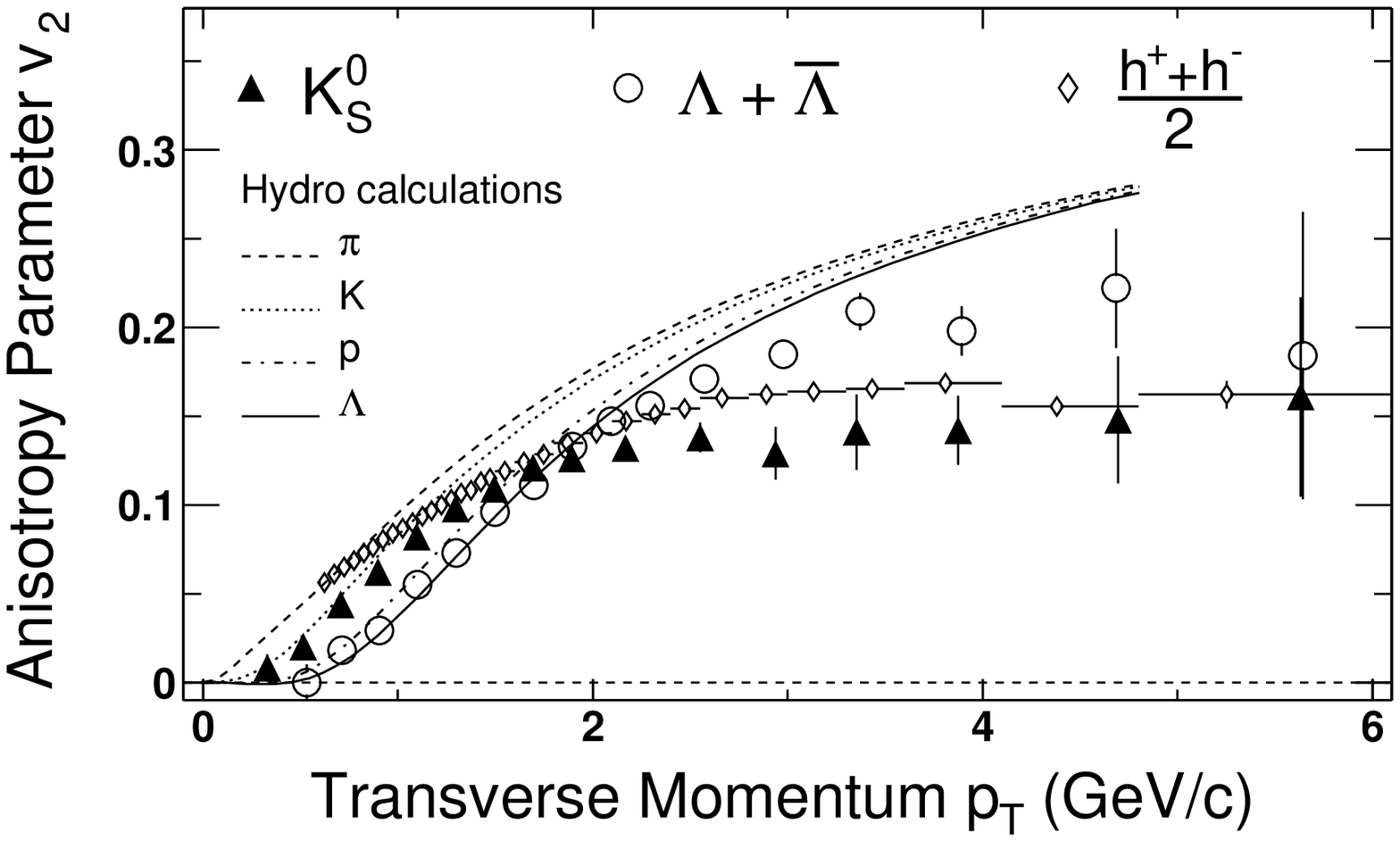}
\vspace*{-0.5cm}
\end{center}
\caption[]{$v_{2}$ as a function of transverse momentum for identified 
particles at RHIC: $\pi^\pm$, $K^\pm$ and $p,\bar{p}$ from PHENIX 
({\it left}), and $K^0_s$ and $\Lambda,\bar{\Lambda}$ from STAR 
({\it right}).}
\label{fig:v2_pT}
\end{figure}

At low $p_{T}$ the strength of the elliptic flow signal is found to be 
large and consistent with hydrodynamics expectations. Above $p_{T}\sim$ 
2 GeV/$c$ where the contribution from collective behaviour is negligible, 
$v_{2}$ is found to be still a sizeable signal which saturates and/or 
slightly decreases as a function of 
$p_{T}$~\cite{Adler:2002ct,Adler:2003kt,Adams:2003am}. The large value 
$v_{2}(p_{T}>$ 2 GeV/$c) \sim$ 0.15 implies unrealistically large 
parton densities and/or cross-sections according to standard parton 
transport calculations~\cite{Molnar:2003ff}. Various interpretations 
have been proposed to account for such a large $v_{2}$ parameter within 
different physical scenarios. In jet quenching models~\cite{Gyulassy:2000gk} 
the resulting momentum anisotropy results from the almond-like density 
profile of the opaque medium (see, however,~\cite{Shuryak:2001me}). 
Calculations based on gluon saturation~\cite{Kovchegov:2002nf} yield a 
(``non-flow'') azimuthal asymmetry component from the fragmentation of 
the released gluons from the initial-state saturated wave functions of 
the colliding nuclei. Finally, quark recombination 
effects~\cite{Molnar:2003ff} can naturally enhance the elliptic flow of 
the produced hadrons compared to that of partons. The measured 
$v_{2}(p_{T})$ for mesons and baryons 
shows a distinct pattern (Fig.~\ref{fig:v2_pT}): $v_{2}^{m}>v_{2}^{b}$ at 
low $p_{T}$, $v_{2}^{m}\approx v_{2}^{b}$ at $p_{T}\approx$ 2 GeV/$c$, and
$v_{2}^{m}<v_{2}^{b}$ at higher $p_{T}$'s; which further constraints the
proposed theoretical explanations.

\begin{figure}[htbp]
\begin{center}
\includegraphics[height=5.cm]{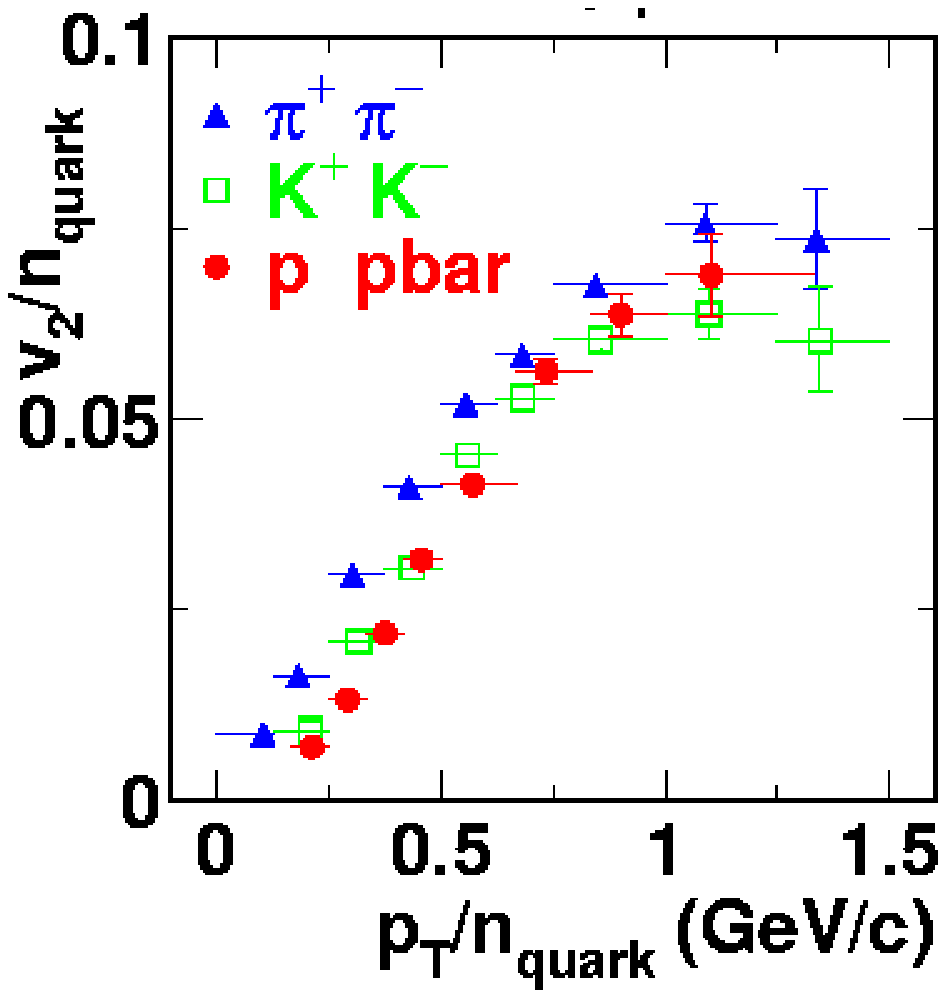}
\hspace*{1.0cm}
\includegraphics[height=5.2cm]{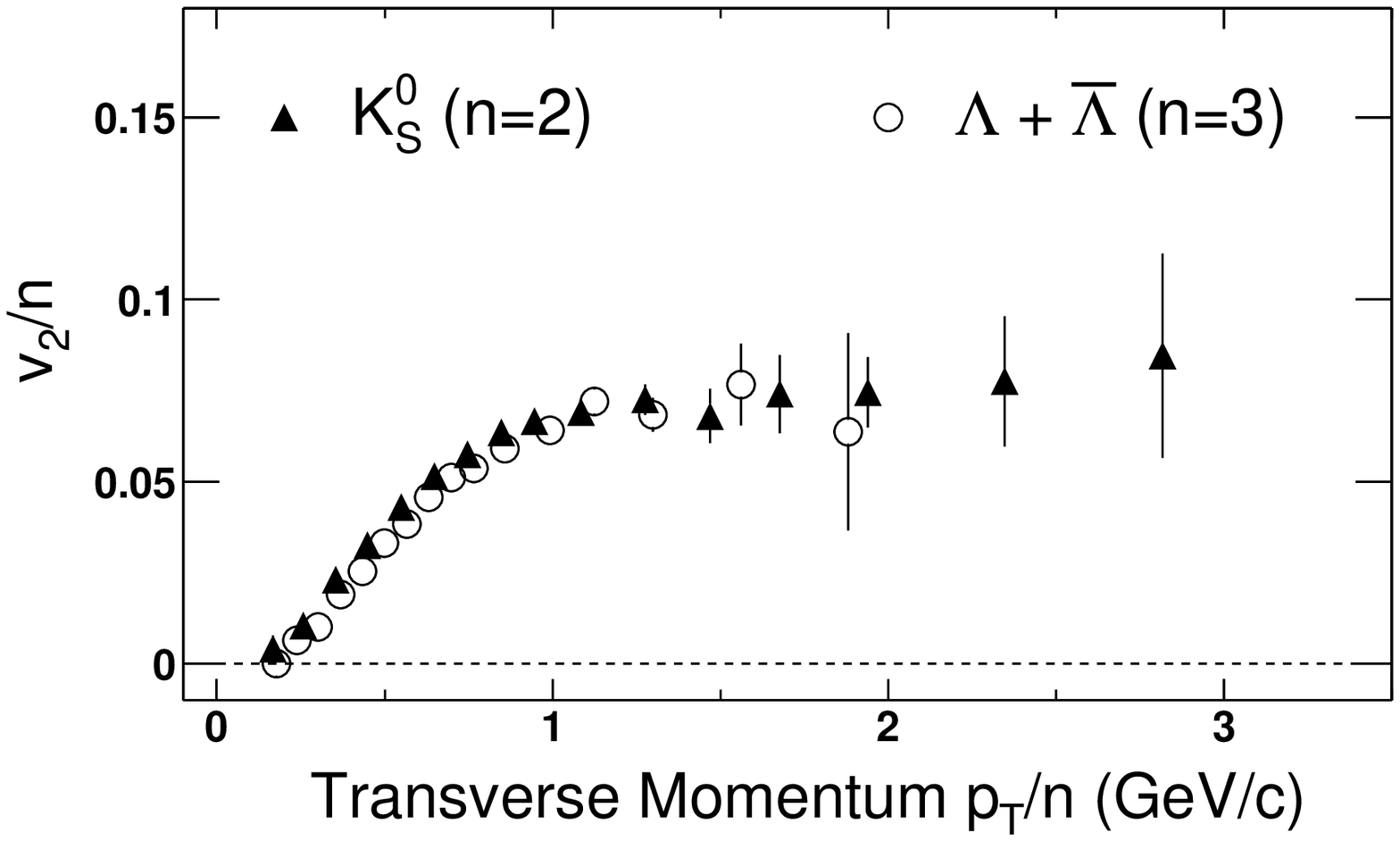}
\vspace*{-0.3cm}
\end{center}
\caption[]{The $v_{2}$ parameter scaled by the number of constituent
quarks ($n$) versus $p_{T}/n$ for $\pi^\pm$, $K^\pm$ and $p,\bar{p}$ 
(PHENIX \cite{Adler:2003kt}, {\it left}) and $K^0_s$ and 
$\Lambda,\bar{\Lambda}$ (STAR \cite{Adams:2003am}, {\it right}).}
\label{fig:v2_n_pT}
\end{figure}

Quark coalescence models~\cite{Molnar:2003ff} naturally lead to weaker
baryon flow than meson flow at low $p_{T}$, while the opposite holds
above 2 GeV/$c$. This simple mass ordering expectation 
recombination models is confirmed by the identified particle data from 
PHENIX and STAR (Fig.~\ref{fig:v2_n_pT}). The fact that the $v_{2}$ 
parameters scaled by the number of constituent quarks ($n$ = 2 for mesons, 
$n$ = 3 baryons) versus $p_{T}/n$, globally fall in a single curve, 
supports the scenario where hadrons at moderate $p_{T}$'s form by 
coalescence of co-moving quarks.

\subsection{High-$p_{T}$ Hadron Production in d+Au Collisions}

\begin{figure}[htbp]
\vspace*{-0.1cm}
\begin{center}
\begin{tabular}{cc}
   \includegraphics[height=4.7cm]{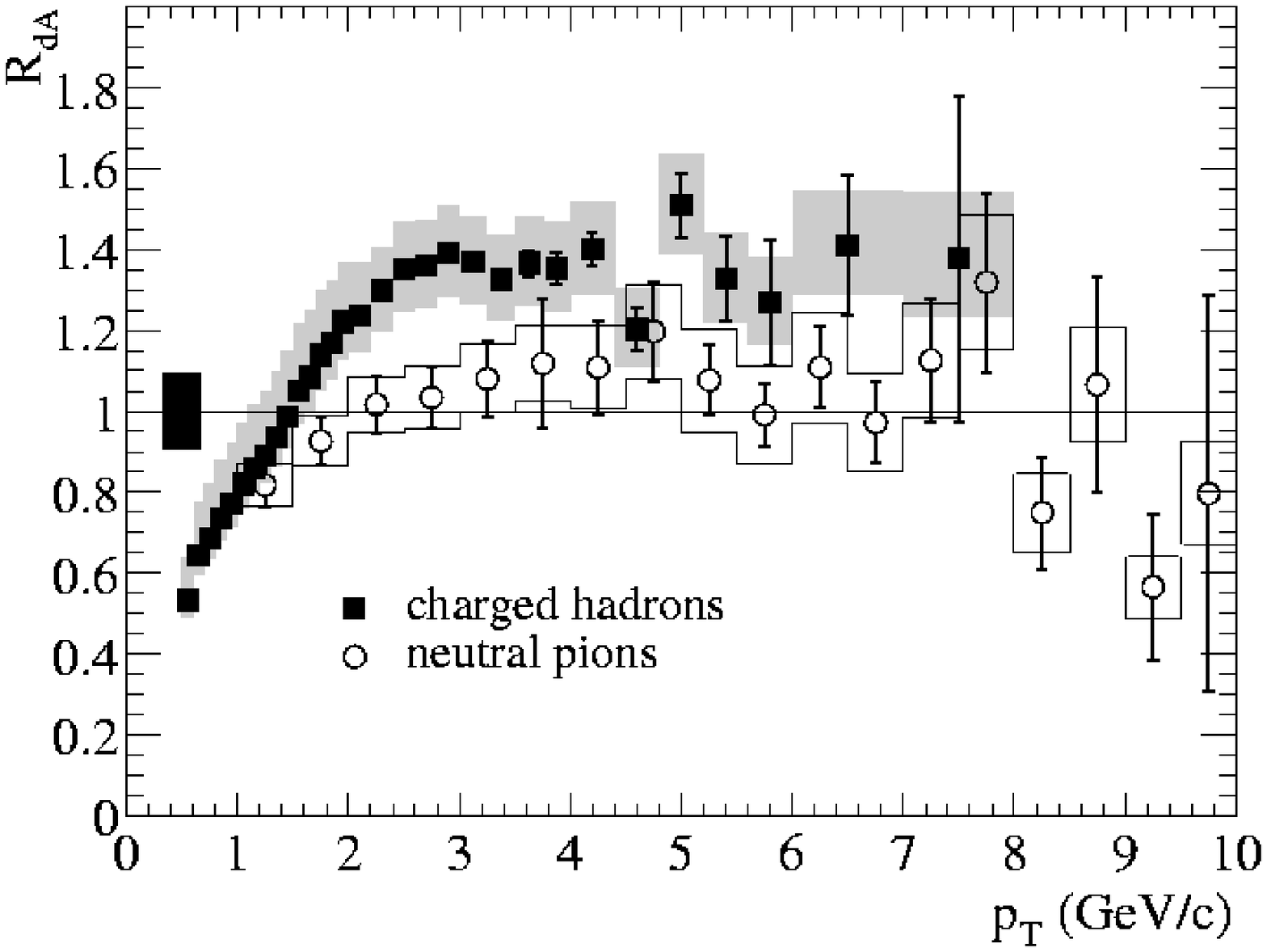} &
   \includegraphics[height=4.7cm]{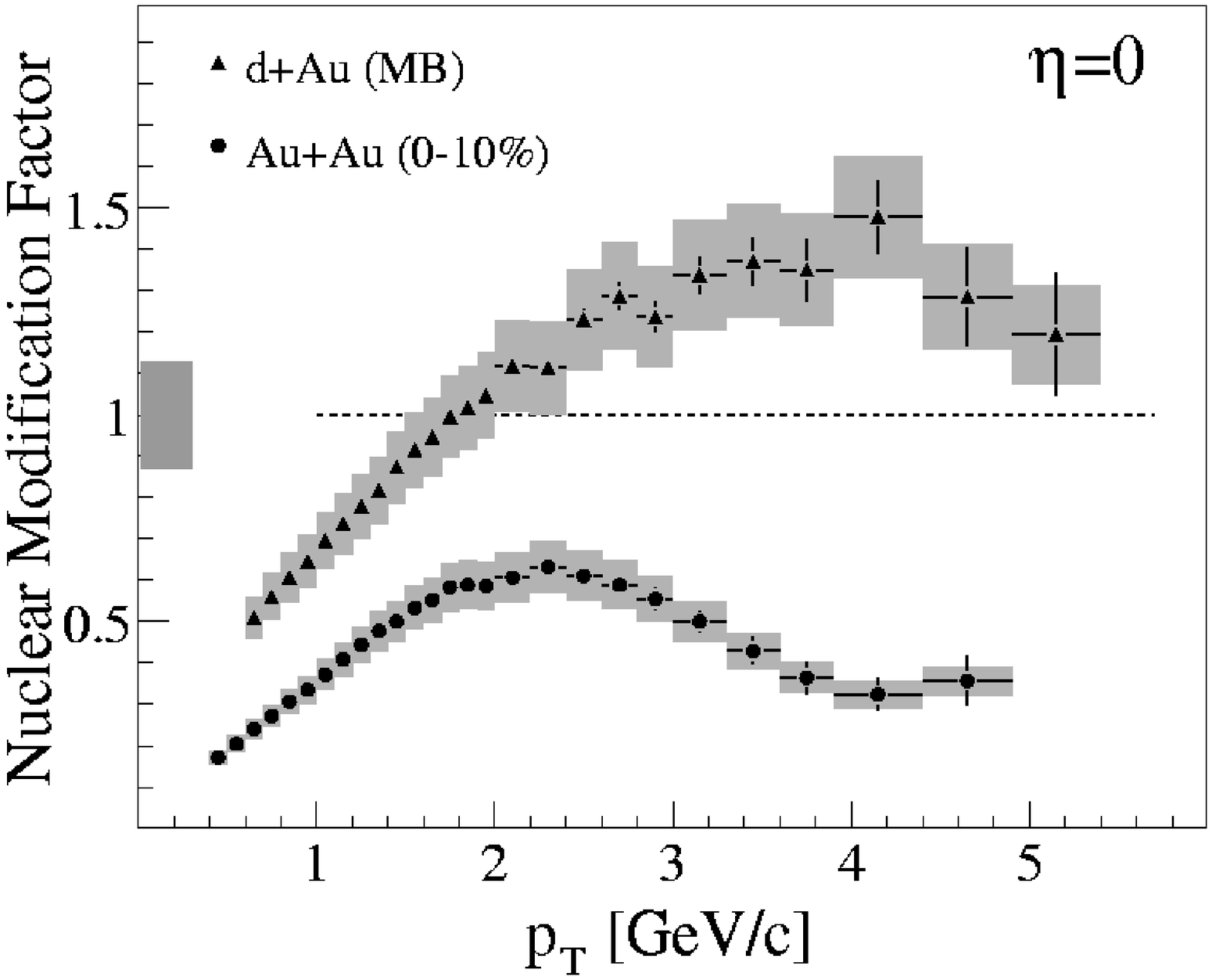} \\
   \includegraphics[height=5.5cm]{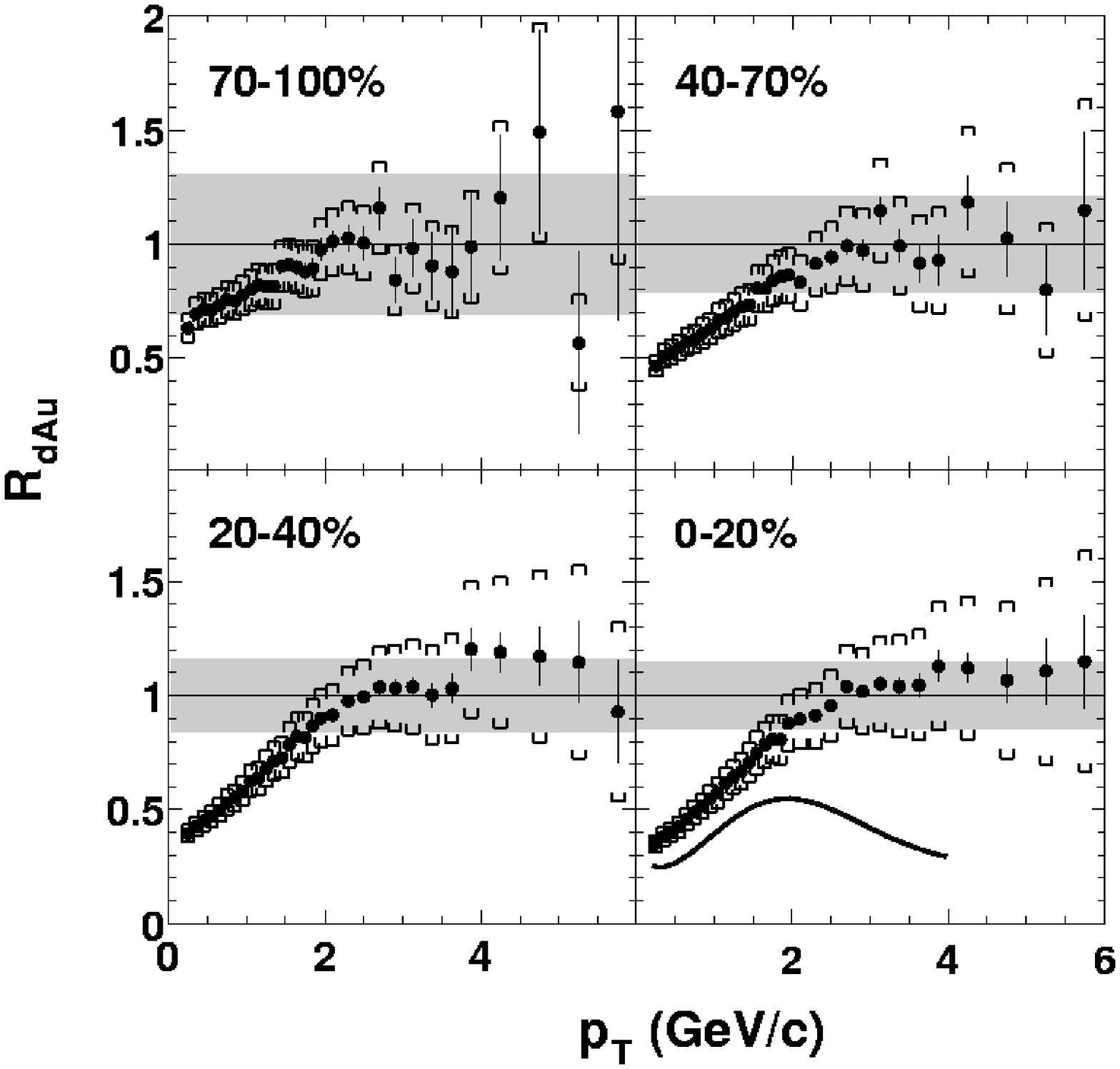} &
   \includegraphics[height=4.5cm]{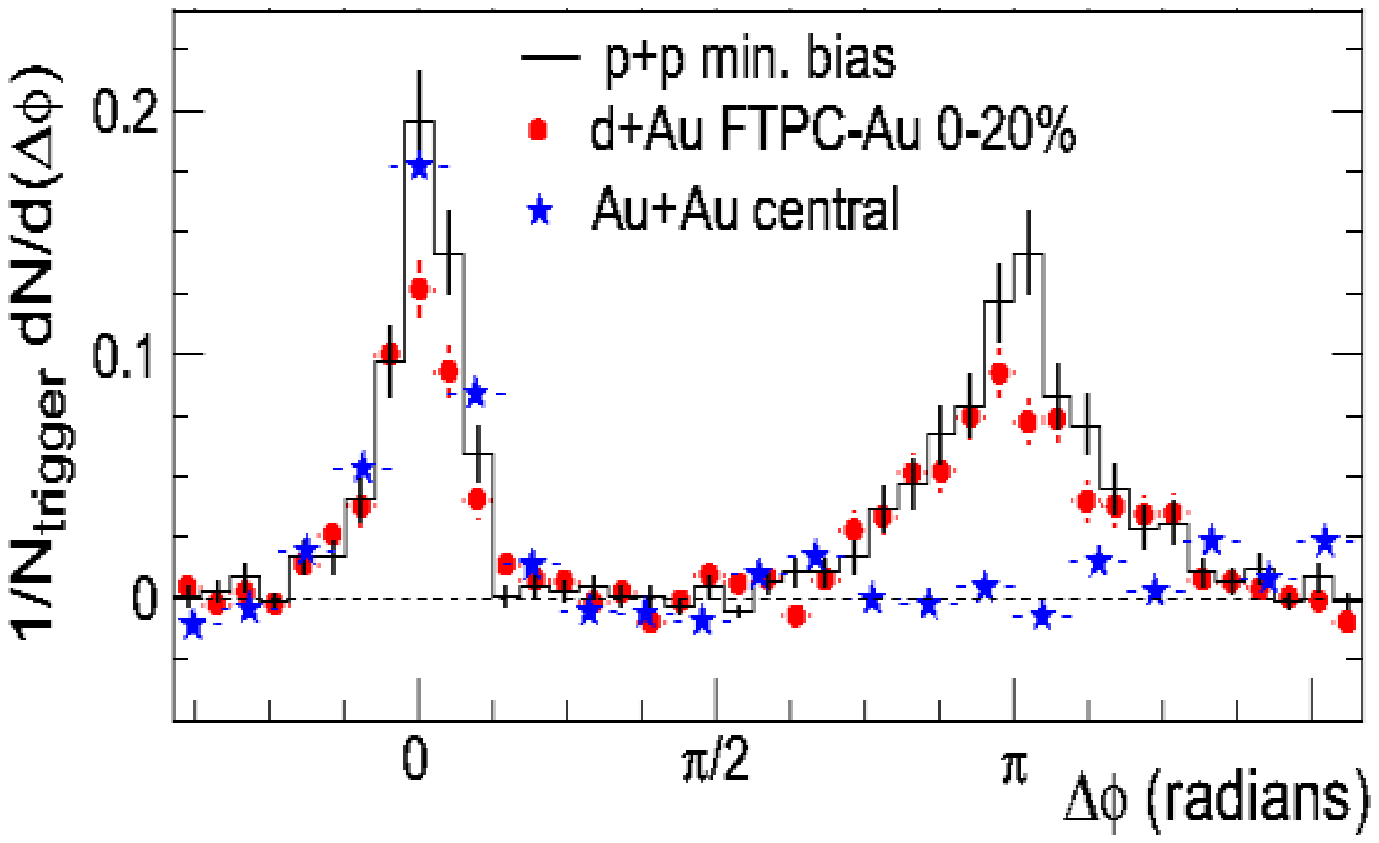} \\
\end{tabular}
\end{center}
\vspace*{-0.5cm}
\caption[]{Top: Nuclear modification factor $R_{dAu}$ for MB d+Au: 
$h^\pm$ and $\pi^0$ (PHENIX, {\it left}), $h^\pm$ (BRAHMS, {\it right}). 
Bottom: $R_{dAu}$ for $h^\pm$ measured by PHOBOS in 4 different d+Au 
centralities ({\it left}), and comparison of two-particle azimuthal 
distributions for central d+Au, p+p, and central Au+Au collisions 
(STAR, {\it right}).}
\label{fig:dAu}
\end{figure}

Proton- (or deuteron-) nucleus collisions constitute a reference 
``control'' experiment needed to determine the influence of {\it cold} 
nuclear matter effects in high $p_{T}$ hadro-production. Since final-state 
medium effects are marginal in p,d+Au collisions, they are basic tools to 
ascertain whether models based on initial- or final- state QCD effects 
can explain the distinct hard scattering behaviour observed in Au+Au 
collisions at RHIC. During the third year of RHIC operation, the 4
experiments collected data from d+Au collisions at 
$\sqrt{s_{_{NN}}}$ = 200 GeV. The resulting high $p_{T}$ results at 
mid-rapidity from PHENIX~\cite{Adler:2003qs}, STAR~\cite{Adams:2003im}, 
PHOBOS~\cite{Back:2003ns}, and BRAHMS~\cite{Arsene:2003yk} consistently 
indicate the following:

\begin{itemize}
\item High $p_{T}$ inclusive 
$h^\pm$~\cite{Adler:2003qs,Adams:2003im,Back:2003ns,Arsene:2003yk} 
and $\pi^0$~\cite{Adler:2003qs} spectra from d+Au minimum bias (MB) 
collisions are not suppressed but are {\it enhanced} compared to p+p 
collisions ($R_{ dAu}$ plots in Fig.~\ref{fig:dAu}), in a way very 
much reminiscent of the ``Cronin effect'' observed in fixed-target p+A 
collisions at lower $\sqrt{s}$~\cite{Antreasyan:cw}. As a matter of fact, 
p+Au collisions (from neutron-tagged d+Au events~\cite{Adler:2003qs}) 
show a similar behaviour as minimum bias d+Au collisions.
\item Above $p_{T}\sim$ 2.5 GeV/$c$ the nuclear modification factor of 
inclusive charged hadrons in MB d+Au collisions saturates 
at~\footnote{
$R_{dAu}^{_{\mbox{\tiny{PHENIX}}}}(p_{T}=2-7$ GeV/$c)\sim$ 1.35, 
$R_{dAu}^{_{\mbox{\tiny{STAR}}}}(p_{T}=2-6$ GeV/$c)\sim$ 1.45, 
$R_{dAu}^{_{\mbox{\tiny{BRAHMS}}}}(p_{T}=2-5$ GeV/$c)\sim$ 1.3.} 
$R_{dAu} \sim$ 1.4. Above 6 GeV/$c$, STAR $h^\pm$ and PHENIX $\pi^0$ 
results seem to indicate that $R_{dAu}$ decreases as a function of 
$p_{T}$, becoming consistent with 1 at around 8 GeV/$c$. 
\item The ``Cronin enhancement'' for unidentified hadrons at high $p_{T}$ 
($R_{dAu}^{h^\pm}\approx$ 1.35) is larger than for neutral pions 
($R_{dAu}^{\pi^0}\approx$ 1.1)~\cite{Adler:2003qs}.
\item The degree of ``enhancement'' in d+Au compared to p+p {\it increases} 
with collision centrality~\cite{Back:2003ns,Adams:2003im}, an opposite 
trend to Au+Au results. 
\item The azimuthal correlations in MB and central d+Au collisions are 
very similar to that of p+p and do not show the significant suppression 
of the away-side peak observed in central Au+Au reactions~\cite{Adams:2003im}.
\end{itemize}

All these results lead to the conclusion that no ``cold'' nuclear matter 
(or initial-state) effects, -like a strong saturation of the nuclear 
parton distribution functions in the relevant ($x,Q^{2}$) kinematical 
region probed by the current experimental setups-, can explain the high 
$p_{T}$ behaviour in central Au+Au. The data suggest, instead, that
final-state interactions are responsible of the high $p_{T}$ suppression 
and the disappearance of back-to-back dijet correlations observed at 
mid-rapidity in central Au+Au reactions.


In summary, these data put strong experimental constraints on the 
properties of the underlying QCD medium produced in Au+Au reactions 
at collider energies. Comparison of the energy spectra and angular 
correlations data to the theoretical calculations globally supports 
pQCD-based models of final-state parton energy loss in a dense medium, 
although non-perturbative effects (like e.g. quark coalescence) are 
needed in order to explain the baryon-meson differences in yield and 
$v_{2}$ in the intermediate $p_{T}$ window ($p_{T}\approx$ 2 -- 5 GeV/$c$). 
Theoretical predictions of a strong saturation of the nuclear wave function 
at high energies are also in agreement with most of the data but do not 
seem to explain consistently Au+Au and d+Au RHIC results at midrapidity. 
Coming ion-ion runs at RHIC and, in the mid-term, Pb+Pb collisions 
at LHC energies will help to further strengthen our current understanding 
of the physics of QCD media at high energy densities.

\section{EXPERIMENTAL CAPABILITIES AT LHC}
\label{sec5}
\subsection{Jet Physics with the ALICE Detector}
\label{sec51}
{\em A.~Morsch}
%
%

ALICE is a multipurpose heavy ion experiment with excellent tracking 
and secondary vertex capabilities, electron and muon detection and a 
high resolution $\gamma$-spectrometer \cite{TP}. In the barrel part of the 
experiment ALICE will measure the flavor content and phase-space 
distribution, event by event, for a large number of particles whose 
momenta and masses are of the order of the typical energy scale involved 
(temperature $\approx \Lambda_{\rm QCD} \approx 200 \, \mbox{${\rm MeV}$}$). 
However, tracking and particle identification capabilities at central 
rapidity reach also far into the transverse momentum region in which 
particle production is expected to be dominated by hard processes, the 
production and fragmentation of high transverse momentum partons, i.e. jets.

Since in its present design complete measurements are restricted to 
charged particles, the ALICE detector has only limited capabilities 
to measure the jet energy. However, energy is not the only jet observable. 
On the contrary, it is likely that the most interesting observables 
which can reveal the presence and kind of interactions of partons with 
deconfined partonic matter and the associated radiation of additional gluons 
(jet quenching) are mainly related to the structure of the jets, 
i.e. the phase space distribution of particle 
around the jet axis, fragmentation function and jet shape.
Recall also that historically the measurement of particle transverse momenta 
relative to the jet-axis has been used to show first evidence for 
gluons radiated from quark jets 
produced in $\mathrm{e^+e^-}$ collisions \cite{Brandelik:1979bd}.
Similar distributions can be measured by ALICE down to very low particle 
momenta and for identified particles. This analysis will be performed
as a function of the energy density by varying the centrality and the 
size of the AA collision system. The study of pA collisions will establish
the reference for comparison with cold nuclear matter.
Moreover, the observables can be studied as a function of the distance 
of the jet or leading particle direction to the reaction plane.  

The STAR experiment at RHIC has shown \cite{STAR} that the combination of an 
electromagnetic calorimeter with a TPC tracking system is functionally 
equivalent to an ideal jet detector in a heavy ion collision environment. 
An electromagnetic calorimeter for ALICE (EMCAL) \cite{EMCAL} has been 
proposed by a group of collaborating US institutes. This would provide an 
opportunity to measure jet energy and the jet structure in heavy 
ion collisions.

\subsubsection{Tracking Particle Identification}

ALICE has been designed to measure single- and multi-particle distributions 
at the highest anticipated charged particle multiplicities for Pb--Pb 
reactions ($dN/d\eta = 8000$). The central tracking devices measure particles 
in the pseudo-rapidity range $|\eta | < 0.9$ with full azimuthal coverage. 
The inner tracker (ITS) \cite{ITS}  provides secondary vertex reconstruction 
of charm 
and hyperon decays, particle identification and tracking of low-momentum 
particles ($\ensuremath{p_{\mathrm{T}}} > 100 \, \mbox{${\rm MeV}$}$). 
The main tracking system is a Time Projection Chamber (TPC)  \cite{TPC} 
providing robust 
tracking capability, good momentum resolution and two particle separation, 
in a high particle density environment. The Transition Radiation Detector 
(TRD) \cite{TRD} for electron identification adds additional tracking 
information at high radii improving the momentum resolution of 
high-\ensuremath{p_{\mathrm{T}}} particles.   
The momentum resolution $\Delta p/ p$ for particle momenta 
$< 2 \, \mbox{${\rm GeV}$}$ is below 1\%. It rises almost linearly 
to up to 16\% for $p = 100 \, \mbox{${\rm GeV}$}$. Tracking efficiency 
reaches almost 100\% for momenta above $1\, \mbox{${\rm GeV}$}$. 
A geometrical inefficiency for high-\ensuremath{p_{\mathrm{T}}} particles 
of 10\% results from dead zones between the TPC read-out chambers.

A Time of Flight System (TOF) \cite{TOF} based on multi-gap resistive 
plate chambers 
(RPCs) extends particle identification in the barrel region to 
$2 \, \mbox{${\rm GeV}$}$ and $3.5 \, \mbox{${\rm GeV}$}$ for 
${\rm K} / \pi $ and ${\rm K / p}$ separation, respectively. Recent studies 
with an improved TPC tracking algorithm have shown that particle 
identification on a statistical basis via $dE/dx$ can be extended into the 
relativistic rise up to $\approx 50 \, \mbox{${\rm GeV}$}$.
The HMPID (High Momentum Particle IDentificationP) system \cite{HMPID} 
using a single-arm Ring Imaging Cerenkov Counter
is devoted to the discrimination of hadrons in the hard part of the 
momentum spectrum. 
It will enhance the PID capability of ALICE by allowing to identify 
particles beyond the momentum 
interval covered by the energy loss measurements and the TOF. The 
useful range for the identification of 
$\pi/{\rm K}$ and K/p of a track by track basis is extended up to 
$3 \; {\rm GeV}$ and $5 \; {\rm GeV}$, respectively. 
The PHOS (PHOton Spectrometer) \cite{PHOS} 
is a single-arm high-resolution electromagnetic calorimeter of high 
granularity.
It is optimised for measuring photons, $\pi^0$'s and $\eta$ mesons 
in broad energy ranges. 
The TRD provides excellent $e/ \pi$-separation in the 
$1 - 100 \, \mbox{${\rm GeV}$}$ momentum range. 

\subsubsection{Event Geometry}
The combined measurement of energy in the zero-degree 
calorimeters (ZDC) \cite{ZDC}
and the forward electromagnetic calorimeter (ZEM) allows ALICE to 
determine the collision centrality in five impact parameter bins 
between 0 and $16 \, {\rm fm}$. Non-zero impact parameters lead to 
pressure gradients in the reaction volume producing directed flow that 
can be observed by ALICE as an anisotropy in the azimuthal distribution 
of final state particles. The measurement of directed flow allows an 
unambiguous determination of the event plane. Thus jet observables can 
be studied as a function of the centrality and the distance of the jet 
axis to the reaction plane.

\subsubsection{Inclusive Transverse Momentum Spectra}
As described in section 3.43 transverse momentum spectra are a mean to 
study final state radiative energy loss via the nuclear modification 
factor $R_{\rm AA}(p_{\rm T})$. Using a sample of $10^7$ Pb--Pb minimum 
bias events and in the absence of quenching ALICE will be able to measure 
the charged particle transverse momentum spectrum up to 
$\approx 100 \, {\rm GeV} $ and down to $\approx 100 \; {\rm MeV}$.
${\cal O} (10)$ events are in the highest momentum bins where jet quenching 
is expected to reduce the charged particle yield by a factor $2-3$. 
ALICE is planning to use a High Level Trigger (HLT) to increase the 
statistics in the high momentum region. Moreover, neutral pions will 
be identified on an event-by-event basis in the high-momentum range from 
$30-100\, {\rm GeV}$.

\begin{figure}
\begin{center}
\begin{tabular}{cc}
\mbox{\includegraphics[width=0.49\linewidth]{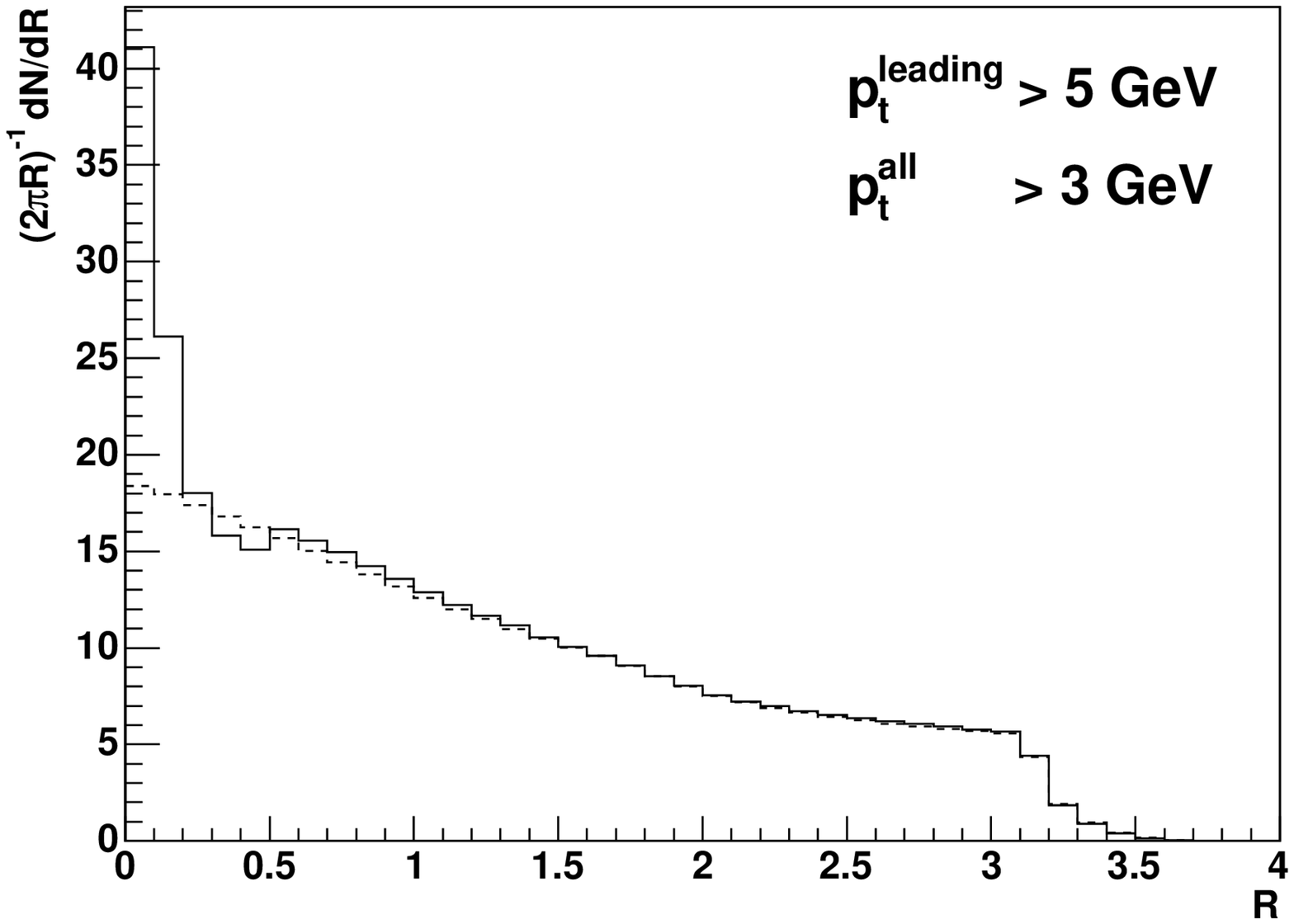}}
&
\mbox{\includegraphics[width=0.49\linewidth]{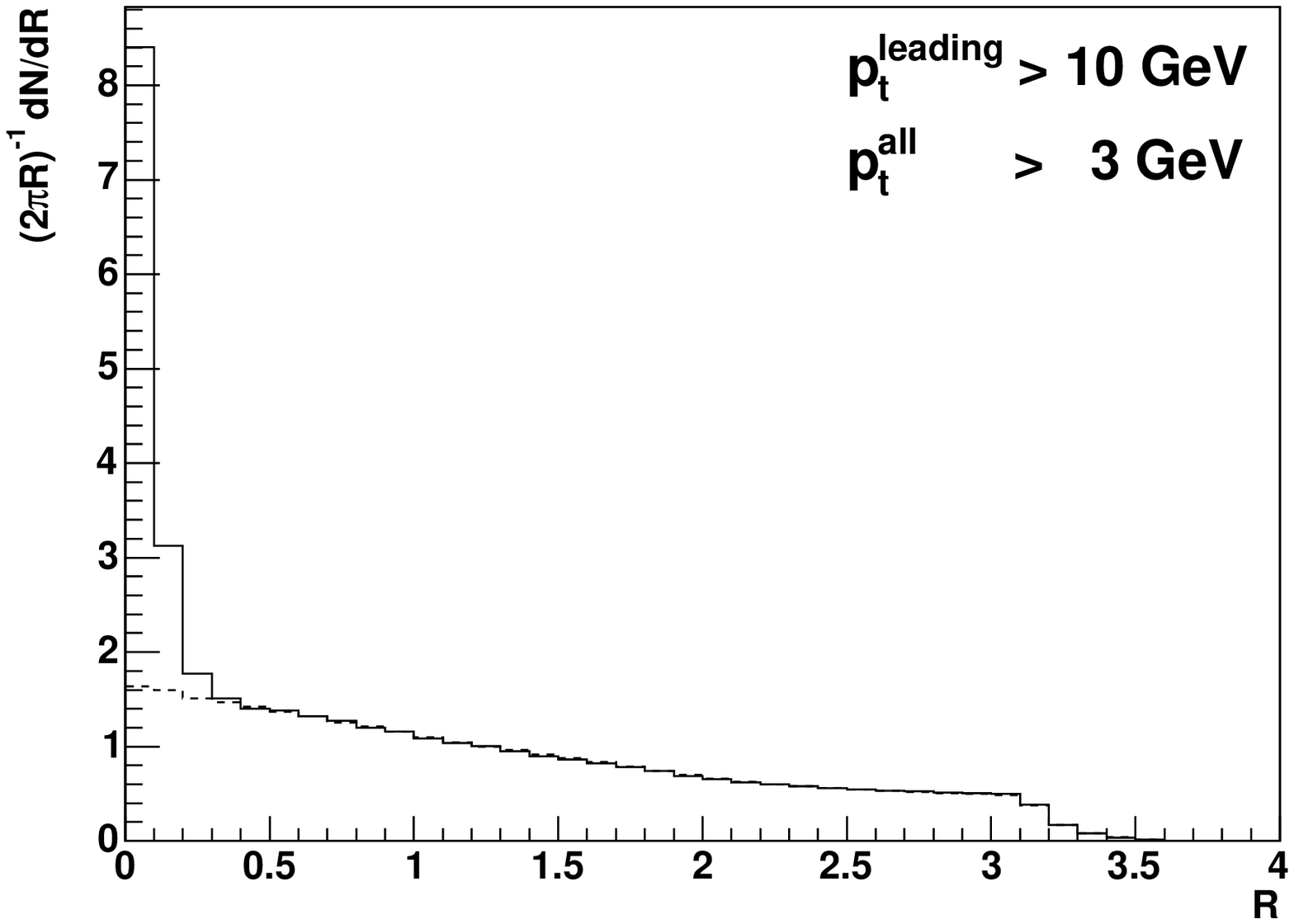}} 
\end{tabular}
\end{center}
\caption{
Distance $R$ of particles with \ensuremath{p_{\mathrm{T}}}\mbox{$>$}3~GeV/c 
to leading particles with  \ensuremath{p_{\mathrm{T}}}\mbox{$>$}5~GeV/c 
(left) and \ensuremath{p_{\mathrm{T}}}\mbox{$>$}10~GeV/c (right). The dashed 
line describing the uncorrelated background is obtained using randomized 
leading particle directions. 
} 
\label{fig:lowet_1}
\end{figure}

\subsubsection{Rates and the Need for Triggering} 
ALICE wants to study the whole spectrum of jet production ranging from 
mini-jets ($\ensuremath{E_{\mathrm{T}}} > 2 \, {\rm GeV}$)
to high-$\ensuremath{E_{\mathrm{T}}}$ jets of several hundred~GeV. 
Concerning the experimental capabilities one has to distinguish four 
energy regions, which are here discussed for central ($b < 5 \, {\rm fm}$) 
Pb--Pb collisions at $\sqrt{s_{\rm NN}} = 5.5 \; {\rm TeV}/{\rm nucleon}$:

\begin{itemize}
\item{}
In the region $E_{\rm T} < 20 \, {\rm GeV}$ several jets of these energies 
overlap in one event within the ALICE acceptance. This means jet 
identification in the traditional sense is not possible. However, their 
presence reveals in inclusive studies of particle correlations.
\item{} 
For $E_{\rm T} < 100 \, {\rm GeV}$ the jet rate of $>  1 \, {\rm Hz}$ is 
high enough so that even with a read-out rate limited by the TPC to 
$40 \, {\rm Hz}$ an event sample of $> 10^5$ jets can be collected in one 
effective month of running ($10^6 \, {\rm s}$).
\item{}
For $E_{\rm T} > 100 \, {\rm GeV}$ triggering will be necessary to collect 
enough data. 
\item{}
Assuming that for a fragmentation function analysis one needs about 
$10^4 - 10^5$ events the statistics limit is reached at about 
$250 \, {\rm GeV}$.
\end{itemize}

In the absence of calorimetry, triggering on jets is only possible using 
a High Level Trigger (HLT). Presently ALICE us studying the possibility 
to trigger on event topologies where two or more 
high-\ensuremath{p_{\mathrm{T}}} tracks are found in a small area of the 
$\eta - \phi $ plane. The search has to be performed on track candidates 
which are themselves the result of a HLT fast clustering and tracking 
procedure.

\subsubsection{Inclusive Jets at Low $\ensuremath{E_{\mathrm{T}}}$}
In proton-antiproton collisions at $1.8 \, {\rm TeV}$ 
evidence for low energy charged particle clusters has been 
seen~\cite{Affolder:2002zg}. These charged particle jets become 
apparent event by event at a charged jet energy of about 
$2 \, {\rm GeV}$ with, on the average, two charged particles with 
$\ensuremath{p_{\mathrm{T}}} > 0.5\, {\rm GeV}$ and grow to, on the average, 
about 10 charged particles at a charged jet energy of $50 \, {\rm GeV}$.

In central Pb--Pb collisions at $5.5 \, \mbox{${\rm TeV}$} $ with an 
anticipated charged particle multiplicity of up to 8000 charged particles 
per unit of rapidity the situation will be completely different. 
It is expected, that in one central collision about 100 jets with 
$\ensuremath{p_{\mathrm{T}}} > 5 \, \mbox{${\rm GeV}$}$, are produced 
within the acceptance of the ALICE central tracking system. The jet 
multiplicity decreases to an average of one, for 
$\ensuremath{p_{\mathrm{T}}} > 20 \, \mbox{${\rm GeV}$}$. The individual 
structures of these low-$\ensuremath{p_{\mathrm{T}}}$ jets dissolve in 
the overall event structure and are not distinguishable event by event. 
Nevertheless, their properties can be studied using inclusive 
particle correlation distributions as will be shown 
in the following.

This effort has two main objectives. First the study of the underlying 
event properties is important for the understanding of the limits of the 
energy resolution for the reconstruction of 
high-$\ensuremath{E_{\mathrm{T}}} $ jets in HI collisions, i.e. the underlying event 
fluctuations. Second, it is expected that in-medium modifications of 
the jet structure will be stronger for low jet energies. ALICE wants 
to study changes of the particle momenta parallel to the jet axis 
(jet quenching) and in the transverse direction (transverse heating).

In order to study inclusive particle correlations, three principle 
methods can be used:
\begin{itemize}
\item Event by event and region by region fluctuations of number of 
particles or energy;
\item Correlations with {\it leading particles};
\item Spatial spectrum analysis (autocorrelation).
\end{itemize}

\begin{figure}
\begin{center}
\begin{tabular}{cc}
\mbox{\includegraphics[width=0.49\linewidth]{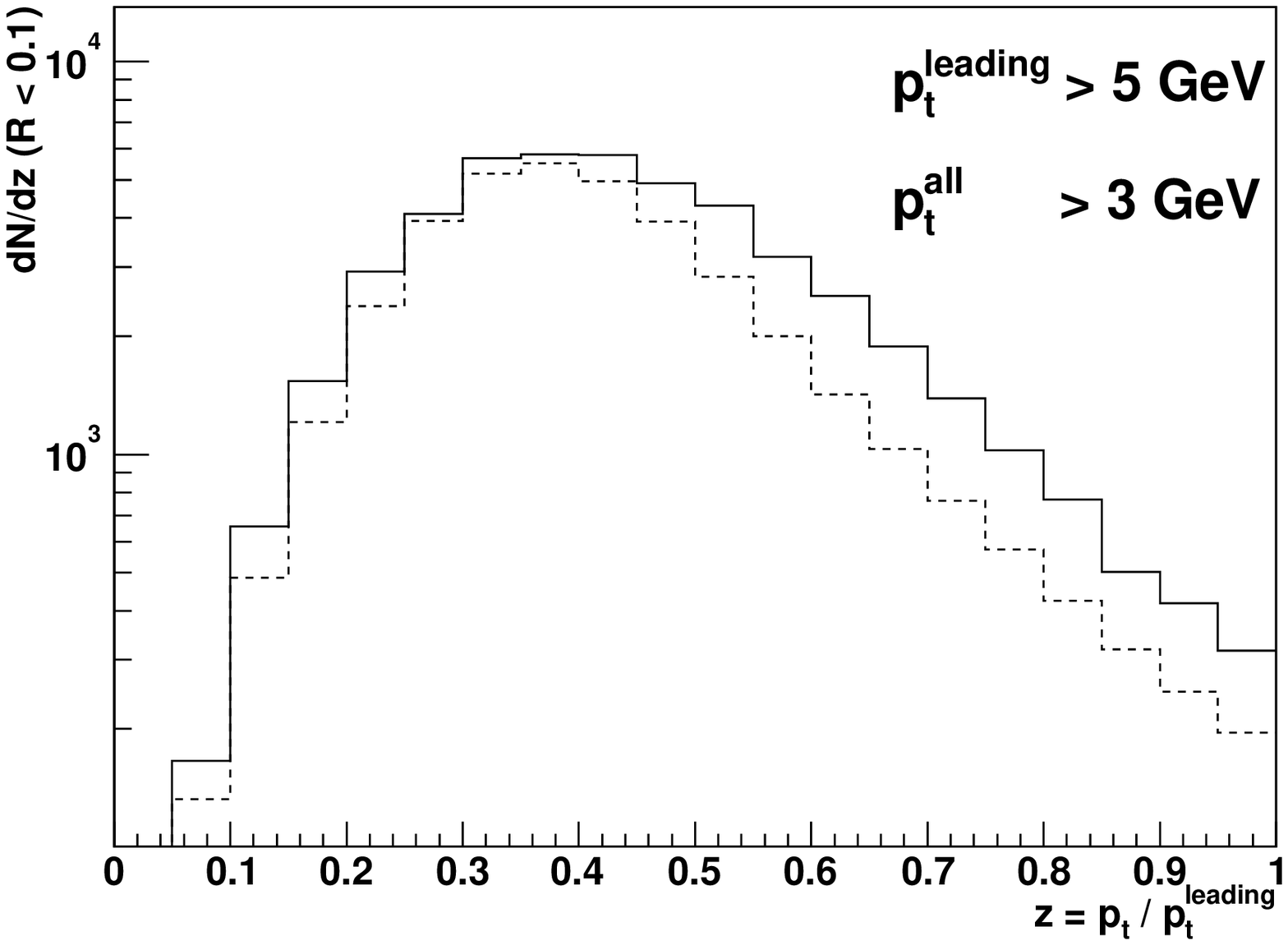}}
&
\mbox{\includegraphics[width=0.49\linewidth]{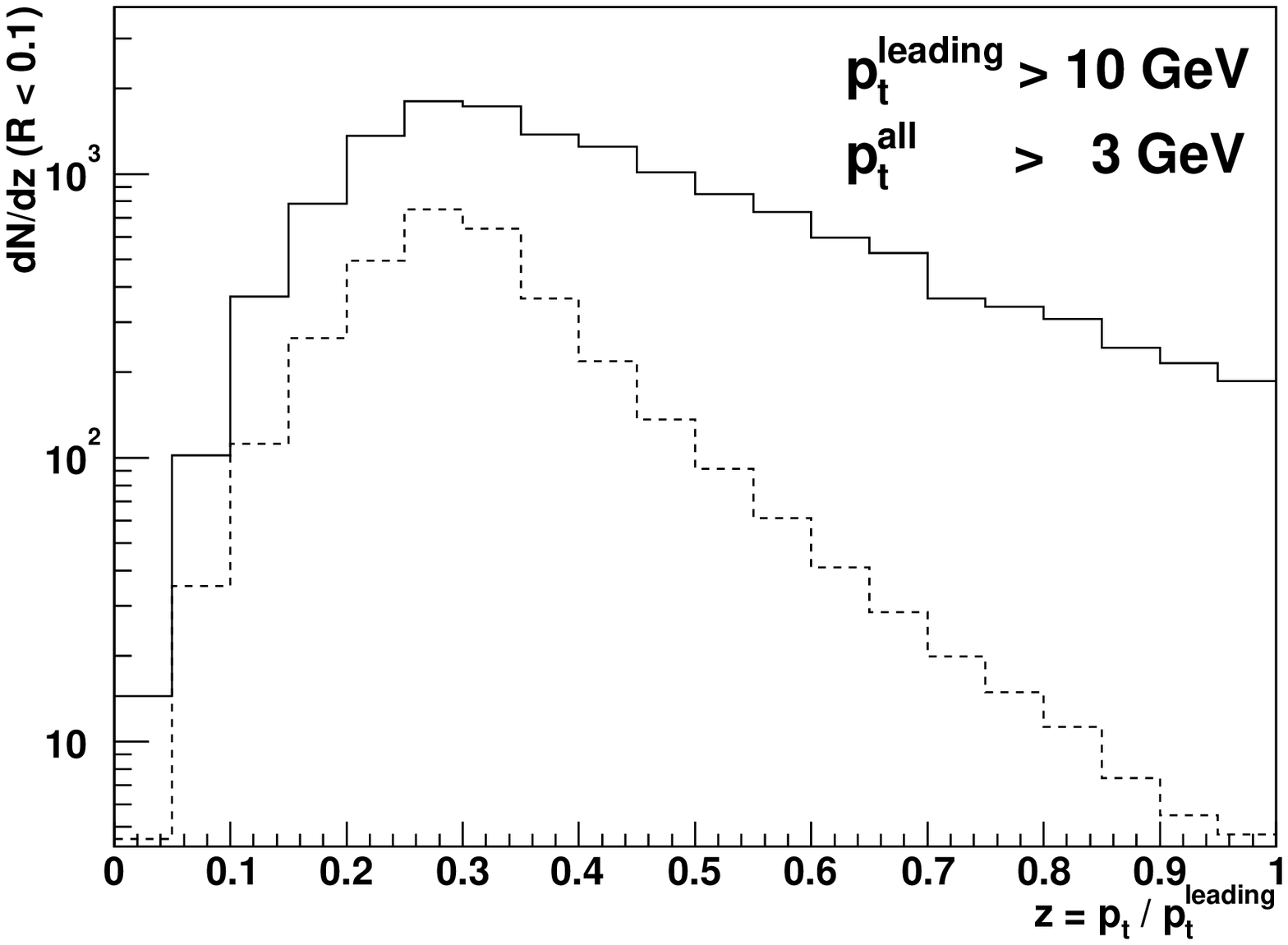}} 
\end{tabular}
\end{center}
\caption{Distribution of   
$z = \ensuremath{p_{\mathrm{T}}} / \ensuremath{p_{\mathrm{T}}}^{\rm leading}$ 
for particles within  $R$\mbox{$<$}0.1 of the leading particle direction 
(solid line). The contribution of uncorrelated particles obtained from a 
guard band  1\mbox{$<$}$R$\mbox{$<$}2 has been subtracted and is shown as a 
dashed line.} 
\label{fig:lowet_2}
\end{figure}

These three methods will be outlined in the following and some 
feasibility studies will be presented.

\paragraph{Underlying event fluctuations}
Here, we will consider fluctuations of the energy contained in a cone of 
size $R_0$. These fluctuations limit the energy resolution of high-energy 
jets obtained with the cone algorithm. For uncorrelated particle production,
the relation between the {\it rms} energy variation $\Delta E$ and the 
number of particles, $N$, the mean transverse momentum, 
$<\ensuremath{p_{\mathrm{T}}} >$, and the {\it rms}  of the 
$\ensuremath{p_{\mathrm{T}}} $-distribution, 
$\Delta \ensuremath{p_{\mathrm{T}}}$, is 
$\Delta E = \sqrt{N} \sqrt{<\ensuremath{p_{\mathrm{T}}}>^2 + 
\Delta \ensuremath{p_{\mathrm{T}}} ^2}$. As an example, this value 
increases by a factor $\sqrt{2}$, if the multiplicity $N$ results from 
$N/2$ clusters of multiplicity 2. In fact, in central Pb--Pb collisions 
simulated with HIJING \cite{Wang:1991ht,Gyulassy:ew}, we observe for  $R_0 = 0.3$ fluctuations that are 
by factors of $1.3-1.5$ higher than expected for uncorrelated particles, 
the exact value depending on the $\ensuremath{p_{\mathrm{T}}}$-cut.

\begin{figure}
\begin{center}
\begin{tabular}{cc}
\mbox{\includegraphics[width=0.49\linewidth]{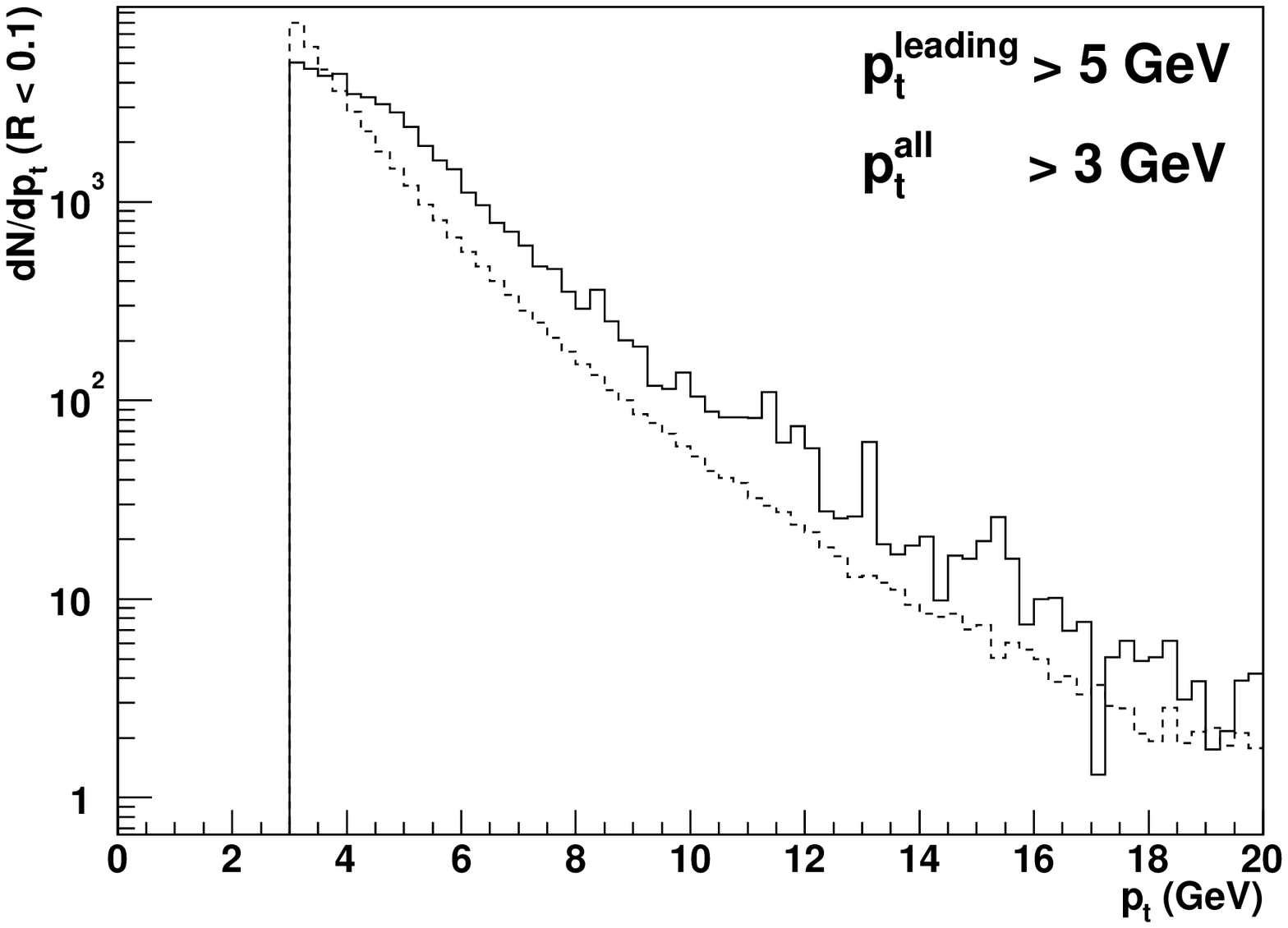}}
&
\mbox{\includegraphics[width=0.49\linewidth]{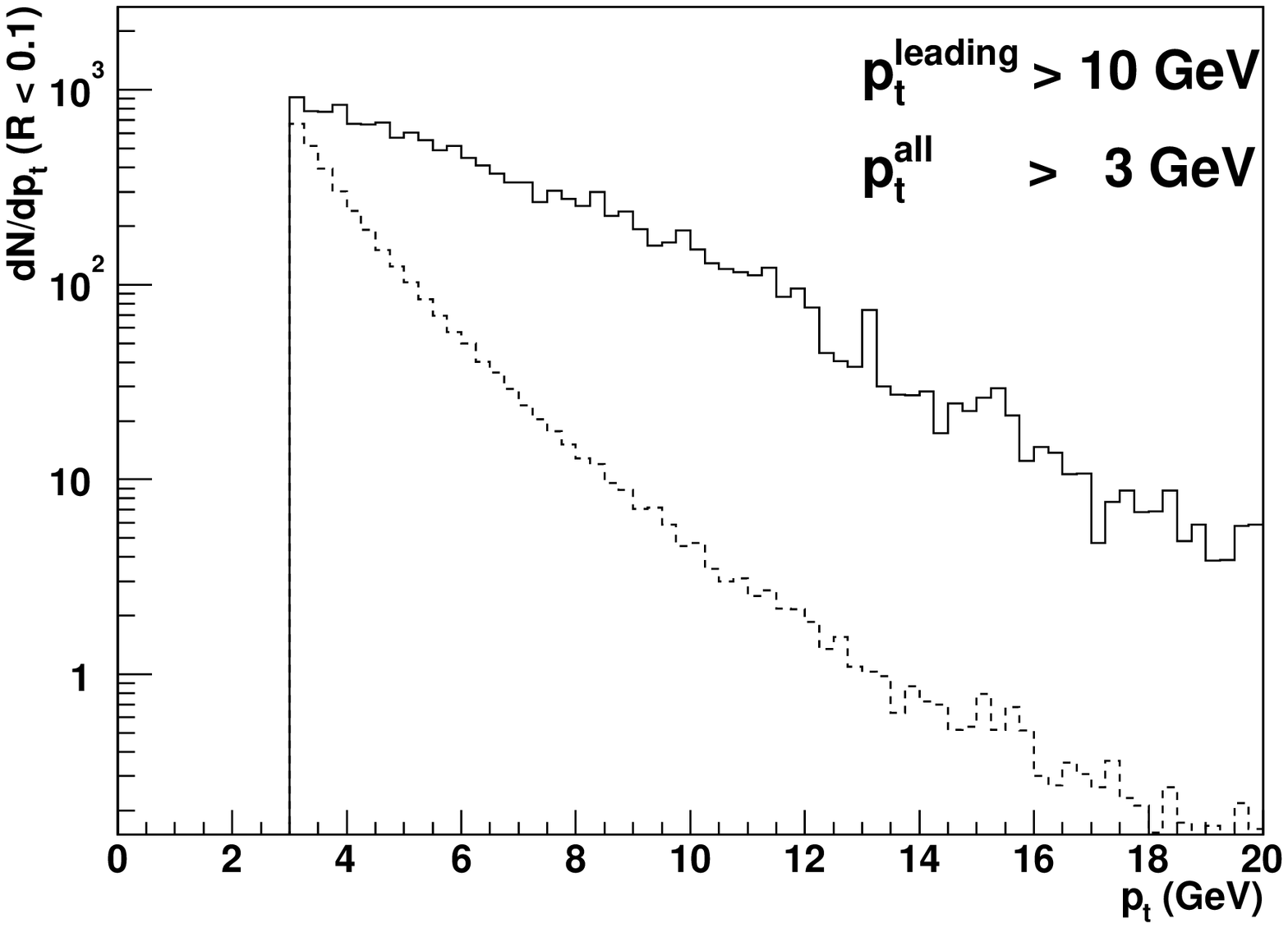}} 
\end{tabular}
\end{center}
\caption{
Charged particle transverse momentum distribution for particles within  
$R$\mbox{$<$}0.1 of the leading particle direction (solid line). The 
contribution of uncorrelated particles obtained from a guard band  
1\mbox{$<$}$R$\mbox{$<$}2 has been subtracted and is shown as a dashed line.
} 
\label{fig:lowet_3}
\end{figure}

\paragraph{Correlations with leading particles}
In order to study correlations with leading particles we apply an 
algorithm similar to the one used for the CDF charged jet analysis. All 
particles with a 
$\ensuremath{p_{\mathrm{T}}} > \ensuremath{p_{\mathrm{T}}} ^{\rm seed}$ 
are leading particle candidates  and are ordered according to their 
$\ensuremath{p_{\mathrm{T}}}$. We start with the highest 
$\ensuremath{p_{\mathrm{T}}}$ candidate  and record the distances $R$ in the 
$\eta  - \phi $-plane between all charged particles and the leading particle.
If another candidate is found within a distance $R < R_{\rm sep}$ it is 
eliminated from the list of leading particles. The procedure continues 
with the next leading particle until none is left.

This algorithm is a natural extension of the cone algorithm used for jet 
reconstruction to inclusive studies in the low-$\ensuremath{p_{\mathrm{T}}}$ 
jet region for heavy ions collisions. To see possible angular correlations 
we plot the particle density $(2\pi R)^{-1} {\rm d} N/{\rm d}R$.
Uncorrelated particle production should result in a flat distribution 
close to the leading particle direction $R=0$. 

HIJING simulations of central Pb--Pb collisions at the LHC centre of 
mass energy have been performed to study these leading particle correlations.
Fig.~\ref{fig:lowet_1} shows such distributions for 
$\ensuremath{p_{\mathrm{T}}}^{\rm seed} = 5$ and $10 \, \mbox{${\rm GeV}$}$   
with a cut on all other particles of $\ensuremath{p_{\mathrm{T}}}^{\rm all} 
> 3 \, \mbox{${\rm GeV}$}$ and  $R_{\rm sep} = 0.5$. 
The dashed lines are the corresponding 
distributions for randomized leading particle directions. 
Clear correlation signals are visible for $R < 0.3$. As expected, the 
significance of the signal increases with the transverse momentum cut.

\paragraph{Fragmentation function}
The distributions of the correlated particle transverse momentum normalized 
by the leading particle transverse momentum 
$z = \ensuremath{p_{\mathrm{T}}} / \ensuremath{p_{\mathrm{T}}}^{\rm leading}$ 
is related to the jet fragmentation function. In Fig.~\ref{fig:lowet_2} we 
show this distribution for particles with $R < 0.1$. The background 
distribution obtained from particles in a guard band $1 < R < 2$ has been 
subtracted and is shown as a dashed line. The signal dominates at high $z$ 
values. The corresponding ``raw'' $\ensuremath{p_{\mathrm{T}}}$ distributions 
are shown in Fig.~\ref{fig:lowet_3}.

$\ensuremath{p_{\mathrm{T}}}^{\rm leading}$ is only a very poor estimator 
of the jet energy and the $dN/dz$ distribution can only be interpreted 
as a smeared out "pseudo"-fragmentation function. This explains the rather 
modest decrease as $z$ approaches 1. 
Nevertheless, since in-medium modifications of the fragmentation function 
are expected to be strong for low energy jets, we expect that the effect 
can be observed in the measured distribution by comparing pPb to Pb--Pb data.

\begin{figure}
\begin{center}
\begin{tabular}{cc}
\mbox{\includegraphics[width=0.49\linewidth]{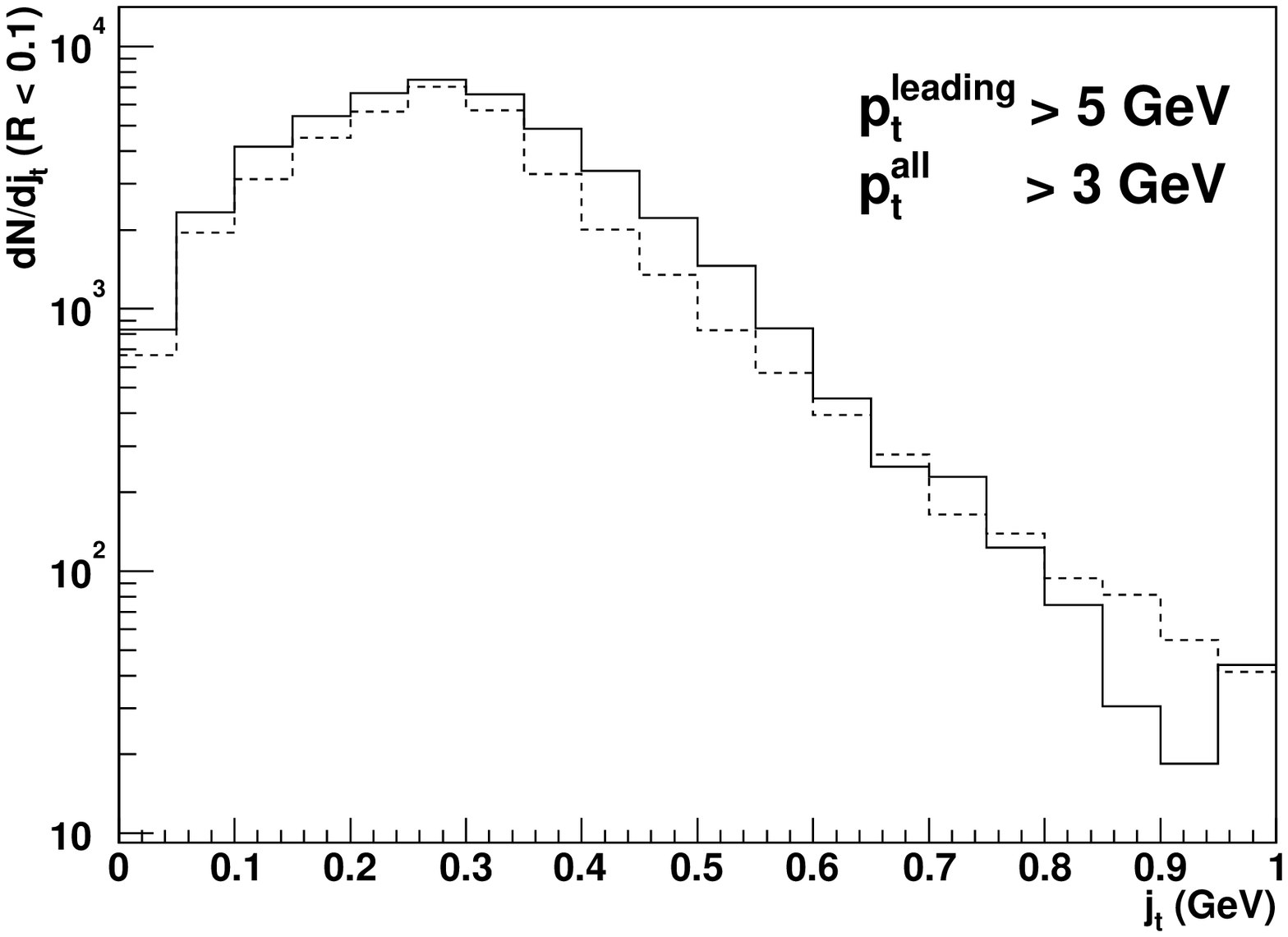}}
&
\mbox{\includegraphics[width=0.49\linewidth]{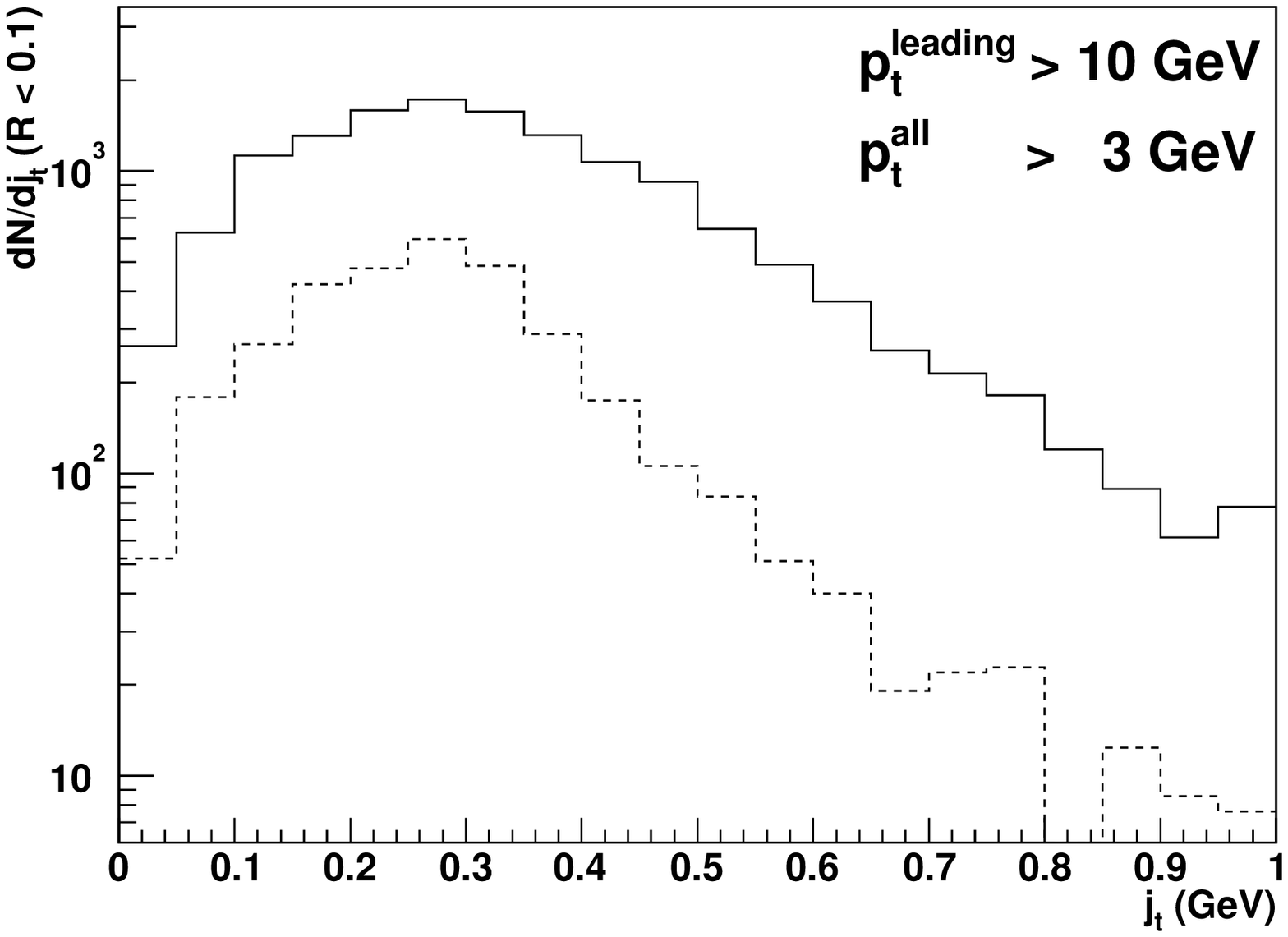}} 
\end{tabular}
\end{center}
\caption{Distributions of the correlated particle momentum perpendicular 
to the jet axis $j_{\rm T} = p  \cdot \sin(\theta({\rm particle,\; 
leading \; particle}))$ obtained from particles within  $R$\mbox{$<$}0.1 
of the leading particle direction (solid line). The contribution of 
uncorrelated particles obtained by randomizing the leading particle 
direction has been subtracted and is shown as a dashed line.} 
\label{fig:lowet_4}
\end{figure}

\paragraph{Momentum transverse to the jet axis}
The distributions of the correlated particle momentum perpendicular to the 
jet axis $j_{\rm T} = p  \cdot 
\sin(\theta({\rm particle,\; leading\; particle}))$ 
is shown in Fig.~\ref{fig:lowet_4}. The background shown as dashed lines 
is obtained using randomized leading particle direction and has been 
subtracted. Signal and background have a similar distribution. 
However, this has to be interpreted in the light of the results 
of the previous paragraph, where we saw that the signal has a harder 
$\ensuremath{p_{\mathrm{T}}}$-spectrum. Ordering of 
$\ensuremath{p_{\mathrm{T}}}$ in the jet fragmentation leads to limited 
$j_{\rm T}$ with mean value of about $250 \, \mbox{${\rm MeV}$}$. 
Higher $\ensuremath{p_{\mathrm{T}}}$ particles are, on the average, 
closer to the jet axis. It is expected that in-medium effects can 
significantly broaden the $j_{\rm T}$ distribution of the particles 
inside the jet. For the expected $S/B > 1$ such changes can be 
easily measured.

\begin{figure}
\begin{center}
\includegraphics[width=8cm]{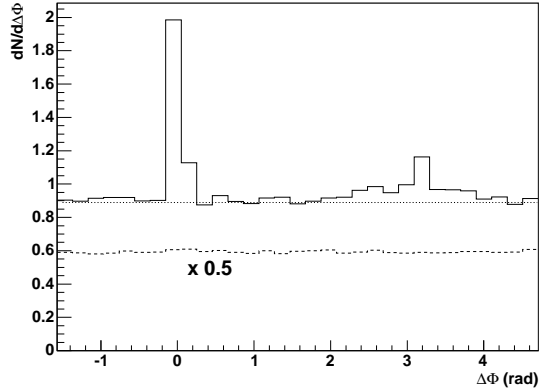}
\end{center}
\caption{
$\Delta\phi$ distribution for particles with 
\ensuremath{p_{\mathrm{T}}}\mbox{$>$}5~GeV/c with respect to 
the leading particle direction with 
\ensuremath{p_{\mathrm{T}}}$^{\rm leading}$\mbox{$>$}10~GeV/c (solid line).
The dashed lines represent the uncorrelated background and has been obtained 
by randomizing the leading particle direction. The dotted line indicates the uncorrelated 
background level below the signal and has been obtained from fit to the 
region between the peaks.
} 
\label{fig:lowet_5}
\end{figure}

\paragraph{Forward-backward correlations}
The backward jets observed through forward-backward $\phi$-correlations 
are more difficult. This is due to the limited $\eta$-acceptance of the 
experiment which reduces the number of two-jet events within the acceptance, 
especially for low-$\ensuremath{p_{\mathrm{T}}}$ jets. Higher statistics 
or harder cuts are needed than for the $R$-correlation to obtain significant 
signals.  
In Fig. ~\ref{fig:lowet_5} we show the $\Delta \phi$ distribution for 
particles with $\ensuremath{p_{\mathrm{T}}} > 5 \, \mbox{${\rm GeV}$}$ 
with respect to the leading particles direction with 
$\ensuremath{p_{\mathrm{T}}}^{\rm leading} > 10 \, \mbox{${\rm GeV}$}$. 
Already for $10^4$ HIJING events a significant backward peak is observed 
indicating that such an analysis is possible at the LHC.

\paragraph{Spatial spectrum analysis}
Jets reveal their presence through repeating length scales. It is well 
known from signal processing that hidden periodicities can be detected 
using Fourier analysis techniques e.g. the spectral power density 
{\it (SPD)} or its back-transformation the autocorrelation function 
{\it(ACF)}. 
For jet analysis the discretized energy density distribution in the 
$\eta - \phi$ plane is used as an input. The transformations are 
performed using a two-dimensional discrete fast Fourier transformation 
algorithm ({\it DFFT}). In heavy ion collisions the length scales not 
only repeat inside the same event but also from event to event. In order 
to profit from this fact, the spectral power densities are averaged 
over many events before the back-transformation is performed to obtain 
the event averaged auto-correlation function.

\subsubsection{Reconstructed Jets}

\paragraph{Charged Jets}

Leading particle analysis provides only a very poor energy measurement. 
It is known from pp data that leading charged particles carry on the 
average 24\% of the charged jet energy. 
The distribution of this energy fraction 
has a {\it rms} of $\approx 50\% $ limiting the jet energy measurement to 
approximately the same value. Nevertheless, there exists some limited 
selection power for jet energy classes, since the jet energy can not be 
lower than the leading particle energy. Due to the steeply falling jet 
production spectrum a leading particle cut of for example 
$\ensuremath{p_{\mathrm{T}}} > 5 \, \mbox{${\rm GeV}$}$ will select 
low-energy jets mainly in the range  $5 < \ensuremath{E_{\mathrm{T}}} < 
25 \, \mbox{${\rm GeV}$} $. 
In the low-energy region underlying event fluctuation are of the order 
of the reconstructible jet energy, i.e. the energy of reconstructed 
charged particles from the jet in a fixed cone after 
\ensuremath{p_{\mathrm{T}}}-cuts. Hence, true jet reconstruction is 
only possible in the high energy region, $\ensuremath{E_{\mathrm{T}}} > 
50 \, \mbox{${\rm GeV}$}$ being our current estimate.

Since the energy resolution will be limited due to the incomplete 
reconstruction, a simple reconstruction algorithm is sufficient. 
The cone-algorithm similar to the one developed by the UA1 collaboration \cite{Arnison:1983gw} 
has been used to evaluate the jet energy resolution in Pb--Pb collisions 
using charged particles only. In this algorithm, charged particles above 
a threshold 
$\ensuremath{p_{\mathrm{T}}} > \ensuremath{p_{\mathrm{T}}} ^ {\rm seed}$ 
are considered as seeds. A jet is defined by all charged particles within 
a cone $R < R_{\rm max}$ with respect to the jet axis, which for the first 
iteration is given by the leading particle direction. In the next iteration 
a new jet direction is obtained from the sum of the momentum vectors of these jet 
particles. The procedure is repeated until convergence is reached, i.e. the 
difference between new and old direction falls below a minimum value.
The energy from the underlying event is determined from the particle 
outside the jet cone and is subtracted from the cone energy.
In central Pb--Pb collisions, for a jet \ensuremath{E_{\mathrm{T}}} of  
50~GeV the energy resolution is similar to that obtained from the leading 
particle analysis ($\Delta \ensuremath{E_{\mathrm{T}}} /
\ensuremath{E_{\mathrm{T}}} \approx 50\%$). 
For jet energies above 100~GeV the resolution improves to 40 \%. 
The resolution is mainly limited by the fluctuation in the
part of the jet energy carried by charged particles (30\%) and the small 
cone-size ($R < 0.3$) that is needed to reduce the energy from the 
underlying event. The latter reduces the reconstructed jet energy and 
thus increases the ratio $\Delta E /E$.

\paragraph{Energy resolution with EMCAL}

The proposed EMCAL adds additional energy information from neutral 
particles. A linear combination of the energies of charged particles and 
calorimeter cells that minimizes the energy dispersion has been
found from MC studies and is used to reconstruct the jet energy. In this 
case jets with $\ensuremath{E_{\mathrm{T}}} > 50 \, \mbox{${\rm GeV}$}$ 
can be reconstructed. The expected resolution is 34\% for  
$\ensuremath{E_{\mathrm{T}}} = 50 \, \mbox{${\rm GeV}$} $ and 29\% for   
$\ensuremath{E_{\mathrm{T}}} = 100 \, \mbox{${\rm GeV}$} $ 
(preliminary results).

\paragraph{Fragmentation function with reconstructed jets}
The fragmentation function represents the average distribution of jet 
energy among the particles produced in the fragmentation of the initial 
parton. The effect of additional gluon radiation when the particle is 
propagating through a deconfined partonic medium will change the 
fragmentation function in two ways: the relatively low-energetic radiated 
gluons will increase the number of particle carrying a small fraction $z$ of 
the jet energy. On the contrary, the high-$z$ part will be depleted. 
It is important to observe both effects experimentally. The challenge 
is to obtain a reasonable $S/B$  in the low-$z$ region and enough 
statistics in the high-$z$ region.

\begin{figure}
\begin{center}
\includegraphics[width=8.0cm]{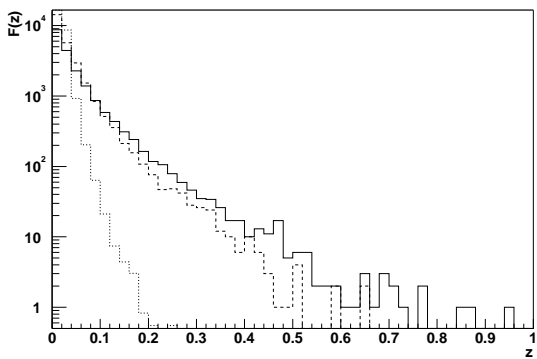}
\end{center}
\caption{
Fragmentation function for jets that have a
reconstructed jet energy in the range $95 < E_{\rm T} < 105 \, {\rm GeV}$. 
The solid line is without jet quenching. The dashed line is for a quenching 
scenario which has been simulated using the superposition of two jets with 
$E_{\rm T} = 80 \, {\rm GeV}$ and $E_{\rm T} = 20 \, {\rm GeV}$, respectively.
The dotted line is the $z$-distribution for charged particles from the 
underlying event.} 
\label{fig:lowet_7}
\end{figure}

As an example we show in Fig.~\ref{fig:lowet_7} the fragmentation function
for jets that have a reconstructed jet energy in the range 
$95 < \ensuremath{E_{\mathrm{T}}} < 105 \, \mbox{${\rm GeV}$}$. The solid 
line is without jet quenching. The dotted line is for a quenching scenario 
which has been simulated using the superposition of two jets with 
$\ensuremath{E_{\mathrm{T}}} = 80 \, \mbox{${\rm GeV}$}$ and 
$\ensuremath{E_{\mathrm{T}}} = 20 \, \mbox{${\rm GeV}$}$, respectively.
The dashed line is the underlying event.

\paragraph{Fragmentation function with identified leading particles}
Reconstructed D-mesons and high-\ensuremath{p_{\mathrm{T}}} electrons 
from semi-leptonic heavy-quark decays with large impact parameter are 
leading particles per se. 
They tag the production and fragmentation of charm and beauty quarks, 
respectively. In the past, experiments like UA1 without secondary vertex 
detection capability for charm and beauty hadrons have used non-isolation 
cuts as an additional criterion to select these heavy particles,
where non-isolation is defined as the energy in a cone around the particle.

In ALICE, the secondary vertex reconstruction capability using the ITS 
allows to select these particles using impact parameter cuts. $D^0$ mesons 
decaying into $\pi K$ can be reconstructed in the transverse momentum range
$0 < \ensuremath{p_{\mathrm{T}}} < 15 \, \mbox{${\rm GeV}$}$. Electrons from 
b-hadron decays cover the transverse momentum range 
$ \ensuremath{p_{\mathrm{T}}} > 2 \, \mbox{${\rm GeV}$}$. Additional 
particles close to the leading particle come from the fragmentation 
and decay of the heavy quark. The $j_{\rm T}$ and $p_{\rm T}$ distribution 
of these particles can be studied in pp, pPb and Pb--Pb collisions. 
Again, differences between these systems are the result of interactions 
between partons and the medium. They are expected to be weaker for heavy 
quarks than for light quarks, which makes this measurement especially 
interesting. 

Moreover, for identified charmed mesons the inclusive 
$\ensuremath{p_{\mathrm{T}}}$-spectra can be studied in the same way as
for example charged light hadrons. Comparing 
$R_{\rm AA}(\ensuremath{p_{\mathrm{T}}})$ of light and heavy mesons will show
whether energy loss is similar for both particle types. Theory expects a 
smaller energy loss for heavy quarks (dead-cone effect).

\paragraph{Photon-tagged jets}
Photons produced at the earliest stage of the collisions,
preserve almost all their energy after traversing through the dense medium. 
Hence, the attenuation of the jet energy can be
measured via comparison of the prompt photon and jet kinematics.
Prompt photons accompanied by a jet produced in the opposite direction
at high $p_{\rm T}$ can be studied as a probe of the dense medium formed in
heavy-ion collisions. ALICE will be able to detect and identify
prompt photons using the PHOS detector, while hadrons from jets will be
detected by the TPC and, optionally, by the EMCAL.

\subsection{Jet Physics with the CMS Detector}
{\em O.L.~Kodolova, I.P.~Lokhtin, S.V.~Petrushanko, C.~Roland, L.I.~Sarycheva, 
S.V.~Shmatov, I.N.~Vardanian} 

\subsubsection{CMS Detector}

The Compact Muon Solenoid (CMS) is a general purpose detector 
designed primarily to 
search for the Higgs boson in proton-proton collisions at LHC~\cite{CMS:1994}. 
The detector is optimized for 
accurate measurements of the characteristics of high-energy leptons and 
photons as well as hadronic jets in a large acceptance, providing unique
capabilities for ``hard probes'' in both $pp$ and $AA$ collisions. In 
particular, jet quenching studies of hard jets, high-mass 
dimuons and high-$p_T$ 
hadrons are primary goals of the CMS Heavy Ion Programme~\cite{Baur:2000}.  

\vskip 0.7 cm 

\begin{figure}[htb]
\begin{center} 
\resizebox{120mm}{120mm} 
{\includegraphics{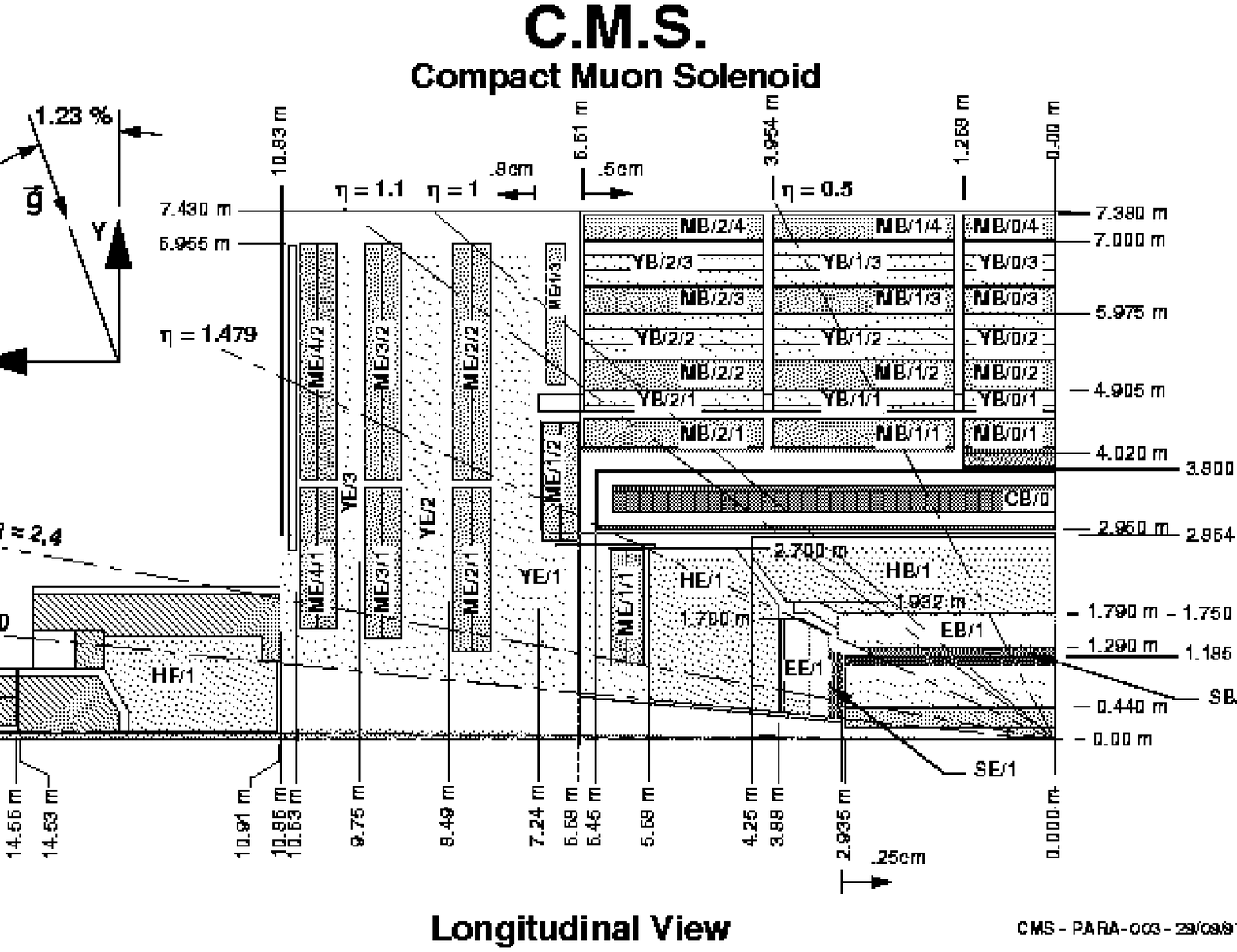}} 
\caption{\small  The longitudinal view of CMS detector.}
\label{cmsjet:fig1}
\end{center} 
\end{figure}

A detailed description of the
detector elements can be found in the corresponding 
Technical Design Reports~\cite{CMSHCAL,CMSMUON,CMSECAL,CMSTRACK}. 
The longitudinal view of the CMS detector is presented in 
Fig.~\ref{cmsjet:fig1}. The central element of CMS is the magnet, 
a $13$ m long solenoid with an internal radius $\approx 3$ m, 
which will provide 
a strong $4~T$ uniform magnetic field. The $4\pi$ detector consists of a $6$ m 
long and $1.3$ m radius central tracker, electromagnetic (ECAL) and hadronic
(HCAL) calorimeters inside the magnet and muon stations outside. 
The tracker and muon chambers cover the pseudorapidity region 
$|\eta|<2.4$, while the 
ECAL and HCAL calorimeters reach $|\eta|=3$. A pair of quartz-fiber 
very forward (HF) 
calorimeters, located $\pm 11$ m from the interaction point, cover the region 
$3<|\eta|<5$ and complement the energy measurement. The tracker is 
composed of pixel 
layers and silicon strip counters. The CMS muon stations consist of 
drift tubes in the barrel
region (MB), cathode strip chambers in the end-cap regions (ME), 
and resistive plate 
chambers in both barrel and endcap dedicated to triggering. 
The electromagnetic 
calorimeter is made of almost 76000 scintillating PbWO$_4$ crystals.  The 
hadronic calorimeter consists of scintillator sandwiched between brass 
absorber plates. 
The main characteristics of the calorimeters, such as energy resolution 
and granularity are 
presented in Table 5. 

\begin{table}[htb]
\begin{center}
{\small Table 5: Energy resolution, $\sigma /E$, and granularity of the 
CMS calorimeters in the barrel (HB, EB), endcap (HE, EE) and very forward 
(HF) regions. The energy resolution is shown for the total energy of 
electrons and photons (ECAL) and transverse energy of hadronic jets 
(HCAL, HF).} 
\label{cmsjet:tab1}

\medskip 

\begin{tabular}{|l|c|c|c|c|c|} \hline  
Rapidity  & \multicolumn{2} {c|} {$0<\mid\eta\mid<1.5$} & 
\multicolumn{2} {c|} 
 {$1.5<\mid\eta\mid<3.0$} 
  & {$3.0<\mid\eta\mid<5.0$} \\ 
 coverage & \multicolumn{2} {c|} {} & 
  \multicolumn{2} {c|} {} &   \\ 
 \hline & & & & & \\
Subdetector & HCAL (HB) & ECAL (EB) & HCAL (HE) & ECAL (EE) & HF \\ 
& & & & & \\ \hline 
$\frac{\sigma}{E}= \frac{a}{\sqrt{E}} \bigoplus b$ & & & & & \\
$a$ & 1.16 & 0.027 & 0.91 & 0.057 & 0.77 \\
$b$ & 0.05 & 0.0055 & 0.05 & 0.0055 & 0.05 \\
\hline
granularity &  &  &  &  &
\\
$\Delta\eta \times \Delta\varphi$ & $0.087 \times 0.087$ & $0.0174 \times
0.0174$ & $0.087 \times 0.087$ & $0.0174 \times 0.0174$ 
& $0.175 \times 0.175$ \\
& & & & to $0.05 \times 0.05$ & \\ \hline 
\end{tabular}
\end{center}
\end{table}

\subsubsection{Dijet, $\gamma + $jet and $Z + $jet Production at CMS}

The following signals of jet quenching by medium-induced 
parton energy loss have 
been identified as being measurable in heavy ion 
collisions with CMS~\cite{Baur:2000}. 

Jet pairs are produced in the initial scattering processes in 
$pp$ and $AA$ collisions 
through reactions such as   
$$gg \rightarrow gg~,~~~~qg \rightarrow qg~,~~~~qq\rightarrow qq~,~~~~~gg
\rightarrow q \overline q~,$$ 
where the $gg \rightarrow gg$ process is dominant. High $p_T$ 
jet pairs produced in 
nucleus-nucleus collisions can be suppressed relative to their 
production in independent 
nucleon-nucleon interactions~\cite{Gyulassy:1990ye}. 
The dijet rates depend on impact parameter~\cite{Lokhtin:2000wm} and 
may exhibit azimuthal 
anisotropy in non-central collisions~\cite{Lokhtin:2001kb}. 

Single jets may be produced opposite a gauge boson in 
$\gamma +$jet~\cite{Wang:1996yh,Wang:1996pe} 
and $Z +$jet~\cite{Kartvelishvili:1996} final states, 
dominantly through processes such as 
$$q g \rightarrow q \gamma~, ~~~~~~q g \rightarrow q Z \, \, .$$ 
In heavy ion collisions, the relative $p_T$ between the jet and the boson
becomes imbalanced due to interactions of the jet within the medium.  The
$Z$ is detected in the dimuon channel, $Z \rightarrow \mu^+ \mu^-$. 

In this section, we will consider mainly production and measurement of 
hard jets, the two points described above. Although not discussed in detail
here, CMS can also measure parton energy loss in two other channels, leading 
particles and heavy quarks. Leading particles in a jet may have their 
momentum suppressed due to medium modifications of the jet fragmentation 
functions~\cite{Wang:xy}.  The capability of the
CMS tracker to measure the momenta of charged particles in heavy ion 
collisions is discussed later. Heavy quark energy loss, particularly $b$ 
quark loss, can be 
measured in two channels.  Semileptonic $B$ and $D$ decays contribute to 
the high mass dimuon
spectra and hadronic $B$ decays to $J/\psi$ are a substantial contribution to
secondary charmonium production~\cite{Lin:1998bd,Lokhtin:2001nh}.

By considering ``jet energy loss'' here, we concentrate on the energy that 
falls outside the jet cone and is truly lost to the jet, see 
Refs.~\cite{Lokhtin:1998ya,Baier:1998yf}. In fact, since coherent 
radiation induces 
a strong dependence of the radiative energy loss of a jet on the angular cone 
size, it will soften particle energy distributions inside the jet, 
increase the multiplicity of secondary particles, and, to a lesser degree, 
affect the total jet energy. On the other hand, collisional energy loss 
turns out to be practically independent of jet cone size and causes the 
loss of total jet energy. 
Moreover, the total energy loss of a jet will be sensitive to the experimental 
capabilities for low-p$_T$ particles, products of soft gluon 
fragmentation. In CMS, most of these low-$p_T$ hadrons may be cleared out of
the central calorimeters by the strong magnetic field.  

Table 6 
presents the event rates for various channels, including hard
jets, in a one month Pb$-$Pb run (assuming two weeks of data taking), 
$T= 1.2 \times 10^6$ s, with luminosity 
$L = 5 \times 10^{26}~$cm$^{-2}$s$^{-1}$ so that 
$$N_{\rm ev}= T \sigma^{h}_{AA} L \, \, .$$ The production cross sections in
minimum bias nucleus-nucleus collisions were obtained from those in 
$pp$ interactions 
at the same energy, $\sqrt{s} = 5.5$ TeV, using the simple parameterization 
$\sigma^h_{AA}= \sigma^h_{pp}A^2$. The $pp$ cross sections were evaluated
using PYTHIA $6.1$~\cite{Sjostrand:1993yb} with the CTEQ5L parton 
distributions. The jet production cross sections in the CMS
acceptance will be large enough to carefully study the dijet rate 
as a function of impact parameter as well as the azimuthal angle and 
rapidity distributions of jet pairs. 
The estimated statistics for $\gamma +$jet production are 
satisfactory for studying 
the $E_T$-imbalance of the process  but the large background from 
jet$+$jet$(\rightarrow \pi^0)$ is still under investigation. The 
corresponding statistics for $Z (\rightarrow \mu^+ \mu^-) +$jet 
are rather low, but the background is less than $10\%$ in this case.  

\begin{table}[htb]
\begin{center}
{\small Table 6: Expected rates for jet production channels 
in a one month Pb$-$Pb run.} 
\label{cmsjet:tab2}

\medskip 

\begin{tabular}{|l|c|c|} \hline  
Channel & Barrel & Barrel+Endcap  \\ \hline 
jet$+$jet, $E_T^{\rm jet}>100$ GeV & 2.1$\times$10$^6$ & 4.3$\times$10$^6$ 
 \\ \hline   
$\gamma +$jet, $E_T^{{\rm jet},\gamma}>100$ GeV & 1.6$\times$10$^3$ & 
3.0$\times$10$^3$ 
\\ \hline 
$Z (\rightarrow \mu^+ \mu^-) +$jet, $E_T^{\rm jet},p_T^{Z}>100$ GeV 
& 30 & 45 \\  
$Z (\rightarrow \mu^+ \mu^-) +$jet, $E_T^{\rm jet},p_T^{Z}>50$ GeV 
& 180 & 300 \\ \hline 
\end{tabular}
\end{center}
\end{table}

Of course, there are some theoretical uncertainties in the absolute 
jet rates in $pp$ collisions due to the choice of parton distribution 
functions, the importance of next-to-leading order of $\alpha_s$ corrections, 
etc. Thus jet measurements in $pp$ or $dd$ 
collisions at the same or similar energies per nucleon as in the heavy ion 
runs are strongly desirable to determine the baseline rate precisely. One
complementary way to reduce uncertainties in the analysis of jet quenching 
is the introduction of a reference process unaffected by medium-induced 
energy loss and with a production cross section 
proportional to the number of nucleon-nucleon collisions, such as 
$Z(\rightarrow \mu^+\mu^-)$ production~\cite{Baur:2000,Kartvelishvili:1996}.   

\subsubsection{Jet Reconstruction} 

The main difficulty of jet recognition in heavy ion collisions arises from the 
``false'' jet background -- transverse energy fluctuations coming from the 
high multiplicity of ``thermal'' secondary particles in the 
event~\cite{Kruglov:1997}. Predictions give between $3000$ to $8000$ 
charged particles per unit rapidity in central Pb$-$Pb collisions at the LHC. 
In these circumstances, reconstruction of ``true'' QCD jets resulting from 
hard parton-parton scatterings is very important. 
The definition of an object like a jet is quite non-trivial even in $pp$
collisions. In particular, the energy and spatial resolution of hard jets are
sensitive to the parameters of the jet finding algorithm, see Ref.~\cite{UA1}. 

\begin{figure}[hbtp] 
\begin{center} 
\resizebox{75mm}{75mm} 
{\includegraphics{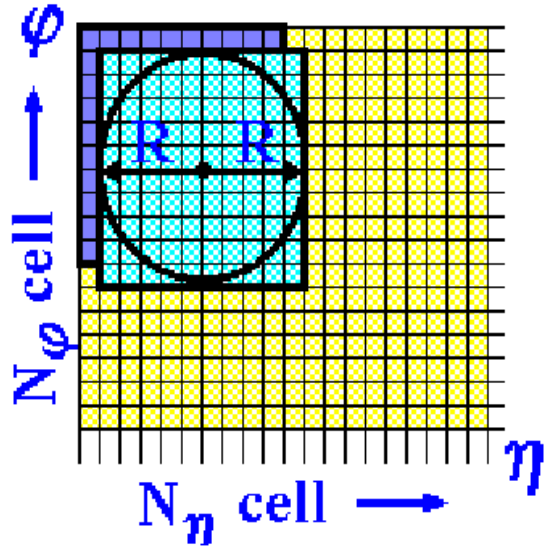}} 
\vskip - 2 mm 
\caption{\small Sliding window-type jet finding algorithm.} 
\label{cmsjet:win}
\end{center} 
\end{figure} 

In the CMS heavy ion physics programme, the modified sliding window-type jet 
finding algorithm has been developed to search for ``jet-like'' clusters 
above the average energy and to subtract the background from the underlying 
event~\cite{Baur:2000,Kodolova:2002}, see Fig.~\ref{cmsjet:win} for an 
illustration. The algorithm is described step-by-step here.

\noindent 
$\bullet$ As a function of pseudorapidity $\eta$, one calculates for each 
event the average transverse energy, $\overline{E_T^{\rm cell} (\eta)}$, 
and the dispersion, $D_T^{\rm cell}(\eta) =$ $\sqrt{\overline{(E_T^{\rm 
cell}(\eta))^2} - \Big(\overline{E_T^{\rm cell}(\eta)}\Big)^2}$, 
in all the calorimeter cells. 
The superscript ``cell'' means averaging
over the calorimeter cells in the given event.\\ 
$\bullet$ All possible rectangular windows, with overlaps, in the calorimeter 
map of $\eta-\varphi$ space are constructed. Each window consists of an
integer number of calorimeter cells. The numbers of cells per window in 
$\eta$, $N_\eta^{\rm wind}$, and $\varphi$, $N_\varphi^{\rm wind}$, are 
calculated separately,
\begin{eqnarray}
N_\eta^{\rm wind} & = & R N_\eta^{\rm tot}/\eta_{\rm max}, \nonumber \\
N_\varphi^{\rm wind} & = & R N_\varphi^{\rm tot}/2\pi, \nonumber
\end{eqnarray}
where $N_\eta^{\rm tot}$ and $N_\varphi^{\rm tot}$ are the total number of 
calorimeter cells in $\eta$ and $\varphi$, $\eta_{\rm max}$ is the maximum 
pseudorapidity and $R$ is the jet cone radius, an external parameter of 
the algorithm. \\ 
$\bullet$ The window energy is calculated as a sum of the call energies 
exceeding background, which is $1 \sigma$, $D_T^{\rm cell}(\eta)$, 
above the averaged energy, $E_T^{\rm cell}$. If the transverse energy 
of the cell is negative after background subtraction, it is set to zero. \\
$\bullet$ The search for jets and the evaluation of their energies is 
started from the window with the maximum transverse energy. \\ 
$\bullet$ Non-overlapping windows with energy $E_T^{\rm wind} > 
E_T^{\rm cut}$ are considered to be jet candidates. \\ 
$\bullet$ The center of gravity of the window is considered as a 
center of the jet. \\
$\bullet$ For correction of the jet axis a cell with maximum transverse 
energy in cone is found and considered as a new geometrical center of 
this jet. Cells within radius $R$ around the new geometrical center are 
collected and center of gravity of jet is recalculated. \\ 
$\bullet$ Cells in a cone with radius $R$ around jet center are collected. \\ 
$\bullet$ The values of $\overline{E_T^{\rm cell}(\eta)}$ and 
$D_T^{\rm cell}(\eta)$ are recalculated using cells which were not 
included in the jets. \\ 
$\bullet$ The jet energy is then the difference between energies in 
collected cells, $E_T^{\rm cell}$, and the background energy per cell, 
$$E_T^{\rm jet}  = \sum _{n_c}\{E_T^{\rm cell}  - 
[\overline{E_T^{\rm cell}(\eta)} + D_T^{\rm cell}(\eta)]\}.$$

\begin{figure} 
\begin{minipage}[t]{78mm}
\resizebox{78mm}{78mm} 
{\includegraphics{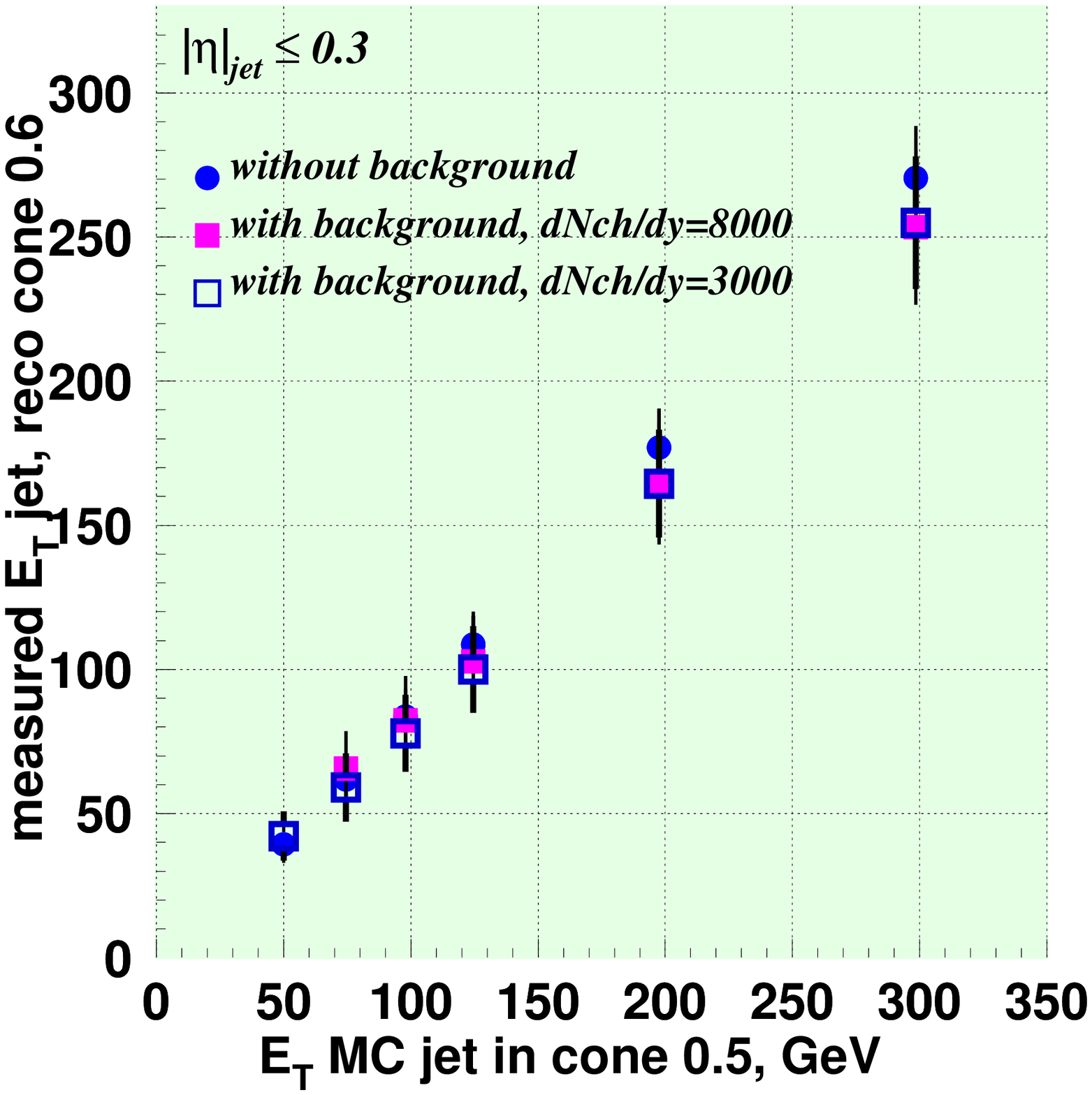}} 
\caption{\small Correlation between reconstructed and generated 
jet transverse energies in Pb$-$Pb and $pp$ events.}
\label{cmsjet:fig2}
\end{minipage}
\hspace{\fill}
\begin{minipage}[t]{78mm}
\resizebox{78mm}{78mm} 
{\includegraphics{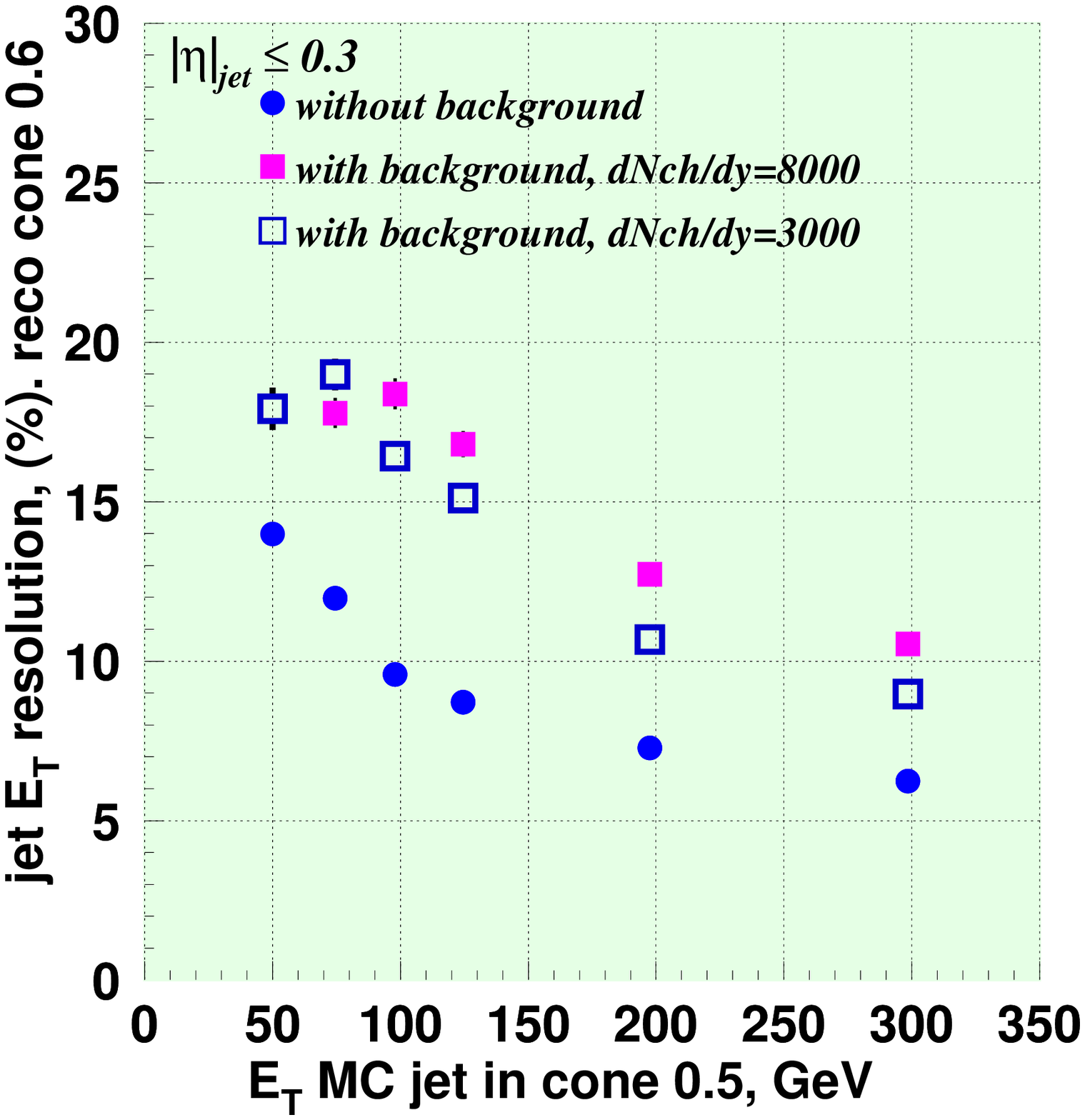}} 
\caption{\small Jet energy resolution in Pb$-$Pb and $pp$ events. (See text for
explanation).} 
\label{cmsjet:fig3}
\end{minipage}
\end{figure} 

\begin{figure} 
\begin{minipage}[t]{77mm}
\resizebox{77mm}{77mm} 
{\includegraphics{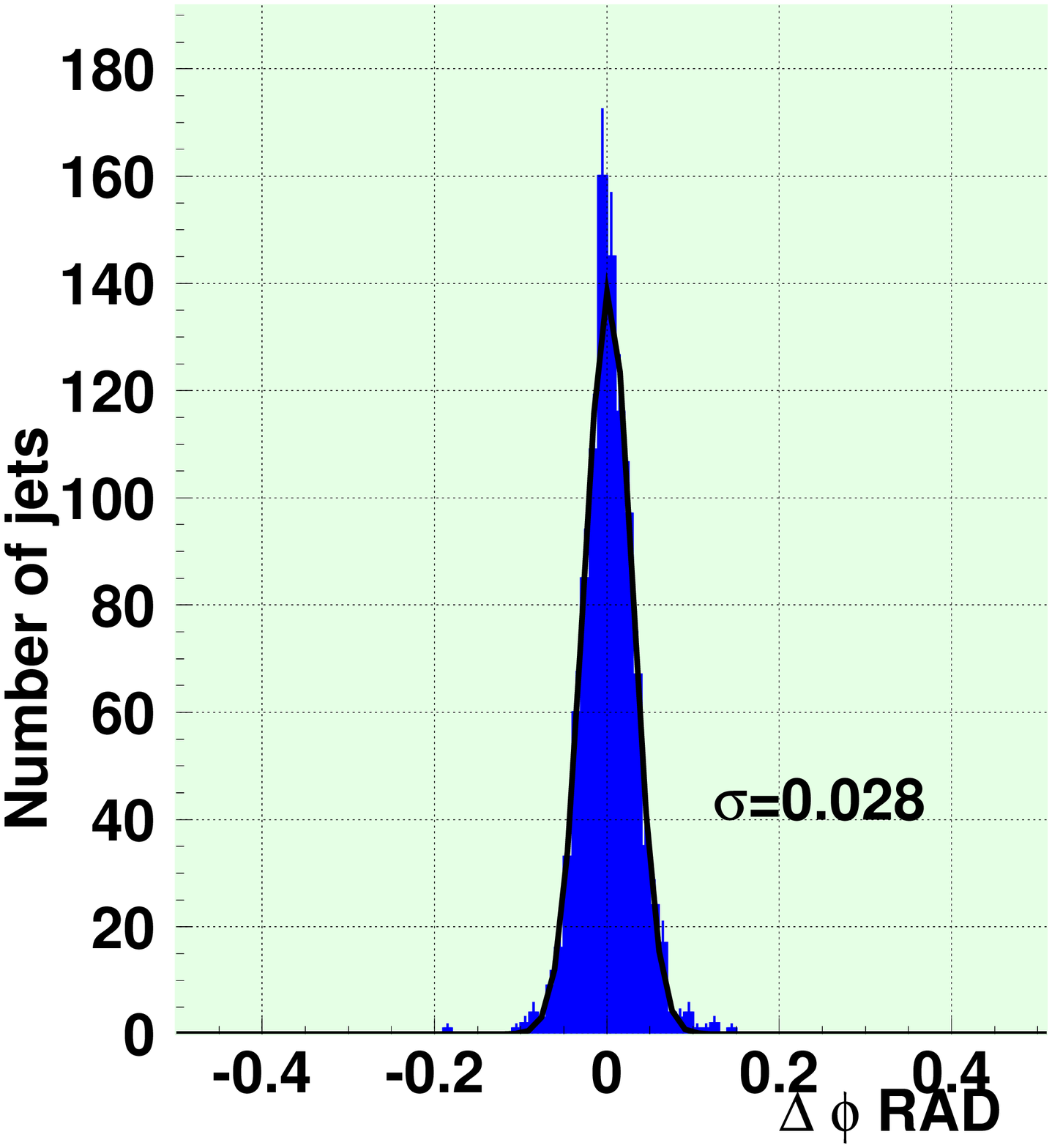}} 
\caption{\small Distribution of differences in azimuthal angle $\varphi$ 
between 
generated and reconstructed jets with $E_T=100$ GeV in $pp$ events.} 
\label{cmsjet:fig4}
\end{minipage}
\hspace{\fill}
\begin{minipage}[t]{77mm}
\resizebox{77mm}{77mm} 
{\includegraphics{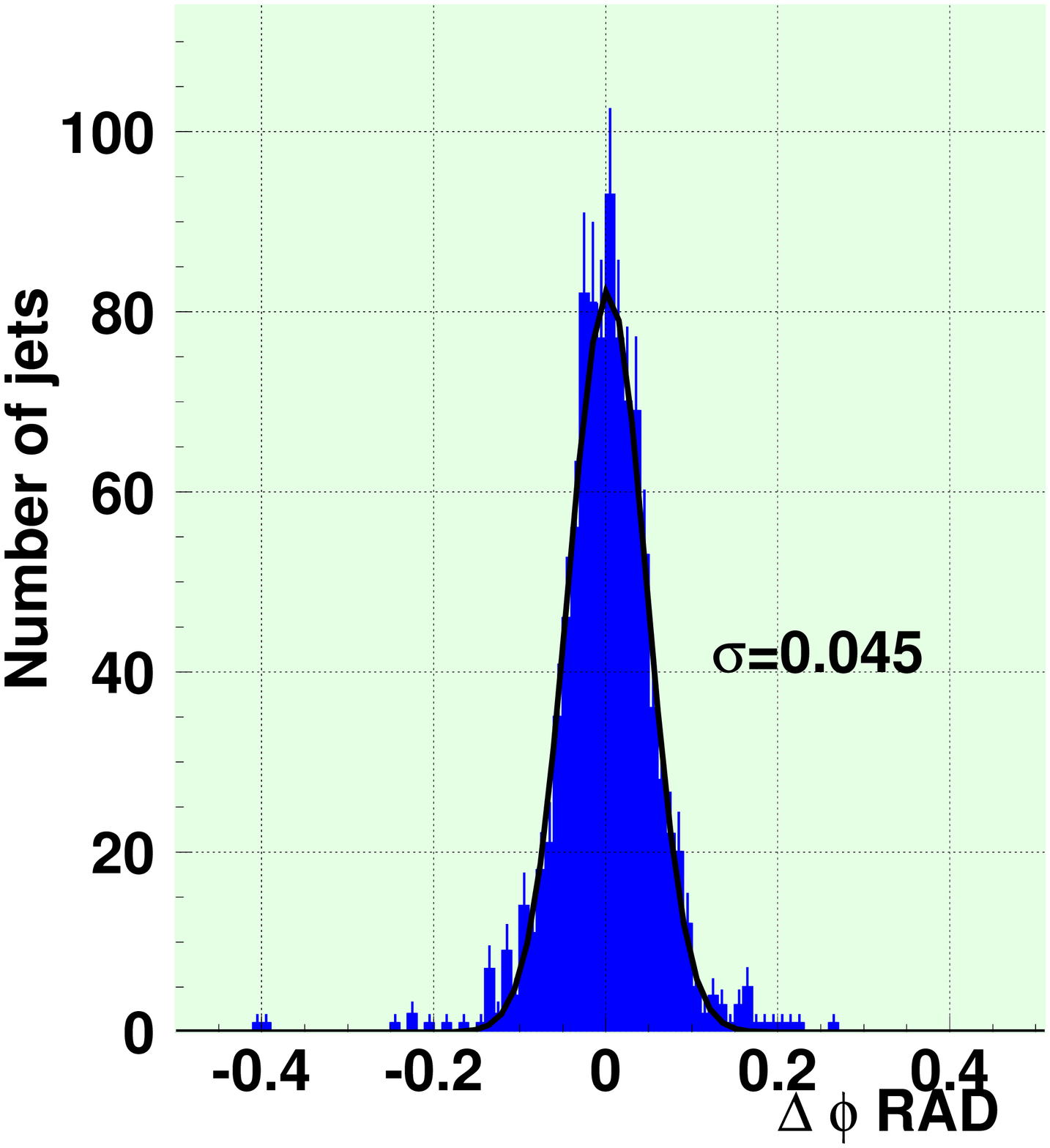}} 
\caption{\small The same as in Fig.~\ref{cmsjet:fig4} but for Pb$-$Pb events 
at $dN^{\pm}/dy (y=0) = 8000$.} 
\label{cmsjet:fig5}
\end{minipage}

\vskip 1cm 

\begin{minipage}[t]{77mm}
\resizebox{77mm}{77mm} 
{\includegraphics{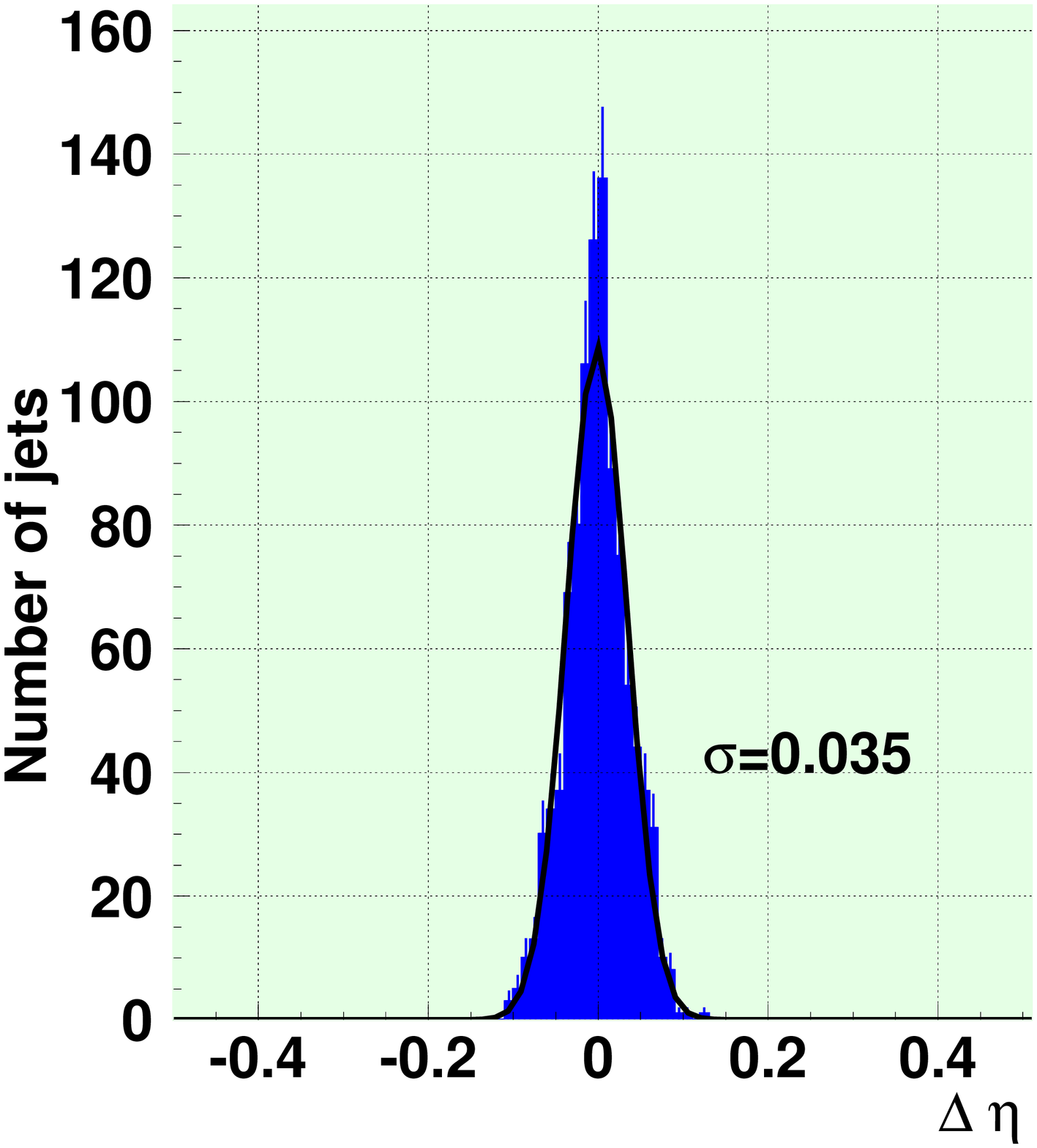}} 
\caption{\small Distribution of differences in pseudorapidity $\eta$ between 
generated and reconstructed jets with $E_T=100$ GeV in $pp$ events.} 
\label{cmsjet:fig4-1}
\end{minipage}
\hspace{\fill}
\begin{minipage}[t]{77mm}
\resizebox{77mm}{77mm} 
{\includegraphics{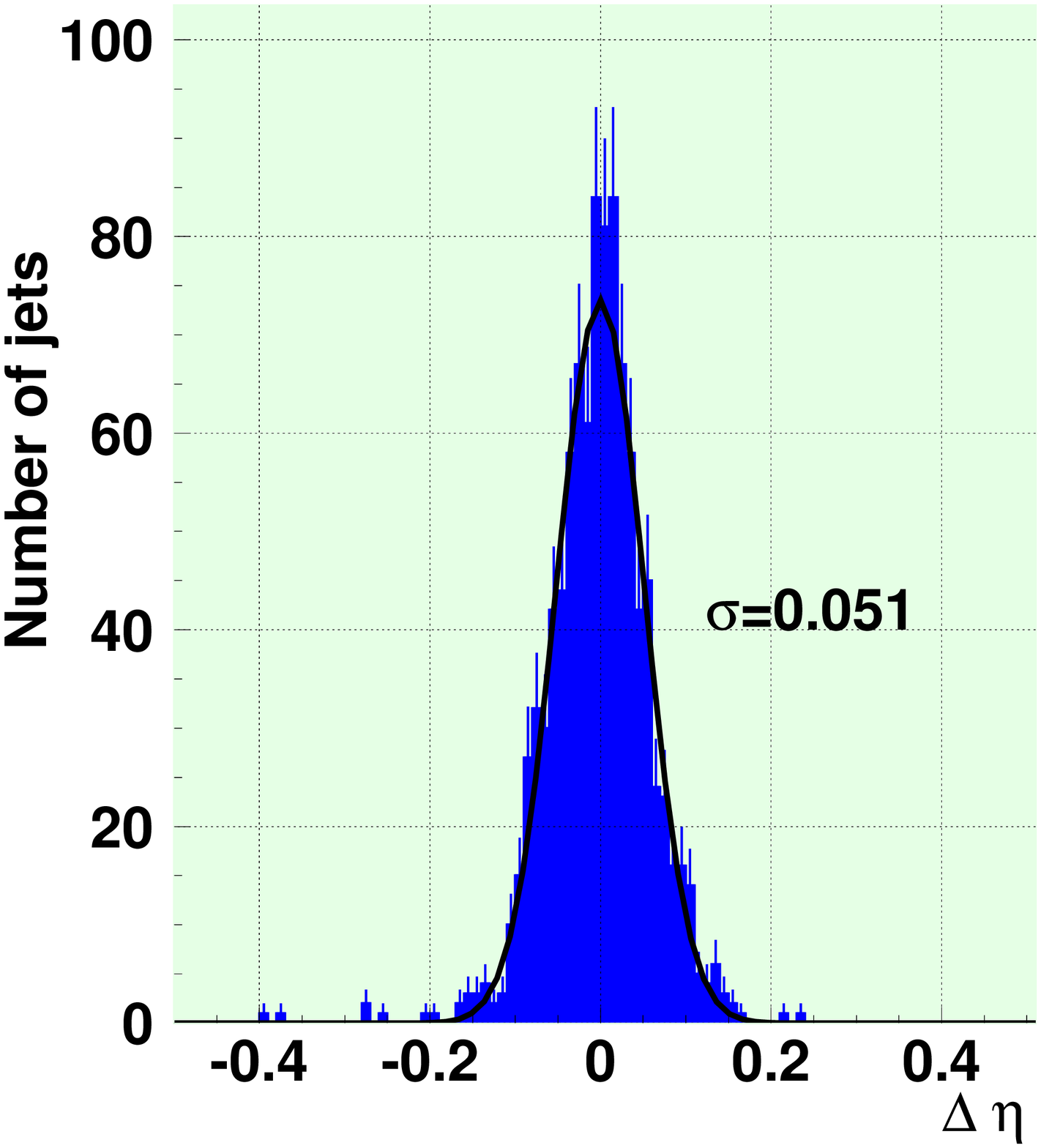}} 
\caption{\small The same as in Fig.~\ref{cmsjet:fig4-1} but for 
Pb$-$Pb events at $dN^{\pm}/dy (y=0) = 8000$.} 
\label{cmsjet:fig5-1}
\end{minipage}
\end{figure}

\begin{figure} 
\begin{minipage}[t]{78mm}
\resizebox{78mm}{78mm} 
{\includegraphics{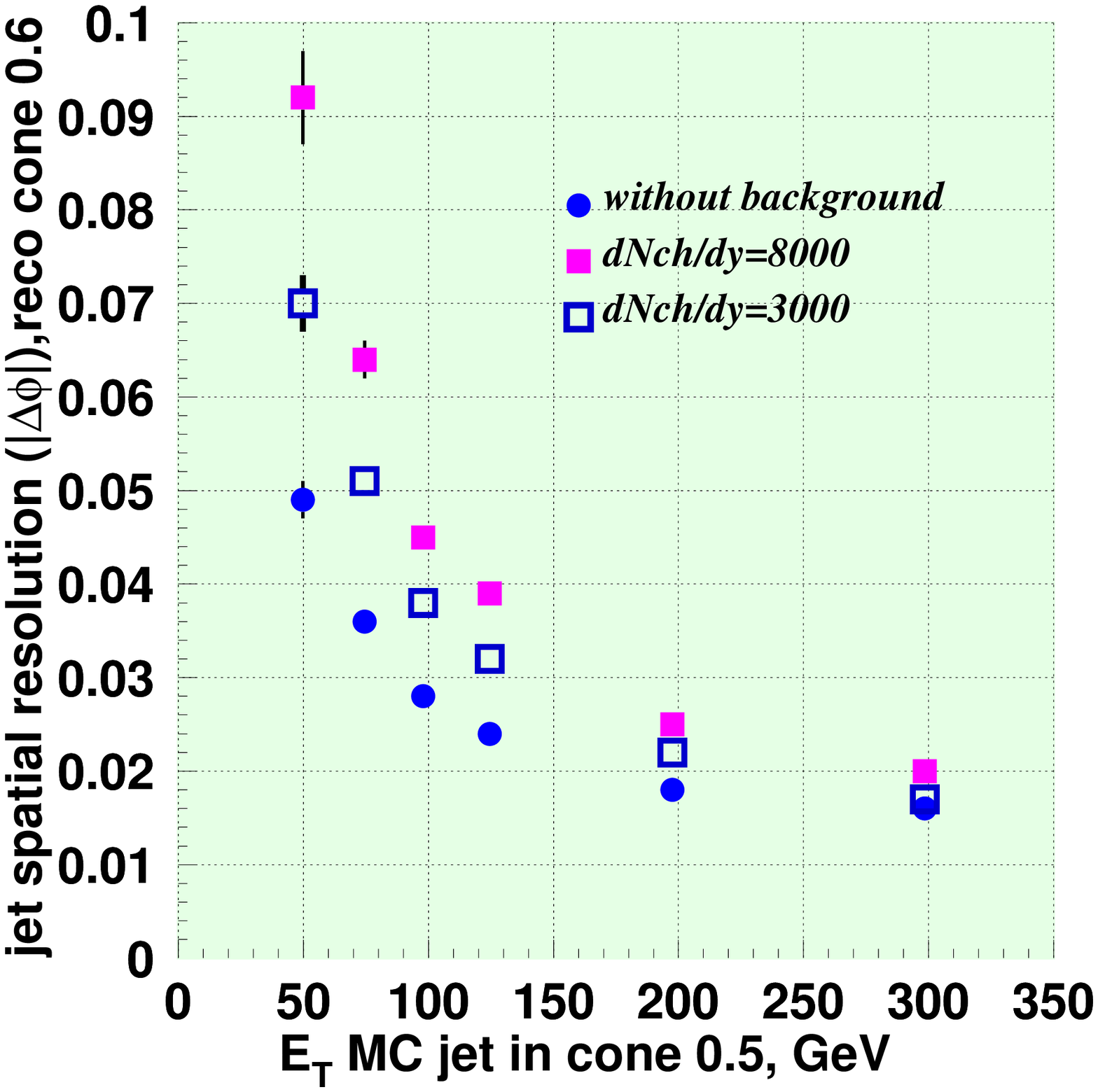}} 
\caption{\small Energy dependence of jet azimuthal angle resolution in Pb$-$Pb and 
$pp$ events.}
\label{cmsjet:phi-et}
\end{minipage}
\hspace{\fill}
\begin{minipage}[t]{78mm}
\resizebox{78mm}{78mm} 
{\includegraphics{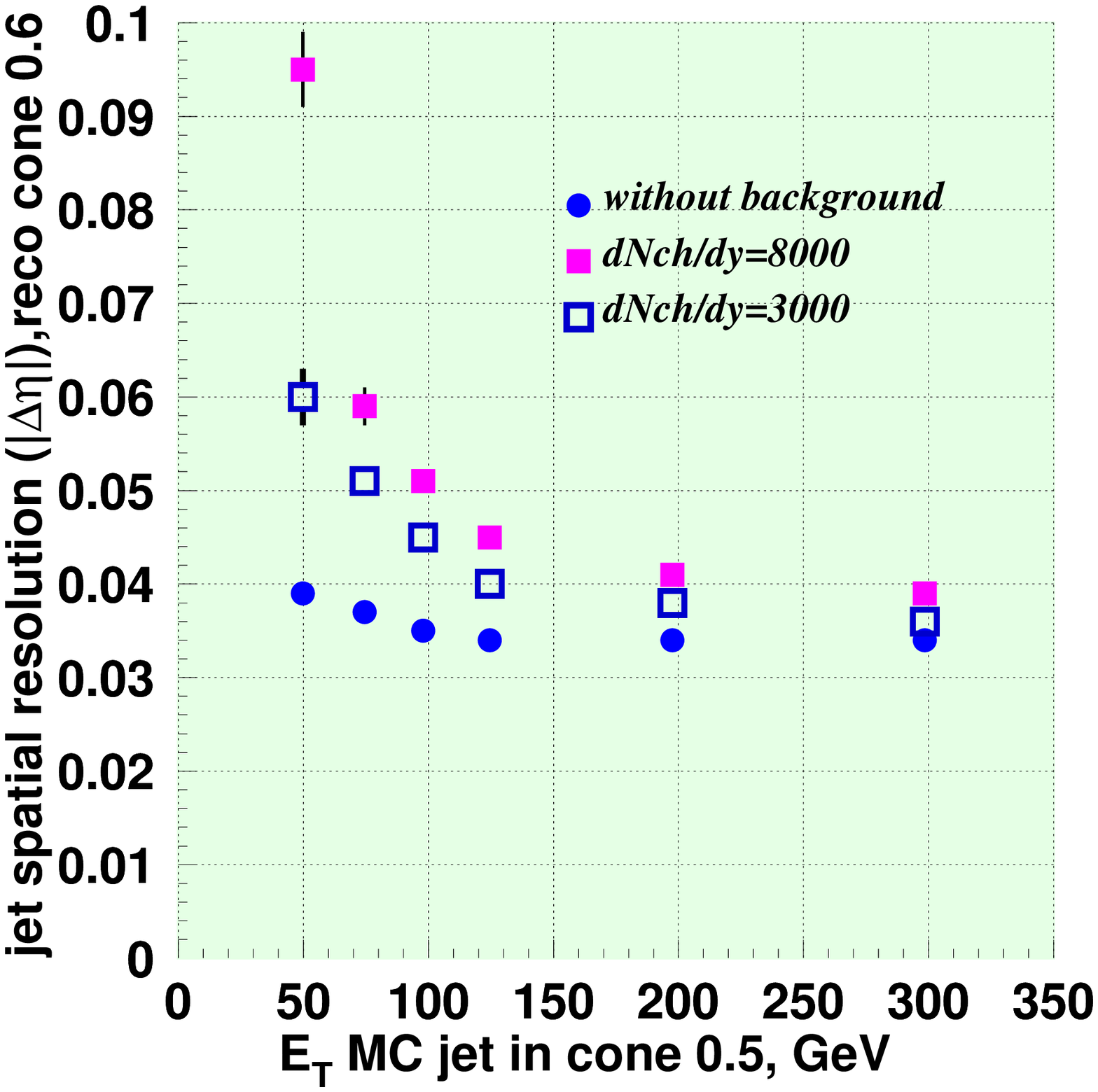}} 
\caption{\small Energy dependence of jet pseudorapidity resolution in Pb$-$Pb and $pp$ 
events.} 
\label{cmsjet:eta-et}
\end{minipage}
\end{figure} 

\begin{table}[htb]
\begin{center} 
{\small Table 7: Purity, noise 
(contamination levels, false jets / generated jets) 
and jet transverse energy resolution in central Pb$-$Pb collisions with 
$dN^{\pm}/dy (y=0) = 3000$ and $8000$.} 
\label{cmsjet:tab3}

\medskip 

\begin{tabular}{|l||c|c||c|c||c|c|} \hline  
$E_T~{\rm min}$ & \multicolumn{2} {c||} {Purity} & \multicolumn{2} {c||} 
{Noise} & \multicolumn{2} {c|} {$\sigma(E_T)/E_T(\%)$} \\ \hline 
(GeV) & $3000$ & $8000$ & $3000$ & $8000$ & $3000$ & $8000$ \\ 
\hline \hline
75 & 0.96 $\pm$ 0.03 & 0.88 $\pm$ 0.03 & 0.021 $\pm$ 0.006 & 0.083 $\pm$ 0.009 
& 19.0 & 17.8 \\ \hline 
100 & 0.99 $\pm$ 0.03 & 0.97 $\pm$ 0.03 & 0.002 $\pm$ 0.001 & 0.011 $\pm$ 0.003 
& 16.4 & 18.4 \\ \hline 
125 & 1.00 $\pm$ 0.03 & 0.99 $\pm$ 0.03 & 0.000 $\pm$ 0.000 & 0.004 $\pm$ 0.002 
& 15.1 & 16.8 \\ \hline 
200 & 1.00 $\pm$ 0.03 & 0.99 $\pm$ 0.03 & 0.000 $\pm$ 0.000 &  0.001 $\pm$ 0.001
& 10.7 & 12.7 \\ \hline 
\end{tabular}
\end{center}
\end{table}

Jet reconstruction was studied in the barrel calorimeters, $|\eta| < 1.5$, with
the GEANT-based package CMSIM$\_$123 (CMS simulation package, version 123) 
adapted to heavy ion collisions. The initial jet distributions in the
nucleon-nucleon sub-collisions at $\sqrt{s} = 5.5$ TeV were generated using 
PYTHIA 6.1~\cite{Sjostrand:1993yb}, as described before. This dijet event 
is then
superimposed on the Pb-Pb event, obtained using a hydrodynamical model of the
hadron spectrum~\cite{Kruglov:1997} as a superposition of thermal hadron 
distributions and collective flow.  The average hadron transverse momentum, 
$\langle p_T^h \rangle = 0.55$ GeV/$c$, was fixed in the model.  The analysis 
was done for two estimates of the charged particle multiplicity at $y=0$ in 
central collisions, $dN^\pm /dy(y=0) = 3000$ and $8000$. 
The calorimeter occupancy is high enough in these cases: mean energy 
per tower is about $4.4$ ($1.7$) GeV for $dN^\pm /dy(y=0) = 8000$ ($3000$) in
the barrel part, and it increases by a factor $\sim 2$ in endcaps. 
Fig.~\ref{cmsjet:fig2} shows the correlation between reconstructed and 
generated transverse energies of jets in Pb$-$Pb and $pp$ events.  The 
generated jet cone has a radius $R=0.5$ while the radius of the
reconstructed cone is larger, $R=0.6$. Since the average measured jet 
energy in Pb$-$Pb collisions is the same as in $pp$, the $pp$ interactions 
are a baseline for jet physics in heavy ion collisions. However, the jet 
transverse energy resolution is degraded by a factor 
$\sim 2$ in the high multiplicity central Pb$-$Pb collisions compared to  
$pp$ interactions, as shown in Fig.~\ref{cmsjet:fig3}. The jet energy 
resolution is defined here as $\sigma(E_T^{\rm reco}/E_T^{\rm gen})/<E_T^{\rm 
reco}/E_T^{\rm gen}>$, where $E_T^{\rm reco}$ is the reconstructed transverse 
energy, and $E_T^{\rm gen}$ is the transverse energy of all generated 
particles inside the given cone radius $R$. The resolution of $75$  
GeV jets with $dN^{\pm}/dy (y=0) = 8000$  is smaller than for 
$100$  GeV jets 
(as well as the resolution of 50 GeV jets with $dN^{\pm}/dy (y=0) = 3000$
is smaller than for 75 GeV jets)
since the background is not fully subtracted and, as a result, the 
reconstructed jet energy is larger in Pb$-$Pb than $pp$. We can define the 
purity of jet reconstruction as the number of events with a true QCD jet 
divided by the number of events with reconstructed jets. Then, for example, 
the purity is $\sim 50\%$ for $50$ GeV jets for $dN^{\pm}/dy (y=0) = 8000$, 
because the 
average energy of the false jets in the background events is also $\sim 50$ 
GeV. The purity increases rapidly with increasing 
$E_T^{\rm reco}$ and becomes $\sim 100\%$ at $100$ GeV.  
Table 7 
summarizes the purity, contamination levels 
(false jets /generated jets) and jet transverse energy resolution in central 
Pb$-$Pb collisions with $dN^{\pm}/dy (y=0) = 3000$ and $8000$.    

Since the azimuthal angle and rapidity distribution of jets are of 
particular interest for jet quenching observables, the angular resolution is
important. Figures~\ref{cmsjet:fig4} and~\ref{cmsjet:fig5} show the 
differences in azimuthal angle $\varphi$ between generated and reconstructed 
$100$ GeV jets in events without and with Pb$-$Pb background. Even in the 
most pessimistic case, $dN^{\pm}/dy (y=0) = 8000$, the $\varphi$ resolution 
is degraded only by a factor of 
$\sim 1.6$ in Pb$-$Pb compared to $pp$ collisions. The resolution is still 
less than the azimuthal size of a calorimeter tower, $\Delta \varphi = 0.087$.
A similar result is also found for the $\eta$ resolution 
(shown in Figs.~\ref{cmsjet:fig4-1} and~\ref{cmsjet:fig5-1}), which
is however somewhat worse than the $\varphi$ resolution.
The reason for that is the vertex was not fixed
here and there was not correction of $\eta$ due to fluctuation of 
$Z$ coordinate in ``pile up'' subtraction algorithm (the latest 
version of the algorithm included this facility).

Figures~\ref{cmsjet:phi-et} 
and~\ref{cmsjet:eta-et} shows the energy dependence of the spatial resolution 
for pp and Pb$-$Pb events with $dN^{\pm}/dy (y=0) = 3000$ and $8000$.
Thus the spatial position of a hard jet can be reconstructed 
in heavy ion collisions at CMS with high enough accuracy for analysis of jet 
production as a function of azimuthal angle and pseudorapidity. 

The Level 1 single jet and electron/photon trigger rates in Pb$-$Pb 
collisions have been estimated using the trigger algorithms developed for 
$pp$ collisions with a parameterization of HIJING results for the 
background~\cite{Baur:2000}. The dominant 
contribution to the trigger rate comes from the single jet trigger 
which uses the transverse energy sums (electromagnetic + hadronic) 
computed in the calorimeter regions 
($4 \times 4$ trigger cells) $\Delta \eta \times \Delta \phi = 0.348 \times 
0.348$. For a threshold of 40$-$50 GeV it gives an acceptable output rate 
of about 400$-$200 Hz 
and is fully efficient for most central collisions. Assuming that with the 
high level trigger full jet reconstruction is possible, the rate can be 
further reduced to a level lower than 10 Hz for jets with reconstructed 
transverse energies greater than $100$ GeV. The 
rate of the single photon trigger is less than 1 Hz for a 
threshold of 50 GeV. With such a threshold, the trigger efficiency is 
close to $100 \%$ for $\gamma +$jet events 
useful for off-line analysis.

\subsubsection{Tracking}

Track finding in heavy ion collisions is difficult due to the large 
number of tracks in an event. We consider heavy ion multiplicities 
up to the worst case, $8000$ charged particles per unit rapidity in 
a central Pb$-$Pb event. In addition to the 
primary tracks, the CMS tracker is occupied by secondaries produced by 
interactions with the detector material. The CMS track reconstruction 
algorithm, originally developed for $pp$ collisions, is based on Kalman 
Filtering and includes seed generation, track propagation, 
trajectory updating and smoothing. However, this track finder, optimized
for the highest efficiency at low density, fails when dealing with central
Pb$-$Pb events due to the large hit combinatorics. Therefore, modification 
of the algorithm is necessary~\cite{cern2003/003}. 

The essential difference for heavy ion collisions is that the primary 
vertex can be determined with dispersion of $200$ $\mu$m using only two 
barrel layers before any tracking. Using this constraint reduces the 
combinatorial background during track seeding. Thus these two features, 
primary vertex finding and restriction of the vertex region, 
were added to the standard package. 

In order to achieve good rejection power to fake tracks from random hit 
combinations we need to require the track to have as many hits as possible.
Only tracks that leave the tracker through the outermost layer are 
considered. This requirement leads to a minimum transverse momentum 
cutoff of $p_T>1$ GeV for the track to be considered reconstructable.

Given this constraint the modified $pp$ reconstruction algorithm gives
about $\approx 80\% $ geometrical acceptance and close to $\approx 100\% $
algorithmic reconstruction efficiency in a low multiplicity environment.
The acceptance varies slightly $\pm 10\% $ over the $\eta$-coverage of 
the tracker and is independent of $p_{T}$

We have tested track propagation algorithms for the case of high occupancy
using Monte Carlo tracks seeds. The estimated efficiency under these
conditions appears high enough, $\approx 80\% $.  The momentum 
resolution is less than $1\%$ at $p_T<100$ GeV. Thus track propagation 
at high density is quite effective. The reconstruction time is around 
$1500$ s/per event. We have also investigated using a regional track 
reconstruction in a limited $\eta - \varphi$ area. This option is 
essentially needed for jet finding and correcting. The efficiency is 
very high, $\ga 95\%$, but we see a large number of fake tracks. 

Thus we believe that pattern recognition is possible in heavy ion 
collisions with the CMS tracking system. The existing $pp$ track 
reconstruction package has been shown to be a valuable framework, 
but needs major restructuring to reduce combinatorics and 
computation time.

\subsubsection{Impact Parameter Determination} 

It is important to study hard jet and high-$p_T$ hadron production in 
heavy ion collisions as a function of centrality. In CMS, the best way 
to determine the impact parameter, $b$, event-by-event is the transverse 
energy deposited in the calorimeters, $E^{\rm tot}_T$, which strongly 
decreases from central to peripheral collisions~\cite{Baur:2000}. The jet 
production rates, $N^{\rm jet}$, can be measured in 
bins of $E^{\rm tot}_T$.  Then the $b-$ and $E^{\rm tot}_T-$ dependencies of 
$N^{\rm jet}$ can be related by the $E^{\rm tot}_T-b$ correlation functions 
$C_{AA}$,  
\begin{eqnarray} 
   N^{\rm jet} (E^{\rm tot}_T) 
   = \int d^2b N^{\rm jet}(b) C_{AA} (E_T^{\rm tot}, b),  
   ~~ C_{AA} (E_T^{\rm tot}, b) 
   = \frac{1}{\sqrt{2\pi}\sigma_{E_T}(b)} \exp {\left( - 
     \frac{\left( E_T^{\rm tot}-\overline{E_T^{\rm tot}}(b) 
                  \right)^2}{2 \sigma_{E_T}^2(b)}\right) } 
  , \nonumber \\  
  N^{\rm jet} (b) = \int d E_T^{\rm tot} N^{\rm jet}(E_T^{\rm tot}) 
     C_{AA} (b, E_T^{\rm tot}),  
     ~~ C_{AA} (b, E_T^{\rm tot})
   = \frac{1}{\sqrt{2\pi}\sigma_{b}(E_T^{\rm tot})} 
     \exp {\left( - \frac{\left( b-\overline{b}(E_T^{\rm tot}) 
                \right)^2}{2 \sigma_{b}^2(E_T^{\rm tot})}\right) }
. \nonumber  
\end{eqnarray} 

\begin{figure}[hbtp]
\begin{center} 
\resizebox{120mm}{80mm} 
{\includegraphics{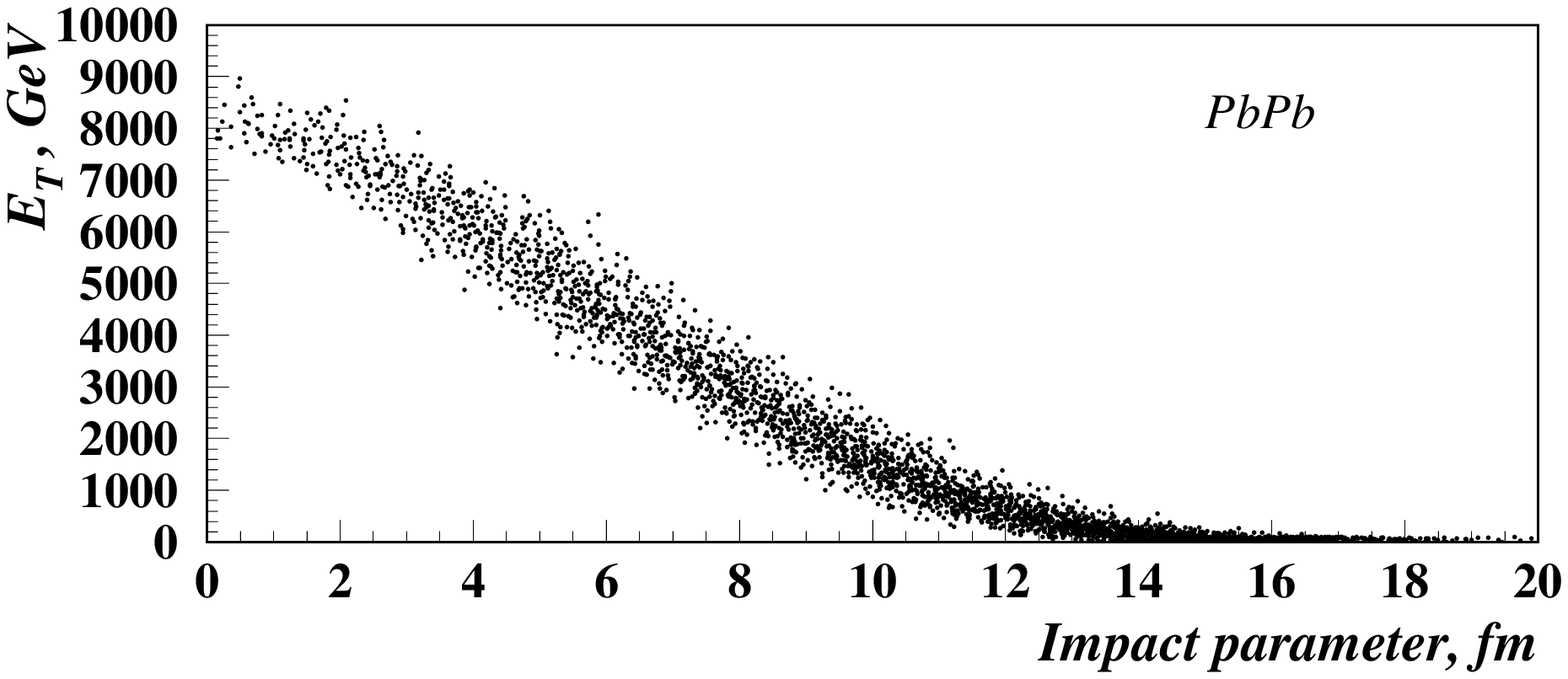}} 
\caption{\small Correlation between the transverse energy flow, $E_T$, 
and impact parameter
$b$ at very forward rapidities, $3\le |\eta| \le 5$, for    
$10000$ minimum bias Pb$-$Pb collisions~\cite{Damgov:et}.} 
\label{cmsjet:fig6}
\resizebox{120mm}{120mm} 
{\includegraphics{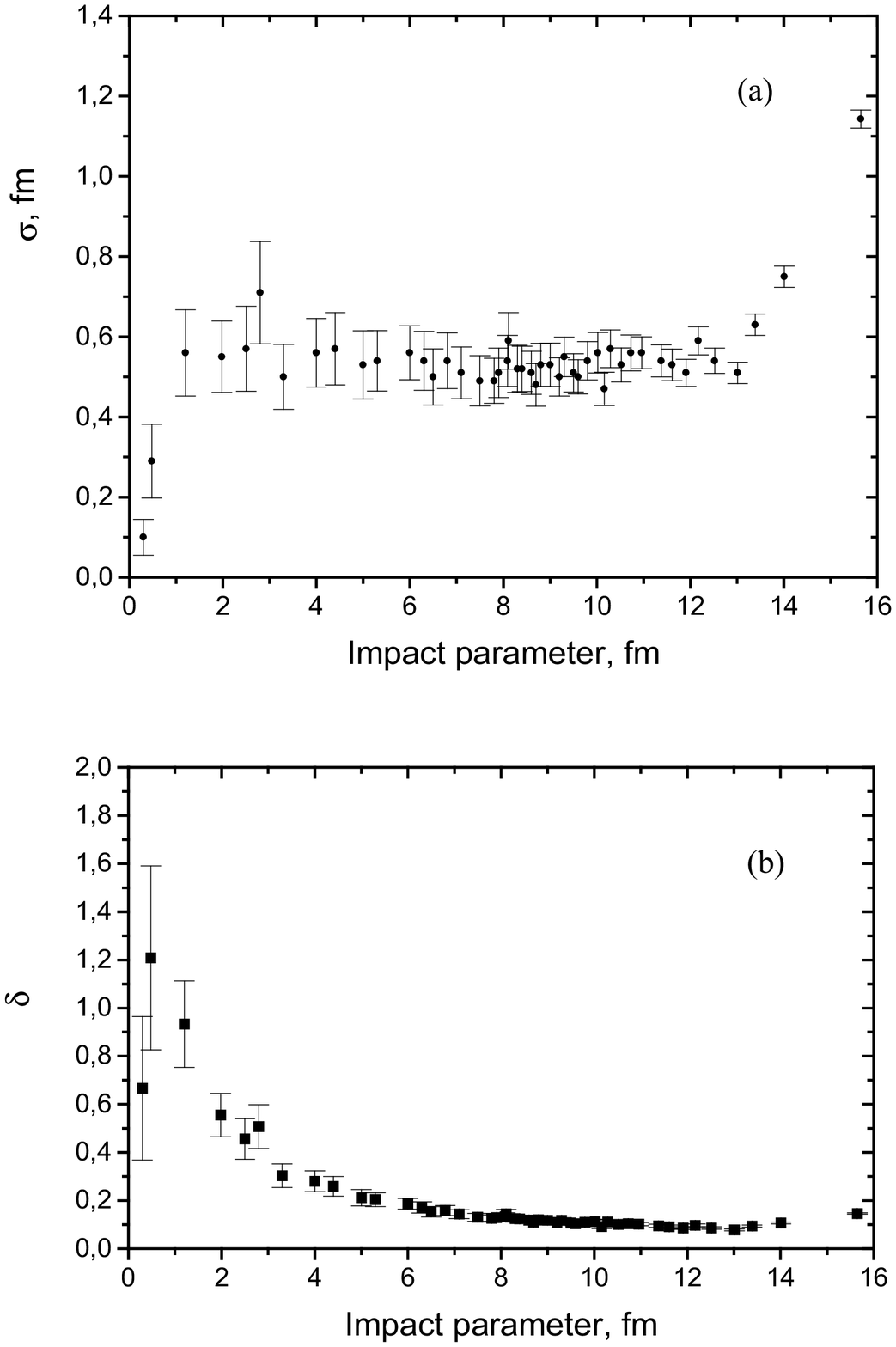}} 
\caption{\small Impact parameter dependence of (a) the Gaussian width 
$\sigma _b$ and (b) the relative error $\delta=2\sigma _b/b$ for Pb$-$Pb 
collisions~\cite{Damgov:et}.}
\label{cmsjet:fig7}
\end{center} 
\end{figure} 

Since very forward rapidity region, $|\eta| \ga 3$, is almost free of final 
state interactions, the (transverse) energy deposition in HF is determined 
mainly by the initial nuclear geometry of a collision rather than by final 
state dynamics. Determining the impact parameter via (transverse) energy 
deposition in the HF calorimeter thus avoids some possible 
uncertainties~\cite{Baur:2000,Damgov:et}. The impact
parameter dependence of the total transverse energy produced in the 
pseudorapidity interval $3 \le |\eta| \le 5$ obtained with 
HIJING~\cite{Wang:1991ht,Gyulassy:ew} is presented in Fig.~\ref{cmsjet:fig6} 
from~\cite{Damgov:et} for Pb$-$Pb 
interactions. The $E_T - b$ correlation is diffuse due to fluctuations in the 
nucleus-nucleus collision dynamics, including fluctuations in the number of 
nucleon-nucleon sub-collisions at a given $b$ and fluctuations of transverse 
energy flow in each nucleon-nucleon interaction. The correlation of the total 
energy flow is of similar shape. Most of the energy produced in the very 
forward direction is between $10$ and $100$ TeV. It is then possible to 
measure the total energy with high accuracy, 
reducing uncertainties in $b$. The impact parameter distribution at 
fixed values of $E_T$ is Gaussian-like with a width, $\sigma _b$, 
dependent on impact parameter, see 
Fig.~\ref{cmsjet:fig7} from Ref.~\cite{Damgov:et}. The absolute accuracy is 
defined here as $\pm 2 \sigma _b$, about $\sim 1$ fm for Pb$-$Pb collisions 
with $1 \mathrel{\mathpalette\fun <} b \mathrel{\mathpalette\fun <} 13$ fm. It is degraded by a factor of $\sim 2$ for very 
peripheral events, $b \ga 13$ fm, 
due to the diminution of the produced energy in the pseudorapidity region. 
At the same time, we see that the relative error is minimal for peripheral 
collisions since the statistics are increased. 

To summarize, we note that although these results were obtained on the 
particle level, similar conclusions are expected to be valid when 
detector effects are taken into account. It has been shown that the finite 
energy and spatial resolution of the HF 
calorimeter do not substantially degrade the accuracy 
of the impact parameter determination in heavy ion 
collisions~\cite{Damgov:et}. 

\subsubsection{Reconstruction of Nuclear Reaction Plane}

\begin{figure} 
\begin{minipage}[t]{78mm}
\resizebox{78mm}{78mm} 
{\includegraphics{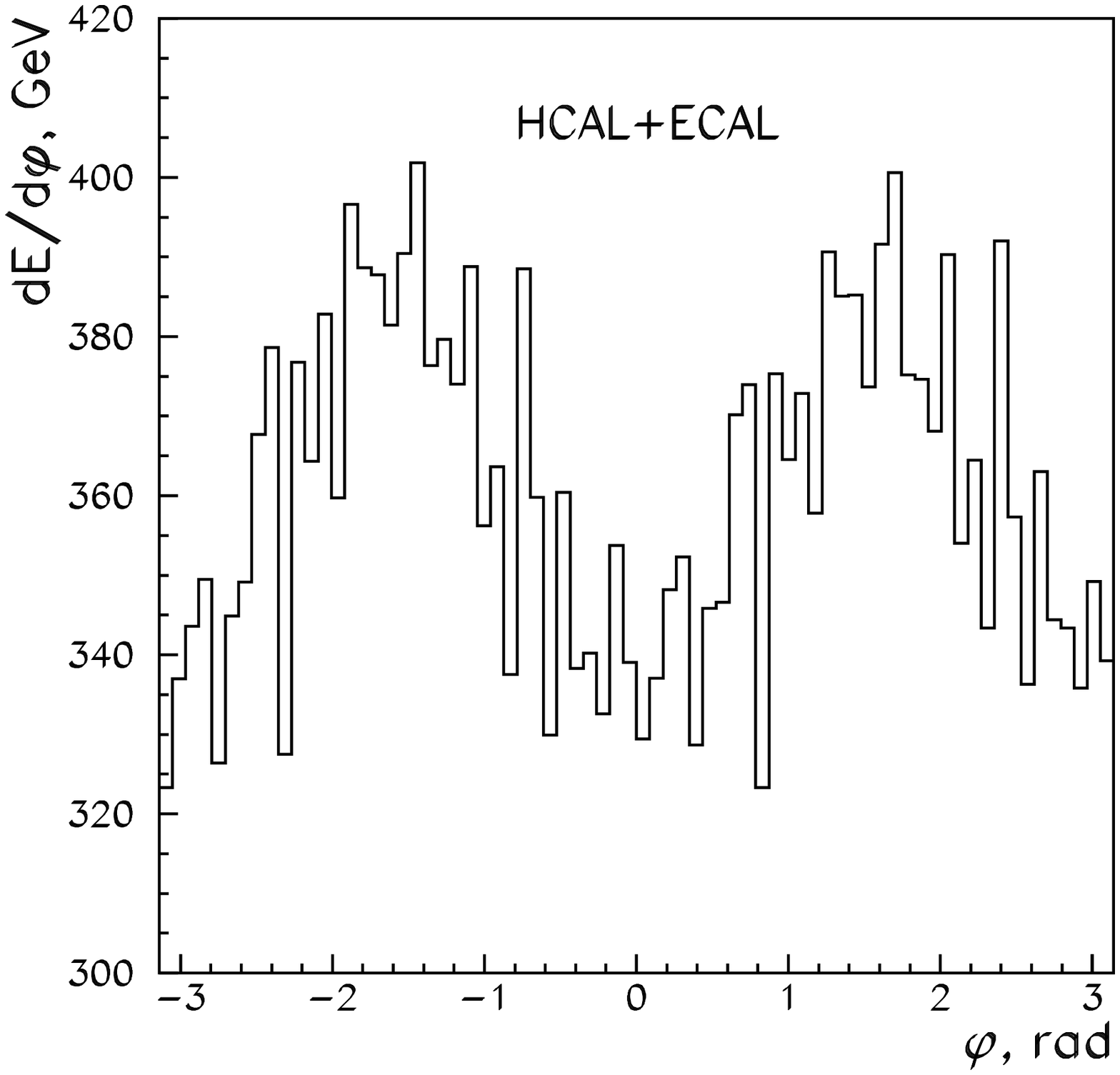}} 
\caption{\small Energy deposition in the barrel and endcap regions of the 
CMS hadronic (HCAL) and electromagnetic (ECAL) calorimeters for Pb$-$Pb 
collisions at $b=6$ fm (hydrodynamics with CMSIM$\_$125).} 
\label{cmsjet:fig8}
\end{minipage}
\hspace{\fill}
\begin{minipage}[t]{78mm}
\resizebox{78mm}{78mm} 
{\includegraphics{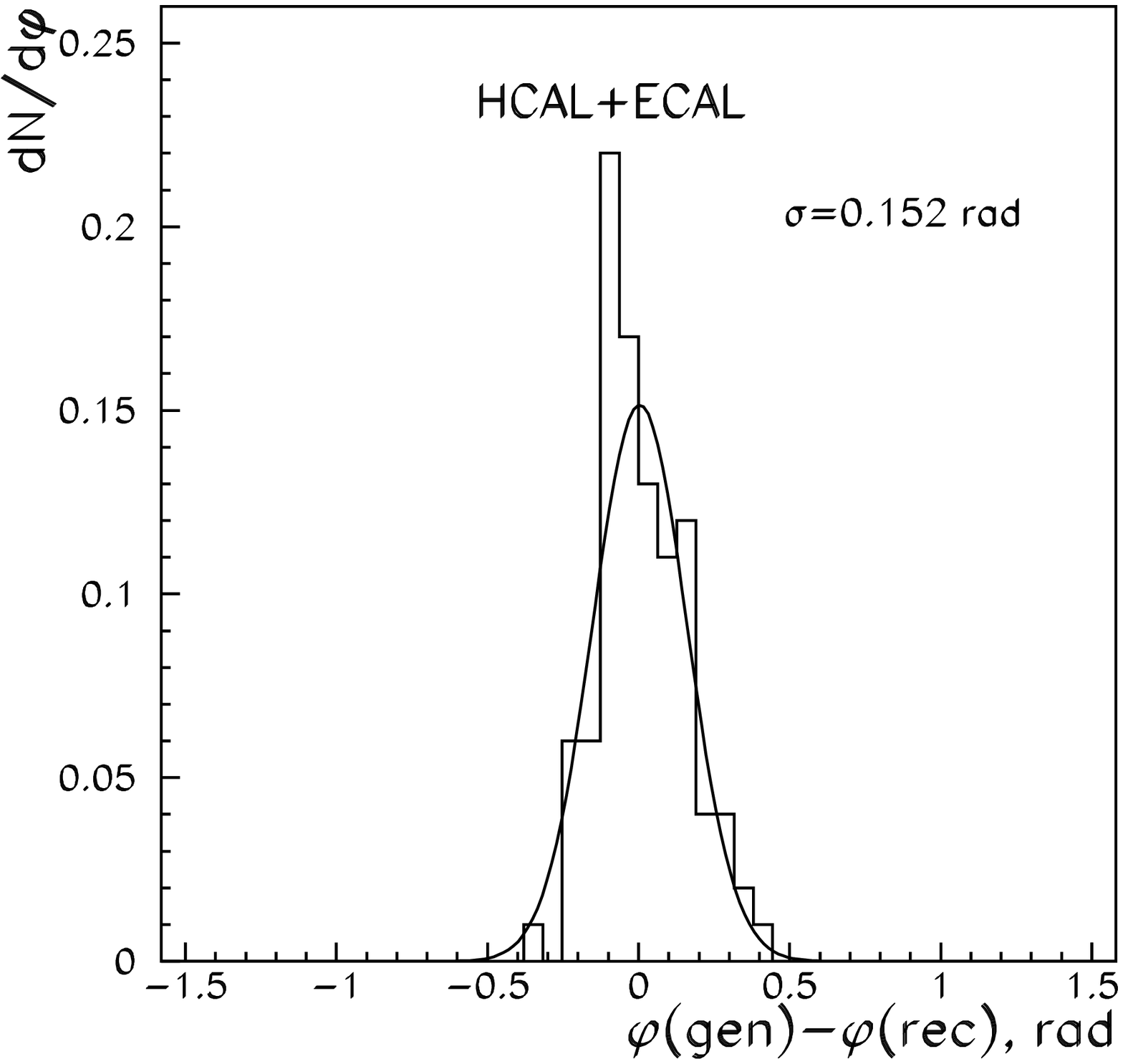}} 
\caption{\small Distribution of differences in the generated and reconstructed 
azimuthal angles of the Pb$-$Pb reaction plane for $b=6$ fm 
(hydrodynamics with CMSIM$\_$125).} 
\label{cmsjet:fig9}
\end{minipage}
\end{figure}

The azimuthal anisotropy of jet and high-$p_T$ particle production in 
semi-central heavy ion collisions is predicted to be a signal of partonic 
energy loss in an azimuthally asymmetric volume of quark-gluon 
plasma~\cite{Lokhtin:2001kb,Wang:2000fq,Gyulassy:2000fs,Gyulassy:2000er}. 
The advantage of azimuthal jet observables is that one needs to reconstruct 
only the azimuthal position of jet, not its total energy. It can be done 
easily and with high accuracy while reconstruction of the jet energy is more 
ambiguous. The  methods summarized in 
Ref.~\cite{Voloshin:1994mz,Poskanzer:1998yz} present ways 
to determine the reaction plane. 
They are applicable to the study of anisotropic flow of soft and semi-hard 
particles in the current dedicated heavy ion experiments at the 
SPS~\cite{Appelshauser:1997dg} and 
RHIC~\cite{Ackermann:2000tr} 
and may also be used at the LHC~\cite{Lokhtin:2001kb}. 
When the azimuthal distribution of particles is described by the elliptic form,
$$\frac{dN}{d \varphi} = \frac{N_0}{2\pi}~ [1+2v_2\cos{2 
(\varphi -\varphi_{\rm reac})}]~,~~~~
N_0 = \int\limits_{-\pi}^{\pi}d\varphi~\frac{dN}{d \varphi}~,$$    
the nuclear reaction plane angle, $\varphi_{\rm reac}$, is 
$$\tan{(2 \varphi_{\rm reac})} = \frac{\sum _i \omega _i \sin{2 \varphi_i}}
{\sum _i \omega _i \cos{2 \varphi_i}}~,$$
where the weights, $\omega_i$, are selected to optimize the resolution. The 
coefficient 
$v_2$ of the azimuthal anisotropy of particle flow is an average over
$\cos(2 \varphi)$.  In CMS, the weights can be introduced~\cite{cern2003/019} 
as energy deposition in calorimeter sector $i$ of position 
$\varphi_i$, $\omega_i = E_i (\varphi_i)$. 
Fig.~\ref{cmsjet:fig8} illustrates the energy deposition 
in sectors of the barrel and endcap regions of the CMS hadronic and 
electromagnetic calorimeters, $|\eta|<3$, for generated with 
hydro-code~\cite{Kruglov:1997,Lokhtin:2002vq} Pb$-$Pb events at $b=6$ fm. The 
detector response is obtained with CMSIM$\_$125. The estimated resolution 
of the $\varphi_{\rm reac}$ determination, $\sigma _{\varphi~\rm reac}=0.15$ 
rad, see Fig.~\ref{cmsjet:fig9}, allows measurement of the coefficient of the 
jet azimuthal anisotropy with $\sim 90\%$ accuracy, defined as the ratio 
of the average of $\langle \cos{2\varphi _{\rm jet}} \rangle$ over all 
events ``measured'' to its ``true'' value. The 
estimates obtained are quite optimistic because hydrodynamic models 
give rather large values of elliptic flow at high-$p_T$. On the other hand, 
the majority of microscopic Monte Carlo models underestimate flow. For 
example, under the same conditions HIJING~\cite{Wang:1991ht,Gyulassy:ew} 
with jet quenching predicts much less anisotropic flow and yields 
$\sigma _{\varphi~\rm reac}=0.8$ with only $\sim 20\%$ accuracy. 

Recently a method was also suggested~\cite{Lokhtin:2002vq} for measuring 
the jet 
azimuthal anisotropy coefficients without event-by-event reconstruction of the 
reaction plane. This technique is based on the correlations between the 
azimuthal position of the jet axis and the angles of particles not 
incorporated in the jet. Then  
$$v^{\rm jet}_{2} \equiv \langle \cos{2\varphi _{\rm jet}} 
\rangle_{\rm event} = 
\left< \frac{\left< \cos{2(\varphi 
_{\rm jet}-\varphi)}~\omega (\varphi) 
\right>} {\sqrt{\left< \cos{2(\varphi_1-\varphi_2)}~\omega _1(\varphi_1)~
 \omega_2(\varphi_2) \right>}} \right>_{\rm event}~,$$ 
where the weights $\omega_i$ are defined as before.  In some sense, this 
represents the development and generalization of the well-known method for 
measuring the azimuthal anisotropy of particle flow originally considered in 
a number of works, see for example 
Refs.~\cite{Voloshin:1994mz,Poskanzer:1998yz,Wang:qh,Ollitrault:bk,Ollitrault:ba}. The 
accuracy of the method improves with increasing multiplicity and particle 
(energy) azimuthal anisotropy and is practically independent of the absolute 
value of the azimuthal anisotropy of the jet itself~\cite{Lokhtin:2002vq}. The 
accuracy of $v^{\rm jet}_{2}$ achieved by such a method are estimated to be 
$94\%$ for hydrodynamics 
and $30\%$ for HIJING, somewhat better than those obtained from direct 
reconstruction of the reaction plane angle.

 \subsection{Jet Physics with the ATLAS Detector}
{\em S. Aronson, K. Assamagan, B.Cole, M. Dobbs, J. Dolesji, H. Gordon,
F. Gianotti, S.Kabana, M.Levine, F. Marroquim, J.Nagle, P. Nevski, 
A. Olszewski, L.Rosselet, P. Sawicki, H.Takai, S. Tapprogge, A. Trzupek, 
M.A. B. Vale, S.White, R. Witt, B. Wosiek and K. Wozniak}

The ATLAS detector is designed to study high $p_T$ physics in
proton-proton collisions at high LHC machine luminosity.  Most of
the detector subsystems will be available for the study of heavy ion
collisions. One of the highlights of the ATLAS detector is its calorimeter
subsystem. Both the electromagnetic and hadronic compartments are
finely segmented and well suited for jet quenching studies. RHIC results
suggest that partons may  radiate gluons in the dense matter formed
in heavy ion collisions. This phenomena can be certainly be well
explored in the ATLAS detector. We report on early assessment of
the detector capabilities in the heavy ion environment.

\subsubsection{The ATLAS Detector}

\begin{figure}[h]\epsfxsize=12.7cm 
\centerline{\epsfbox{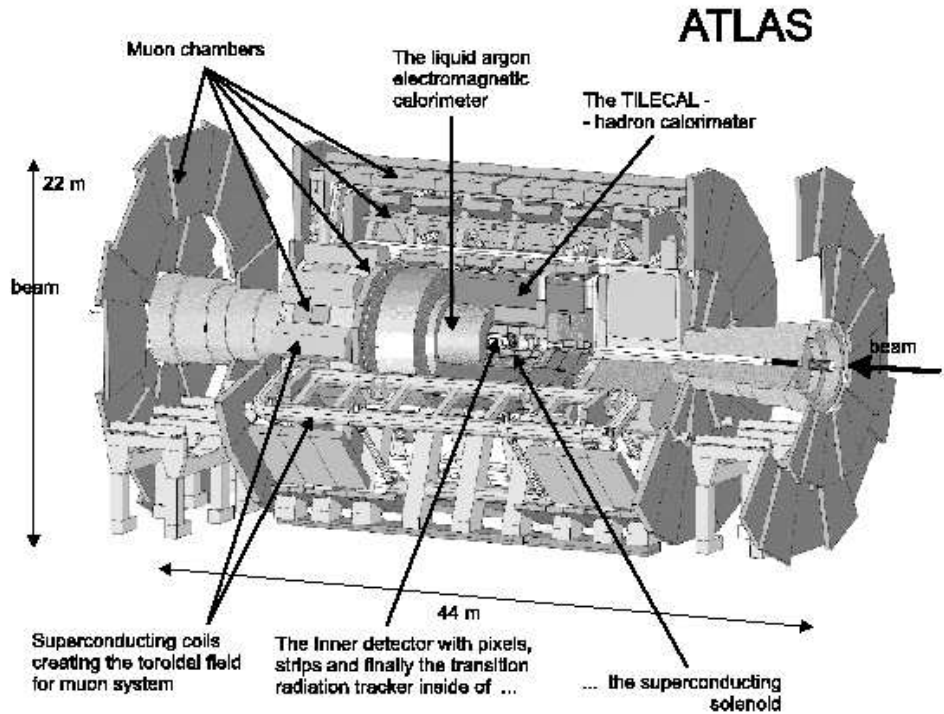}}
\vspace{0.5cm}
\caption{The overall layout of the ATLAS detector
}\label{fig:label}
\end{figure}

The ATLAS detector is designed to study proton-proton collisions at
the LHC design energy of 14~TeV in the center of mass.  The physics
pursued by the collaboration is vast and includes: Higgs boson search,
searches for SUSY, and other scenarios beyond the Standard Model, as
well as precision measurements of process within (and possibly beyond)
the Standard Model. To achieve these goals at a full machine
luminosity of $10^{34} cm^{-2}s^{-1}$. ATLAS will have a precise
tracking system (Inner Detector) for charged particle measurements, an
as hermetic as possible calorimeter system, which has an extremely
fine grain segmentation, and a stand-alone muon system. An overview of
the detector is shown in Fig.~\ref{fig:label}.

%

The Inner Detector is composed of (1) a finely segmented silicon pixel
detector, (2) silicon strip detectors (Semiconductor Tracker (SCT))
and (3) the Transition Radiation Tracker (TRT).  The segmentation is
optimized for proton-proton collisions at design machine luminosity.
The inner detector system is designed to cover a pseudo-rapidity of
$\mid \eta \mid < 2.5$ and is located inside a 2~T solenoid magnet.

The calorimeter system in the ATLAS detector surrounds the solenoid
magnet is divided into electromagnetic and hadronic sections and
covers pseudo-rapidity $\mid \eta \mid < 4.9$. The EM calorimeter is
an accordion liquid argon device and is finely segmented
longitudinally and transversely for $\mid \eta \mid \le 3.1$.  The first
longitudinal segmentation has a granularity of 0.003 x 0.1 $(\Delta
\eta \times \Delta \phi)$ in the barrel and slightly coarser in the
endcaps.  The second longitudinal segmentation is composed of $\Delta
\eta \times \Delta \phi = 0.025 \times 0.025$ cells and the last
segment $\Delta \eta \times \Delta \phi = 0.05 \times 0.05$ cells.  In
addition a finely segmented $(0.025 \times 0.1)$ pre-sampler system is
present in front of the electromagnetic (EM) calorimeter.  The overall
energy resolution of the EM calorimeter determined experimentally is
$10\%/\sqrt{E} \oplus 0.5\%$. The calorimeter also has good pointing
resolution, $60 mrad/\sqrt{E}$ for photons and timing resolution
better than 200 ps for showers of energy larger than 20 GeV.

The hadronic calorimeter is also segmented longitudinally and
transversely.  Except for the endcaps and the forward calorimeters,
the technology utilized for the calorimeter is a lead-scintillator
tile structure with a granularity of $\Delta \eta \times \Delta \phi =
0.1 \times 0.1$.  In the endcaps the hadronic calorimeter is
implemented in liquid argon technology for radiation hardness with the
same granularity as the barrel hadronic calorimeter.  The energy
resolution for the hadronic calorimeters is $50\%/\sqrt{E} \oplus 2\%$
for pions.  The very forward region, up to $\eta =4.9$ is covered by
the Forward Calorimeter implemented as an axial drift liquid argon
calorimeter.  The overall performance of the calorimeter system is
described in [1].

The muon spectrometer in ATLAS is located behind the calorimeters,
thus shielded from hadronic showers. The spectrometer is implemented
using several technologies for tracking devices and a toroidal magnet
system, which provides a field of 4~T strength to have an independent
momentum measurement outside the calorimeter volume.  Most of the
volume is covered by MDTs, (Monitored Drift tubes). The forward region
where the rate is high, Cathode Strip Chamber technology is chosen.
The stand-alone muon spectrometer momentum resolution is of the order
of $2\%$ for muons with $p_T$ in the range 10 - 100 GeV. The muon
spectrometer coverage is $\mid \eta \mid < 2.7$.

The trigger and data acquisition system of ATLAS is a multi-level
system, which has to reduce the beam crossing rate of 40~MHz to an
output rate to mass storage of $\mathcal{O}(100)$~Hz. The first stage
(LVL1) is a hardware based trigger, which makes use of coarse 
granularity calorimeter data and dedicated muon trigger chambers only,
to reduce the output rate to about 75~kHz, within a maximum latency
of 2.5~$\mu$s.

The performance results mentioned have been obtained using a detailed
full simulation of the ATLAS detector response with GEANT and have 
been validated by an extensive program of testbeam measurements of 
all components.

\subsubsection{Jet Physics and ATLAS}

The ATLAS calorimer coverage and its fine segmentation will
be an asset for jet studies in the heavy ion environment. Signatures
of jet quenching in central heavy ion collisions could manifest
as a larger jet cone (as compared to proton-proton collision) and/or
modifications in the jet fragmentation function. The finely segmented
(longitudinally and transversely) electromagnetic calorimeter
will allow us to reconstruct EM clusters in the jet environment and
in particular measure the $\pi^0$ containt in the jet.

There is excellent opportunity in ATLAS to measure $\gamma - jet$,
jet-jet and Z-jet events where one can more fully characterize the
modified fragmentation functions. In particular, the $\gamma$(or Z) in
$\gamma(or Z)-jet$ processes provides a ``control'' over the away-side
jet energy and direction that will allow the physics of quenching to
be studied quantitatively and in great detail  \cite{Wang:1996pe}. 
The effects of
hard gluon radiation on the photon/jet energy imbalance and angular
distribution can be studied in great detail using the high-statistics
p-p data set. The $\gamma - jet$ channel requires the identification
of a photon.  In proton-proton collisions the rejection of $\gamma /
\pi^0$ is about a factor of three up to a $p_T$ of 50 GeV. However,
the heavy ion environment presents considerable more challenge.  $Z_0$
production rates have been estimated by Wang and Huang \cite{Wang:1996pe}.  
For $p_T$ larger than 40 GeV, we expect of the order of $\sim$500 $Z_0
\rightarrow \mu^+\mu^-$ events for one month-run.  Therefore multiple
runs may be required to extract relevant information on jet
fragmentation.

\subsubsection{Expected Detector Performance for Jet Studies}

The ATLAS calorimeter granularity is shown in Table 1.
The calorimeter system is not only segmented in $\eta$
and $\phi$, but also longitudinally.  The calorimeter is fully segmented
to $\eta = 5$.
The electromagnetic calorimeter is in
most places $25X_0$ deep and designed to fully contain a 1 TeV
electron or photon. The hadronic calorimeter is more than $10 \lambda_{int}$ 
deep and contains
all of the hadrons in typical pp high luminosity runs. 

\begin{center}
Table 1: ATLAS calorimeter system segmentation. Listed are the number of 
longitudinal
and the size of the transverse segmentation in the different calorimeters. 
\vskip 0.2cm
\begin{tabular} {l|c|c|l}
\hline
Calorimeter System  & $\eta$ coverage & Long. & Transversal segmentation\\
\hline\hline
LAr Electromagnetic & $0.0<\eta<3.2$ & 3 & $0.003\times0.1$, $0.025\times0.025$, and $0.05\times0.05$\\
LAr Hadronic &  $1.5< \eta < 3.2$&4 & $0.1\time0.1$ for $1.5<\eta<2.5$ and $0.2 \times 0.2$ otherwise.\\
Hadronic Tile & $0.0<\eta<1.7$ & 3 & $0.1\times0.1$, $0.1\times 0.1$, and $0.2\times0.1$ \\
Forward Calorimeter &  $3.1 < \eta <4.9$ & 4&$0.2 \times 0.2$ \\
\hline\hline
\end{tabular}
\end{center}

The jet energy resolution for high luminosity proton-proton run is
$50\%/\sqrt{E} \oplus 2\%$ and the jet reconstruction threshold
approximately $E_T \sim 20 GeV$ for proton-proton run. These numbers
are expect to be different for the heavy ion environment. Preliminary
simulations indicate that the detection thresholds should be at around
40$\sim$50 GeV.  We expect a worsening of the energy resolution
because of the soft background but the energy scale should remain
untouched, unless the ratio of EM to Hadronic components in the jet
changes substantially. 

Preliminary figures for energy deposition in the EM calorimeter
from full HIJING central events indicate that approximately
4 GeV of transverse energy is deposited in a tower $\eta \times \phi=
0.1 \times 0.1$. This number is consistent with reported by CMS. 
However, due to the longitudinal segmentation and predominantly low
$p_T$ ($<1~GeV$)nature of the particles in the background, we do expect that
most of the energy to be deposited in the first compartament of the
calorimeter.  Thus jets could be reconstructed on the basis of
the remaining compartaments. Detailed studies are under way.

To study jet quenching  in a direct way is to measure its
fragmentation function and possible changes in the jet cone radius.  The
fragmentation function can only be measured if particles are
identified within the jet. 
The segmentation of the ATLAS calorimeter is such that allows,
in principle, for the
identification of $\pi_0$'s.  In spite of the soft background preliminary
studies shows encouraging results. However, we have observed a
significant worsening of the EM cluster energy resolution. Studies
performed in proton-proton collisions do have not addressed the
issue of EM cluster reconstruction, specially for $\pi^0$s at
low energy.

\section*{Acknowledgements}

\noindent The following sources of funding are acknowledged:

\noindent~~~ Alexander von Humboldt Foundation: R. Fries;

\noindent~~~ CERN TH Division, Visitor Programme: 

 A. Accardi, F. Arleo, R. Baier, R.J. Fries, J.W. Qiu, I. Vitev, R. Vogt;


\noindent~~~ CICYT of Spain under contract AEN99-0589-C02: N. Armesto;


\noindent~~~ European Commission IHP program,
Contract HPRN-CT-2000-00130: A. Accardi;


\noindent~~~ German Science Foundation (DFG),
Contract FOR 329/2-1: R. Baier;


\noindent~~~ Marie Curie Fellowship of the European Community programme TMR,

Contract HPMF-CT-2000-01025: C.A. Salgado; 


\noindent~~~ United States Department of Energy,
 
Contract No. DE-FG02-93ER40764: A. Accardi, I. Vitev;

Contract no. DE-FG02-96ER40945: R.~Fries;

Contract No. DE-FG02-87ER40371: J.W.~Qiu, I. Vitev;

Contract No. DE-AC03-76SF00098: R.~Vogt;

Contract No. DE-AC03-76SF00098: X.N.~Wang;


\noindent~~~ United States NSFC project Nos. 19928511 and 10135030: X.N. Wang;


\noindent~~~ Universidad de C\'ordoba: N. Armesto.



\end{document}